\documentclass[12pt,twoside]{book} 
\usepackage{a4wide}
\usepackage{graphicx} 
\usepackage[dvips]{epsfig}
\usepackage{times} 
\usepackage{latexsym} 
\usepackage{amsmath}
\usepackage{amsfonts}
\usepackage{amssymb}
\usepackage{amsbsy}
\usepackage{amsthm} 
\usepackage{amscd}      % commutative diagrams ?
\usepackage{pstricks} 
\usepackage{pst-text} 
\usepackage{pst-node} 
\usepackage{fancyvrb}

\RecustomVerbatimEnvironment%
  {Verbatim}{Verbatim}
  {numbers=left,numbersep=2mm,frame=lines,framerule=0.4pt,%
   firstnumber=last,numberblanklines=false,fontfamily=cmtt,%
   commandchars=\\\{\},framesep=4pt}

\def\date{January 25, 2002}

\def\a{\alpha}%GreekGreekGreek
\def\b{\beta}
\newcommand{\g}{\gamma}
\def\d{\delta}
\newcommand{\D}{\Delta} 
\def \Del{\Delta}
\newcommand{\e}{\eta}
\newcommand{\ep}{\epsilon}
\newcommand{\G}{\Gamma}

\newcommand{\la}{\lambda}
\newcommand{\La}{\Lambda}
\newcommand{\om}{\omega}

\newcommand{\vp}{\varphi}

\newcommand{\si}{\sigma}

\def\th{\theta}

\def\cA{{\cal A}}
\def \A{{\cal A}}
\def\cD{{\cal D}}
\def\cC{{\cal C}}
\def\cF{{\cal F}}
\def\F{{\cal F}}
\def\cG{{\cal G}}
\def\cH{{\cal H}}
\def\cI{{\cal I}}
\def\cM{{\cal M}}
\def\cR{{\cal R}}
\def \RR {{\cal R}}
\def \TR{\tilde{\cal R}}

\def\cG{{\cal G}}
\def\cU{{\cal U}}
\def\cV{{\cal V}}
\def\cW{{\cal W}}
\def\cN {{\cal N}}
\def\cH {{\cal H}}

\def \S{{\cal S}^2}

\def\R{{\mathbb R}}
\def\C{{\mathbb C}}
\def\N{{\mathbb N}}
\def\Z{{\mathbb N}}

\def\dag{^{\dagger}}

\def\mg {{\mathfrak g}}
\def\mgh {\widehat{\mathfrak g}}
\def\mk {{\mathfrak k}}
\def\GcG{\cG_L \tens^\cR \cG_R}
\def\gcg{U_q(\mg_L \times \mg_R)_\cR}
\def\eps{\varepsilon}
\def\gcheck{g^{\vee}}

\newcommand \one{{\bf 1}}
\def\haf{\frac 12}

\def\kbar{{\mathchar'26\mkern-9muk}}  
\def\wm{\mathbin{*}}

\def\ket#1{\vert #1 \rangle}

\def \tp{\tilde\phi}
\def \[{\left[}
\def \]{\right]}
\newcommand{\ttr}{\tilde\triangleright}
\def \wtilde{\widetilde}
\def \dl{\partial}
\def \p{\varphi}

\def\dim{{\rm dim}}%%%%%%%%%%%%%%%%% roman
\def\diag{{\rm diag}}
\def\det{{\rm det}}

\def\tr{{\rm tr}}
\def\tw {{\om}}

\def\exp{{\rm exp}}

\def\Mo{M^{(0)}}
\def\del{\partial}

\def\id{\rm id}

\def \Tr{\mbox{Tr}}

\def\rep{representation }
\def\reps{representations }

\newcommand{\beq}{\begin{equation}}
\newcommand{\eeq}{\end{equation}}
\newcommand{\be}{\begin{equation}}
\newcommand{\ee}{\end{equation}}
\newcommand{\beqa}{\begin{eqnarray}}
\newcommand{\eeqa}{\end{eqnarray}}
\def \berr{\begin{eqnarray}}
\def \err{\end{eqnarray}}
\newcommand{\barr}{\begin{array}}
\newcommand{\earr}{\end{array}}
\newcommand{\ben}{\begin{enumerate}}
\newcommand{\een}{\end{enumerate}}
\newcommand{\bit}{\begin{itemize}}
\newcommand{\eit}{\end{itemize}}

\def\llap#1{\hbox to 0pt{\hss#1}}%definition of llap
\def\pola{a\llap{\hbox{\char'30\kern-1.2pt}}}
\def\pole{e\llap{\hbox{\char'30\kern-.8pt}}}
\newcommand{\hf}{\hspace{\fill}}

\newcommand{\non}{\nonumber\\}

\def\nn{\nonumber}

\def\obar{\overline}
\def\tens{\otimes}

\def\trr{\triangleright}
\def\smash{\mbox{$\,\rule{0.3pt}{1.1ex}\!\times\,$}}

\def\half{{\mbox{\small  $\frac{1}{2}$}}}

\newcommand{\refeq}[1]{(\ref{#1})}

\newcommand{\lb}{\left(}
\newcommand{\rb}{\right)}

\def\rara{\rangle\rangle}

\newcommand{\matR}[4]{R{}^{#1}{}_{#2}\,{}^{#3}{}_{#4}\,}

\newtheorem{prop}{Proposition}[section]
\newtheorem{theorem}[prop]{Theorem}
\newtheorem{lemma}[prop]{Lemma}

\begin{document}
\pagestyle{empty}
\mbox{}
\vskip 0.5truecm
\begin{center}
\rule{\textwidth}{3pt}
\vskip 1.5truecm
\huge\bf\sf
Field theoretic models \\[2ex] 
on covariant quantum spaces
\end{center}
\vspace{1truecm}
\rule{\textwidth}{3pt}
\vspace{2truecm}
\begin{center}
{\Large\bf\sf
Habilitationsschrift}

\vspace{1truecm}

{\large\bf\sf
Harold Steinacker}
\end{center}
\vspace{3truecm}
\parbox{0.5\textwidth}{%
\hfil \black
\includegraphics[height=3.5cm]{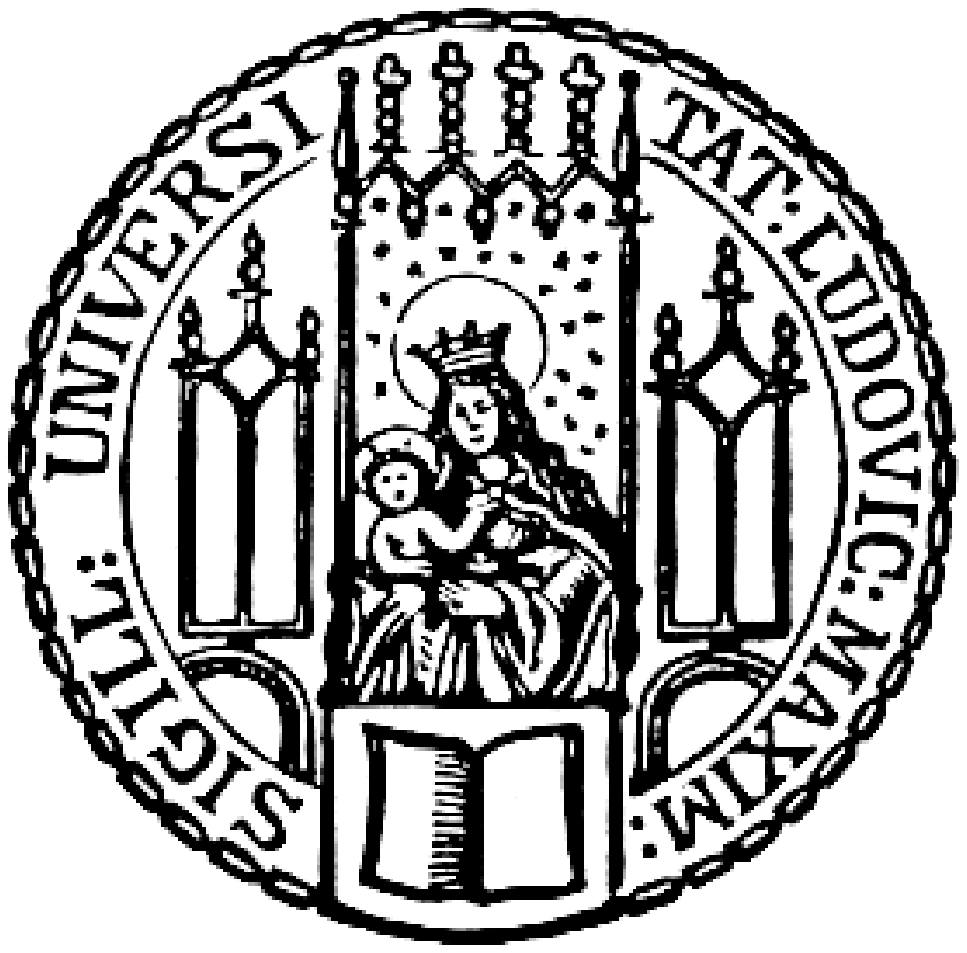}
\hfil}
\parbox{0.5\textwidth}{%
\begin{center}
\large\sf
Ludwig-Maximilians Universit\"at M\"unchen\\
Sektion Physik%\\
%80333 M\"unchen
\end{center}
}
\vspace{1truecm}
\begin{center}
\large\bf\sf
Oktober 2002
\end{center}

%\newpage
\mbox{}
\newpage
\mbox{}
\vskip0.3\textheight
%\begin{center}
%\Large\sf
%To \\[2ex]
%\end{center}
%\newpage
\mbox{}
\newpage

%%\setcounter{page}{1}
%%\def\thepage{\Roman{page}}
%%\pagestyle{myheadings}
%%\newpage
%%%%%%%%%%%%% once more%%%%
%\pagestyle{empty}
%\mbox{}
%\vskip 0.5truecm
%\begin{center}
%\rule{\textwidth}{3pt}
%\vskip 1.5truecm
%\huge\bf\sf
%Field theoretic models \\[2ex] 
%on covariant quantum spaces
%\end{center}
%\vspace{1truecm}
%\rule{\textwidth}{3pt}
%\vspace{2truecm}
%\begin{center}
%{\Large\bf\sf
%Habilitationsschrift}

%\vspace{1truecm}

%{\large\bf\sf
%Harold Steinacker}
%\end{center}
%\vspace{3truecm}
%%\parbox{0.5\textwidth}
%%\parbox{0.5\textwidth}{%
%\begin{center}
%\large\sf
%Ludwig-Maximilians Universit\"at M\"unchen\\
%Sektion Physik%\\
%%80333 M\"unchen
%\end{center}
%\vspace{1truecm}
%\begin{center}
%\large\bf\sf
%Oktober 2002
%\end{center}
%%}
%%%%%%%%%%%%%%%%%%%%%%%%%%%%%
%\addcontentsline{toc}{chapter}{Abstract}
%\input{abstract.tex}

\setcounter{page}{1}
\def\thepage{\Roman{page}}
\pagestyle{myheadings}
\newpage
\relax
\newpage

%\addcontentsline{toc}{chapter}{Table of Contents}
\tableofcontents
%
%\newpage\mbox{}
\newpage
%%% delete if necessary
%\newpage
%\addcontentsline{toc}{chapter}{Preface}
%\input{preface.tex}
%
%\newpage\mbox{} %% if ends on a odd page
\newpage
\setcounter{page}{1}
\def\thepage{\arabic{page}}
%% Intro

\chapter*{Introduction}
\addcontentsline{toc}{part}{Introduction}

The topic of this thesis is the study of field theory on 
a particular class of quantum spaces with quantum group symmetries.
These spaces turn out to describe certain $D$-branes on 
group manifolds, which were found in string theory.

It is an old idea going back to Heisenberg \cite{Heisenberg:1996bv} 
that the ultraviolet (UV)--divergences 
of quantum field theory (QFT) are a reflection of our 
poor understanding of the physics of spacetime 
at very short distances, and should disappear if 
a quantum structure of spacetime is 
taken into account. One expects indeed
that space-time does not behave like a classical manifold at or below the
Planck scale, where quantum gravity should modify its structure. 
However, the scale where a quantum structure of spacetime 
becomes important might as well be much larger than the Planck scale. 
There has been considerable effort to 
study field theory on various non-commutative spaces,
sparked by the development of noncommutative geometry \cite{Connes:book}.
With very little experimental guidance, finding the 
correct description is of course very difficult.

In recent years, quantized spaces have attracted a lot of attention
in the framework of string-theory \cite{Douglas:1998fm,Chu:1998qz,Seiberg:1999vs,Schomerus:1999ug,Connes:1998cr,Alekseev:1999bs,Myers:1999ps}, 
for a somewhat different 
reason. Even though this thesis is not about string theory, we 
want to take advantage of this connection, 
and therefore we need to explain it briefly. This is independent of
whether or not string theory is a theory of nature: it is certainly
a rich mathematical laboratory, and provides new insights
into physical aspects of noncommutative field theories.
The connection with noncommutative spaces is through
$D$-branes, which are submanifolds of the target space,
on which open strings end. It turns out that these $D$-branes become 
non-commutative (=quantized) spaces if there is a non-vanishing $B$-field,
i.e. a particular 2-form field in the target space.
The important point is that these 
$D$-branes carry certain induced
field theories, arising from the open strings which end on them.
This discovery gave a boost to the study of  field theories
on such non-commutative spaces, and led to important new insights. 
Their realizations in string theory is strong support for the existence of 
``physically interesting'' field theories on quantized spaces. 
In general, it is far from trivial to 
study QFT on noncommutative spaces, and results which originated from
string theory such as the Seiberg-Witten map
were very helpful to establish a certain intuitive understanding.

The space which is usually considered in this context
is the so-called quantum plane $\R^n_\theta$, defined by
the algebra of coordinate ``functions'' 
\beq
[x_i, x_j] = i \theta_{ij}. 
\eeq
Here $\theta_{ij}$ is a constant
antisymmetric tensor, which is related to the background $B$ field
mentioned above. 
Field theory has been studied in considerable detail 
on this space, and a formulation of 
gauge theory was achieved using the so-called 
Seiberg-Witten map \cite{Seiberg:1999vs,Jurco:2001rq,Jurco:2000dx}.
On a formal level, this can be generalized to spaces with a 
Moyal-Weyl star product 
in the context of deformation quantization \cite{Jurco:2001my}. 

While $\R^n_\theta$ appears to be one of the simplest quantum spaces,
it has several drawbacks:
\begin{itemize}
\item rotation invariance is lost.

\item the desired regularization of the UV divergences does not occur.
Worse yet, there is a new phenomenon known as UV/IR mixing, 
which appears to destroy perturbative renormalizability.

\item there are serious mathematical
complications which are related to the use of deformation quantization.
\end{itemize}
In view of this, it seems that the apparent simplicity of the
quantum space $\R^n_\theta$ is not borne out, and we want to look for other,
``healthier'' quantum spaces. The first problem to overcome
is the lack of symmetry. We therefore insist
on quantum spaces with some kind of generalized symmetry; in our context,
this will be a symmetry under a quantum group. 
Next, we insist that no divergences occur at all. This may seem
too much to ask for, but it will be satisfied since the spaces
under consideration here
admit only a finite (but sufficiently large) number of modes. 
This finiteness property is extremely helpful, 
because no intuition is available at this stage 
%of studying quantum spaces 
which could allow to circumvent the difficult 
technical aspects of infinite-dimensional non-commutative algebras.
Nevertheless, one can obtain e.g.  $\R^2_\theta$ 
from these spaces by a scaling procedure, as we will see. 

In this thesis, we study a class of quantum spaces which satisfy
the above requirements. Moreover, 
we will argue that these spaces are realized in string theory, 
in the form of $D$-branes on (compact) group manifolds $G$. 
Indeed, $G$ always carries a non-vanishing $B$-field for consistency reasons, 
due the WZW term.  Therefore these branes are noncommutative spaces.
They have been studied from various points of views
% branes have been studied in considerable detail recently 
\cite{Alekseev:1999bs,Alekseev:1998mc,Alekseev:2002rj,Pawelczyk:2000ah,
Bordalo:2001ec,Schweigert:2000iv,Bachas:2000ik,Felder:1999ka,Maldacena:2001xj,Fredenhagen:2000ei,Figueroa-O'Farrill:2000kz,Stanciu:2000fz}, leading to a nice and coherent picture.
On the quasi-classical level they\footnote{we consider only 
the simplest type of $D$-branes here} are adjoint orbits in $G$, 
for example 2-spheres, or higher-dimensional analogs thereof.
The exact description on the world-sheet level is given by a WZW model,
which is a 2-dimensional conformal field theory (CFT).
This is one of the few situations where string theory on a curved space
is well under control. It is known that 
without $D$-branes, the WZW model leads to a chiral algebra of 
left and right currents on $G$, which satisfy affine Lie algebras 
$\widehat \mg_{L,R}$ at level $k \in \Z$. 
These $\widehat \mg_{L,R}$ in turn are closely related to the 
quantum groups (more precisely quantized universal enveloping algebras)
$U_q(\mg_{L,R})$ for $q = exp(\frac{i \pi}{k+\gcheck})$.
However, this quantum group structure disappears once the left and right 
currents are combined into the full currents of {\em closed} string theory.
This changes in the presence of $D$-branes, which
amount to a boundary condition for the 
{\em open} 
strings ending on them, relating the left with the right chiral current.
Then only one copy of the affine Lie algebras
(the ``vector'' affine Lie algebra $\widehat \mg_{V}$) survives, 
which preserves the (untwisted) branes $D$, 
reflecting the geometric invariance of $D$ under the adjoint action of $G$.
Now the corresponding 
quantum group $U_q(\mg_V)$ is manifest, and the
brane turns out to be a $q$-deformed space.

We present here a simple and compact description 
of these quantized $D$-branes on the classical
(compact, simple) matrix groups $G$ of type
$SU(N), SO(N)$ and $USp(N)$, 
and study field theory on them in some simple cases.
%These quantized branes are certain ``finite'' quantizations of 
%adjoint orbits, given by finite operator algebras, with additional structure 
%related to quantum groups which determine their geometrical properties.
%More explicitly, 
We first give an algebraic description of the quantized
group manifold $G$, different from the one proposed by Faddeev, Reshetikhin 
and Takhtajan \cite{Faddeev:1990ih} and 
Woronowicz \cite{Woronowicz:1987vs}. Instead, it is based on 
the so-called reflection equation \cite{Mezincescu:1991ui,Kulish:1992qb}.
This quantized group manifold turns out to admit an array
of discrete quantized adjoint orbits, which are
quantum sub-manifolds described by finite operator algebras. 
Their position in the group manifold and hence their size
is quantized in a particular way. 
This array of quantized adjoint orbits turns out to match precisely 
the structure of $D$-branes on group manifolds 
as found in string theory.
Such non-commutative $D$-branes were 
first found - in a somewhat different formulation - 
in the work \cite{Alekseev:1999bs} of Alekseev, Recknagel and Schomerus.
We proposed more generally in \cite{PS:G}  
the quantum-algebraic description of these $D$-branes.
%in analogy to the Moyal-Weyl quantum spaces which
%describe $D$-branes in the background of a constant $B$ field.
The essential new feature of our approach is the covariance under
a quantum analog of the full group of motions $G_L \times G_R$.
This allows to address global issues on $G$, 
rather than just individual branes.

In the simplest case of the group $SU(2)$, these quantized orbits are
$q$-deformations of the fuzzy spheres $S^2_N$
introduced by John Madore \cite{Madore:1992bw},
embedded in $G$. In the present context
$q$ is necessarily a root of unity, and 
is related to the ``radius'' of the group $G$:
The limit $q \to 1$ corresponds to infinite radius or 
zero curvature of $G$, where the $D$-branes coincide with the 
``standard'' fuzzy spheres. Keeping $q\neq 1$ basically describes
the effects of curvature on the group manifold, in a very remarkable 
way which will be discussed in detail.

This thesis is organized as follows. 
We start by giving in Chapter \ref{chapter:qDbranes} the
general description of quantized ajoint orbits
on quantized group manifolds, including a brief review of the main results 
from string theory (more precisely conformal field theory).
This first chapter is probably the most difficult one; however,
it is not indispensible for the remainder of this thesis. We chose this
approach because the algebraic description of the quantized group
manifolds is strikingly simple and compact, 
once the mathematical background has been digested. 
Many characteristic quantities of these quantum spaces will be 
calculated and compared with results from string theory, 
with very convincing agreement.
Of course, no knowledge of string theory is needed to 
understand the mathematical constructions of these quantum spaces,
and the remaining chapters can be read independently.

In Chapter \ref{chapter:qFSI}, we proceed to study field theory
on the simplest of these space, the $q$-deformed fuzzy spheres $S^2_{q,N}$,
on the first-quantized level.
While scalar fields offer little surprise, the formulation 
of gauge theories naturally leads to Yang-Mills and Chern-Simons type actions. 
The gauge fields contain an additional scalar degree of freedom,
which cannot be disentangled from the ``usual'' gauge fields. 
This is matched by the corresponding analysis of gauge theory on 
fuzzy 2-branes on $SU(2)$ \cite{Alekseev:2000fd}, 
which is possible in a certain limit. 
The kinetic term of the Yang-Mills action arises dynamically,
du to a beautiful mechanism 
which is similar to spontaneous symmetry breaking. This can also be 
related to the ``covariant coordinates'' introduced in 
\cite{Madore:2000en}.
Indeed, the formulation of gauge theories on this (and other) noncommutative
space is in a sense much simpler than the classical description in terms of
connections on fiber bundles. 
This suggests that if properly understood, 
noncommutative gauge theories should be simpler than
commutative ones, rather than more complicated.

There are other aspects of field theory
which are intrinsically related to 
the quantum nature of the space. For example, we find
transitions between different geometric phases of gauge theory on 
$S^2_N$, which are parallel to instantons in ordinary gauge theories. 
These are described briefly in Chapter \ref{chapter:fuzzyinst}.

The next step is to study quantized field theories. We first 
consider the undeformed fuzzy sphere $S^2_N$, putting $q=1$. 
Chapter \ref{chapter:floops} is a review of 
the first one-loop calculation for QFT on $S^2_N$,
which was done in \cite{fuzzyloop} for a scalar $\varphi^4$ theory. 
This was originally motivated by trying to understand 
the so-called ``UV/IR mixing'', which is a rather mysterious 
and troublesome phenomenon 
in quantum field theories on the quantum plane $\R^n_\theta$.
It turns out that $\R^2_\theta$ can be approximated by ``blowing up''
$S^2_N$ near the north pole, and 
we are able to understand how the UV/IR mixing arises as the 
limit of an intriguing ``non-commutative anomaly'',
which is a discontinuity of the quantized (euclidean) field theory
as the deformation is sent to zero. 

This application nicely illustrates the
long-term goal of this line or research, which is to take advantage 
of the rich mathematical structure of fuzzy $D$-branes
in order to get new handles on quantum field theory.
One might of course wonder whether adjoint orbits on Lie groups are  
of great physical interest. 
We claim that they could be of considerable interest:
There exist 4-dimensional versions such as  $\C P^2$,
and various limits of them can be taken.
Furthermore, one could also consider non-compact
groups, and look for branes with Minkowski signature. 
Knowing that there exist Yang-Mills type gauge theories on these
branes \cite{Alekseev:2000fd}, this should lead to new ways of formulating
physically interesting gauge theories, and quite possibly 
to new insights into old problems. 

Finally, we should point out  that 
quantizations of {\em co}-adjoint orbits have been considered 
for a long time 
\cite{Kirillov-book}, \cite{Kostant-book}. 
This is {\em not} what we do here. The quantized adjoint orbits 
considered here are intimately related to the WZW 3-form 
on a group manifold, and their description in terms
of generalized Poisson-structures
is quite involved \cite{Alekseev:2002km}.

This thesis is based on several publications 
\cite{PS:G,qFSI,fuzzyinst,fuzzyloop,qFSII} which were written 
in collaboration with Chong-Sun Chu,
Harald Grosse, John Madore, Marco Maceda and 
Jacek Pawelczyk. I have considerably rewritten and 
adapted them to make this thesis a coherent work, 
and added additional material throughout 
to make it more accessible. Nevertheless, 
some familiarity with Hopf algebras and quantum groups is assumed.
There are many good textbooks available on this subject 
\cite{Chari:1994pz,Klimyk:1997eb,Majid:1996kd}, and 
a short summary of the main aspects needed here is given in Chapter 
\ref{chapter:qDbranes}.
Chapters \ref{chapter:fuzzyinst} and \ref{chapter:floops} 
do not involve quantum groups at all.

\section*{Acknowledgements}

I am indebted to Julius Wess for letting me join his group in Munich
allowing me to pursue research which is not always
in line with the fashion, and for sharing his deep insights into  
theoretical physics.
Part of the work related to this thesis was done 
at the Laboratoire de Physique Th\'{e}orique in Orsay,
where I enjoyed being hosted for one year by John Madore,
and during several useful visits at the Erwin Schr\"odinger Institut in Vienna.

\vskip5pt

My special thanks go to
Chong-Sun Chu, Harald Grosse, John Madore, Marco Maceda 
and Jacek Pawelczyk for very pleasant   and inspiring
collaborations on much of the material included here. 

\vskip5pt

Many fruitful discussions helped me understand various 
aspects of this subject, in particular with 
Paolo Aschieri,
Christian Blohmann,  Gaetano Fiore, 
Branislav Jur\v{c}o,  
Peter Schupp,  Christoph Schweigert, Raimar Wulkenhaar,
and many others. 

\vskip5pt

Finally, the ground of this thesis has been laid many years ago in Berkeley,
where Bruno Zumino sparked my interest in quantum groups, which 
has not decreased since then. 
 
%% some basic Math (q and no q)
%\input{math.tex} 
%%  Quantized D-branes on group manifolds

\chapter{An algebraic description of quantized D-branes on group
  manifolds}
\label{chapter:qDbranes}

In this chapter\footnote{This chapter is based on the joint 
work \cite{PS:G} with J. Pawelczyk.}
we exhibit a certain class of quantum spaces,
namely quantized adjoint orbits on (simple, compact) Lie groups $G$.
Some of them will be studied in more detail in the later chapters. 
We shall not only describe these spaces, but also their 
embedding in the quantized group manifold, which is an essential 
part of the mathematical construction.  
This will allow us to relate them in a convincing way to 
certain quantized $D$-branes on $G$ in the framework of string theory.
There, $D$-branes are by definition submanifolds in the target space
(which is $G \times M$ for some suitable manifold $M$),
which are expected to be noncommutative spaces. 

We shall in fact argue that our quantum spaces 
are {\em the} appropriate description of these $D$- branes. 
This provides valuable physical insights; 
in particular, the open strings ending on $D$- branes are expected to
induce a gauge theory on them. 
One can therefore expect that 
gauge theories on these quantum spaces exist and are physically 
meaningful. 

The simplest class of $D$- branes on group manifolds $G$
corresponds to quantized adjoint orbits. 
We will see that their structure and properties can be described in a
compact and simple way in terms of certain $q$-deformed quantum algebras,
which have been considered in various contexts before. 
``$q$-deformed'' here refers to the fact that they are covariant under
the quantized universal 
enveloping algebra $U_q(\mg)$ of the Lie algebra $\mg$ of $G$, as defined by
Drinfeld and Jimbo \cite{Jimbo:1985zk,Drinfeld:1988in}. Our
description will reproduce essentially all 
known characteristics of stable branes as found in the framework of
WZW models on $G$, in particular their configurations in $G$,
the set of harmonics, and their energies. It covers
both generic and degenerate branes. 

The quantized $D$-branes constructed here are hence
quantizations of adjoint orbits on $G$.
This might be somewhat puzzling to the reader:
It is well-known that there exists a canonical Poisson structure
on co-adjoint orbits, which live in the dual $\mg^*$ of the Lie algebra
of $G$; the quantization of these is in fact related to ours.
However, this is not what we do, and
there is no canonical Poisson structure on adjoint orbits in
$G$. Rather, there is a more complicated so-called ``twisted Dirac
structure''
on $G$, where the twisting is basically given by the 
Wess-Zumino 3-form on the group. These differential-geometric aspects 
have been investigated in 
\cite{Alekseev:2002km}.
Here we decide to circumvent these rather involved classical considerations 
by starting directly with the full quantum version, which 
turns out to be much simpler than the semi-classical one. 
I find this quite gratifying: since the real world is quantum,
one should start with the quantum theory and then derive its
classical limit, rather than the other way round.

Before discussing the quantum spaces, we first recall
some general aspects of noncommutative $D$-branes in string theory. 
We then provide some mathematical background on adjoint orbits, and
review the main properties of the corresponding $D$-branes in WZW models
in Section \ref{sec:cft}. The account
on conformal field theory (CFT)
is not self-contained, since it is not required to understand the 
remainder of this thesis. Rather, these results are used as a guideline,
and to connect our algebraic framework with string theory.
In order to make this chapter more readable, we will sometimes postpone
the technical details to a ``local'' appendix in Section 
\ref{sec:technical1} .

\section{Noncommutative $D$-branes in string theory}
\label{subsec:string-dbranes}

We recall some aspects of noncommutative $D$-branes in string theory. 
This is not intended as an introduction to this topic,
which has become a very active field of research in recent years.

$D$-branes in string theory are by definition subvarieties of the target 
space (which is $T = G \times M$ here for some suitable manifold $M$)
on which open strings can end. The properties of these strings are
governed by the action 
\beq
S = \int_\Sigma g^{ij} \del_i X^\mu \del_j X^\nu G_{\mu \nu} +
        i \eps^{ij} \del_i X^\mu \del_j X^\nu B_{\mu \nu}
\label{sigma-action}
\eeq
\begin{figure}[htpb]
\begin{center}
\epsfxsize=3in
%  \vspace{-1in} 
   \hspace{0.1in}
   \epsfbox{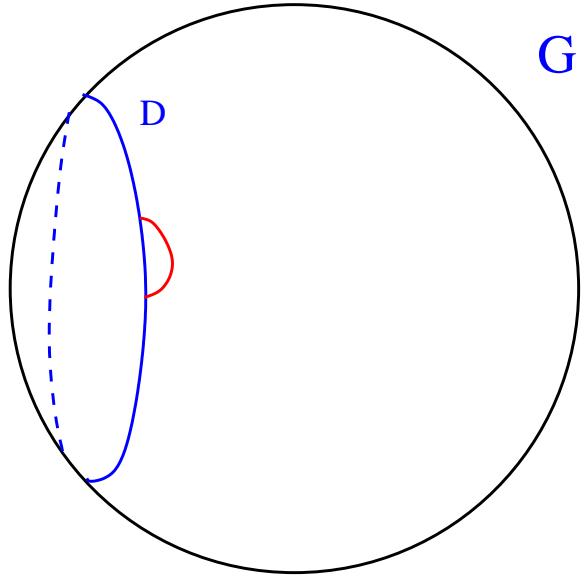}
\end{center}
 \caption{String ending on $D$-brane in $G$}
\label{fig:strings-branes}
\end{figure}
Here $\Sigma$ is the wordsheet of the string which has $D$ as a boundary, 
$X: \Sigma \to T$ is the embedding of the string in $T$, 
$G_{\mu \nu}$ is the metric on the target space $T$, 
and $B_{\mu \nu}$ is the crucial antisymmetric 2-form field in $T$. 
The structure of such  $D$-branes with a nonvanishing $B$ field background 
has attracted much attention, because they turn out to be
noncommutative spaces \cite{Douglas:1998fm,Chu:1998qz,Schomerus:1999ug}, 
at least in a suitable limit. 
Indeed, the $B$ field is topological in the bulk (if $dB = 0$),
and it can be shifted to the boundary of the worldsheet,
which is the $D$-brane. There it induces a Poisson-structure,
and the brane becomes a quantization of this Poisson structure.
The noncommutative algebra of functions on the brane
can be extracted from the algebra of boundary vertex operators.
The simplest case of flat branes in a 
constant $B$ background has been studied extensively 
(see e.g. \cite{Douglas:2001ba} for a review), and leads to quantum spaces
with a Moyal-Weyl star product corresponding to the constant Poisson 
structure. This was later generalized to
non-constant, closed $B$  \cite{Jurco:2001my}. 

A more complicated situation arises on group manifolds $G$ which
carry $D$-branes, because the $B$ field is not closed any more,
but satisfies $dB = H = const \neq 0$ where $H$ is the WZW term. 
One should therefore expect that the resulting quantization of 
the branes is more ``radical'' than just a deformation quantization.
Indeed, they turn out to be some generalizations of ``fuzzy'' 
spheres with finitely many degrees of freedom.
The best approach to study this situation is BCFT, which is an 
exact description of the worldsheet theory in the presence of the $D$-brane.
Other approaches include the Dirac-Born-Infeld (DBI) 
action, which can be used in the limit of
large radius of $G$. 
Using these methods, it has been shown in 
\cite{Alekseev:1998mc,Felder:1999ka,Bachas:2000ik,Pawelczyk:2000ah} 
that stable branes can 
wrap certain conjugacy classes in the group manifold. 
This will be explained in more detail in Section \ref{sec:cft}.
On the other hand, matrix model \cite{Myers:1999ps} and again CFT calculations 
\cite{Alekseev:2000fd} led to an intriguing picture where, in a 
special limit, the macroscopic branes arise as 
bound states of $D0$-branes, i.e. zero-dimensional branes.

Attempting to reconcile these various approaches, we proposed
in \cite{PS:su2} a  matrix description of $D$ - branes on
$SU(2)$. This led to a ``quantum'' algebra based on quantum group symmetries,
which reproduced all static properties of stable
$D$-branes on $SU(2)$. 

In the following chapter, we generalize the approach of \cite{PS:su2},
and present a simple and convincing description 
of all (untwisted) D-branes on group manifolds $G$ as
noncommutative, i.e. quantized spaces, in terms of certain 
``quantum'' algebras related to $U_q(\mg)$. This is analogous to 
the quantization of flat branes in a constant background $B$ using star 
products. As in the latter case, our description is based
on exact results of boundary conformal field theory (BCFT). 
We shall show that a simple algebra which was 
known for more than 10 years as reflection equation (RE)
reproduces exactly the same
branes as the exact, but much more complicated CFT description. 
It not only reproduces their configurations in $G$, 
i.e. the positions of the corresponding conjugacy classes, 
but also the (quantized) algebra of functions on the branes is
essentially the one given by the CFT description. 
Moreover, both generic and degenerate branes are predicted,
again in agreement with the CFT results. For example, 
we identify branes on $SU(N+1)$ which are quantizations of $\C P^N$,
and in fact $q$-deformations of the fuzzy $\C P^N$ spaces
constructed in \cite{Alexanian:2001qj,Grosse:1999ci}. 

We will not attempt here to recover all known branes on $G$, such as
twisted branes or ``type B branes'' \cite{Maldacena:2001ky},
but concentrate on the untwisted branes. Given the success and simplicity
of our description, it seems quite possible, however, that these other
branes are described by RE as well.

\section{The classical geometry of $D$--branes on group manifolds}

\subsection{Some Lie algebra terminology}
\label{subsec:liealg}

We collect here some basic definitions, in order to fix the notation.
Let $\mg$ be a (simple, finite-dimensional) Lie algebra, with 
Cartan matrix $A_{ij}=2\frac{\a_i\cdot\a_j}{\a_j\cdot\a_j}$.
Here $\a_i$ are the simple roots, and $\cdot$ is the
Killing form which is defined for arbitrary weights.
The generators $X^{\pm}_i, H_i$ of $\mg$ satisfy the relations
\beqa
\;[H_i, H_j]      &=& 0,  \quad
\;[H_i, X^{\pm}_j] = \pm A_{ji} X^{\pm}_j,     \\
\;[X^+_i, X^-_j]   &=& \d_{i,j} H_i
\label{UEA-0}
\eeqa
The length of a root (or weight) $\a$ is defined by
$d_\a = \frac{\a\cdot\a}2$.
We denote the diagram automorphisms by $\g$, which extend to $\mg$ by
$\g(H_i) = H_{\g(i)},\; \g(X^\pm_i) = X^\pm_{\g(i)}$.
For any root $\a$, the reciprocal root is denoted by 
$\a^{\vee} = \frac {2\a}{(\a \cdot \a)}$.
The dominant integral weights are defined as 
\beq
P_+ = \{\sum n_i \La_i;\;\; n_i \in \N \},
\eeq 
where the fundamental weights $\La_i$ satisfy
$\a_i^{\vee}\cdot\La_j = \d_{ij}$.
The Weyl vector is the sum over all positive roots, 
$\rho = \frac 12 \sum_{\a>0} \a$.
For any weight $\la$, we define $H_\la \in \mg$ 
to be the Cartan element which takes the value
$H_\la v_\mu =(\la\cdot\mu)\;v_\mu$
on a weight vectors $v_{\mu}$ in some representation.
In particular, $H_i = H_{\a_i^{\vee}}$.

For a positive integer $k$, 
one defines the ``fundamental alcove'' in weight space as
\beq \fbox{$
P_k^+ = 
 \{\la \in P^+; \; \la\cdot\theta \leq k\} $}
\label{fundalcove}
\eeq
where $\theta$ is the highest root of $\mg$.
It is a finite set of dominant integral weights.
For $G=SU(N)$, this is explicitly
$P_k^+ = \{\sum n_i \La_i;\;\sum_i n_i \leq k\}$. 
We shall normalize the Killing form such that
$d_\theta = 1$, so that the dual Coxeter number is given by
$\gcheck = (\rho + \frac 12 \theta)\cdot \theta$,
which is $N$ for $SU(N)$. 

We will consider only finite-dimensional representations of $\mg$.
$V_{\la}$ denotes the irreducible
highest-weight module of $G$ with highest weight $\la \in P_+$, and
$V_{\la^+}$ is the conjugate module of $V_{\la}$.
The defining \rep of $\mg$ for the classical matrix
groups $SU(N)$, $SO(N)$, and $USp(N)$ will be denoted by 
$V_N$, being $N$-dimensional.

\subsection{Adjoint orbits on group manifolds}
\label{sec:c-branes}

Let $G$ be a compact matrix group of type $SU(N)$, $SO(N)$ or $USp(N)$,
and $\mg$ its Lie algebra. For simplicty we shall concentrate on $G=SU(N)$,
however all constructions apply to the other cases as well,
with small modifications that will be indicated when necessary.

(Twisted) adjoint orbits on $G$ have the form 
\beq\fbox{$
\cC(t) = \{g t \g(g)^{-1}; \quad g \in G\}. $}
\label{conj-classes}
\eeq
Here $\g$ is an auto-morphism of $G$.%, which will appear again in\refeq{twinbc}. 
In this work we shall study only untwisted branes corresponding to
$\g = id$, leaving the twisted case for future investigations.
One can assume that $t$ belongs to a maximal torus $T$ of $G$, i.e. 
that $t$ is a diagonal matrix for $G = SU(N)$.
As explained in Section \ref{subsec:harmonic-brane}, 
$\cC(t)$ can be viewed as homogeneous space:
\beq
\cC(t) \cong G/K_t.
\label{coset}
\eeq
Here $K_t = \{g \in G:\; [g,t] = 0\}$ is the stabilizer of $t \in T$.
``Regular'' conjugacy classes are those with $K_t = T$, 
and are isomorphic to $G / T$. In particular, their dimension is 
$dim(\cC(t)) = dim(G) - rank(G)$.
``Degenerate'' conjugacy classes have a larger stability group $K_t$,
hence their dimension is smaller. Examples of degenerate conjugacy classes
are $\C P^N \subset SU(N+1)$, and the extreme case is a point $\cC(t=1)$.

\paragraph{Symmetries.}

The group of motions on $G$ is the product $G_L \times G_R$,
which act on $G$ by left resp. right multiplication. 
It contains the ``vector'' subgroup $G_V \hookrightarrow G_L \times G_R$,
which is diagonally embedded and acts on $G$ via conjugation. 
In general, the motions rotate the conjugacy classes on the group manifold.
However all conjugacy classes are invariant under the adjoint action
of $G_V$,
\beq\label{adj-action}
G_V \cC(t) G_V^{-1} = \cC(t).
\eeq
%This reflects the breaking $\mgh_L\times\mgh_R \to \hat\mg_V$.
We want to preserve this symmetry pattern
in the quantum case, in a suitable sense.

\paragraph{The space of harmonics on $\cC(t)$}

A lot of information about the spaces $\cC(t)$ can be obtained using
harmonic analysis, i.e. by decomposing functions on $\cC(t)$
into irreps under the action of the (vector) symmetry $G_V$. 
This is particularly useful here, because 
quantized spaces are described in terms of their
algebra of functions. The decomposition of this 
space of functions $\cF(\cC(t))$ into harmonics can be calculated explicitly
using \refeq{coset}, and it must 
be preserved after quantization, at least up to some cutoff.
Otherwise, the quantization would not be admissible.
One finds (see Section \ref{subsec:harmonic-brane} and \cite{Felder:1999ka}) 
\beq
\cF(\cC(t)) \cong  \bigoplus_{\la \in P_+} \; mult_{\la^+}^{(K_t)}\; V_{\la}.
\label{FCt-decomp}
\eeq
Here
$mult_{\la^+}^{(K_t)}\; \in \N$ is the dimension of the subspace of 
$V_{\la^+}$ which is invariant under $K_t$.

\subsection{Characterization of the stable $D$--branes}

\begin{figure}[htpb]
\begin{center}
\epsfxsize=3in
%  \vspace{-1in} 
   \hspace{0.1in}
   \epsfbox{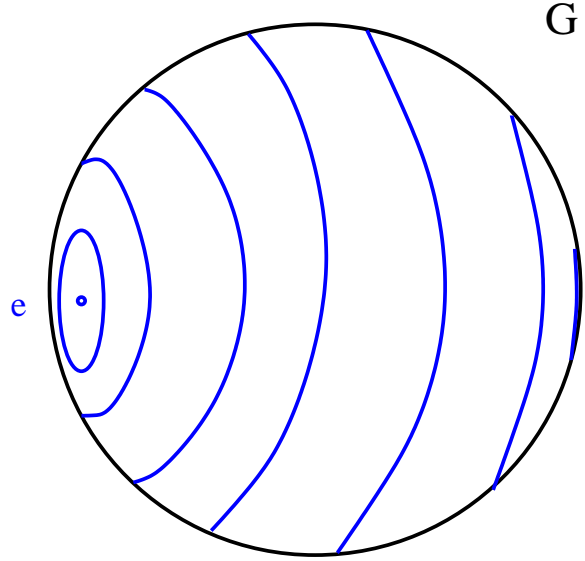}
\end{center}
 \caption{$D$-branes in $G$}
\label{Gbranes-eps}
\end{figure}

From the CFT \cite{Alekseev:1998mc,Felder:1999ka} 
considerations as reviewed in Section \ref{sec:cft},
it follows that there is only a finite
set of stable $D$--branes on $G$ (up to global motions),
one for each integral weight $\la\in P_k^+$ in 
the fundamental alcove \refeq{fundalcove}. 
They are given by $\cC(t_{\la})$ for 
\beq\fbox{$
t_{\la} = \exp(2 \pi i \frac{H_\la + H_\rho}{k + \gcheck})$}
\label{stable-branes}
\eeq
where $\rho = \frac 12 \sum_{\a>0} \a$ is the Weyl vector. 
The restriction to $\la\in P_k^+$ follows from the fact that
different integral $\la$ may label the same conjugacy class,
because the exponential in \refeq{stable-branes} is periodic. This 
happens precisely if
the weights are related by the affine Weyl group, which is 
generated by the ordinary Weyl group together with translations of the form
$\la \rightarrow \la + (k+\gcheck) \frac{2 \a_i}{(\a_i\cdot \a_i)}$.
Hence one should restrict the weights to be in
the fundamental domain of this affine Weyl group, which is precisely the 
fundamental alcove $P_k^+$ \refeq{fundalcove}.
The resulting pattern of stable branes is sketched in Figure 
\ref{Gbranes-eps}
for the $G = SU(2)$. This case is discussed in more detail in
Section \ref{subsec:su2}.

We are interested not only in the branes themselves, but also in their
location in $G$.
Information about the location of these (untwisted) branes in $G$
is provided by the quantities 
\beq
s_n = \tr(g^n) = \tr(t^n), \quad g \in \cC(t)
\eeq
which are invariant under the adjoint action \refeq{adj-action}.
The trace is over the defining \rep $V_N$ 
($=V_{\La_1}$ in the case of $SU(N)$, 
where $\La_1$ is the fundamental weight) of the matrix group $G$,
of dimension $N$. For the conjucacy classes  $\cC(t_{\la})$,
they can be calculated easily:
\beq\fbox{$
s_n =  \tr_{V_N}\; (q^{2n (H_\rho +H_\la)}) = 
   \;\sum_{\nu \in V_N}\; 
   q^{2 n (\rho +\la)\cdot \nu}$}
\label{sn}
\eeq
where 
$$
q = e^{\frac{i \pi}{k+\gcheck}}.
$$
The $s_n$ are independent functions 
of the weight $\la$ for $n = 1,2,..., rank(G)$,  which
completely characterize the class $\cC(t_{\la})$.
These functions have the great merit that their quantum analogs 
(\ref{central2-n}) can also be calculated explicitly.

An equivalent characterization of these conjugacy classes is provided by 
a characteristic equation: for any $g \in \cC(t_\la)$, the relation 
$
P_\la(g) = 0
$
holds in $Mat(V_N,\C)$, 
where $P_\la$ is the polynomial 
\beq
P_\la(x) = \prod_{\nu \in V_N} (x-q^{2(\la+\rho)\cdot\nu}).
\label{charpoly-class}
\eeq
This follows immediately from (\ref{stable-branes}), since
$t_\la$ has the eigenvalues $q^{2(\la+\rho)\cdot\nu}$
on the weights $\nu$ of the defining \rep $V_N$.
Again, we will find analogous characteristic equations in the 
quantum case.

We should perhaps point out that co-adjoint orbits on Lie groups
carry a natural symplectic structure, and the quantization 
of these spaces has been considered before 
\cite{Kirillov-book}, \cite{Kostant-book}. As individual manifolds, 
the quantized $D$-branes we shall find are
related to these quantized (co)adjoint orbits, which by itself would
perhaps not be too exciting. The point is, however, that
our description not only gives isolated quantized orbits, but
a quantization of the entire group manifold, and the 
quantized orbits are embedded at precisely the positions as predicted
by string theory. This means that our construction will {\em global} 
and reproduces the orbits correctly not only near the origin,
but also reproduces correctly the shrinking of the branes 
beyond the ``equator'' of $G$, as they approach the point-liks
brane at $-id \in G$. This works only for $q$ a root of unity,
and is the major result of our construction.

\section{The CFT description of D-branes in the WZW model}
\label{sec:cft}

This section reviews some results from the 
CFT description of branes in WZW models on $G$, and
is {\em not} self-contained. It serves as inspiration to
the algebraic considerations below,
and to establish the relations with string theory. 
All results presented here are well known. They are not necessary to 
understand the rest of this chapter, and the reader who is not
familiar with CFT and string theory may skip this section and go
directly to the Section \ref{sec:LR-symm}

%\subsection{Branes in the WZW model}
%\label{sec:wzw-b}

The WZW model (see e.g. \cite{Fuchs:1992nq,DiFrancesco:1997nk})
is a CFT with a Lie group $G$ as target space, and
action \refeq{sigma-action}
\beq
S = \int_\Sigma g^{-1} dg g^{-1} dg  + B, \qquad dB = k (g^{-1} dg)^3.
\label{WZW-action}
\eeq
It is specified by a level $k$ of the affine Lie algebra $\mgh$ 
whose horizontal subalgebra is $\mg$, the Lie algebra of $G$.
We shall consider only simple, compact groups $G$, 
so that the level $k$ must be a positive integer. 
On an algebraic level, the WZW branes can  be
described by boundary states $\vert B\rara\in \cH^{\rm closed}$ respecting  a
set of boundary conditions. 
A large class of boundary conditions 
is of the form
\beq
 \biggl( J_{n} + \g(\tilde J)_{-n} \biggr) \vert
B \rangle \rangle =0 \qquad n\in {\bf Z} 
\label{twinbc}
\eeq
where $\g$ is an auto-morphism of the affine Lie algebra $\mgh$.
Here  $J_n$ are the modes of the left-moving currents, and $\tilde
J_n$ are the modes of the right-moving currents.  Boundary states
with $\g=1$ are called ``symmetry-preserving branes'' or "untwisted branes":
these are the object of interest here.
The untwisted ($\g=1$) boundary condition \refeq{twinbc} breaks half of the
symmetries  $\mgh_L\times\mgh_R$ of the WZW model down to the vector part
$\mgh_V$. This will be important for the choice of  the relevant
quantum symmetry $\gcg$ (or $\GcG$) in the algebraic considerations of
Section \ref{sec:LR-symm}.

The condition \refeq{twinbc} alone does not define a good boundary state:
one also has to impose open-closed string duality of the 
amplitude describing interactions of branes. This leads to so called
Cardy (boundary) states. 
For the untwisted case they are labeled by
$\lambda\in P_k^+$ corresponding to integrable irreps of  
$\widehat \mg$, which are
precisely the weights in the ``fundamental alcove'' \refeq{fundalcove}.
Therefore 
the stable branes in the WZW model are in one-to-one correspondence to
the weights $\la \in  P_k^+$.

The CFT description yields furthermore an important formula for the energy 
of the brane corresponding to $\la$:
{\large
\beq\label{crdymss}\fbox{$ 
E_\la=\prod_{\a > 0}
\frac{\sin\bigl(\pi \frac{\alpha\cdot (\lambda+ \rho)}{k+\gcheck}\bigr)}
     {\sin\bigl(\pi \frac{\alpha\cdot  \rho}{k+\gcheck}\bigr)}.$}
\eeq}
For $k\gg N $, one can expand the denominator in \refeq{crdymss} to obtain a
formula which compared with the
DBI description \cite{Maldacena:2001xj} shows that the leading $k$-dependence
fits perfectly with the interpretation of a brane wrapping 
once a conjugacy class $\cC(t_{\la})$
given by the element $t_\la$ \refeq{stable-branes}
of the maximal torus of $G$.

The BCFT is also the basis for the description of 
branes as noncommutative spaces, in complete analogy to 
the case of flat branes in a constant
$B$ field as explained in Section \ref{subsec:string-dbranes}. 
The properties of the D-branes are determined by open strings 
ending on them, which are described by boundary vertex operators.
The relevant operators here are the primary fields of the BCFT
which transform under the unbroken symmetry algebra 
$\mgh_V \subset \mgh_L\times\mgh_R$.
Their number is finite for any compact WZW model. 
They satisfy an operator-product algebra (OPE), which has the form
\beq
    Y^I_i(x) \; Y^J_j(x') \sim \sum_{K,k} \; 
     (x-x')^{h_I+h_J - h_K}  
    \left[\begin{array}{ccc} I&J&K\\ i&j&k \end{array} \right]
    \left\{\begin{array}{ccc} I&J&K\\ \la&\la&\la \end{array} \right\}_q 
    Y^K_k(x') + ... 
\eeq
where we assume $G = SU(2)$ for simplicity\footnote{We use the 
$su(2)$ convention here that the weight $\la$ corresponds to half-integers.
This should not cause any confusion, because the $6j$ symbols depend
on \reps rather than numbers.}.
Here $h_I = I(I+1)/(k+\gcheck)$  
are the conformal dimension of the primaries $Y^I_i$, and
the sum over $K$ has a cutoff
$K_{\rm max}= \min(I+J,k-I-J,2\la,k-2\la)$. It involves the $3j$ symbols
$[...]$ of $su(2)$, and the $6j$ symbols $\{...\}_q$  of $U_q(su(2))$,  
the quantized universal enveloping algebra of $su(2)$.
The variables $x, x'$ are on the boundary of the world sheet $\Sigma$,
which is usually taken to be the upper half plane.
The dots denote operators (descendants) which do not enter the description 
of the brane as a manifold. 
For higher groups, the OPE has a similar form \cite{Felder:1999ka}
involving the corresponding $3j$ and
$6j$ symbols of $\mg$ and $U_q(\mg)$, respectively. 
This indicates the deep relation between 
affine Lie algebras and quantum groups at roots of unity 
\cite{Alvarez-Gaume:1990aq,Chari:1994pz,KL}.

For $k \to \infty$, the conformal weights $h_I$ become zero, and
this OPE reduces to the algebra of functions
on the fuzzy sphere $S_N^2$ in the case $G=SU(2)$
(see also Chapter \ref{chapter:floops}). 
The $Y^I_i$ then play the role of the spherical harmonics.
This interpretation was extended in \cite{Felder:1999ka} 
to branes on arbitrary $G$.
For finite $k$, this interpretation 
is less clear, because the conformal dimensions are
nonvanishing, and the truncation of the OPE may seem unjustified. 
We shall nevertheless consider the brane with label $\la$ 
as a quantized manifold, whose space of function is encoded in
the finite set $\{Y^I_i \}$ of primaries. 
%The other massless modes (in particular the gauge fields) 
%should be included in a further development of our approach.

To extract an algebra of functions on the branes, we note that
the variables $x,x'$ in the above OPE are only world-sheet variables
and not physical. It was therefore suggested in \cite{Alekseev:1999bs}
in the case 
$G=SU(2)$ to associate with this OPE the  ``effective'' algebra
\beq
  Y^I_i Y^J_j  = {\sum}_{K,k}
       \left[\begin{array}{ccc} I&J&K\\ i&j&k \end{array} \right]
   \left\{\begin{array}{ccc} I&J&K\\ \la&\la&\la \end{array} \right\}_q 
    Y^K_k,
\label{quasiass-ope}
\eeq
interpreted as algebra of functions on the brane,
which is covariant under $su(2)$.
Unfortunately this algebra is only quasi-associative,
due to the curious mixing of undeformed and $q$-deformed
group theoretical objects. 
However, as explained in Sections \ref{subsec:drinfeld_twist} and 
\ref{subsec:6j-alg},
it is equivalent by a Drinfeld-twist to the associative algebra
\beq\fbox{$
  Y^I_i \star Y^J_j  = {\sum}_{K,k}
       \left[\begin{array}{ccc} I&J&K\\ i&j&k \end{array} \right]_q
   \left\{\begin{array}{ccc} I&J&K\\ \la&\la&\la \end{array} \right\}_q 
    Y^K_k,  $}
\label{ass-ope}
\eeq
which is covariant under $U_q(su(2))$.
Associativity of course makes this twisted algebra much easier to work with,
and appears to be the most natural description. 
The twisting of the algebra 
could presumably be realized on the CFT level by introducing
some kind of ``dressed'' vertex operators, similar as in \cite{Moore:1989ni}.
%However, we will see that 
%\refeq{ass-ope} is a $U_q(su(2))$--module algebra, while 
%\refeq{quasiass-ope} is a $U(su(2))$--module algebra. 
For other groups $G$, the relevant algebras are entirely analogous,
involving the generalized $3j$ and $6j$ symbols of $U_q(\mg)$.

We will recover exactly the same algebra \refeq{ass-ope} 
from the quantum-algebraic approach below, based on symmetry principles
involving quantum groups. This is 
the basis of our algebraic description in the following sections. 
It was indeed known for a long time \cite{Alvarez-Gaume:1990aq} that 
the $6j$ symbols of $U_q(\mg)$ enter the chiral OPE's of WZW models,
but they become ``invisible'' in the full CFT. 
The remarkable thing here is that on the $D$-branes, the $q$-deformation
becomes manifest, because the chiral symmetry
$\mgh_L\times\mgh_R$ is broken down to $\mgh_V$. 
Before proceeding, we must therefore pause and collect some relevant facts 
about $U_q(\mg)$ and related algebraic structures.

\section{Some quantum group technology}
\label{subsec:q-groups}
\label{a:basics}

We recall here the basic definitions and properties
of quantum groups which are needed below. Some the topics
will be elaborated further in later chapters.
This section 
should also serve as some kind of ``background check'' for the reader:
while it might help as a first crash course to 
get some basic understanding, the presentation 
is definitely not self-contained.
For a more detailed account, the reader is referred to some
of the existing monographs, for example 
\cite{Majid:1996kd,Chari:1994pz,Fuchs:1992nq}.

\subsection{$U_q(\mg)$}

Given a (simple, finite-dimensional)
Lie algebra $\mg$ as in Section \ref{subsec:liealg}, 
the quantized universal enveloping 
algebra\footnote{we will ignore subtle mathematical issues of
topological nature, being interested only in 
finite-dimensional \reps
where all series terminate and convergence issues are trivial.}
\cite{Drinfeld:1988in,Jimbo:1985zk} $U_q(\mg)$ is generated by 
the generators $X^{\pm}_i, H_i$ satisfying the relations
\beqa
\;[H_i, H_j]      &=& 0,  \quad
\;[H_i, X^{\pm}_j] = \pm A_{ji} X^{\pm}_j,     \\
\;[X^+_i, X^-_j]   &=& \d_{i,j}
                      \frac{q^{d_i H_i} -q^{-d_i H_i}}{q^{d_i}-q^{-d_i}}
%                    = \d_{i,j} [H_i]_{q_i},
\label{UEA}
\eeqa
plus analogs of the Serre relations. 
Here $d_i = \frac{\a_i\cdot\a_i}2$ is the length of the root $\a_i$.
The Hopf algebra structure of $U_q(\mg)$ is given by the 
comultiplication and antipode 
\beqa
\Delta(H_i)          &=& H_i \tens 1 + 1 \tens H_i,  \qquad
\Delta(X^{\pm}_i) =  X^{\pm}_i \tens q^{d_i H_i/2} + q^{-d_i H_i/2}
\tens  X^{\pm}_i, \nn\\
S(H_i)    &=& -H_i, \qquad
S(X^{\pm}_i)  = -q^{\pm d_i} X^{\pm}_i.
\label{coproduct-X}
\eeqa
The coproduct is conveniently written in Sweedler-notation as
$\Delta(u) = u_1 \tens u_2$, for $u\in U_q(\mg)$,
where a summation is implied. 
Generators $X_\a^\pm$ corresponding to the other roots can be defined 
as well. It is easy to verify the important relation
$S^2(u) = q^{2 H_\rho}\; u\; q^{-2 H_\rho}$ 
which holds for all $u \in U_q(\mg)$.
%This is used in the definition of the quantum traces 
%\refeq{central2-n},\refeq{qdim}.
The {\em quasitriangular structure} of $U_q(\mg)$ is given by the 
universal $\cR \in U_q(\mg) \tens U_q(\mg)$, which
satisfies the properties
\beqa
(\Delta\tens id) \cR  &=& \cR_{13} \cR_{23},  \quad
(id \tens \Delta) \cR = \cR_{13} \cR_{12}  \label{qtr_2}\\
\Delta'(x) &=& \cR \Delta(x) \cR^{-1} \label{qtr_3}
\eeqa
where $\Delta'$ denotes the reversed coproduct.
This implies the Yang-Baxter equation
\beq
\label{YBE}
\cR_{12} \cR_{13} \cR_{23} = \cR_{23} \cR_{13} \cR_{12}. 
\eeq
Introducing the matrices 
\beqa
L^{+} & = & (id \tens \pi)(\cR ), \nn\\
SL^{-} &=& (\pi \tens id) (\cR )
\eeqa
where $\pi$ is a \rep of $U_q(\mg)$, the 
relations \refeq{UEA} can be written in the form
\beqa
R_{12} L^+_2 L^+_1 &=& L^+_1 L^+_2 R_{12}, \nn\\
R_{12} L^-_2 L^-_1 &=& L^-_1 L^-_2 R_{12},\nn\\
R_{12} L^+_2 L^-_1 &=& L^-_1 L^+_2 R_{12}
\label{RLL-relations}
\eeqa
which follow easily from \refeq{YBE}.
Explicitly, $\cR$  has the form
\beq
\cR = q^{H_i (B^{-1})_{ij} \tens H_j} \; (1 \tens 1 + \sum U^+ \tens U^-)
\eeq
where $B$ is the symmetric matrix $d_j^{-1} A_{ij}$, and
$U^+, U^-$ stands for terms in the Borel sub-algebras of
rising respectively lowering operators.
Therefore $L^+$ are lower triangular matrices with 
$X^+_\a$'s 
below the diagonal, and $L^-$ are upper triangular matrices with  
$X^-_\a$'s 
above the diagonal. For $\mg = sl(2)$ one has explicitly
\beq
L^+=\left(\barr{cc} q^{H/2}& 0\\ q^{-\half}\la X^+& ~~~q^{-H/2}
   \end{array}\right), \quad 
L^-=\left(\barr{cc} q^{-H/2}& ~~~ - q^{\half}\la X^- \\ 0& q^{H/2}
\end{array}\right)
\eeq
The above definitions are somewhat sloppy mathematically, 
$q$ being unspecified. This is justified here because 
we are only interested in certain
finite-dimensional \reps of $U_q(\mg)$, and we will in fact
only consider the case where $q$ is a root of unity, 
$$\fbox{$
q = e^{\frac{i \pi}{k+\gcheck}}. $}
$$
Hence we are dealing with (some version of the) ``finite quantum group''
$u_q^{fin}(\mg)$ at roots of unity \cite{Chari:1994pz}. 

\subsection{Representations of $U_q(\mg)$}
\label{subsec:reps}

For generic (or formal) 
$q$, the \rep theory of $U_q(\mg)$ is completely parallel to
that of $U(\mg)$. The situation is very different at roots of unity, 
many non-classical types of representations arise. Here we consider 
only the simplest ones, which are the irreducible highest weight 
\reps $V_\la$ with highest weight $\la \in P_k^+$ in the fundamental alcove
\refeq{fundalcove}. We shall call them 
{\em regular} representations.

A useful concept at roots of unity is 
the {\em quantum dimension} of a \rep $V$, 
\beq
\dim_q(V) = \tr_{V}(q^{2 H_\rho})
\label{qdim-def}
\eeq
which for highest-weight \reps $V_\la$ with $\la \in P_k^+$ can be 
calculated using Weyls character formula
\beq
\dim_q(V) = \prod_{\a > 0} 
  \frac{\sin(\pi\frac{\a\cdot(\la + \rho))}{k + \gcheck})}
       {\sin(\pi\frac{\a\cdot\rho}{k + \gcheck})}.
\label{qdim}
\eeq
Clearly $\dim_q(V_\la)>0$ for $\la \in P_k^+$, i.e. for regular 
representations. 
These representations are in one-to-one correspondence 
with the integrable modules of the affine Lie algebra $\mgh$ at level $k$,
see e.g. \cite{Fuchs:1992nq}. 
This is part of a deep relation between 
affine Lie algebras and quantum groups at roots of unity 
\cite{Chari:1994pz,KL}.
This relation extends to tensor products as follows:
In general, the tensor product 
$V_\la \tens V_{\la'}$ is in general not completely reducible. Rather,
it decomposes as 
$V_\la \tens V_{\la'} = (\oplus_\mu V_\mu) \oplus T$ where 
$V_\mu$ is regular, and 
$T$ denotes indecomposable ``tilting modules'' \cite{Chari:1994pz}.
It turns out that $\dim_q(T) = 0$, and one can define a 
``truncated tensor product'' $\obar{\tens}$ 
by simply omitting all the modules $T$ in the 
tensor product which have vanishing quantum dimension. It turns out
that $\obar{\tens}$ is associative, and coincides precisely with the 
fusion rules of integrable modules of $\mgh$ at level $k$.
This gives exactly the modes which occur in the rhs of \refeq{quasiass-ope}. 

Furthermore, one can show that 
they are unitary \reps with respect to the star-structure
\beq
H_i^* = H_i, \quad (X_i^\pm)^* = X_i^\mp,
\eeq
see \cite{UnitaryReps} for a proof. 
For $\la\not\in P_k^+$, this is no longer the
case in general, 
and the corresponding highest weight \reps become non-classical.

\paragraph{Invariant tensors}

Consider some irreducible \rep $\pi^a_b(u)$ of $U_q(\mg)$.
Then $g^{ab}$ is an 
invariant 2- tensor if
\beq
\pi(u_1)^{a}_{a'} \pi(u_2)^{b}_{b'} g^{a' b'} = g^{ab}\; \eps(u),
\eeq
and similarly for higher-order tensors. There exist as many invariant tensors
for a given \rep as in the undeformed case, because
they are in one-to-one correspondence with trivial components in the 
tensor product of the representation, whose decomposition
is the same as classically for generic $q$. For roots of unity, of course,
there may be additional ones.
Invariant 2-tensors can be used as usual 
to rise and lower indices, however one has to be careful with the ordering 
of indices. 
For example, if $f^{abc}$ is an invariant 3-tensor, one can define
\beq
f^{ab}{}_c = f^{abc'} g_{c' c},  \; 
f^{a}{}_{bc} = f^{ac' b'} g_{b' b} g_{c' c},\; 
f_{abc} = f^{c' b' a'} g_{a' a} g_{b' b} g_{c' c}.
\eeq
The invariance of the tensor with lower indices is then as follows:
\beq
f_{a'b'c'}\;
\pi(Su_3)^{a'}_a \pi(Su_2)^{b'}_b \pi(Su_1)^{c'}_c
= f_{abc}\; \eps(u).
\eeq

\subsection{Dual Hopf algebras and $Fun_q(G)$}
\label{subsec:dualHA}

Two Hopf algebras $\cU$ and $\cG$ 
are {\em dually paired}
if there exists a non-degenerate pairing $<\;,\;>:$
$\cG \tens \cU \rightarrow \C$, such that
\beqa
<a b,u> & = & <a \tens b, \Delta (u)>, \qquad
<a,u v>  =  <\Delta (a),u \tens v>, \nn \\
<S(a),u> & = & <a,S(u)>, \quad
<a,1>  = \eps (a),\quad <1,u>=\eps (u),
\label{hopf_dual}
\eeqa
for $a,b \in \cG$ and $u,v \in \cU$.
The dual Hopf algebra of $U_q(\mg)$ is $Fun_q(G)$,
as defined in \cite{Faddeev:1990ih}. It is generated by
the elements of a $N \times N$ matrix 
$A = ({A^i}_j) \in Mat(N, Fun_q(G))$, which can be interpreted as 
quantized coordinate functions on $G$.
The dual evaluation is fixed by  
${\pi^i}_j(u) = <{A^i}_j,u>$, where $\pi$ is the defining  
\rep of $U_q(\mg)$. 
The coalgebra structure on $Fun_q(G_q)$ comes out as classically:
\beqa
\Delta A &=& A \dot{\tens} A, \quad\mbox{i.e.}\quad\Delta ({A^i}_j)=
{A^i}_k\tens{A^k}_j. \label{coprod_A} \nn\\
S(A) &=& A^{-1},\quad \ep(A^i_j) = \d^i_j, \quad \ep(A^i_j) = \d^i_j.
\eeqa
The inverse matrices $A^{-1}$ are well-defined after 
suitable further constraints on $A$ are 
imposed, as in \cite{Faddeev:1990ih}: for example, there can be an invariant tensor 
$(\eps_q)_{i_1 ... i_N}$ of $U_q(\mg)$ which defines 
the quantum determinant of $Fun_q(G_q)$,
\beq
\det_q(A) = 1
\eeq
where
\beq
\det_q(A)\; (\eps_q)_{j_1 .... j_N} = 
(\eps_q)_{i_1 .... i_N}\; A^{i_1}_{j_1} .... A^{i_N}_{j_N}  \label{det_1}
\eeq
and other constraints for the orthogonal and symplectic cases,
which can be found in the literature.
One also finds that the matrix elements of $A$ 
satisfy the commutation relations
\beq
{R^{ij}}_{kl} {A^k}_m {A^l}_n  =   {A^j}_s {A^i}_r {R^{rs}}_{mn},
\eeq 
which can be written more compactly in tensor product notation as
\beq
R_{12} A_1 A_2  =    A_2 A_1 R_{12}. \label{RTT}
\eeq

\subsection{Covariant quantum algebras}  
\label{subsec:cov-alg}

Assume that $\cG,\ \cU$ are two Hopf algebras. Then   
an algebra $\cM$ is called a {\em (left) $\cU$-module algebra} if  
there is an action $\trr$ of $\cU$ on $\cM$ such that 
\beq 
u \trr (xy) = (u\trr x)\trr y,\quad u\trr 1 =  \eps(u) 1  
\eeq  
for $u \in \cU$. Similarly, $\cM$ is a {\em (right) $\cG$-comodule algebra}
if there is a coaction $\nabla: \cM \to \cM \tens \cG$ 
of $\cG$ on $\cM$ such that 
\beq 
(id \tens \Delta)\nabla = (\nabla \tens id)\nabla,\quad 
(\eps \tens id )\nabla = id. 
\eeq
It is easy to see that if $\cG,\ \cU$ are dually paired Hopf algebras,
then a (right) $\cG$ - comodule algebra $\cM$ 
is automatically a (left) $\cU$ -- module algebra 
by  
\beq 
u \trr M = \langle\nabla(M),u\rangle 
\eeq 
where $\langle m\tens a,u\rangle = m\langle a,u\rangle$, 
and vice versa.  

%\subsection{Invariant tensors}

%In the adjoint representation, there are
%invariant tensors $g_{ab}, d_{abc}$ and $f_{abc}$, 
%which are deformations of the usual ones. For example,
%\beq
%f_{a'b'c'}\; 
%\pi_{(ad)}(u_1)^{a'}_a \pi_{(ad)}(u_1)^{b'}_b \pi_{(ad)}(u_3)^{c'}_c  
% = f_{abc}\; \eps(u).
%\eeq
%The ``Killing form'' $g_{ab}$ satisfies
%\beq
%g_{ab} = g^{ab}, \; g_{ab} g^{bc} = \d_a^c = g^{cb} g_{ba},
%\eeq
%(this follows e.g. using 
%the universal Weyl element, see e.g. \cite{ads-space})
%and can be used to rise and lower indices (note that $g_{ab}$ 
%is not symmetric). For example, 
%\beq
%f_{ab}{}^c = f_{abc'} g^{c' c},  \; 
%f_{a}{}^{bc} = f_{ac' b'} g^{c' c} g^{b' b},\; 
%f^{abc} = f_{c' b' a'} g^{c' c} g^{b' b}g^{a' a}.
%\eeq
%The invariance property of the upper-index tensor is 
%\beq
%\pi_{(ad)}(Su_3)_{a'}^a \pi_{(ad)}(Su_2)_{b'}^b 
%  \pi_{(ad)}(Su_1)_{c'}^c\; f^{a'b'c'} = f^{abc}\; \eps(u)
%\eeq

\subsection{Drinfeld twists}
\label{subsec:drinfeld_twist}

A given Hopf algebra $\cU$ can be {\em twisted} 
\cite{Drinfeld:quasi,Drinfeld:quasiGal}
using an invertible element
$$
\F = \F_1\tens \F_2 \; \in \; \cU \tens \cU
$$
(in a Sweedler notation, where a sum is implicitly understood) 
which satisfies
\beqa
&&(\varepsilon\tens id)\F=\one=(id\tens \varepsilon)\F,
%\label{cond2bis}
\eeqa
This works as follows: Let $\cU_\F$ be the Hopf algebra which 
coincides with $\cU$ as algebra, but has the twisted coalgebra structure
\beqa
\Delta_\F(u)&=& \F\Delta(u)\F^{-1},        \label{defd}\\
S_{\F}(u) &=& \g^{-1} S(u) \g
\eeqa
where $\g = S(\F_1^{-1}) \F_2^{-1}$.
$\F$ is said to be a cocycle if the  
coassociator
\beq
\phi:=[(\Delta\tens \id)\F^{-1}](\F^{-1}\tens \one)(\one \tens \F)
[(id \tens \Delta)\F] \qquad \in\cU^{\tens 3}
\label{defphi-0}
\eeq
is equal to the unit element. In that case, 
one can show that $\Delta_\F$ is coassociative,
and moreover if $\cU$ is quasitriangular, then so is $\cU_\F$ with
\beq
\RR_\F = \F_{21} \RR \F^{-1}.       % \label{defR}
\eeq
The same twist $\F$ can also be used to twist the corresponding (co)module 
algebras: Assume that $\cM$ is a left $\cU$-module algebra. Then 
\beq
a \star b := (\F^{-1}_1\trr a) \; (\F^{-1}_2\trr b) 
\label{star-0}
\eeq
defines a new product on the same space $\cM$, which turns it into a 
left $\cU_\F$-module algebra; see 
Section \refeq{sec:twisting} for more details.
Associativity of $\star$ is a consequence of
the cocycle condition on $\cF$, provided $\cM$ is associative.

There is also a more general notion of twisting where $\F$ is not a cocycle,
which relates $U_q(\mg)$ to $U(\mg)$ 
(or more precisely a related quasi--Hopf algebra. The relation between 
$U(\mg)$ and $U_q(\mg)$ will be discussed in more
detail in Section \ref{sec:drinfeld_twist}). 
A fundamental theorem by Drinfeld  (Proposition 3.16
in Ref. \cite{Drinfeld:quasi}) states that there exists an algebra 
isomorphism
\beq
\varphi: U_q(\mg) \rightarrow U(\mg)[[h]]
\label{phi-0}
\eeq
(considering $q=e^h$ as formal for the moment),
and a twist $\cF$ which relates the undeformed coproduct on $U(\mg)$
to the $q$-deformed one by the formula \refeq{defd}. 
This twist $\F$ (which is not a cocycle) 
can now be used to twist the quasiassociative 
$U(\mg)$ -module algebra \refeq{quasiass-ope}, and turns it into the
associative $U_q(\mg)$ -module algebra \refeq{ass-ope}.
Indeed, from \refeq{defd} it follows that $\cF$ relates the
undeformed Clebsch--Gordan coefficients to the $q$-deformed ones:
\beq
\left[\begin{array}{ccc} I&J&K\\ i&j&k' \end{array} \right] (g^{(K)})^{k' k}= 
    \left[\begin{array}{ccc} I&J&K\\ i'&j'&k' \end{array} \right]_q 
     (g_q^{(K)})^{k' k}\;\; \pi^{i'}_i(\cF^{(1)}) \pi^{j'}_j(\cF^{(2)}).
\eeq
Here we have raised indices using $(g_q^{(K)})^{k' k}$, which 
is the $q$--deformed invariant tensor. Both 
$(g_q^{(K)})^{k' k}$ and its underformed counterpart $(g^{(K)})^{k' k}$ 
will sometimes be suppressed, i.e. absorbed in the Clebsches.
It should also be noted that even though the abstract element $\F$ exists 
only for formal $q$, 
the representations of $\F$ on the {\em truncated} tensor product
of ``regular'' \reps $V_\la$ does exist at roots of unity,
because the decomposition into irreps as well as 
the Clebsches are then analytic in $q$.
Hence the twisted multiplication rule \refeq{star-0}
for the generators $Y^I_i$ is precisely \refeq{ass-ope}, which defines
an associative algebra\footnote{there is a subtle point here:
the twist $\cF$ is only defined up to ``gauge invariance''
\refeq{F_gauge}, which amounts to 
the ambiguity of the Clebsch-Gordan coefficients up to a phase.
One can fix this ambiguity by requiring the ``standard'' normalization,
which then yields an associative algebra.}. 
Associativity can be verified either using the
pentagon identity for the $q$-deformed $6j$ symbols
similar as in Chapter \ref{chapter:floops}, or simply by an
explicit construction of 
this algebra in terms of operators on an irreducible representation 
space of $U_q(\mg)$. This is explained in Section \ref{subsec:6j-alg}.

\section{Quantum algebras and symmetries for $D$-branes}
\label{sec:LR-symm}

We return to the problem of describing the quantized $D$-branes on $G$.
To find the algebra of quantized funcitons on $G$ and
its $D$-branes, we shall make an ``educated guess'', and then 
verify the desired properties. 

The quantum spaces under consideration should certainly
admit some appropriate quantum versions of the symmetries
$G_L, G_R$ and $G_V$. 
In technical terms, this suggests that they should be 
module algebras $\cM$ under some quantum groups.
In view of the results in Section \ref{sec:cft}, 
we expect that these symmetry algebras are essentially
$U_q(\mg)$ \cite{Chari:1994pz}, whose representations are known to be 
parallel to those of $\mgh$ of the WZW model \cite{KL}.

Since we are considering matrix groups $G$, 
we further assume that the appropriate 
(module) algebra $\cM$ is generated by 
matrix elements $M^i_j$ with indices $i,j$ in the defining \rep $V_N$ of $G$,
subject to some commutation relations and constraints. 
With hindsight, we claim that these relations are given by
the so-called {\em reflection equation} (RE) 
\cite{Mezincescu:1991ui,Kulish:1992qb}, 
which in short-hand notation reads
\beq \fbox{$
R_{21} M_1 R_{12} M_2 = M_2 R_{21} M_1 R_{12}. $}
\label{re}
\eeq
Here $R$ is the $\cR$ matrix  of 
$U_q(\mg)$ in the defining representation. 
Displaying the indices explicitly, this means 
\beq
({\rm RE })^{\ i\ k}_{\ j\ l}: \qquad
\matR kaib\ M^b_c\ \matR cjad\ M^d_l = M^k_a\ \matR abic\ M^c_d\ \matR djbl.
\label{re-i}
\eeq
The indices $\{i,j\}$, $\{k,l\}$ correspond to the first (1) and the
second (2) vectors space in \refeq{re}. 
Because $\cM$ should describe a quantized group
manifold $G$, we need to impose constraints which ensure that the
branes are indeed embedded in such a quantum group manifold. 
In the case $G=SU(N)$, these are
$\det_q(M)=1$ where $\det_q$ is the so-called quantum 
determinant \refeq{qdet}, and
suitable reality conditions imposed on the generators
$M^i_j$. Both will be discussed below.

One can also think of
\refeq{re} as being analogs of the boundary condition \refeq{twinbc}.  
As we shall see, RE has indeed similar symmetry properties. 
This is the subject of the the following subsection.
We should mention here that RE 
appeared more then 10 years ago in the context of the boundary integrable
models, and is sometimes called boundary YBE \cite{Mezincescu:1991ui}.

\subsection{Quantum symmetries of RE}

There are 2 equivalent ways to look at the symmetry of $\cM$, 
either as a (right) comodule algebra or as a (left) module algebra
under a suitable quantum group.
We recall the concepts of Section \ref{subsec:cov-alg} here.
The symmetry algebras of $\cM$ are
$\cG_L \tens^\cR \cG_R$ and $U_q(\mg_L \times \mg_R)_\cR$, 
which are dual Hopf algebras. 

\paragraph{1) $\cM$ as comodule algebra.}
The easiest way to find the full symmetry
is to postulate that the matrix $M$ transforms as
\beq\label{st-coaction}\fbox{$
M^i_j \to (s^{-1}M t)^i_j $}
\eeq
where $s^i_j$ and $t^i_j$ generate the algebras $\cG_L$ and 
$\cG_R$ respectively, which both coincide with the 
well--known quantum groups \cite{Faddeev:1990ih} $Fun_q(G)$ as described in 
Section
\ref{subsec:dualHA}. 
For example, $s_2s_1 R =R s_1s_2 ,\ t_2t_1 R=R  t_1t_2.
$\footnote{$R$ with suppressed indices means $R_{12}$.}
In  \refeq{st-coaction} matrix multiplication is understood.
It is easy to see that this is a symmetry of the RE if we impose that 
(the matrix elements of) $s$ and $t$ commute with $M$, and additionally satisfy
$ s_2 t_1  R= R t_1s_2$. Notice that \refeq{st-coaction} is a quantum analog
of the action of the classical isometry group $G_L\times G_R$ on classical
group element $g$ as in Section (\ref{sec:c-branes}). 

Symmetries become powerful only because they have a group-like
structure,
i.e. they can be iterated. In mathematical language this means that 
we have a Hopf algebra denoted by $\GcG$:
\beqa\label{gg-algebra}
&&s_2s_1\ R =R\  s_1s_2 , \quad  t_2t_1\  R=R\  t_1t_2 , \quad s_2 t_1\ 
R= R\
t_1s_2 \\ 
&&\D s=s\otimes s,\quad \Delta t=t\otimes t,\\
&&S(s)=s^{-1},\quad \ep(s^i_j)=\d^i_j,\quad S(t)=t^{-1},\quad 
\ep(t^i_j)=\d^i_j
\eeqa
(here $S$ is the antipode, and $\ep$ the counit).
The inverse matrices $s^{-1}$ and $t^{-1}$ are defined after 
suitable further (determinant-like) constraints on $s$ and $t$ are
imposed, as explained in Section \ref{subsec:dualHA}, \cite{Faddeev:1990ih}.
Formally, $\cM$ is then a right $\cG_L \tens^\cR \cG_R$ - comodule algebra;
see Section \ref{subsec:cov-alg} for further details.

Furthermore, $\cG_L \tens^\cR \cG_R$ can be mapped to a 
vector Hopf algebra $\cG_V$ with generators $r$, by 
$s^i_j\otimes 1\to r^i_j$ and $1\otimes t^i_j\to r^i_j$ (thus basically 
identifying $s=t=r$ on the rhs). 
The (co)action of $\cG_V$ on the $M$'s is then
\beq\label{v-coaction}
M^i_j \to (r^{-1}M r)^i_j.
\eeq

\paragraph{2) $\cM$ as module algebra.}

Equivalently, we can consider $\cM$ as a left module algebra under
$U_q(\mg_L \times \mg_R)_\cR$, which is dual Hopf algebra to
$\cG_L \tens^\cR \cG_R$. As an algebra, it is the usual tensor product
$U_q(\mg_L) \tens U_q(\mg_R)$, generated by 2 copies of $U_q(\mg)$.
The dual evaluation $\langle, \rangle$ between $\cG_L \tens^\cR \cG_R$
and $U_q(\mg_L \times \mg_R)_\cR$ is defined componentwise,
using the standard dualities of $\cG_{L,R}$ with $U_q(\mg_{L,R})$.
The Hopf structure turns out to be
\beqa
\Delta_{\cR}: U_q^L \otimes U_q^R &\rightarrow&
    (U_q^L \otimes U_q^R) \otimes 
     (U_q^L \otimes U_q^R), \non
   u^L \otimes u^R &\mapsto& 
    {\cF}(u^L_1 \otimes u^R_1) \otimes (u^L_2 \otimes u^R_2) {\cF}^{-1}
\label{coprod-so(4)-new}
\eeqa
with ${\cF} = 1\tens \cR{}^{-1}\tens 1$.
This is a special case of a Drinfeld twist as discussed in 
Section \ref{subsec:drinfeld_twist}.
The action of $U_q(\mg_L \times \mg_R)_\cR$ on $\cM$ which is dual to 
\refeq{st-coaction} comes out as
\beq
(u_l\tens u_R) \trr M^i_j %= M^k_l \langle S(s^i_k) t^l_j,u_l\tens u_R\rangle
= \pi^i_l(S u_L) M^l_k \pi^k_j(u_R),
\label{UU-action}
\eeq
where $\pi()$ is the defining \rep $V_N$ of $U_q(\mg)$.
This is a symmetry of $\cM$ 
in the usual sense, because the rhs is again an element in $\cM$.
Moreover, there is a Hopf-algebra map 
$u \in U_q(\mg_V) \to \Delta(u) \in U_q(\mg_L \times \mg_R)_\cR$
where $\Delta$ is the usual coproduct. This
defines the vector sub-algebra $U_q(\mg_V)$. 
It induces on $\cM$ the action 
\beq\label{q-adj}
 u \trr M^i_j = \pi^i_k(S u_1) M^k_l \pi^l_j(u_2),
\eeq
which is 
again dual to the coaction \refeq{v-coaction}.
At roots of unity, these dualities are somewhat subtle. We will 
not worry about this, because covariance of
the reflection equation under $U_q(\mg_L \times \mg_R)_\cR$ can also be 
verified directly.

The crucial points in our construction is the existence of 
a vector sub-algebra $U_q(\mg_V)$ of $\gcg$ (or the analogous
notion for the dual $\GcG$). We will see that the
central terms of $\cM$ which characterize its \reps is
invariant only with respect to that $U_q(\mg_V)$ (or $\cG_V$).
This will allow to interpret
these sub-algebras as isometries of the quantum D-branes. 
In other words, the RE imposes very similar
conditions on the symmetries and their breaking as the original BCFT WZW
model described in section \ref{sec:cft} does. 
This is  not the case for other conceivable quantized algebras of
funcitons on $G$, such as  $Fun_q(G)$.

\subsection{Central elements of RE}

Next we discuss some general properties of the algebra defined by
\refeq{re}. We need to find the central elements,
which are expected to characterize its irreps. This problem was
solved in the second paper of \cite{Kulish:1993ep,Kulish:1992qb,Kulish:1993pb}.
The (generic) central elements of the algebra \refeq{re} are 
\beq\label{central2-n}
c_n=\tr_q(M^n)\equiv \tr_{V_N}(M^n\ v) \; \in\cM,
\eeq
where the trace is taken over the 
defining representation $V_N$, and
\beq
v = \pi(q^{-2H_\rho})
\eeq
is a numerical matrix which satisfies $S^2(r)=v^{-1} r v$
for the generator $r$ of $\cG_V$.
These elements $c_n$ are independent for $n = 1, 2,..., rank(G)$.
A proof of centrality can be found e.g. in the book \cite{Majid:1996kd}, 
see also Section \ref{a:cov}.
Here we verify only invariance under \refeq{v-coaction}:
\beqa
c_n\to&& \tr_q(r^{-1} M^n r)=(r^{-1})^i_j (M^n)^j_k r^k_l v^l_i\nn\\
&=&S(r^i_j) (M^n)^j_k v^k_l S^2(r^l_i)=(M^n)^j_k v^k_l S(S(r^l_i)r^i_j)=
(M^n)^j_k v^k_j=c_n
\label{cn-inv}
\eeqa
as required.  As we shall see, these 
$c_n$ for $n=1,...rank(G)-1$
determine the position of the branes on the group manifold: They 
are quantum analogs of the $s_n$ \refeq{sn}.

There should be another central term, which is the 
quantum analog of the ordinary determinant. 
It is known as the quantum determinant, $\det_q(M)$
(which is of course different from the quantum determinant of $Fun_q(G)$). 
While it can be expressed as a polynomial in $c_n$'s ($n=1,..., rank(G)$),
$\det_q(M)$ is invariant under the full chiral symmetry algebra.
Hence we impose the constraint
\beq
1 = \det_q(M).
\label{qdet}
\eeq
For other groups such as $SO(N)$ and $SP(N)$, 
additional constraints (which are also invariant 
under the full chiral quantum algebra) must be imposed. 
These are known and can be 
found in the literature \cite{Schupp:1993mt}, but their explicit form is not
needed for the forthcoming considerations.
Section \ref{a:qdet} contains details about how to calculate 
$\det_q(M)$
and provides some explicit expressions. 

\subsection{Realizations of RE}
\label{subsec:solutions}

In this section we find realizations of the RE \refeq{re}
in terms of other known algebras. 
It can be viewed as an intermediate step towards finding 
representations. We use a technique generating new solutions
out of constant solutions (trivial \reps).
Thus first we consider 
 \beq
R_{21} \Mo_1 R_{12} \Mo_2 = \Mo_2 R_{21} \Mo_1 R_{12}.
\label{re-o}
\eeq
where the entries of the matrices $\Mo$ are c-numbers.
Then one easily checks that
\beq\label{m-sol}
M=L^+ M^{(0)} S(L^-). 
\eeq
solves \refeq{re} if $L^\pm$ respects 
\beqa\label{l-alg}
&&R L^\pm_2L^\pm_1=L^\pm_1 L^\pm_2 R, \quad R L^+_2L^-_1=L^-_1L^+_2 R.
\eeqa
In fact
this is the same calculation as checking $\GcG$ invariance of the RE.
Notice  that $\det_q(M)=\det_q(\Mo)$, due to chiral invariance of the
q-determinant.  Thus we have trade our original problem to
the problem of finding matrices $L^\pm$
respecting \refeq{l-alg}. Of course there is a well-known answer
due to the famous work of Faddeev, Reshetikhin and Takhtajan 
\cite{Faddeev:1990ih}: the
relations \refeq{l-alg} are exactly the same as those of the generators
of $U_q(\mg)$, \refeq{RLL-relations}. 
This implies that there is an algebra map 
\beqa
\cM &&\to \; U_q(\mg), \nn\\
M^i_j &&\mapsto (L^+ M^{(0)} S(L^-))^i_j
\label{M-U-subalgebra}
\eeqa
determined by the constant solution $M^{(0)}$,
which will be very useful below. However, 
this map is not an isomorphism.

From now on, we will concentrate on the case where $\cM$ is realized as
a sub-algebra of $U_q(\mg)$ via \refeq{M-U-subalgebra}. 
The sub-algebra depends of course on 
$\Mo$. We will not discuss the most general $\Mo$ here
(see e.g. \cite{Kulish:1993ep,Kulish:1992qb,Kulish:1993pb}), 
but consider only the most obvious solution, which is a
diagonal matrix. The specific values of the diagonal entries do
not change the algebra generated by the elements of $M$. 
To be explicit, we give the solution for $\mg = sl(2)$ and $\Mo=\diag(1,1)$:
\beq
M=L^+ S(L^-)=\left(\begin{array}{cc} q^{H} & q^{-\half}\la q^{H/2} X^-\\
q^{-\half}\la X^+\ q^{H/2} & ~~~~q^{-H}+q^{-1}\la^2X^+X^- \end{array}\right)
\eeq
where $\la = q-q^{-1}$.
One can verify that $\det_q(M)=1$, according to \refeq{qdet2}.
As we will see, choosing a definite \rep of $U_q(\mg)$ then
corresponds to choosing a brane
configuration, and determines the algebra of functions on the brane.

The other, non-diagonal solutions for
$\Mo$'s presumably also correspond to some branes. We hope to
come back to this subject in a future paper.

\subsection{Covariance}
\label{subsec:cov}

We show in Section \ref{a:cov} that for any solutions
of the form $M=L^+ M^{(0)} S(L^-)$ where $M^{(0)}$ is a constant solution
of RE, the ``vector'' rotations \refeq{q-adj} 
can be realized as quantum adjoint action:
\beq
 u \trr M^i_j = \pi^i_k(S u_1) M^k_l \pi^l_j(u_2)
              = u_1\; M^i_j\; S u_2\;\; \in \cM 
\label{cov}
\eeq
for $u \in \cM$, where $\pi()$ is the defining \rep $V_N$ of $U_q(\mg)$.
Here we consider $\cM \subset U_q(\mg)$, so that 
$\Delta(u) = u_1 \tens u_2$ is defined in $U_q(\mg) \tens U_q(\mg)$;
nevertheless the rhs is in $\cM$. 
The proof for $M^{(0)} = \one$ is very simple and well-known.
This is as it should be in a quantum theory: the action of a symmetry
is implemented by a conjugation in the algebra of operators.
It will be essential later to perform the harmonic analysis on the branes.

\subsection{Reality structure}
\label{sec:star}

An algebra $\cM$ can be considered as a
quantized (algebra of complex-valued functions on a space
only if it is equipped with a $*$-structure, i.e. an anti-linear 
(anti)-involution. 
For classical unitary matrices, the 
condition would be $M^{\dagger} = M^{-1}$.
To find the correct quantum version is a bit tricky here, 
and strictly speaking 
the star given below can only be justified after going to 
representations\footnote{so that certain inverses etc. exist; this
This will be understood implicitly}.
We determine it by requiring that on finite-dimensional representations
of $M = L^+ S L^-$, the $*$ will become the usual matrix adjoint.
In term of the generators of $U_q(\mg)$, this means that 
$(X_i^\pm)^* = X_i^\mp$, $H_i^* = H_i$. 
In the $SU(2)$ case, this leads to
\beq
\left(\begin{array}{cc} a^* & b^* \\ c^* & d^* \end{array}\right) = 
\left(\begin{array}{cc} a^{-1} & -q c a^{-1}\\ -q a^{-1} b\;\; & 
q^2 d + a-q^2 a^{-1}\end{array}\right);
\label{su2-star}
\eeq
$a^{-1}$ indeed exists on the irreps of $\cM$ considered here.
A closed form for this star structure for general $\mg$ 
is given in Section \ref{subsec:reality}.

\section{Representations of $\cM$ and quantum $D$--branes}

$\cM$ should be considered as quanization of the manifold $G$
in the spirit of non-commutative geometry. 
However, we are interested here in the quantization of the 
orbits $\cC(t_\la)$, which are submanifolds of $G$. 
We will now explain how these arise in terms of $\cM$, and then show
that the functions 
on these submanifolds decompose into the same harmonics as
on the classical  $\cC(t_\la)$,
under the action of the vector symmetry 
algebra \refeq{cov}. This correspondence hold only up to some cutoff. 
In other words, the harmonic analysis on the classical 
and quantum $\cC(t_\la)$ turns out to be the same up to the cutoff.

\paragraph{Submanifolds in noncommutative geometry.}

To clarify the situation, we recall the definition of a submanifold
in (non)commutative geometry. A submanifold $V\subset W$ of a classical
manifold $W$ is characterized by an embedding map
\beq
\iota: \quad V \hookrightarrow W
\eeq
which is injective. Dualizing this, one obtains a surjective algebra map
\beq
\iota^*: \quad \cW \to \cV,
\eeq
where $\cV \cong \cW / \cI$ where $\cI = \rm Ker(\iota^*)$ is a 2-sided ideal.
Clearly $\cV$ describes the internal structure of the submanifold,
while $\cI$ describes the embedding, hence the location of $V$ in $W$.
Explicitly, 
\beq
\cI = \{f:W \to \C;\;\; f(V) = 0 \}
\eeq
in the classical case.
For noncommutative spaces, we define subspaces as quotients
of the noncommutative algebra of functions by some 2-sided ideal $\cI$,
which describes the location of the subspaces.

\subsection{Fuzzy $D_\la$}

To identify the branes, 
we should therefore look for algebra maps $\cM \to \cV$
with nontrivial kernel and image. There are obvious candidates for such 
maps, given by the (irreducible) representations
\beq
\cM \to U_q(\mg) \to  Mat(V_\la)
\label{M-reps}
\eeq
of $U_q(\mg)$, extending the map \refeq{M-U-subalgebra}
for $M^{(0)} = \one$.
On such an irreducible representation, $\cM$ becomes 
the matrix algebra $Mat(V_\la)$, and 
the Casimirs $c_n = \tr_q(M^n)$ \refeq{central2-n} of $\cM$ 
take distinct values $c_n(\la)$ which can be calculated. This defines 
the ideal $\cI = \cap_n \langle c_n - c_n(\la)\rangle_\cM$.
Hence every \rep of  $U_q(\mg)$ defines a quantum submanifold.
Here we consider only the realization $M = L^+ SL^-$ \refeq{m-sol}. 
Then the Casimirs $c_n$ are invariant under (vector) rotations
as shown in \refeq{cn-inv}. In view of their form, 
this suggests that the matrix algebra $Mat(V_\la)$
should be considered as 
quantization of (the algebra of functions on) some untwisted $D$--brane,
the position of which is determined by the value $c_n(\la)$.
One should realize, however, that 
there exist other \reps of $\cM$ which are {\em not} \reps 
of $U_q(\mg)$, which may describe different branes. We shall not consider
such possibilities here.

Now we need the \rep theory of $U_q(\mg)$, which is largely understood,
although quite complicated at roots of unity.
Not all \reps of $U_q(\mg)$ define admissible branes. 
For example, they should be $\star$-representations of $\cM$ with 
respect to a suitable star structure; this is because 
we want to describe real manifolds with complex-valued functions. 
We claim that the admissible representations are those
$V_\la$  with $\la\in P_k^+$, because they have the following properties
(see Section \ref{subsec:reps}):
\begin{itemize}
\item
they are unitary, i.e. $*$ -\reps of $\cM$
with respect to the $*$ structure of 
Section \ref{sec:star} (see \cite{UnitaryReps})
\item their quantum-dimension $\dim_q(V_{\la})  = \tr_{V_\la}(q^{2 H_\rho})$
given by \refeq{qdim} is positive 
\item $\la$ corresponds precisely to the integrable 
modules of the affine Lie algebra $\mgh$ which governs the 
CFT.
\end{itemize}
We will show that these irreps of $\cM$ 
describe precisely the {\em stable} D-branes of string theory.
Since the algebra $\cM$ is\footnote{more precisely the semi-simple quotient
of $\cM$, see Section \ref{subsec:harmonics}} 
the direct sum of the corresponding 
representations, 
the whole group manifold is recovered
in the limit $k \rightarrow \infty$ where the branes become dense.
To confirm this interpretation, we will calculate
the position of the branes on the group manifold, and study their
geometry by performing the
harmonic analysis on the branes, i.e. by determining the set of harmonics.
The \reps belonging to the boundary of $P_k^+$ will correspond to the
degenerate branes.

To summarize, we propose that
{\bf the representation \refeq{M-reps} of $\cM$ for $\la \in P_k^+$
is a quantized or ``fuzzy'' D--brane, denoted by $D_\la$}. 
It is an
algebra of maps from $V_{\la}$ to $V_{\la}$ which transforms under 
the quantum adjoint action \refeq{cov} of $U_q(\mg)$.
For ``small'' weights\footnote{see Section \ref{subsec:harmonics}} $\la$, 
this algebra coincides with $Mat(V_{\la})$.
There are some modifications for ``large'' weights
 $\la$  because $q$ is a root of 
unity, which will be discussed in Section \ref{subsec:harmonics}. The reason
is that $Mat(V_{\la})$ then contains unphysical degrees of freedom which 
should be truncated. 
Moreover, we claim that the {\bf $D_\la$ correspond
precisely to the stable $D$-branes on $G$.}

A first support for this claim
is that there is indeed a one--to--one correspondence
between the (untwisted) branes in string theory and these quantum branes
$D_\la$, since both are labeled by $\la \in P_k^+$.
To give a more detailed comparison, we
calculate the traces (\ref{central2-n}), derive a characteristic equation,
and then perform the harmonic analysis on $D_\la$.
Furthermore, the energy \refeq{crdymss} of the branes in string theory 
will be recovered precisely in terms of the quantum dimension.

\subsection{Location of $D_\la$ and values of the Casimirs}
\label{subsec:pos}
 
The values of the Casimirs $c_n$ on  $D_\la$ are  
calculated in Section \ref{a:casimirs}:
\beqa
%c_0 &=& \tr_{V_N}\; (q^{-2 H_\rho}) %=  \tr_{V_N}\; (q^{- 2 H_\rho}) 
%        =\dim_q(V_N), \\
c_1(\la) &=& \tr_{V_N}\; (q^{2 (H_\rho +H_\la)}), \label{c1}\\
c_n(\la) &=& \sum_{\nu \in V_N;\; \la + \nu \in P_k^+} 
   q^{2n((\la + \rho)\cdot \nu - \la_N\cdot\rho)} \; 
   \frac{\dim_q(V_{\la+\nu})}{\dim_q(V_{\la})},\quad n \geq 1.  
\label{cn}
\eeqa
Here $\la_N$ is the highest weight of the defining \rep $V_N$, and 
the sum in \refeq{cn} goes over all $\nu \in V_N$ 
such that $\la + \nu$ lies in $P_k^+$. 
%$c_0$ is $\la$-independent uninteresting number.

The value of $c_1(\la)$ agrees precisely with the corresponding
value \refeq{sn}
of $s_1$ on the classical conjugacy classes $\cC(t_{\la})$.
For $n \geq 2$, the values of $c_n(\la)$ agree only approximately
with $s_n$ on $\cC(t_{\la})$,
more precisely they agree
if $\frac{\dim_q(V_{\la+\nu})}{\dim_q(V_{\la})} \approx 1$,
which holds provided $\la$ is large 
(hence $k$ must be large too).
In particular, this holds for branes which are not ``too close'' to
the unit element. This slight 
discrepancy for small $\la$ is perhaps not too surprising, since the
higher--order Casimirs are defined in terms of non-commutative 
coordinates and are therefore subject to  operator--ordering
ambiguities.

Finally, we show in Section \ref{a:chareq} that
the generators of $\cM$ satisfy the following characteristic 
equation on $D_{\la}$:
\beq
P_\la(M) = \prod_{\nu \in V_N} 
                      (\; M -q^{2 (\la + \rho)\cdot\nu-2 \la_N\cdot\rho}) = 0.
\label{char-poly-Y}
\eeq 
Here the usual matrix multiplication of the $M^i_j$ is understood.
Again, this (almost) matches with the classical version \refeq{charpoly-class}.

Hence we see that the positions and the ``size'' of the branes
essentially agree with the results from string theory. 
In particular, their size shrinks to zero
if $\la$ approaches a corner of $P_k^+$,
as can be seen easily in the $SU(2)$ case \cite{PS:su2}: 
as $\la$ goes from $0$ to $k$,
the branes start at the identity $e$, grow up to the equator,
and then shrink again around $-e$.
We will see that the algebra of functions on $D_\la$ precisely reflects 
this behavior; 
however this is more subtle and will be discussed below.
All of this is fundamentally 
tied to the fact that $q$ is a root of unity.

It is worth emphasizing here that the agreement of the 
values of $c_n$ with their
classical counterparts \refeq{sn} shows 
that the $M$'s are very reasonable variables to describe the branes.

\subsection{Energy of $D_\la$ and quantum dimension}

Following \cite{PS:su2}, the $M^i_j$'s can be thought of as some matrices 
(as in Myers model \cite{Myers:1999ps}) 
out of which one can form an action which is invariant under
the relevant quantum groups. The action should have the structure 
$S=\tr_q(1+...)$, where dots represent some expressions in the $M$'s.
The point of \cite{PS:su2} was that  for some
equations of motion, the "dots"- terms vanish
on classical configurations. 
We postulate that the equations of motion for $M$
are given by RE \refeq{re}. If so, then their energy is equal to
\beq\label{energy}
E=\tr_q(1).
\eeq
This energy is not just a constant as might be suggested by
the notation, but it depends on the representation $V_\la$ of the 
algebra, where it becomes the quantum dimension \refeq{qdim}
\beq
\dim_q(V_\la) = \tr_{q}(1) = \tr_{V_\la}(q^{2 H_\rho}) = 
   \prod_{\a > 0} \frac{\sin(\pi\frac{\a\cdot(\la + \rho))}{k + \gcheck})}
                     {\sin(\pi\frac{\a\cdot\rho}{k + \gcheck})} = E_\la.
\label{qdim-2}
\eeq
The equality follows from Weyl character formula.
This is indeed the value of the energy of the corresponding
$D$-brane in the BCFT description, \refeq{crdymss}.

\subsection{The space of harmonics on $D_\la$.}
\label{subsec:harmonics}

As discussed in Section \ref{sec:c-branes}, 
we must finally match the space of functions
or harmonics on $D_\la$ with the ones on $\cC(t_{\la})$,
up to some cutoff. Using covariance \refeq{cov}, 
this amounts to calculating the decomposition of 
the \rep $Mat(V_\la)$ of $\cM$ 
under the quantum adjoint action of $U_q(\mg)$ \refeq{q-adj} 
for $\la \in P_k^+$.
However, recall from Section \ref{subsec:reps} that the full tensor product 
$Mat(V_\la) =  V_{\la} \tens V_{\la}^*$
is problematic since $q$ is a root of unity.
To simplify the analysis, we
assume first that $\la$ is not too 
large\footnote{roughly speaking if $\la = \sum n_i \La_i$, 
then $\sum_i n_i < \frac 12 (k+\gcheck)$.}, so that 
this tensor product is completely reducible. Then
$D_\la$ coincides with the matrix algebra acting
on $V_\la$,
\beq
D_\la \cong Mat(V_\la) =  V_{\la} \tens V_{\la}^*
 \cong \oplus_{\mu} N_{\la \la^+}^{\mu} \; V_{\mu},
\label{little-rich}
\eeq
where $N_{\la \la^+}^{\mu}$ are the usual fusion rules of $\mg$
which can be calculated explicitly using formula \refeq{racah-speiser}.
Here $\la^+$ is the conjugate weight to $\la$, so that 
$V_{\la}^* \cong V_{\la^+}$.
This has a simple geometrical meaning if 
$\mu$  is small enough (smaller than all {\em nonzero}
Dynkin labels of $\la$, roughly speaking; 
see Section \ref{subsec:harmonic-brane} for details):
 then
\beq
N_{\la \la^+}^{\mu} =  mult_{\mu^+}^{(K_{\la})}
\label{Dla-decomp}
\eeq
where $K_{\la}\subset G$ is the stabilizer group\footnote{which acts
by the (co)adjoint action on weights} of $\la$,
and $mult_{\mu^+}^{(K_{\la})}$ is the dimension of the subspace of
$V_\mu^*$ which is invariant under $K_{\la}$.
This is proved in Section \ref{subsec:harmonic-brane}.
Note in particular that the mode structure (for small $\mu$)
does not depend on the particular value of $\la$, only 
on its stabilizer $K_{\la}$. 
Comparing this with the decomposition \refeq{FCt-decomp} 
of $\cF(\cC(t_\la))$, we see that indeed
\beq \fbox{$
D_\la \cong \cF(\cC(t_{\la}'))  $}
\label{D-C-match}
\eeq
up to some cutoff in $\mu$, 
where $t_{\la}' = \exp(2 \pi i \frac{H_\la}{k + \gcheck})$.
This differs slightly from  \refeq{stable-branes}, by a shift
$\la \rightarrow \la + \rho$. It implies that 
degenerate branes do occur in the our quantum algebraic description, 
because $\la$ may be invariant under a nontrivial subgroup
$K_{\la} \neq T$.
These degenerate branes have smaller dimensions than 
the regular ones. An example for this is fuzzy 
$\C P^N$, which will be discussed in some detail below.
We want to emphasise that it is only the result \refeq{D-C-match}
which allows to identify these quantized spaces with classical ones.

Here we differ from \cite{Felder:1999ka}
who identify only regular $D$--branes in the CFT description,
arguing that $\la + \rho$ is always regular.
This is due to a particular limiting procedure for 
$k \rightarrow \infty$ which was chosen in \cite{Felder:1999ka}. 
We assume $k$ to be large but finite, 
and find that degenerate branes do occur. This is in agreement with
the CFT description of harmonics on $D_\la$,
as will be discussed below.
Also, note that \refeq{D-C-match}
reconciles the results \refeq{c1}, \refeq{sn} on the position of the branes
with their mode structure as found in CFT.

Now we consider the general case where 
the tensor product $Mat(V_\la) = V_{\la} \tens V_{\la}^*$
may not be completely reducible. Then
$Mat(V_\la) = V_{\la} \tens V_{\la}^*$
contains non-classical \reps with vanishing quantum dimension,
which have no obvious interpretation.
However, there is a well--known remedy: one can replace the 
full tensor product by the so-called ``truncated tensor product'' 
\cite{Fuchs:1992nq},
which amounts to discarding\footnote{note that the calculation
of the Casimirs in Section \ref{subsec:pos} is still valid, because
$V_\la$ is always an irrep} the \reps with $dim_q = 0$.
This gives a decomposition into irreps
\beq
D_\la  \cong V_{\la} \obar{\tens} V_{\la}^*
 \cong \oplus_{\mu \in P_k^+} \obar{N}_{\la \la^+}^{\mu} \; V_{\mu}
\label{little-rich-trunc}
\eeq
involving only modules $V_{\mu}$ of positive quantum dimension. 
These 
$\obar{N}_{\la \la^+}^{\mu}$
are known to coincide with the fusion rules for integrable modules
of the affine Lie algebra $\mgh$ at level $k$, and
can be calculated explicitly.
These fusion rules in turn coincide (see e.g. \cite{Felder:1999ka}) with
the multiplicities of harmonics on the D-branes
in the CFT description, 
i.e. the primary (boundary) fields. 

We conclude that the structure of
harmonics on $D_\la$, \refeq{little-rich-trunc}
is in complete agreement with the CFT results. 
Moreover, it is known (see also \cite{Felder:1999ka})
that the  structure constants of the corresponding boundary operators
are essentially given by the $6j$ symbols of $U_q(\mg)$, which in turn are 
precisely the structure constants of the algebra of functions 
on $D_\la$, as explained in the next subsection.
Therefore our quantum algebraic description not only 
reproduces the correct set of boundary fields, but also 
essentially captures their algebra in (B)CFT.

Finally, it is interesting to note 
that branes $D_\la$ which are  ``almost'' degenerate
(i.e. for $\la$  near some boundary of $P_k^+$) have only few
modes $\mu$  in some directions\footnote{this is just the condition on
$\mu$ discussed before \refeq{Dla-decomp}}
and should therefore be interpreted
as degenerated branes with ``thin'', but finite walls. 
They interpolate between branes of different dimensions.

\subsection{$6j$ symbols and the algebra on $D_\la$}
\label{subsec:6j-alg}

Finally we show that the structure constants of the algebra of functions 
on $D_\la$ coincide precisely with those of the (twisted) algebra of
boundary operators \refeq{ass-ope} on the branes, which are
given by the $6j$ symbols of $U_q(\mg)$.
This is done using the explicit realization of $D_\la$ as matrix algebra
$Mat(V_\la)$, or more precisely 
$D_\la  \cong V_{\la} \obar{\tens} V_{\la}^*$ \refeq{little-rich-trunc}.
For simplicity, we assume that $\mg = su(2)$ here, but all
arguments generalize to the general case.

Let us denote the branes on $SU(2) \cong S^3$ as $\S_{\la} = D_\la$,
being 2-spheres.
The decomposition (\ref{little-rich}) of $\S_{\la}$ is then explicitly
\beq
 \S_{\la} = D_\la = (1) \oplus (3) \oplus (5) \oplus ... \oplus (2N+1).
\label{A_decomp-0}
\eeq
Here $(k)$ denotes the $k$-dimensional \rep of $su(2)$.

\begin{figure}[htpb]
\begin{center}
\epsfxsize=5in
%  \vspace{-1in} 
   \hspace{0.1in}
   \epsfbox{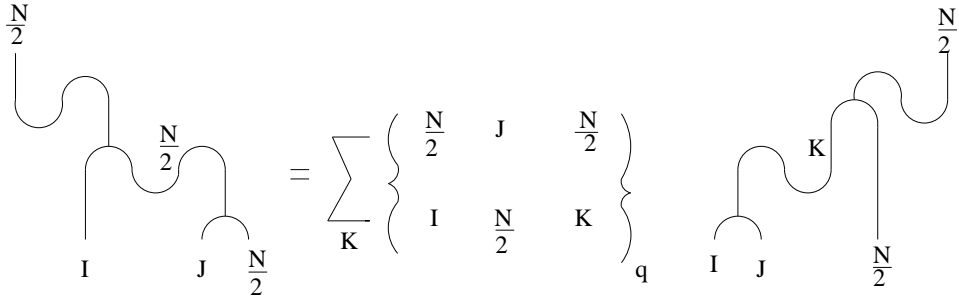}
\end{center}
%   \vspace{-2.2in}
 \caption{Derivation of the algebra (\ref{ass-ope})}
\label{fig:6jcalc}
\end{figure}

Let $Y^I_i \in Mat(V_\la)$ be a ``function'' on $\S_{\la}$, which
transforms in the spin $I$ 
representation\footnote{for higher groups, it will carry an additional
degeneracy label} of $U_q(su(2))$. 
Denote with $\pi(Y^I_i)^r_s$ the matrix which represents $Y^I_i$ 
on $V_\la$, in a weight basis of $V_\la$.
Because it transforms in the adjoint, 
it must be proportional to the Clebsch--Gordan coefficient
of the decomposition $(2I+1) \tens V_\la \rightarrow V_\la$,
after lowering the index $r$.
Hence in a suitable normalization of the basis $Y^I_i$, we can 
write\footnote{recall the convention of Section \ref{sec:cft} 
that the weight $\la$ corresponds to $\frac N2$}.
\beq
\pi(Y^I_i)^r_s =  
 (g_q^{(N/2)})^{r r'} \left[\begin{array}{ccc} N/2&I&N/2\\ r'&i&s \end{array} 
\right]_q
 = \left[\begin{array}{ccc}I&N/2&N/2\\i&s&r'\end{array} \right]_q 
 (g_q^{(N/2)})^{r' r}.
\eeq
Therefore the matrix representing the operator 
$Y^I_i Y^J_j$ is given by 
\beqa
 \pi(Y^I_i)^r_s \!\!\!\!\!\!\!\!\!\!\!\! && \pi(Y^J_j)^s_t =
  (g_q^{(N/2)})^{r r'} \left[\begin{array}{ccc} N/2&I&N/2\\ r'&i&s 
       \end{array} \right]_q
  (g_q^{(N/2)})^{s s'} \left[\begin{array}{ccc} N/2&J&N/2\\ s'&j&t 
       \end{array} \right]_q  \nn\\
 &=& \sum_K \left\{\begin{array}{ccc} N/2&J&N/2\\I&N/2&K \end{array}\right\}_q
     \left[\begin{array}{ccc} I&J&K\\ i&j&k' \end{array} \right]_q 
      (g_q^{(K)})^{k' k}
   \left[\begin{array}{ccc} K&N/2&N/2\\ k&t&r'\end{array} \right]_q 
       (g_q^{(N/2)})^{r' r}   \nn\\
 &=& \sum_K \left\{\begin{array}{ccc} I&J&K\\ N/2&N/2&N/2\end{array}\right\}_q
    \left[\begin{array}{ccc} I&J&K\\ i&j&k' \end{array} \right]_q 
                (g_q^{(K)})^{k' k}\; \pi(Y^K_k)^r_t.  \nn\\
\eeqa
Here we used the identity
\beq
\left\{\begin{array}{ccc} N/2&J&N/2\\I&N/2&K \end{array}\right\}_q
 = \left\{\begin{array}{ccc} I&J&K\\ N/2&N/2&N/2\end{array}\right\}_q,
\eeq
which is proved in \cite{Kirillov:1991ec}. This calculation 
is represented graphically in Figure \ref{fig:6jcalc}, which shows that it 
simply boils down to the definition of the $6j$--symbols.
Associativity could also be verified using the 
$q$-Biedenharn Elliott identity \cite{Kirillov:1991ec}.
Therefore the algebra of $\S_{\la}$ is precisely 
(\ref{ass-ope}), which is a twist of the algebra (\ref{quasiass-ope}). 

\section{Examples}
\subsection{Fuzzy $\C P^{N-1}_q$}

Particularly interesting examples of degenerate conjugacy classes are  
the complex projective spaces $\C P^{N-1}$.
We shall demonstrate the scope of our general results by extracting some 
explicit formulae for this special case.
This gives a $q$-deformation of the fuzzy $\C P^{N-1}$ discussed 
in \cite{Alexanian:2001qj,Grosse:1999ci}.

We first give a more explicit description of branes on $SU(N)$. 
Let us parameterize the matrix $M$ ($= L^+ SL^-$ acting on 
$V_\la$) as
\beq
M = \xi_\a \la^\a  = \sum_a \xi_a \la^a + \xi_0 \la^0
\label{M-param}
\eeq
where $\a = (0,a)$ and $a = 1,2, ..., N^2 -1$.
The $\xi_\a$ will be generators of a non-commutative algebra,
and $\la^a = (\la^a)^\a{}_{\dot{\b}}$ for $a = 1, 2, .., N^2-1$
are $q$-deformed Gell-Mann matrices;
we set $\la^0 \equiv {\bf 1}$.
Using covariance \refeq{cov}, $M$ transforms as
\beq
M \rightarrow u \trr M = \xi_\a\; \pi(Su_1) \la^\a \pi(u_2)
 = (u_1 \xi_\a Su_2)\; \la^a.
\eeq
We will hence choose a basis such that
\beq
\xi_a \rightarrow u\trr \xi_a = u_1 \xi_a Su_2 = \xi_b\; (\pi_{(ad)}(u))^a_b
\eeq 
transforms under the adjoint of $U_q(su(N)_V)$, and 
\beq
\pi(Su_1) \la^a \pi(u_2) = 
    (\pi_{(ad)}(u))^a_b\; \la^b
\eeq
can be viewed as the right adjoint action of $U_q(su(N)_V)$ on 
$Mat(N,\C)$. 
Therefore the $\la^a$ are the intertwiners
$(N) \tens (\obar{N}) \to (N^2-1)$ under the right action of $U_q(su(N)_V)$.
They satisfy the relation
\beq
\la^a \la^b = \frac1{\dim_q(V_N)} g^{ab} + 
                 (d^{ab}{}_c  +  f^{ab}{}_c) \la^c
\label{gell-mann-id}
\eeq
where $g^{ab}$, $d^{ab}{}_c$ and  $f^{ab}{}_c$ 
are (right) invariant tensors in a suitable
normalization, and $\tr_q(\la^a) = 0$ (for $a\neq 0$).
We can now express the Casimirs $c_n$ \refeq{cn} in terms of the
new generators:
\beqa
c_1 &=& \tr_q(M) = \xi_0\; \dim_q(V_N), \label{brane-pos}\\
c_2 &=& g^{ab}\; \xi_a \xi_b + \xi_0^2\; \dim_q(V_N), \label{brane-radius}\\
\eeqa
etc, which are numbers on each $D_\la$. 
An immediate consequence of \refeq{brane-pos} is
\beq
[\xi_0, \xi_a] = 0
\label{brane-CR-1}
\eeq
for all $a$. One can show furthermore
that the reflection equation \refeq{re}, which is equivalent 
to the statement that the ($q$-)antisymmetric part of $MM$ vanishes, 
implies that
\beq
f^{ab}{}_{c}\; \xi_a \xi_b = \a \; \xi_0 \xi_c.
\label{brane-CR}
\eeq
On a given brane $D_\la$, $\xi_0$ is a number determined by \refeq{brane-pos}, 
while $\a$ is a (universal) constant 
which can be determined explicitly, as indicated below.

\refeq{brane-CR-1} and \refeq{brane-CR} hold for all branes $D_\la$.
Now consider $\C P^{N-1} \cong SU(N)/ U(N-1)$, which
is the conjugacy class through $\la = n \La_1$
(or equivalently $\la = n \La_N$) where $\La_i$ 
are the fundamental weights; 
indeed, the stabilizer group for $n\La_1$ is $U(N-1)$.
The quantization of $\C P^{N-1}$ is therefore the brane $D_{\la}$.
It is characterized by a further relation among the generators $\xi_a$,
which has the form
\beq
d^{ab}{}_{c}\; \xi_a \xi_b = \b_n\; \xi_c
\label{CPN-rel}
\eeq
where the number $\b_n$  can be determined explicitly as indicated below.
For $q=1$, these relations reduce to the ones given in \cite{Alexanian:2001qj}.
\refeq{CPN-rel} can be quickly derived using the results in 
Section \ref{subsec:harmonics}:
It is easy to see using \refeq{racah-speiser} that 
\beq
D_{n \La_1} \cong \oplus_n (n,0,..., 0,n)
\eeq
up to some cutoff,
where $(k_1, ..., k_N)$ denotes the highest-weight \rep with Dynkin labels
$k_1, ..., k_N$. 
The important point is that all multiplicities are one. It follows that 
the function $d^{ab}{}_{c} \xi_a \xi_b$ on $D_{n \La_1}$ must 
be proportional to $\xi_c$, because it transforms as $(1,0,..., 0,1)$
(which is the adjoint). Hence \refeq{CPN-rel} holds.

The constant $\a$ in \refeq{brane-CR} can be calculated
either by working out RE explicitly, or 
by specializing \refeq{brane-CR} for $D_{\La_1}$. 
We shall only indicate this here:
On $D_{\La_1}$, $\xi_a = c \la_a$ for some $c \in \C$. 
Plugging this into \refeq{brane-CR}, one finds
$c\, f^{ab}{}_{c}\; \la_a \la_b = \a \;\xi_0 \la_c$,
and  $c^2\; g^{ab}\; \la_a \la_b + \xi_0^2\; \dim_q(V_N) = c_2$. 
Calculating $\xi_0$ and the  Casimirs  explicitly on $D_{\La_1}$,
one obtains $\a$ which vanishes as $q\rightarrow 1$. 
Similarly using the explicit value of $c_3$ given in Section \ref{subsec:pos},
one can also determine $\b_n$.
Alternatively, they be calculated using creation - and
annihilation operator techniques of \cite{Grosse:1996ar}, \cite{qFSI,AspqFS}.

In any case, we recover the relations of fuzzy $\C P^{N-1}$
as given in \cite{Alexanian:2001qj} in the limit $q \to 1$. As an algebra, 
it is in fact identical to it,
as long as $k$ is sufficiently large.

\subsection{$G=SU(2)$ model}
\label{subsec:su2}

\begin{figure}[htpb]
\begin{center}
\epsfxsize=3in
%  \vspace{-1in} 
   \hspace{0.1in}
   \epsfbox{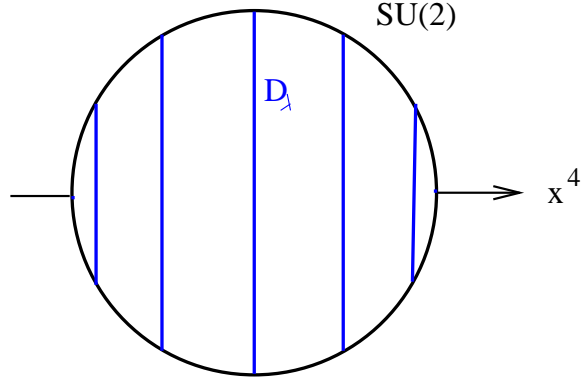}
\end{center}
 \caption{Position of branes on $SU(2)$}
\label{S3branes-eps}
\end{figure}

In this section we shall show how one can recover the results 
of \cite{PS:su2} from the 
general formalism we discussed so far.
The solution to RE given by $L^\pm$ operators and $\Mo=\diag(1,1)$ is
\beq
M=L^+\Mo S(L^-)=\left(\barr{cc}{q^{H}}&{q^{-\half}\la q^{H/2} X^-}\\
{q^{-\half}\la X^+\ q^{H/2}}& {~~~~q^{-H}+q^{-1}\la^2X^+X^- }\earr\right)
\eeq
Let us parameterize the $M$ matrix as
\beq\label{m-para}
M =\left(\barr{cc} 
M_4- i M_0& -i q^{-3/2}\sqrt{[2]}M_+\\ iq^{-1/2}\sqrt{[2]}M_-& M_4+iq^{-2}M_0 
\earr\right)
\eeq
as in \refeq{M-param}, then RE is equivalent to 
\beqa\label{eq-m}
[M_4, M_l] = 0,\quad
\ep^{ij}_l \; M_i M_j = i (q-q^{-1}) M_4 M_l 
\label{su2-branes}
\eeqa
which will be studied in great detail in Section ... .
In order to calculate the central terms 
we need 
\beq
v=\pi(q^{-2H_\rho}) = \pi(q^{-H}) =\diag(q^{-1},q)
\eeq
so that using \refeq{central2-n},\refeq{qdet2}
\beqa
c_1&=&\tr_q(M)= [2]\,M_4\\
c_2 &=&\tr_q(M^2)= [2]\;(M_4^2 - q^{-2} g^{ij} M_i M_j)  \\
\det_q(M)&=&M_4^2+M_0^2-q^{-1}M_+M_- -q\,M_- M_+
 =  M_4^2 +g^{ij} M_i M_j.
\eeqa
Only $\det_q(M)$ is invariant under $U_q(su(2)_L \times su(2)_R)_\cR$.
The explicit value of $M_4=c_1/[2]$ is obtained from
$$
M_4=\frac1{[2]}(q^{H-1}+q^{-(H-1)}+ \la^2 X^+X^-)
$$
which is proportional to the 
standard Casimir of $U_q(su(2))$. 
On the n-th brane $D_n$,  $H$ takes the value $-n$ on the
lowest weight vector, thus
\beq\fbox{$
M_4=\cos(\frac{(n+1)\pi}{k+2})/\cos(\frac{\pi}{k+2}) $}
\eeq
for $n=0,1,.., k$. This leads to the pattern of branes as 
shown in Figure \ref{S3branes-eps}.
If the square of the radius of the quantum $S^3$ is chosen to be $\det_q(M)=k$ 
(which is the value given by the supergravity solution for the background),
$g^{ij} M_i M_j$
leads to the correct formulae for the square of the radius of the n-th branes.

\section{Summary}

We proposed in this chapter a simple and compact description 
of all (untwisted) D-branes on group manifolds $G$ based on
the reflection equation RE. The model 
can be viewed as a finite matrix model in the spirit of the
non-abelian DBI model of D0-branes \cite{Myers:1999ps},
but contrary to the latter it yields results well beyond 
the $1/k$ approximation. 
In fact,  the model properly describes all branes on the group manifold
regardless of their positions. 
This covers an astonishing wealth of data on the configurations
and properties of  branes such as their positions and spaces 
of functions, which are shown to be in very good 
agreement with the CFT data. 
It also shows that $M$ is a very reasonable
variable to describe the branes. 
Our construction also sheds light on the fact that the energies
of these branes are given by so-called quantum dimensions. 

The branes are uniquely given by certain ``canonical'' irreps of the RE
algebra, and their world-volume can be interpreted as quantum manifolds. 
The characteristic feature of our construction is
the covariance of RE under
a quantum analog of the group of isometries $G_L \times G_R$ of $G$.
A given brane configuration breaks it 
to the diagonal (quantum) $\cG_V$, an analog 
of the classical vector symmetry $G_V$.

It should be clear to the reader, however, that 
the present picture does not cover all aspects of branes physics
on group manifolds. For example, we
did not study all \reps of RE, only the most obvious ones
which are induced by the algebra map $RE \to U_q(\mg)$. 
There exist other \reps of RE, some of which can be investigated 
using the  technique in Section \ref{subsec:solutions},
some of which may be entirely different. One may hope that all of 
the known $D$-branes on groups, including those not discussed here such as 
twisted branes
or ``type B''-branes, can be described in this way.
This is an interesting open problem, in particular
in view of the recent progress \cite{Alekseev:2002rj}
made on the string-theoretic side of twisted branes.
Moreover, we did not touch here the dynamical aspects of $D$-branes, 
such as their excitations and interactions. For this it may be necessary to
extend the algebraic content  
presented here, and the well-developed theory of quantum groups
may become very useful.

These considerations have a large number of possible physical 
applications, independent of string theory. 
The general construction of quantized branes
presented here provides a variety of specify examples of 
finite (``fuzzy'') quantum spaces, such as $\C P^N_q$. They may serve
as useful testing grounds for noncommutative field theories,
which are completely finite on these spaces, due to the finite number 
of degrees of freedom.
This is the subject of the remaining chapters.

\section{Technical complements to Chapter 1}
\label{sec:technical1}

\subsection{Reality structure}
\label{subsec:reality}

In term of the generators of $U_q(\mg)$, the star is defined as
$(X_i^\pm)^* = X_i^\mp$, $H_i^* = H_i$. Using \cite{KhoroshkinTol}
\beq
\cR^{*\tens*} = \cR^{-1}_{21}
\label{R-real}
\eeq
which holds for $|q|=1$, it follows that
\beq
M^{\dagger} := M^{T *} = SL^+ SL^- M^{-1} L^+ L^-.
\label{M-real}
\eeq
The rhs is indeed in $\cM$, because the $L$ - matrices
can be expressed on irreps in terms of the $M$ - matrices\footnote{this holds 
only on completely reducible representations}.
This can be cast into a more convenient form using 
the generator of the ``longest Weyl reflection'' 
$\om \in U_q(\mg)$ \cite{Kirillov:1990vb} 
(more  precisely in an extension thereof), which satisfies
\beq
\Delta(\om) = \cR^{-1} \om\tens \om =  \om \tens \om \cR^{-1}_{21}.
\label{del_om_left}
\eeq
Furthermore, it implements the automorphism
$S\theta \g$ of $U_q(\mg)$ as an inner automorphism:
\beqa
\theta(X_i^\pm) &=& X_i^\mp, \; \theta (H_i) = H_i, \nn\\
S\theta\g(u) &=& \om^{-1} u \om 
\eeqa
for any $u \in U_q(\mg)$.
Hence the star can also be written as
\beq
M^{\dagger} = \pi(\om^{-1}) \om^{-1} M^{-1} \om \pi(\om) =
              \pi(\om^{-1}) S\theta\g(M^{-1}) \pi(\om) 
\eeq
where $\pi$ is the defining \rep $V_N$.
This form is useful to verify the consistency of the star
with the relations \refeq{re} and \refeq{qdet}. 
The classical limit $q \rightarrow 1$ of this star structure
is correct, because then
$\pi(\om^{-1}) S\theta\g(M) \pi(\om) \rightarrow M$. 
In the $SU(2)$ case, one recovers the star given in \refeq{su2-star}.

\subsection{The harmonics on the branes}
\label{subsec:harmonic-brane}

\paragraph{The modes on $\cC(t)$ \refeq{FCt-decomp}.}

Consider the map
\beqa
G/K_t &&\rightarrow \cC(t), \nn\\
g K_t  &&\mapsto  g t g^{-1}  \nn
\eeqa
which is clearly well--defined and bijective.
It is also compatible with the group actions, in the sense that 
the adjoint action of $G$ on $\cC(t)$ translates into the {\em left} action
on $G/K_t$. Hence we want to decompose functions on 
$G/K_t$ under the left action of $G$. 

Functions on 
$G/K_t$ can be considered as functions on $G$ which are invariant under the 
{\em right} action of $K_t$, and this correspondence is one-to-one
(because this action is free). Now
the Peter-Weyl theorem states that the 
space of functions on $G$ is isomorphic as a bimodule to 
\beq
\cF(G) \cong  \bigoplus_{\la \in P^+} \; V_{\la} \tens V_{\la}^*.
\label{FG-decomp}
\eeq
Here $\la$ runs over all dominant integral weights, and
$V_{\la}$ is the corresponding highest-weight module. 
Let $mult_{\la^+}^{(K_t)}$ be the dimension of the 
subspace of $V_{\la}^* \equiv V_{\la^+}$ which is
invariant under the action of $K_t$. Then
\beq
\cF(\cC(t)) \cong  \bigoplus_{\la \in P} \; mult_{\la^+}^{(K_t)}\; V_{\la}
\eeq
follows.

\paragraph{The modes on $D_{\la}$ and proof of
\refeq{Dla-decomp}.}

We are looking for the Littlewood--Richardson 
coefficients $N_{\la \la^+}^{\mu}$ in the decomposition
\beq
V_{\la} \tens V_{\la}^* \cong \oplus_{\mu} N_{\la \la^+}^{\mu} \; V_{\mu}
\eeq
of $\mg$ - modules.
Now we use $N_{\la \la^+}^{\mu} = N_{\la \mu^+}^{\la}$
(because $N_{\la \la^+}^{\mu}$ is given by
the multiplicity of the trivial component in 
$V_{\la} \tens V_{\la^+} \tens V_{\mu^+}$,
and so is $N_{\la \mu^+}^{\la}$).
But $N_{\la \mu^+}^{\la}$ can be calculated using the formula 
\cite{Fuchs:1992nq}
\beq
N_{\la \mu^+}^{\la} = \sum_{\sigma  \in W} \; (-1)^{\sigma} \;
        mult_{\mu^+}(\sigma \star \la - \la),
\label{racah-speiser}
\eeq
where $W$ is the Weyl group of $\mg$.
Here $mult_{\mu^+}(\nu)$ is the multiplicity of the weight space $\nu$
in $V_{\mu^+}$, and $\sigma \star \la = \sigma(\la + \rho) - \rho$ 
denotes the action of $\sigma$ 
with reflection center $-\rho$.
Now one can see already that for large, generic $\la$ (so that 
$\sigma \star \la - \la$ is not a weight of $V_{\mu^+}$ unless $\sigma = 1$), 
it follows that $N_{\la \mu^+}^{\la} = mult_{\mu^+}(0) = mult_{\mu^+}^{(T)}$,
which proves \refeq{Dla-decomp} for the generic case. 
To cover all possible $\la$, we proceed as follows:

Let $\mk$ be the Lie algebra of $K_{\la}$, and
$W_{\mk}$ its Weyl group; it is the subgroup of $W$ which 
leaves $\la$ invariant, generated by 
those reflections which preserve $\la$
(the $u(1)$ factors in  $\mk$ do not contribute to $W_{\mk}$).
If $\mu$ is ``small enough'', then 
the sum in \refeq{racah-speiser}
can be restricted to $\sigma \in W_{\mk}$, because otherwise 
$\sigma \star \la - \la$ is too large to be in $V_{\mu^+}$;
this defines the cutoff
in $\mu$. It holds for any given $\mu$ if $\la$ has the form
$\la = n \la_0$ for large $n \in \N$ 
and fixed $\la_0$\footnote{This constitutes 
our definition of ``classical limit''.
For weights $\la$ which do not satisfy this requirement, 
the corresponding $D$--brane $D_{\la}$ cannot be interpreted as 
``almost--classical''. Here we differ from the approach in \cite{Felder:1999ka},
which do not allow degenerate $\la_0$.}.
We will show below that 
\beq
mult_{\mu^+}^{(K_{\la})} = \sum_{\sigma  \in W_{\mk}} \; (-1)^{\sigma} \;
        mult_{\mu^+}(\sigma \star \la - \la)
\label{mult-eq}
\eeq
for all  $\mu$, which implies \refeq{Dla-decomp}.
Recall that the lhs is the dimension of the subspace of
$V_{\mu^+}$ which is invariant under $K_{\la}$. 

To prove \refeq{mult-eq}, first observe the following fact: Let
$V_{\la}$ be the highest weight irrep of some simple Lie algebra $\mk$
with highest weight $\la$. Then 
\beq
\sum_{\sigma \in W_{\mk}}\; (-1)^{\sigma}\;mult_{V_{\la}}(\sigma \star 0)
  = \d_{\la,0}
\label{trivial-rep}
\eeq
i.e. the sum vanishes unless $V_{\la}$ is the trivial 
representation; here $\mk = u(1)$ is allowed as well.
This follows again from \refeq{racah-speiser},
considering the decomposition of $V_\la \tens (1)$.
More generally, assume that  $\mk = \oplus_i \mk_i$ is a direct sum
of simple Lie algebras $\mk_i$, with corresponding 
Weyl group $W_{\mk} = \prod_i W_i$. 
Its irreps have the form $V = \tens_i V_{\la_i}$, where $V_{\la_i}$ 
denotes the highest weight module of $\mk_i$ with highest weight $\la_i$.
We claim that the relation
\beq
\sum_{\sigma \in W_{\mk}}\; (-1)^{\sigma}\;mult_{V}(\sigma \star 0)
  = \prod_i\; \d_{\la_i,0}
\label{trivial-mult}
\eeq
still holds. Indeed, assume that some $\la_i \neq 0$; then 
$$
\sum_{\sigma \in W_{\mk}}\; (-1)^{\sigma}\;mult_{V}(\sigma \star 0) = 
\Big(\sum_{\sigma' \in\; \prod' W_i} (-1)^{\sigma'}\Big)
\Big(\sum_{\sigma \in W_i} (-1)^{\sigma_i}
 mult_{V}(\sigma_i\star (\sigma' \star 0))\Big) = 0
$$
in self--explanatory notation. The last bracket vanishes by 
\refeq{trivial-rep}, since $(\sigma' \star 0)$ has weight $0$ with 
respect to $\mk_i$, while $V$ contains no trivial component of $\mk_i$
(notice that $\rho = \sum \rho_i$, and the operation $\star$ 
is defined component-wise).
Therefore for any (finite, but not necessarily irreducible) $\mk$--module 
$V$, the number of trivial components in $V$ is given by
$\sum_{\sigma \in W_{\mk}}\; (-1)^{\sigma}\;mult_{V}(\sigma \star 0)$.

We now apply this to \refeq{mult-eq}. Since the sum is over 
$\sigma \in W_{\mk}$, we have $\sigma(\la) = \la$ 
by definition, and $\sigma \star \la - \la  = \sigma \star 0$. Hence
the rhs can be replaced by 
$\sum_{\sigma \in W_{\mk}}\; (-1)^{\sigma}\;mult_{\mu^+}(\sigma \star 0)$.
But this is precisely the number of  vectors in $V_{\mu^+}$
which are invariant under $K_{\la}$, as we just proved.  
Notice that we use here the fact that $\mk$ contains the 
Cartan sub-algebra of $\mg$, so that the space of weights of $\mk$
is the same as the space of weights of $\mg$; therefore the 
multiplicities in \refeq{mult-eq} and \refeq{trivial-mult}
are defined consistently. This is why we had to include the case 
$\mk_i = su(1)$ in the above discussion.

To calculate the decomposition \refeq{little-rich-trunc} for all 
allowed $\la$ (with $\dim_q(V_{\la}) > 0$), 
the ordinary multiplicities in \refeq{little-rich} should be replaced with 
with their truncated versions $\obar{N}_{\la \mu^+}^{\la}$
\refeq{little-rich-trunc} corresponding
to $U_q(\mg)$ at roots of unity. 
There exist generalizations of the formula \refeq{racah-speiser}
which allow to calculate $\obar{N}_{\la \mu^+}^{\la}$
 efficiently; we refer here to the 
literature, e.g. \cite{Felder:1999ka}.

\subsection{The quantum determinant}
\label{a:qdet}
\def\tR{{\tilde R}}

Here we quote a formula for the quantum determinant, following 
\cite{Schupp:1993mt,Faddeev:1996ie}.
We need the q-deformed 
totally (q)--antisymmetric tensor $\eps_q^{i_1 ... i_N}$,
which for $U_q(sl(N))$ has the form
\beq
\eps_q^{\si(1) ... \si(N)}=(-q)^{-l(\si)} = \eps^q_{\si(1)
  ... \si(N)} 
\eeq
where $l(\si)$ is the length of the permutation $\si$. The important formula
respected by $\eps_q$ is 
\beq
\hat R_{i,i+1} \eps_q^{1...N}=-q^{-\frac{N+1}N}\;\eps_q^{1...N}.
\eeq
(for $U_q(sl(N))$. The factor differs for other groups.)
The most obvious form of the determinant is 
\beq
det(M) \eps_q^{1...N} = (M_n) (\hat R_n M_n \hat R_n) ... 
(\hat R_2 \hat R_{3} ... \hat R_n M_n \hat R_n ... \hat R_{3}\hat
R_2)
\eps^{12...n}_q
\eeq
where $\hat R_k = \hat R_{k-1,k}$ (ignoring a possible constant factor).
We check invariance inder the chiral coaction $M \to s^{-1} M t$ 
(noting that $t_n \hat R_n s_n^{-1} = s_{n-1}^{-1} \hat R_n t_{n-1}$):
observe first that
\beqa
(M_n) (\hat R_n M_n \hat R_n) &\to&
s_n^{-1} (M_n)t_n (\hat R_n s_n^{-1} M_n t_n \hat R_n) = 
s_n^{-1} M_n (s_{n-1}^{-1} \hat R_n t_{n-1}) M_n t_n \hat R_n \nn\\
&=& s_n^{-1} s_{n-1}^{-1} (M_n \hat R_n M_n) t_{n-1} t_n \hat R_n
= s_n^{-1} s_{n-1}^{-1} (M_n) (\hat R_n M_n\hat R_n) t_{n-1} t_n \nn
\eeqa
let us work out one more step: since 
\beq
t_{n-1} t_n \hat R_{n-1} \hat R_n s_n^{-1} 
= t_{n-1}\hat R_{n-1}s_{n-1}^{-1} \hat R_n t_{n-1}
= s_{n-2}^{-1} \hat R_{n-1} \hat R_n t_{n-2} t_{n-1}
\eeq
etc, it follows that
\beqa
&& (M_n) (\hat R_n M_n \hat R_n) 
(\hat R_{n-1} \hat R_{n} M_n \hat R_{n}\hat R_{n-1})\to \nn\\
&&= s_n^{-1} s_{n-1}^{-1} (M_n) (\hat R_n M_n\hat R_n) t_{n-1} t_n 
(\hat R_{n-1} \hat R_{n} s_n^{-1} M_n t_n \hat R_{n}\hat R_{n-1}) \nn\\
&&= s_n^{-1} s_{n-1}^{-1} (M_n) (\hat R_n M_n\hat R_n)
s_{n-2}^{-1} \hat R_{n-1} \hat R_n t_{n-2} t_{n-1}M_n t_n 
  \hat R_{n}\hat R_{n-1} \nn\\
&&= s_n^{-1} s_{n-1}^{-1}s_{n-2}^{-1}(M_n) (\hat R_n M_n\hat R_n)
\hat R_{n-1} \hat R_n M_n  t_{n-2} t_{n-1} t_n  \hat R_{n}\hat R_{n-1} \nn\\
&&= s_n^{-1} s_{n-1}^{-1}s_{n-2}^{-1}(M_n) (\hat R_n M_n\hat R_n)
(\hat R_{n-1} \hat R_n M_n  \hat R_{n}\hat R_{n-1}) t_{n-2} t_{n-1} t_n
\eeqa
Invariance now follows from the determinant condition
$t_1 ... t_{n-1} t_n \eps^{12...n}_q = \eps^{12...n}_q$,
and similarly for the $s$.
This in turn implies that indeed $det(M) = 1$ (fixing the constant
factor in the definition of $det(M) = 1$) in the realization $M = L^+ SL^-$. 

(another possible, essentially equivalent form of the determinant is  
\cite{Faddeev:1996ie}
\beq
det(M) = (M_n \hat R_2 ... \hat R_{n-1}\hat R_n)^n\eps^{12...n}
\eeq
)

For $N=2$, this becomes
$det(M) = M_2 \hat R_{12} M_2 \hat R_{12}\; \eps^{12} 
= -q^{-\frac{3}2} M_2 \hat R_{12} M_2 \eps^{12}$,
%$(M_1R'_1) (M_1R'_1) \eps_q^{12}=-\frac1q M_1R'_1M_1 \eps_q^{12}$.
which is  proportional to $\eps_q^{12}$ times
$(M^1_1M^2_2-q^2M^2_1M^1_2)$. Hence we find
the quantum determinant 
\beq\label{qdet2}
\det_q(M)=(M^1_1M^2_2-q^2M^2_1M^1_2)
\eeq
as in \cite{Majid:1996kd}.
For other groups such as $SO(N)$ and $SP(N)$, the explicit form for 
$\eps_q^{i_1 ... i_N}$ is different, 
and additional constraints (which are also invariant 
under the chiral symmetries) must be imposed. These are known and can be 
found in the literature \cite{Schupp:1993mt,Faddeev:1996ie}.

\subsection{Covariance of $M$ and central elements}
\label{a:cov}

For any numerical matrix $M^{(0)}$ (in the defining \rep of $U_q(\mg)$), 
consider
\beq
M = L^+ M^{(0)} SL^- = (\pi \tens 1)(\cR_{21})\; M^{(0)}\; 
   (\pi \tens 1)\cR_{12}.
\label{MC-explicit}
\eeq
Let $\cM \subset U_q(\mg)$  be the sub-algebra generated by the 
entries of this matrix. 
First, we note that $\cM$ is a (left) coideal sub-algebra, 
which means that $\Delta(\cM) \in U_q(\mg) \tens \cM$. This is 
verified simply by calculating the coproduct of $M$, 
\beq\label{coid}
\Delta({M}^i_l) = {L^+}^i_s {SL^-}^t_l \tens (M)^s_t.
\eeq
In particular if $M^{(0)}$ is a constant 
solution of the  reflection equation \refeq{re}, it follows by taking the 
defining \rep of \refeq{MC-explicit} that $[\pi({M}^i_j), M^{(0)}] = 0$,
and therefore $[\pi(\cM), M^{(0)}] = 0$. Then  
for any $u \in \cM \subset U_q(\mg)$,
\beqa
((\pi\tens 1)\Delta(u)) M &=& 
           (\pi\tens 1)(\Delta(u)\cR_{21})\; M^{(0)}\; SL^- \nn\\
  &=&(\pi\tens 1)(\cR_{21}\Delta'(u))\; M^{(0)}\; SL^- \nn\\
  &=& L^+\; M^{(0)}\; (\pi\tens 1)(\Delta'(u)\cR_{12})\nn\\
  &=& L^+\; M^{(0)}\; SL^- (\pi\tens 1)\Delta(u) = M\; (\pi\tens 1)\Delta(u).
\eeqa
In the second line we used $\Delta'(u)\equiv u_2\otimes u_1=R \Delta(u) R^{-1}$,
in the third line,  the coideal property \refeq{coid}. Using Hopf algebra identities
(i.e. multiplying from left with $(\pi(S u_0) \tens 1)$ and from the right 
with $(1 \tens Su_3)$), this is equivalent to
$(1\tens u_1) M (1\tens Su_2) = (\pi(Su_1) \tens 1) M (\pi(u_2)\tens 1)$,
or 
\beq
u_1 M S u_2 = \pi(Su_1) M \pi(u_2)
\eeq
for any $u \in \cM$, as desired.
This  implies immediately that 
\beq
 u_1 \tr_q(M^n) Su_2  = \tr_q(\pi(Su_1) M^n \pi(u_2)) = \eps(u)\; \tr_q(M^n),
\eeq
or equivalently 
\beq
[u, \tr_q(M^n)] = 0
\eeq
for any $u \in \cM$. 
This proves in particular that the Casimirs $c_n$ \refeq{central2-n} 
are indeed central.

\subsection{Evaluation of Casimirs}
\label{a:casimirs}

\paragraph{Evaluation of $c_1$}

Consider the fuzzy $D$--brane $D_\la$. Then $c_1$  acts on 
the highest--weight module $V_{\la}$, and has the form
\beq
c_1 =  \tr_q(L^+ SL^-) = (\tr_q \pi \tens 1) (\cR_{21} \cR_{12}).
\eeq
Because it is a Casimir, it is enough to evaluate it on the 
lowest--weight state $|\la_-\rangle$ of $V_\la$, given by 
$\la_- = \sigma_m(\la)$ where 
$\sigma_m$ denotes the longest element of the Weyl group. 
Now the universal $\cR$ has the form
\beq
\cR = q^{H_i (B^{-1})_{ij} \tens H_j} \; (1 \tens 1 + \sum U^+ \tens U^-).
\eeq
Here $B$ is the (symmetric) 
matrix $d_j^{-1} A_{ij}$ where $A$ is the Cartan Matrix, 
$d_i$ are the lengths of the simple roots ($d_i = 1$ for $\mg = su(N)$) and
$U^+, U^-$ stands for terms in the Borel sub-algebras of
rising respectively lowering operators. Hence 
only the diagonal elements of $(SL^-)^i_j$ are non--vanishing on a
lowest--weight state, and due to the trace only the diagonal elements 
of $(L^+)^i_j$ enter. We can therefore write
\beq
c_1\; |\la_-\rangle = 
      (\tr_q \pi \tens 1) 
         (q^{2\; H_i (B^{-1})_{ij} \tens H_j}) |\la_-\rangle 
    = (\tr \;\pi \tens 1) (q^{-2 H_\rho} \tens 1)
              (q^{2\; H_i (B^{-1})_{ij} \tens H_j}) \; |\la_-\rangle
\eeq
Here %$<H_\rho, \mu> = (\rho, \mu)$
$H_\la |\mu\rangle =(\la\cdot\mu)\;|\mu\rangle$
for any weight $\mu$. Therefore the eigenvalue of $c_1$ is
\beq
c_1 = \sum_{\mu \in V_N}\; q^{-2 \mu\cdot\rho + 2 \mu\cdot\la_-}
    = \sum_{\mu \in V_N}\; q^{2 \mu \cdot (-\rho +\la_-)}.
\eeq
Using $\sigma_m(\rho) = -\rho$, this becomes
\beq
c_1 = \sum_{\mu \in V_N}\; q^{2 (\sigma_m(\mu))\cdot(\rho +\la)}
= \tr_{V_N}\; (q^{2 (\rho +\la)})
\eeq
because the weights of $V_N$ are invariant under the Weyl group.

\paragraph{Evaluation of $c_n$ in general}

Since $c_n$ is proportional to the identity matrix on irreps, 
it is enough to calculate
$\tr_q(c_n) = \tr_{V_\la}(c_n\; q^{-2 H_\rho})$ on $V_{\la}$, 
noting that $\tr_q(1) = \dim_q(V_{\la})$
is known explicitly:
\beq
\tr_q(c_n) = (\tr_q \tens \tr_q)((\cR_{21} \cR_{12})^n)
\eeq
where the traces are over $Mat(N)$ and $Mat(V_{\la})$.
Now we use the fact that $\cR_{21} \cR_{12}$ commutes with $\Delta(U_q(\mg))$,
i.e. it is constant on the irreps of $V_{N} \tens V_{\la}$,
and observe that $\Delta(q^{-2 H_\rho}) = q^{-2 H_\rho} \tens q^{-2 H_\rho}$,
which means that the quantum trace factorizes. Hence we can
decompose the tensor product $V_{N} \tens V_{\la}$
into irreps:
\beq
V_{N} \tens V_{\la} = \oplus_{\mu\in P_k^+} V_{\mu}
\label{laLa-decomp}
\eeq
where the sum goes over all $\mu$  which have the form $\mu = \la + \nu$
for $\nu$ a weight of $V_{N}$. The multiplicities are equal
one because $V_N$ is the defining representation.
The eigenvalues of $\cR_{21} \cR_{12}$ on $V_{\mu}$ are 
known \cite{Reshetikhin:1988ix} to be
$q^{c_{\mu} - c_{\la} - c_{\la_N}}$, where $\la_N$
denotes the highest weight of $V_N$ and  $c_\la = \la\cdot(\la+2\rho)$. 
Now for $\mu = \la + \nu$, 
\beq
c_{\mu} - c_{\la} - c_{\la_N} = 2 (\la + \rho)\cdot \nu - 2\la_N\cdot\rho,
\eeq
hence the set of eigenvalues of $\cR_{21} \cR_{12}$ is
\beq
\{q^{2 (\la + \rho)\cdot\nu - 2\la_N\cdot \rho};\;\; \nu \in V_{N}\}.
\eeq
Putting this together, we obtain
\beq
\tr_q(c_n) = c_n\; \tr_{V_{\la}}(q^{-2 H_\rho}) = \sum_{\mu}\;
       q^{2n((\la + \rho)\cdot\nu - \la_N\cdot \rho)}\; 
            \tr_{V_{\mu}}(q^{-2 H_\rho})
\eeq
where the sum is as explained above.
Then \refeq{cn} follows, since 
$\tr_{V_{\mu}}(q^{-2 H_\rho}) = \dim_q(V_{\mu})$.

\subsection{Characteristic equation for $M$.}
\label{a:chareq}

(\ref{char-poly-Y}) can be seen as follows:
On $D_{\la}$, the quantum matrices $M^i_j$ become the operators
\beq
(\pi^i_j \tens \pi_\la)(\cR_{21} \cR_{12})
\label{RR-rep}
\eeq
acting on $V_\la$. As above, the \rep of $\cR_{21} \cR_{12}$ acting
on $V_N \tens V_{\la}$ has eigenvalues 
$\{q^{c_{\mu} - c_{\la} - c_{\la_N}}$
$=q^{2(\la + \rho)\cdot\nu -2 \la_N\cdot\rho}\}$ 
on $V_{\mu}$ in the decomposition 
\refeq{laLa-decomp}. Here $\mu = \la + \nu$ for $\nu \in V_{N}$,
and $\la_N$ is the highest weight of $V_N$.
This proves (\ref{char-poly-Y}). Note that 
if $\la$ is on the boundary of the fundamental Weyl chamber,
not all of these $\nu$ actually occur in the decomposition;
nevertheless, the characteristic equation holds.

%% Classical Field theory on S^2_{q,N}

\chapter{Field theory on the $q$-deformed fuzzy sphere}
\label{chapter:qFSI}

In this section we elaborate the simplest case of 
Chapter \ref{chapter:qDbranes}, 
which are spherical branes on the group $SU(2)$. 
We will ignore here the ``target'' space $G = SU(2)$, and 
study the structure of the individual branes, which are 
$q$--deformed fuzzy spheres $\S_{q,N}$. Those are precisely the 
branes which were briefly exhibited in Section \ref{subsec:su2}.
This chapter is based\footnote{I adapted the conventions 
to those of chapter 1. In particular, the coproduct is reversed
from \cite{qFSI}} 
on the paper \cite{qFSI} which was written in collaboration with Harald 
Grosse and John Madore.

After reviewing the undeformed fuzzy sphere, we give a definition of 
$\S_{q,N}$ in Section \ref{sec:fuzzyqsphere} for both $q\in \R$ and $|q|=1$.
As an algebra, it is simply
a finite--dimensional matrix algebra, equipped with additional structure 
such as an action of $U_q(su(2))$, a covariant differential calculus,
a star structure, and an integral. For $q \in \R$,
this is precisely the ``discrete'' series of Podle\'s spheres 
\cite{Podles:1987wd}.
For $|q|=1$, the algebra \refeq{su2-branes} is reproduced, with $N \equiv n$.
This case, which is relevant to string theory as discussed before, 
has apparently not been studied in detail before. In
Section \ref{sec:diff-calc}, we develop the non--commutative
differential geometry on $\S_{q,N}$, using an approach which is suitable
for both $q \in \R$ and $|q|=1$.  The differential calculus
turns out to be elaborate, but quite satisfactory.
We are able to show, in particular, that 
in both cases there exists a 3--dimensional exterior differential calculus
with real structure and a Hodge star, and we develop a frame 
formalism \cite{Dubois-Violette:1989at,madoreDim,Mad99c}.
This allows us to write Lagrangians for field theories on $\S_{q,N}$.
In particular, the fact that the tangential space is 3--dimensional
unlike in the classical case is very interesting physically,
reflecting the fact that the $D$--branes are embedded 
in a higher-dimensional space.

Using these tools, we study in Section \ref{subsec:scalars}
 actions for scalar fields
and abelian gauge fields on $\S_{q,N}$. The latter case is particularly
interesting, since it turns out that certain actions for gauge 
theories arise in a very natural way in terms of polynomials
of one--forms. In particular, the kinetic terms arise automatically
due to the noncommutativity of the space. 
Moreover, because the calculus is 3--dimensional, the gauge field
consists of a usual (abelian) gauge field plus a (pseudo) scalar
in the classical limit. This is similar to a Kaluza--Klein reduction.
One naturally obtains analogs of Yang--Mills and 
Chern--Simons actions, again because the calculus is
3--dimensional. In a certain limit where $q=1$, 
such actions were shown to arise from open strings ending
on $D$--branes in the $SU(2)$ WZW model \cite{Alekseev:2000fd}.
The gauge theory actions for $q \neq 1$ suggest a new
version of gauge invariance, where the gauge ``group'' is 
a quotient $U_q(su(2))/I$, which can be identified with the space 
of functions on the deformed fuzzy sphere. This is 
discussed in Section \ref{subsec:gauge}.

In this chapter, we shall only consider the
first--quantized situation; the second quantization is discussed
in Chapter \ref{chapter:qFSII}. The latter
turns out to be necessary for implementing the symmetry
$U_q(su(2))$ on the space of fields in a fully satisfactory way.

We should perhaps add a general remark on the type of noncommutative 
spaces considered here. It is customary in the 
literature on noncommutative geometry to define the dimension of 
a noncommutative space in terms of so-called Chern-Connes 
characters \cite{Connes:book}, which are related to the $K$ theory of the 
underlying function algebras. In this framework, all fuzzy spheres
are ``zero-dimensional'', being finite matrix algebras. 
However it will become clear in the later sections that this 
does not correspond to the physical concept of dimension in physics. 
We will see explicitly that the field theories defined on these
spaces do behave as certain ``regularized'' field theories
on ordinary spheres. The reason is the harmonic analysis:
if decomposed under the appropriate symmetry algebra, the space or algebra of 
functions is for ``low energies'' (=for small \reps) exactly the same
as classically. It is the number of degrees of freedom
below a given cutoff $\La$ (more precisely its scaling with $\La$)
which determines the physical dimension of a space,
and enters the important quantities such as entropy, 
spectral action etc. We will therefore
not worry about these K-theoretic issues any further here.

\section{The undeformed fuzzy sphere}
\label{sec:fuzzysphere}

We give a quick review of the ``standard'' fuzzy sphere
\cite{Madore:1992bw,Grosse:1996ar,Grosse:1997pr}.
Much information about the standard unit sphere $S^2$ in ${\R^3}$
is encoded in the infinite dimensional algebra 
of polynomials generated by 
$\tilde x = (\tilde x_1, \tilde x_2, \tilde x_3) \in {\R^3}$ with the 
defining relations
\be
[\tilde x_i, \tilde x_j] = 0, \ \ \ \sum_{i=1}^3 \tilde x_i^2 = r^2
\ee
The algebra of functions on the  fuzzy sphere is defined 
as the finite algebra 
${\cal S}^2_N$ generated by $\hat{x} = (\hat{x}_1, \hat{x}_2, \hat{x}_3)$, with
relations 
\beqa
[\hat{x}_i, \hat{x}_j ] &=& i \lambda_N \e_{ijk} \hat{x}_k \label{defl1}\\
\sum_{i=1}^3 \hat{x}_i^2 &=& r^2 
\label{fuzzy_x}
\eeqa
The real parameter $\lambda_N > 0$ characterizes the non-commutativity,
and has the dimension of a length.
The radius $r$ is quantized in units of $\la_N$ by
\be\label{defl3-0}
\frac {r}{\la_N} = \sqrt{\frac{N}{2} \left( \frac{N}{2} +1\right)
} \; , 
\quad \mbox{$N = 1,2,\cdots$ } 
\ee
This quantization can be easily understood. Indeed \refeq{defl1} is
simply the Lie algebra $su(2)$, whose irreducible representation
are labeled by the spin $N/2$. The Casimir 
of the spin-$N/2$ representation is quantized, and related to $r^2$ by 
\refeq{defl2}.
The fuzzy sphere $S_N^2$ is thus characterized by its radius $r$ and the
``noncommutativity parameters'' $N$ or $\la_N$. 
The integral of a function $F \in S_N^2$ over the fuzzy sphere is given by
\be
\int F  = \frac{4 \pi R^2}{N+1} \tr[ F(x)],
\ee
It agrees with the usual integral on $S^2$ in the large $N$ limit. 
Invariance of the integral under  the rotations 
$SU(2)$ amounts to invariance of the trace under adjoint action.

The algebra of functions is  
most conveniently realized using the Wigner-Jordan realization of the
generators $\hat{x}_i \ , i=1,2,3$, in terms of two pairs of annihilation and
creation operators 
$A_{\alpha} , {A^+}^{\alpha} , \alpha = \pm \frac 12$, which satisfy
\be
[A_{\a} , A_{\b} ] = [{A^+}^{\a} , {A^+}^{\b} ] = 0 \ ,\ \
[A_{\a} , {A^+}^{\b} ] = \delta_{\a}^{\b} \ ,
\label{A_cr}
\ee
and act on the Fock space $\cal{F}$ spanned by the vectors
\be
|n_1 ,n_2 \rangle =
\frac{1}{\sqrt{n_1 ! n_2 !}} ({A^+}^{\haf})^{n_1} ({A^+}^{-\haf})^{n_2} 
|0 \rangle \ .
\ee
Here $|0 \rangle$ is the vacuum defined by $A_i |0 \rangle = 0$. 
The operators $\hat{x}_i$ take the form
\be
\hat{x}_i = \frac{\lambda_N}{\sqrt{2}} 
           {A^+}^{\a'} \eps_{\a' \a} \sigma_i^{\a \b} A_{\b}. 
\label{JW}
\ee
Here $\eps_{\a \a'}$ is the antisymmetric tensor 
(spinor metric), and $\sigma_i^{\a \b}$ are the  
Clebsch-Gordan coefficients, 
that is rescaled Pauli--matrices.
The number operator is given by 
$\hat N = \sum_{\a} {A^+}^{\a} A_{\a}$. When restricted to
the $(N+1)$-dimensional subspace
\be
{\cal{F}}_N = \{\sum {A^+}^{\a_1} ... \; {A^+}^{\a_N}|0\rangle  \; 
       (N\; \mbox{creation operators})\}.
\ee
it yields for any given $N = 0,1,2,...\ \ $ the irreducible unitary
representation in which the parameters $\lambda_N$ and $r$ are related as
\be
\frac {r}{\lambda_N} = \sqrt{\frac{N}{2} \left( \frac{N}{2} +1\right) } \ .
\ee
The algebra ${\cal S}^2_N$ generated by the $\hat{x}_i$ is clearly the
simple matrix algebra $Mat(N+1)$.
Under the adjoint action of $SU(2)$, it decomposes into the 
direct sum $(1) \oplus (3) \oplus (5) \oplus ... \oplus (2N+1)$
of irreducible \reps of $SO(3)$ \cite{Grosse:1996ar,Grosse:1997pr}.

\section{The $q$-deformed fuzzy sphere}
\label{sec:fuzzyqsphere}

\subsection{The $q$--deformed algebra of functions}
\label{subsec:q_fuzzy}

The fuzzy sphere $\S_N$ is invariant under the action of $SU(2)$, 
or equivalently 
under the action of $U(su(2))$. We shall define finite algebras 
$\S_{q,N}$ generated by ${x}_i$ for $i=1,0,-1$,
which have completely analogous properties to
those of $\S_{N}$, but which are
covariant under the quantized universal enveloping algebra
$U_q(su(2))$. This will be done for both $q \in \R$ and $q$ a phase,
including the appropriate reality structure.
In the first case, the $\S_{q,N}$ will turn out to be the  
``discrete series'' %$S^2_{q,c(N-1)}$ 
of Podle\'s' quantum spheres \cite{Podles:1987wd}.
Here we will study them more closely from the above point of view.
However, we also allow $q$ to be a root of unity,
with certain restrictions.
In a twisted form, this case does appear naturally 
on $D$--branes in  the $SU(2)$ WZW model, as was shown in \cite{Alekseev:1999bs}.

In order to make the analogy to the undeformed case obvious,
we  perform a $q$--deformed Jordan--Wigner 
construction, which is covariant under $U_q(su(2))$. To fix the notation, we 
recall the basic relations of $U_q(su(2))$
\be  
[H, X^{\pm}] = \pm 2 X^{\pm}, \quad
[X^+, X^-]   =  \frac{q^{H}-q^{-H}}{q-q^{-1}}  =  [H]_{q}
\label{U_q_rel}
\ee
where the $q$--numbers are defined as 
$[n]_q = \frac {q^n-q^{-n}}{q-q^{-1}}$. 
The action of $U_q(su(2))$ on a tensor product of \reps 
is encoded in the coproduct
\be
\Del(H)       = H \tens 1 + 1 \tens H, \qquad
\Del(X^{\pm}) =  X^{\pm} \tens q^{H/2} + q^{-H/2}\tens X^{\pm}.
\label{coproduct_X} 
\ee
The antipode and the counit are given by
\be
S(H)    = -H, \quad  S(X^\pm)  = -q^{\pm 1} X^\pm,  \qquad
\eps(H) = \eps(X^{\pm})=0.
\ee
The star structure is related to the
Cartan--Weyl involution $\theta(X^{\pm}) = X^{\mp}, \; \theta(H) = H$,
and will be discussed below.
All symbols will now be understood to carry a label ``$q$'',
which we shall omit.

Consider $q$--deformed creation and anihilation operators 
$A_{\a} , {A^+}^{\a}$ for $ \alpha = \pm \haf$, which
satisfy the relations (cp. \cite{Wess:1991vh,Carow-Watamura:1991zp}) 
\berr
{A^+}^{\a} A_{\b} &=& \d^{\a}_{\b} + q \hat{R}^{\a \g}_{\b\d}
                               A_{\g}  {A^+}^{\d} \nn\\
(P^-)^{\a \b}_{\g\d} A_{\a} A_{\b} &=& 0 \nn\\
(P^-)^{\a \b}_{\g\d} {A^+}^{\d} {A^+}^{\g} &=& 0
\label{AA_CR} 
\err  
where $\hat{R}^{\a \g}_{\b\d} = q (P^+)^{\a \g}_{\b\d} - q^{-1}  
(P^-)^{\a \g}_{\b\d}$ is the decomposition of the $\hat R$ --matrix
of $U_q(su(2))$ into the projection operators on the symmetric  
and antisymmetric part. They can be written as  
\berr
(P^-)^{\a \b}_{\g\d}  &=& \frac 1{-[2]_q}\eps^{\a \b} \eps_{\g\d}, \nn\\
(P^+)^{\a \b}_{\g\d}  &=& \sigma_i^{\a \b} \sigma^i_{\g \d}
\label{sigma_completeness}
\err
Here $\eps_{\a \b}$ is the $q$--deformed invariant
antisymmetric tensor (see Section \ref{subsec:reps}), 
and $\sigma^i_{\a \b}$ are the  $q$--deformed
Clebsch--Gordan coefficients; they are given explicitly in 
Section \ref{subsec:inv-su2}. 
The factor $-[2]_q^{-1}$ arises from the relation
$\eps^{\a \b}\eps_{\a\b} = -[2]_q$. 
The above relations are covariant under $U_q(su(2))$,  
and define a left $U_q(su(2))$--module algebra. 
We shall denote the action on the generators with lower indices by  
\be 
u \trr A_{\a} = A_{\b}\; \pi^{\b}_{\a}(u),  
\ee  
so that $\pi^{\a}_{\b}(u v ) = \pi^{\a}_{\g}(u) \pi^{\g}_{\b}(v)$ for  
$u,v \in U_q(su(2))$.
The generators with upper indices transform in the contragredient  
representation, which means that  
\be 
A^+_{\a} := \eps_{\a \b} {A^+}^{\b} 
\ee 
transforms in the same way under $U_q(su(2))$ as $A_{\a}$.

We consider again the corresponding 
Fock space $\cal{F}$ generated by the  ${A^+}^{\a}$ acting on the   
vacuum $|0\rangle$, and its sectors 
\be 
{\cal{F}}_N = \{\sum {A^+}^{\a_1} ... \; {A^+}^{\a_N}|0\rangle  \; 
       (N\; \mbox{creation operators})\}.
\label{F_N}
\ee
It is well--known that these subspaces ${\cal{F}}_N$
are $N+1$--dimensional, as they are when $q=1$, 
and it follows that they form irreducible \reps 
of $U_q(su(2))$ (at root of unity, this will be true due to the restriction 
(\ref{q_phase}) we shall impose). 
This will be indicated by writing ${\cal{F}}_N = (N+1)$, and 
the decomposition of ${\cal F}$ into irreducible representations is
\be
{\cal F} = {\cal{F}}_0 \oplus {\cal{F}}_1 \oplus {\cal{F}}_2 \oplus ... = 
            (1) \oplus (2) \oplus (3) \oplus ... 
\label{F_decomp}
\ee
Now we define 
\be
\hat Z_i = {A^+}^{\a'} \eps_{\a \a'} \sigma_i^{\a \b} A_{\b} 
\label{Z_i}
\ee   
and  
\be
\hat N = \sum_{\a} {A^+}^{\a'}\eps_{\a \a'} \eps^{\a\b}A_{\b}.  
\ee
After some calculations, 
these operators can be shown to satisfy the relations
\berr
&&\eps_k^{i j} \hat Z_i \hat Z_j =
 \frac{q^{-1}}{\sqrt{[2]_q}}(q^{-1} [2]_q - \la\hat N) \hat Z_k 
\label{ZZ_sigma}\\
&&\hat Z^2 := g^{i j} \hat Z_i \hat Z_j =
            q^{-2}\frac{[2]_q + \hat N}{[2]_q} \hat N 
\label{ZZ_square}
\err
Here $\la = (q-q^{-1})$, $g^{ij}$ is the $q$--deformed invariant tensor 
for spin 1 representations, 
and $\eps_k^{ij}$ is the corresponding $q$--deformed Clebsch--Gordan 
coefficient; they are given in Section \ref{subsec:inv-su2}.
Moreover, one can verify that
\berr
\hat N{A^+}^{\a} &=& q^{-3}{A^+}^{\a} + q^{-2} {A^+}^{\a} \hat N, \nn\\
\hat N{A}_{\a}   &=& -q^{-1}{A}_{\a} + q^{2} {A}_{\a} \hat N,
\err
which implies that
$$
[\hat N , \hat Z_i ] = 0.
$$
On the subspace ${\cal F}_N$, the ``number'' operator  $\hat N$ 
takes the value 
\be
\hat N {\cal F}_N = q^{-N-2} [N]_q {\cal F}_N
\label{hat_N}
\ee
It is convenient to introduce also an undeformed number operator
$\hat n$ which has eigenvalues
$$
\hat n \F_N = N \F_N,
$$
in particular $\hat n A_{\a} = A_{\a} (\hat n-1)$.

On the subspaces ${\cal F}_N$, the relations (\ref{ZZ_sigma}) become
\berr
\eps_k^{i j}  x_i  x_j &=& 
            \La_N \; x_k, \label{xx_sigma}\\ 
 x\cdot x &:=& g^{i j} x_i x_j =  r^2. 
\label{xx_square}
\err
Here the variables have been rescaled to $x_i$ with  
$$
x_i = r \; \frac{q^{\hat n +2}}{\sqrt{[2]_q C_N}} \; \hat Z_i.
$$
The $r$ is a real number, and we have defined
\berr
C_N &=& \frac{[N]_q [N+2]_q}{[2]_q^2}, \nn\\
\La_N &=& r \; \frac{[2]_{q^{N+1}}}{\sqrt{[N]_q [N+2]_q}}.
\err
Using a completeness relation (see Section \ref{subsec:inv-su2}),
(\ref{xx_sigma}) can equivalently be written as 
\be
(P^-)^{ij}_{kl}  x_i  x_j  = \frac 1{[2]_{q^2}}\; \La_N  \eps^n_{kl} \; x_n.
\ee
There is no $i$ in the commutation relations, because we use
a weight basis instead of Cartesian coordinates.
One can check that these relations precisely reproduce the 
``discrete'' series %$S^2_{q,c(N+1)}$ 
of Podle\'s' quantum spheres 
(after another rescaling), see \cite{Podles:1987wd}, Proposition 4.II.
Hence we define $\S_{q,N}$ to be 
the algebra generated by the variables $x_i$ acting on ${\cal F}_N$.
Equipped with a suitable star structure and a differential structure, 
this will be the $q$--deformed fuzzy sphere.

It is easy to see that the algebra $\S_{q,N}$ 
is simply the full matrix algebra $Mat(N+1)$, i.e.
it is the same algebra as $\S_{N}$ for $q=1$.
This is because ${\cal{F}}_N$ is an irreducible \rep
of $U_q(su(2))$. To see it, we use complete reducibility \cite{Rosso:1988gg}
of the space of polynomials in $x_i$ of degree $\leq k$ to conclude that 
it decomposes into the direct sum of irreducible \reps
$(1) \oplus (3) \oplus (5) \oplus ... \oplus (2k+1)$.
Counting dimensions and noting that $x_1^{N} \neq 0 \in (2N+1)$,
it follows that $\dim(\S_{q,N}) = (N+1)^2 = \dim Mat(N+1)$, and hence
\be \fbox{$
\S_{q,N} = (1) \oplus (3) \oplus (5) \oplus ... \oplus (2N+1)  $}
\label{A_decomp}
\ee
(as in \refeq{A_decomp-0}).
This is true even if $q$ is a root of unity provided 
(\ref{q_phase}) below holds, a relation which will be necessary 
for other reasons as well. This is the decomposition of the functions 
on the $q$--deformed fuzzy sphere into $q$--spherical harmonics, and 
it is automatically truncated.
Note however that 
not all information about a (quantum) space 
is encoded in its algebra of functions; in addition, one must 
specify for example a differential calculus and symmetries.
For example, the action of $U_q(su(2))$ on $\S_{q,N}$ is different
from the action of $U(su(2))$ on $\S_{N}$.

The covariance of $\S_{q,N}$ under $U_q(su(2))$ can also be 
stated in terms of the
quantum adjoint action. It is convenient to consider the  
cross--product algebra $U_q(su(2)) \smash \S_{q,N}$, which 
as a vector space is equal to $U_q(su(2)) \tens \S_{q,N}$,
equipped with an algebra structure defined by
\be
ux = (u_{(1)}\trr x)u_{(2)}.
\label{smash}
\ee
Here the $\trr$  denotes the action of
$u \in U_q(su(2))$ on $x \in \S_{q,N}$.
Conversely, the action of $U_q(su(2))$ on $\S_{q,N}$ 
can be written as $u \trr x = u_{(1)} x Su_{(2)}$.
The relations (\ref{smash}) of $U_q(su(2)) \smash \S_{q,N}$
are automatically realized on the \rep $\F_N$.

Since both algebras $\S_{q,N}$ and $U_q(su(2))$ act on $\F_N$ and 
generate the full matrix algebra 
$Mat(N+1)$, it must be possible to express the generators 
of $U_q(su(2))$ in terms of the $Z_i$. 
The explicit relation can be obtained by 
comparing the relations (\ref{xx_sigma}) with (\ref{smash}). One finds
\berr
X^+ q^{-H/2} &=& q^{N+3} \;\hat Z_1,     \nn\\
X^- q^{-H/2} &=& -q^{N+1}\; \hat Z_{-1}, \nn\\
q^{-H} &=& \frac{[2]_{q^{N+1}}}{[2]_q} + 
              \frac{q^{N+2}(q-q^{-1})}{\sqrt{[2]_q}}\; \hat Z_0,
\label{U_A_relation}
\err
if acting on $\F_N$. In fact, this defines an algebra map
\be
j:\quad U_q(su(2)) \rightarrow \S_{q,N}
\label{j_map}
\ee
which satisfies
\be\fbox{$
j(u_{(1)}) x j(Su_{(2)}) = u \trr x $}
\label{q_adjoint}
\ee
for $x \in \S_{q,N}$ and $u \in U_q(su(2))$. This is analogous to 
results in \cite{Cerchiai:2000qu,Cerchiai:2000tc}. 
We shall often omit $j$ from 
now on. In particular, $\S_{q,N}$ is 
the quotient of the algebra $U_q(su(2))$ by the relation (\ref{xx_square}).
The relations (\ref{q_adjoint}) and those of $U_q(su(2))$ can be verified
explicitly using (\ref{ZZ_sigma}). Moreover, one can verify that 
it is represented correctly on $\F_N$
by observing that $X^+ (A^+_{1/2})^N |0\rangle = 0$, which means that
$(A^+_{1/2})^N |0\rangle $ is the highest--weight vector of $\F_N$.

\subsection{Reality structure for $q \in \R$ }

In order to define a real quantum space, we must also construct a star
structure, which is an involutive anti--linear anti--algebra map. 
For real $q$, the algebra
(\ref{AA_CR}) is consistent with the following star structure
\berr
(A_{\a})^{\ast} &=& {A^+}^{\a}  \nn\\
({A^+}^{\a})^{\ast} &=& A_{\a}
\label{A_star_real}
\err
This can be verified using the standard compatibility relations
of the $\hat R$ --matrix with the invariant tensor \cite{Faddeev:1990ih}.
On the generators $x_i$, it implies the relation
\be
x_i^{\ast} = g^{ij} x_j,
\label{x_star_real}
\ee
as well as the equality
$$
\hat N^{\ast} = \hat N.
$$
The algebras $\S_{q,N}$ are now precisely 
Podle\'s' ``discrete'' $C^*$ algebras $\tilde{S}^2_{q,c(N+1)}$.
Using (\ref{U_A_relation}), this is equivalent to
\be
H^{\ast} = H, \; (X^{\pm})^{\ast} = X^{\mp},
\label{U_star}
\ee
which is the star--structure for the compact form $U_q(su(2))$.
It is well--known that there is a unique Hilbert space structure on 
the subspaces ${\cal F}_N$ such that they are unitary irreducible \reps of 
$U_q(su(2))$. Then the above star is simply the operator adjoint.

\subsection{Reality structure for $q$ a phase}
\label{subsec:real_form}

When $q$ is a phase, finding the correct star structure is not quite
so easy. The difference with the case $q \in \R$ is that 
$\Delta(u^{\ast}) = (\ast \tens \ast) \Delta'(u)$
for $|q|=1$ and $u \in U_q(su(2))$, 
where $\Delta'$ denotes the flipped coproduct. 
We shall define a star only on the algebra
$\S_{q,N}$ generated by the $x_i$, and not on the full
algebra generated by $A_{\a}$ and ${A_{\a}^+}$.

There appears to be an obvious choice at first sight, namely
$x_i^{*} = x_{i}$, which is indeed 
consistent with (\ref{xx_sigma}). However, it is the
wrong choice for our purpose, because it induces the noncompact 
star structure $U_q(sl(2,\R))$. 

Instead, we define a star--structure on $\S_{q,N}$ as follows. 
The algebra
$U_q(su(2))$ acts on the space $\S_{q,N}$, which generically
decomposes as $(1)\oplus (3) \oplus ...\oplus (2N+1)$. 
This decomposition should be a direct sum of {\em unitary} \reps  
of the compact form 
of $U_q(su(2))$, which means that the star structure 
on $U_q(su(2))$ should be (\ref{U_star}), as it is for real $q$.
There is a slight complication, because
not all finite--dimensional irreducible \reps are 
unitary if $q$ is a phase \cite{Keller:1990tg}.
However, all \reps with dimension $\leq 2N+1$ are
unitary provided $q$ has the form
\be
q = e^{i\pi \vp}, \quad \mbox{with} \;\; \vp < \frac 1{2N}.
\label{q_phase}
\ee
This will be assumed from now on.

As was pointed out before, we can consider the algebra  $\S_{q,N}$
as a quotient of $U_q(su(2))$ via (\ref{U_A_relation}). It acts on
${\cal F}_N$, which is an irreducible \rep of $U_q(su(2))$, and hence 
has a natural Hilbert space structure.  We define the 
star on the operator algebra $\S_{q,N}$ by the adjoint
(that is by the matrix adjoint in an orthonormal basis), hence by the 
star (\ref{U_star}) using the identification (\ref{U_A_relation}).

There is a very convenient way to write down this star structure
on the generators $x_i$, similar as in \cite{Steinacker:1999xu}. 
It involves an element $\tw$ of an extension of $U_q(su(2))$ introduced by 
\cite{Kirillov:1990vb} and \cite{lev-soib},
which implements the Weyl reflection on irreducible representations.
The essential properties are
\berr
\Del(\tw) &=& \RR^{-1}\tw\tens \tw,        \\ 
\tw u \tw^{-1} &=& \theta S^{-1}(u),   \label{S_theta}\\
\tw^2 &=& v \epsilon,
\err
where $v$ and $\epsilon$ are central elements in $\in U_q(su(2))$ 
which take the values $q^{-N(N+2)/2}$ resp. $(-1)^{N}$ on $\F_N$. 
Here $\RR = \RR_1 \tens \RR_2 \in U_q(su(2)) \tens U_q(su(2))$ is the 
universal $R$ element. In a unitary representation of $U_q(su(2))$,
the matrix representing $\tw$ in an orthonormal basis 
is given the invariant tensor 
in a certain normalization, $\pi^i_j(\tw) = - q^{-N(N+2)/4} g^{ij}$,
and $\tw^{\ast} = \tw^{-1}$. 
This is discussed in detail in \cite{Steinacker:1999xu}. 
From now on, we denote
with $\tw$ the element in $\S_{q,N}$ which represents this element 
on $\F_N$.

We claim that the star structure 
on $\S_{q,N}$ as explained above is given by the following 
formula:
\be
x_i^{\ast} = - \tw x_i \tw^{-1} \
        = x_j {L^-}^{j}_{k} q^{-2} g^{k i},
\label{x_star_phase}
\ee
where
\be
{L^-}^{i}_{j} = \pi^{i}_{j}(\RR_1^{-1}) \RR_2^{-1}           \label{L_-}
\ee
as usual \cite{Faddeev:1990ih}; a priori, ${L^-}^{i}_{i} \in U_q(su(2))$, 
but it is understood here as an element of $\S_{q,N}$ via 
(\ref{U_A_relation}).
One can easily verify  using $(\eps^{ij}_k)^* = - \eps^{ji}_k$ (for $|q|=1$)
that (\ref{x_star_phase}) is consistent with the relations 
(\ref{xx_sigma}) and (\ref{xx_square}). In the limit $q \rightarrow 1$, 
${L^-}^{i}_{j} \rightarrow \d^{i}_{j}$, therefore (\ref{x_star_phase})
agrees with (\ref{x_star_real}) in the classical limit.
Hence we define the $q$--deformed fuzzy sphere for $q$ a phase
to be the algebra $\S_{q,N}$ equipped with 
the star--structure (\ref{x_star_phase}).

To show that (\ref{x_star_phase}) is correct in the sense explained above, 
it is enough to 
verify that it induces the star structure (\ref{U_star}) on $U_q(su(2))$, 
since both $U_q(su(2))$ and $\S_{q,N}$ generate 
the same algebra $Mat(N+1)$. This can easily be seen 
using (\ref{S_theta}) and (\ref{U_A_relation}).
A somewhat related conjugation has been proposed in 
\cite{Steinacker:1999xu,Mack:1992tg} using the universal element
$\RR$.

\subsection{Invariant integral}

The integral on $\S_{q,N}$ is defined to be the 
unique functional on $\S_{q,N}$ which is invariant 
under the (quantum adjoint) action of $U_q(su(2))$.
It is given by the projection on the 
trivial sector in the decomposition (\ref{A_decomp}). We claim that it can be 
written explicitly using the quantum trace:
\be
\int\limits_{\S_{q,N}} f(x_i) := 4\pi r^2 \frac 1{[N+1]_q} \Tr_q(f(x_i)) = 
              4\pi r^2  \frac 1{[N+1]_q} \Tr(f(x_i) \;q^{H})
\label{integral}
\ee
for $f(x_i) \in \S_{q,N}$, where the trace is taken on $\F_N$. 
Using $S^{-2} (u) = q^{-H} u q^{H}$
for $u \in U_q(su(2))$, it follows that 
\be
\int\limits_{\S_{q,N}}f g = \int\limits_{\S_{q,N}} S^{-2}(g) f.
\label{cyclic_integral}
\ee
This means that it is indeed 
invariant under the quantum adjoint action\footnote{Note the difference to 
\refeq{cn-inv}: there, the trace is over the {\em indices} of the coordinates
$M^i_j$, not over the representation. This leads to $q^{-H}$
rather than $q^H$.}, 
\berr
\int\limits_{\S_{q,N}} u \trr f(x_i) &=& \int\limits_{\S_{q,N}} 
                                            u_1 f(x_i)S(u_2) \nn\\
    &=& \int\limits_{\S_{q,N}} S^{-1}(u_2) u_1 f(x_i)  
    = \eps(u) \; \int\limits_{\S_{q,N}} f(x_i),
\label{invariance_int}
\err
using the identification (\ref{U_A_relation}).
The normalization constant is obtained from
$$
\Tr_q(1) = \Tr(q^{H}) = q^{N} + q^{N-2} + ... + q^{-N} = [N+1]_q
$$
on $\F_N$, so that $\int\limits_{\S_{q,N}} 1 =4\pi r^2$.

\begin{lemma}
Let $f \in \S_{q,N}$. Then 
\be
\Big(\int\limits_{\S_{q,N}} f\Big)^* = \int\limits_{\S_{q,N}}f^{\ast}
\label{I_star_real}
\ee
for real $q$, and 
\be
\Big(\int\limits_{\S_{q,N}}f\Big)^* = \int\limits_{\S_{q,N}} f^{\ast} q^{-2H}
\label{I_star_phase}
\ee
for $q$ a phase, with the appropriate star structure 
(\ref{x_star_real}) respectively (\ref{x_star_phase}). 
In (\ref{I_star_phase}), we use (\ref{U_A_relation}).
\end{lemma}

\begin{proof}
Assume first that $q$ is real, and consider the functional
$$
I_{q,N}(f) :=  \Tr(f^{\ast}q^{H})^*
$$
for $f \in \S_{q,N}$. Then
\berr
I_{q,N}(u \trr f) &=& \Tr((u_1 f S(u_2))^{\ast}\; q^{H})^* \nn\\ 
   &=& \Tr (S^{-1}((u^{\ast})_2) f^{\ast}
       (u^{\ast})_1 \; q^{H})^* 
      = \Tr (f^{\ast}(u^{\ast})_1 S((u^{\ast})_2) q^{H})^* \nn\\
    &=& \eps(u) \;  I_{q,N} (f),
\err
where $(S(u))^{\ast} = S^{-1}(u^{\ast})$ and 
$(\ast\tens\ast) \Delta(u) = \Delta(u^{\ast})$
was used. Hence $I_{q,N}(f)$ is invariant as well, 
and (\ref{I_star_real}) follows  
using uniqueness of the integral (up to normalization).
For $|q| = 1$, we define 
$$
\tilde I_{q,N}(f) :=  \Tr(f^{\ast}q^{-H})^*
$$
with the star structure (\ref{x_star_phase}). 
Using $(S(u))^{\ast} = S(u^{\ast})$ and 
$(\ast\tens\ast) \Delta(u) = \Delta'(u^{\ast})$,
an analogous calculation shows that $\tilde I_{q,N}$ is invariant under 
the action of $U_q(su(2))$, which again implies (\ref{I_star_phase}).
\end{proof}

For  $|q|=1$, the integral is neither real nor positive, hence
it cannot be used for a GNS construction. Nevertheless, it is clearly the 
appropriate functional to define an action for field theory, since it is
invariant under $U_q(su(2))$. To find a way out, we 
introduce an auxiliary antilinear algebra--map on $\S_{q,N}$ by
\be
\obar{f} = S^{-1}(f^{\ast})
\label{obar}
\ee
where $S$ is the antipode on $U_q(su(2))$, using (\ref{U_A_relation}).
Note that $S$ preserves the relation (\ref{xx_square}), hence it
is well--defined on $\S_{q,N}$. 
This is not a star structure, since
$$
\obar{\obar{f}} = S^{-2} f
$$
for $|q|=1$. Using (\ref{U_A_relation}), one finds in particular
\be
\obar{x_i} = - g^{ij} x_{j}.
\ee
This is clearly consistent 
with the relations (\ref{xx_sigma}) and (\ref{xx_square}).
We claim that (\ref{I_star_phase}) can now be stated as
\be
\Big(\int\limits_{\S_{q,N}}f\Big)^* = \int\limits_{\S_{q,N}} \obar{f}
\qquad \mbox{for } \; |q|=1.
\label{I_star_bar}
\ee
To see this, observe first that 
\be
\Tr(S(f)) = \Tr(f),
\ee
which follows either from the fact that 
$\hat I_{q,N}(f):= \Tr(S^{-1}(f) q^{-H})  = \Tr(S^{-1}(q^{H} f))$
is yet another invariant functional, or using 
$\tw f \tw^{-1} = \theta S^{-1}(f)$ together with the observation that the 
matrix \reps of $X^{\pm}$ in a unitary \rep are real. 
This implies 
\be
\Tr_q (f^{\ast} q^{-2H}) = 
\Tr(f^{\ast} q^{-H}) = \Tr(S(q^{H} S^{-1}(f^{\ast}))) = 
\Tr(q^{H} S^{-1}(f^{\ast})) = \Tr_q( \obar{f}), 
\ee
and (\ref{I_star_bar}) follows.
Now we can write down a positive inner product on $\S_{q,N}$:
\begin{lemma}
The sesquilinear forms
\be
(f,g):=  \int\limits_{\S_{q,N}} f^{\ast} g \quad \mbox{for}\;\;  q\in \R
\ee
and
\be
(f,g):=  \int\limits_{\S_{q,N}} \obar{f} g, \quad \mbox{for}\;\;  |q|=1
\ee
are hermitian, that is $(f,g)^* = (g,f)$, and satisfy
\be
(f, u\trr g) = (u^{\ast} \trr f, g)
\ee
for both $q\in\R$ and $|q|=1$. They are positive definite provided
(\ref{q_phase}) holds for $|q|=1$, and 
define a Hilbert space structure on $\S_{q,N}$. 
\label{inner_product_lemma}
\end{lemma}
\begin{proof}
For $q\in \R$, we have 
\berr
(f, u \trr g) &=& \int\limits_{\S_{q,N}} f^{\ast}u_1 g Su_2 = 
                   \int\limits_{\S_{q,N}} S^{-1}(u_2) f^{\ast} u_1 g = 
                   \int\limits_{\S_{q,N}} (S((u^{\ast})_2)^{\ast} 
                     f^{\ast} ((u^{\ast})_1)^{\ast} g  \nn\\
               &=&  \int\limits_{\S_{q,N}}\left((u^{\ast})_1 f
                    S(u^{\ast})_2\right)^{\ast} g =  (u^*\trr f, g),
\err
and hermiticity is immediate. For $|q|=1$, consider
\berr
(f, u \trr g) &=& \int\limits_{\S_{q,N}} \obar{f}u_1 g Su_2 = 
              % \int\limits_{\S_{q,N}} S^{-1}(f^{\ast}) u_1 g Su_2=
                  \int\limits_{\S_{q,N}} S^{-1}(u_2) S^{-1}(f^{\ast})u_1 g=
                \int\limits_{\S_{q,N}} S^{-1}\left( (u^{\ast})_1 f 
                  S(u^{\ast})_2 \right)^{\ast} g \nn\\
              &=&  (u^*\trr f, g).
\err 
Hermiticity follows using (\ref{I_star_bar}):
$$
(f, g)^* = \int\limits_{\S_{q,N}} \obar{\obar{f}} \obar{g} = 
                \int\limits_{\S_{q,N}} S^{-2}(f) \obar{g} = 
                \int\limits_{\S_{q,N}} \obar{g} f = (g, f).
$$
Using the assumption (\ref{q_phase}) for $|q|=1$, 
it is not difficult to see that they are also positive--definite.
\end{proof}

\section{Differential Calculus}
\label{sec:diff-calc}

In order to write down Lagrangians, it is convenient to use the notion 
of an (exterior) differential calculus \cite{Woronowicz:1989rt,Connes:book}. 
A covariant differential calculus over $\S_{q,N}$ is a 
graded bimodule $\Omega^*_{q,N} = \oplus_n \; \Omega^n_{q,N}$
over $\S_{q,N}$ which is a $U_q(su(2))$--module algebra, 
together with an exterior derivative $d$ which satisfies $d^2=0$ and 
the graded Leibnitz rule. We define the dimension of
a calculus to be the rank of $\Omega^1_{q,N}$
as a free right $\S_{q,N}$--module.

\subsection{First--order differential forms}
 
Differential calculi for the Podle\'s sphere have been studied before
\cite{Podlescalc,Apel:1994du}. 
It turns out that 2--dimensional calculi 
do not exist for the cases we are interested in;
however there exists a unique 3--dimensional module of 1--forms.
As opposed to the classical case, it contains an 
additional ``radial'' one--form.
This will lead to an additional scalar field, 
which will be discussed later.

By definition, it must be possible to write any term 
$x_i d x_j$ in the form $\sum_k d x_k f_k(x)$. Unfortunately the
structure of the module of 1--forms turn out to be not quadratic, 
rather the $f_k(x)$ are polynomials of order up to 3.
In order to make it more easily tractable and to find suitable
reality structures, we will construct this calculus using 
a different basis.
First, we will define the bimodule of 1--forms $\Omega^1_{q,N}$ 
over $\S_{q,N}$ which is covariant under 
$U_q(su(2))$, such that $\{d x_i\}_i$  is a free right 
$\S_{q,N}$--module basis, together with a map 
$d: \S_{q,N} \rightarrow \Omega^1_{q,N}$
which satisfies the Leibnitz rule. Higher--order differential forms 
will be discussed below.

Consider a basis of one--forms $\xi_i$ for $i=-1,0,1$ with the 
covariant commutation
relations\footnote{note that this is not the same as
$u \xi_i = u_{(1)} \trr \xi_i u_{(2)}$.}
\be
x_i \xi_j = \hat{R}^{k l}_{ij} \xi_k x_l,
\label{xxi_braid}
\ee
using the $(3) \tens (3) $  $\hat{R}$--matrix of $U_q(su(2))$.
It has the projector decomposition
\be
\hat{R}^{k l}_{ij} = q^2 (P^+)^{kl}_{ij} - q^{-2} (P^-)^{kl}_{ij} + 
                     q^{-4} (P^0)^{kl}_{ij},
\ee
where $(P^0)^{kl}_{ij} = \frac 1{[3]_q} g^{kl}g_{ij}$, and 
$(P^-)^{kl}_{ij} = \sum_n \frac 1{[2]_{q^2}}\; \eps^{kl}_n \eps^n_{ij}$.
The relations (\ref{xxi_braid}) are consistent with (\ref{xx_square}) and 
(\ref{xx_sigma}), using the braiding  relations 
\cite{Reshetikhin:1988ix,Reshetikhin:1988iw}
\berr
\hat{R}^{k l}_{ij} \hat{R}^{rs}_{lu} \eps^{ju}_n &=& 
                     \eps^{kr}_t \hat{R}^{ts}_{in}, \label{RRC_braid} \\
\hat{R}^{k l}_{ij} \hat{R}^{rs}_{lu} g^{ju} &=& g^{kr} \d^{s}_{i}
\label{RRg_braid}
\err
and the quantum Yang--Baxter equation
$\hat{R}_{12}\hat{R}_{23} \hat{R}_{12} = \hat{R}_{23}\hat{R}_{12}\hat{R}_{23}$,
in shorthand--notation \cite{Faddeev:1990ih}.
We define $\Omega^1_{q,N}$ to be the free right module over $\S_{q,N}$ 
generated by the $\xi_i$. It is clearly a bimodule
over $\S_{q,N}$. To define the exterior derivative, consider
\be
\Theta:= x \cdot \xi = x_i \xi_j g^{ij},
\ee
which is a singlet under $U_q(su(2))$.  
It turns out (see Section \ref{subsec:proofs-I}) that 
$[\Theta, x_i] \neq 0 \in  \Omega^1_{q,N}$.
Hence 
\be\fbox{$
df := [\Theta, f(x)] $}
\label{d_0}
\ee
defines a nontrivial derivation $d: \S_{q,N} \rightarrow \Omega^1_{q,N}$, 
which completes the definition of the calculus up to first order.
In particular, it is shown in Section \ref{subsec:proofs-I} that
\be
dx_i = - \La_N \eps^{nk}_i x_n \xi_k +
               (q-q^{-1})(q x_i \Theta - {r^2} q^{-1} \xi_i).
\label{dx_i}
\ee
Since all terms are linearly independent, this is a 3--dimensional 
first--order differential calculus, and by the uniqueness 
it agrees with the 3--dimensional calculus in \cite{Podlescalc,Apel:1994du}.
In view of (\ref{dx_i}), it is not surprising that the commutation relations 
between the generators $x_i$ and $d x_i$ are very complicated 
\cite{Apel:1994du}; will not write them down here. The meaning of the
$\xi$--forms will become clearer in Section \ref{subsec:frames}. 

Using (\ref{C_g}) and the relation $\xi \cdot x = q^4 x \cdot \xi$, 
one finds that
$$
x \cdot dx = \left(-\La_N^2 +([2]_{q^2} - 2) r^2\right) \Theta.
$$
On the other hand, this must be equal to 
$x_i \Theta x_j g^{ij} - r^2 \Theta$,
which implies that
$$
x_i \Theta x_j g^{ij} = \a r^2 \Theta
$$
with
\be
\a = [2]_{q^2} -1 - \frac{\La_N^2}{r^2} =  1-\frac 1{C_N}.
\ee
Combining this, it follows that 
\be
dx \cdot x = r^2\; \frac 1{C_N}\; \Theta = -  x\cdot dx.
\ee  

Moreover, using the identity (\ref{CC_id}) one finds
\be
 \eps_i^{jk} x_j dx_k = (\a-q^2) r^2  \eps_i^{jk} x_j \xi_k - \La_N r^2 \xi_i
                     + q^2 \La_N x_i \Theta,
\label{sxdx_id}
\ee
which together with (\ref{dx_i}) yields
\be 
\xi_i = \frac{q^2}{r^2}\; \Theta x_i 
       + \frac{q^2 C_N \La_N}{r^4} \eps_i^{jk} x_j dx_k
       - q^2 (1-q^2) \frac{C_N}{r^2} dx_i.
\label{xi_i}
\ee

\subsection{Higher--order differential forms}

Podle\'s \cite{Podlescalc} has constructed an extension 
of the above 3--dimensional calculus including higher--order forms
for a large class of quantum spheres. This class does not include ours,
however, hence
we will give a different construction based on $\xi$--variables,
which will be suitable for $q$ a phase as well.

Consider the algebra
\be
\xi_i \xi_j = -q^2 \hat R^{kl}_{ij} \xi_k \xi_l
\label{xixi}
\ee
which is equivalent to $(P^+)^{ij}_{kl} \xi_i \xi_j = 0$, 
$(P^0)^{ij}_{kl} \xi_i \xi_j = 0$
where $P^+$ and $P^0$ are the projectors on the symmetric
components of $(3) \tens (3)$ as above; hence the product is totally
($q$--) antisymmetric. 
It is not hard to see (and well--known) that the dimension of the 
space of polynomials of order $n$ in the $\xi$ is 
$(3,3,1)$ for $n=(1,2,3)$, and zero for $n>3$, as classically.
We define $\Omega^n_{q,N}$ to be the free right $\S_{q,N}$--module
with the polynomials of order $n$ in $\xi$ as basis; this is covariant
under $U_q(su(2))$. Then $\Omega^n_{q,N}$ is in fact a (covariant) 
$\S_{q,N}$--bimodule, since the commutation relations (\ref{xxi_braid})
between $x$ and $\xi$ are consistent with (\ref{xixi}), which follows from
the quantum  Yang--Baxter equation.  
There remains to construct the exterior derivative.
To find it, we first note that (perhaps surprisingly) $\Theta^2 \neq 0$,
rather
\be
\Theta^2 = -\frac{q^{-2} \La_N}{[2]_{q^2}} \eps^{ijk} x_i \xi_j \xi_k.
\label{theta_2}
\ee
The $\eps^{ijk}$ is defined in (\ref{q_epsilon}).
By a straightforward but lengthy calculation which is sketched 
in Section \ref{subsec:proofs-I}, one can show that
\be
dx_i dx_j g^{ij} + \frac {r^2}{C_N} \Theta^2 = 0. \nn
\ee
We will show below that an extension of the calculus to higher--order forms
exists; then this can be rewritten as
\be
d \Theta - \Theta^2 = 0.
\label{dtheta}
\ee
The fact that $\Theta^2 \neq 0$ makes the construction of the extension
more complicated, since now $\a^{(n)} \rightarrow [\Theta,\a^{(n)}]_{\pm}$
does not define an exterior derivative. To remedy this,
the following observation is useful: the map
\berr
\ast_H: \; \Omega^1_{q,N} &\rightarrow& \Omega^2_{q,N}, \nn\\
     \xi_i &\mapsto&  -\frac{q^{-2}\La_N}{[2]_{q^2}}\; \eps_i^{jk} \xi_j \xi_k
\err
defines a left--and right $\S_{q,N}$--module map; in other words, the 
commutation relations between $\xi_i$ and $x_j$ are the same as between 
$\ast_H(\xi_i)$ and $x_j$. This follows from the 
braiding relation (\ref{RRC_braid}). This is in fact the natural analogue
of the Hodge--star on 1--forms in our context, 
and will be discussed further below. Here we note the important identity
\be
\a (\ast_H \b) = (\ast_H \a) \b
\label{star_adj}
\ee
for any $\a, \b \in \Omega^1_{q,N}$, which is proved 
in Section \ref{subsec:proofs-I}.
Now (\ref{theta_2}) can be stated as
\be
\ast_H(\Theta) =  \Theta^2,
\label{ast_theta}
\ee
and applying $\ast_H$ to $df = [\Theta, f(Y)]$ one obtains
\be
[\Theta^2, f(x)] =  \ast_H df(x).
\label{comm_theta2}
\ee
Now we define the map 
\beq \fbox{$
\begin{array}{rl}
d: \; \Omega^1_{q,N} &\rightarrow \Omega^2_{q,N}, \\
                   \a &\mapsto [\Theta,\a]_+ - \ast_H(\a).
\end{array} $}
\label{d_1}
\eeq
It is easy to see that this defines a graded derivation 
from $\Omega^1_{q,N}$ to $\Omega^2_{q,N}$, and the
previous equation implies immediately that
$$
(d \circ d) f = 0.
$$
In particular,
\be 
d \xi_i = 
(1-q^2) \xi \Theta + \frac{q^{-2} \La_N}{[2]_{q^2}}\eps_i^{jk} \xi_j\xi_k.
\ee
To complete the differential calculus, we extend it to
$\Omega^3_{q,N}$ by
\beq \fbox{$
\begin{array}{rl}
d: \; \Omega^2_{q,N} &\rightarrow \Omega^3_{q,N}, \\
                   \a^{(2)} &\mapsto [\Theta,\a^{(2)}].
\end{array} $}
\label{d_2}
\eeq
As is shown in Section \ref{subsec:proofs-I}, this satisfies indeed
$$
(d \circ d) \a = 0 \quad \mbox{for any} \;\; \a \in \Omega^1_{q,N}.
$$
It is easy to see that the map (\ref{d_2}) is non--trivial.
Moreover 
there is precisely one monomial of order 3 in the $\xi$ variables, given by
\be
\Theta^3 = -\frac{q^{-6}\La_N r^2}{[2]_{q^2}[3]_q}\eps^{ijk} \xi_i \xi_j \xi_k,
\label{om_3}
\ee
which commutes with all functions on the sphere,
\be
[\Theta^3, f] = 0
\label{theta3_f}
\ee
for all $f  \in \S_{q,N}$.
Finally, we complete the definition of the Hodge star operator by 
\be
\ast_H(1) = \Theta^3,
\ee
and by requiring that $(\ast_H)^2 = id$.

\subsection{Star structure}

A $*$--calculus (or a real form of $\Omega^{*}_{q,N}$) is
a differential calculus which is a graded $*$--algebra such that the star 
preserves the grade, and satisfies \cite{Woronowicz:1989rt}
\berr
(\a^{(n)} \a^{(m)})^{\ast} &=& 
          (-1)^{nm} (\a^{(m)})^{\ast} (\a^{(n)})^{\ast}, \nn\\
(d \a^{(n)})^{\ast} &=& d(\a^{(n)})^{\ast}
\label{real_forms}
\err
for $\a^{(n)} \in \Omega^n_{q,N}$; moreover, the action of 
$U_q(su(2))$ must be compatible with the star on $U_q(su(2))$. 
Again, we have to distinguish the cases $q \in \R$ and $|q|=1$.

\paragraph{ \bf 1) $q \in \R$.}
In this case, the star structure must satisfy
\be
(dx_i)^{\ast} =  g^{ij} dx_j, \quad x_i^{\ast} = g^{ij} x_j,
\label{x_calc_real}
\ee
which by (\ref{xx_square}) implies 
\be
\Theta^{\ast} = -\Theta.
\label{theta_real}
\ee
Using (\ref{xi_i}), it follows that
\berr
\!\!\!\xi_i^{\ast} &=& -g^{ij}\xi_j +  
         q^2 (q-q^{-1})\frac{[2]_q C_N}{r^2 } g^{ij} d x_j \nn \\
         &=& -g^{ij}\xi_j  - q^2 (q-q^{-1})\frac{[2]_q C_N}{r^2 }
                   g^{ij}\left( \La_N \eps_j^{kl} x_k \xi_l
              - (q-q^{-1})(q x_j \Theta - q^{-1} r^2 \xi_j )\right). 
\label{xi_real}
\err
To show that this is indeed compatible with (\ref{xxi_braid}),
one needs the following identity
\be
q^2(q -q^{-1})\frac{[2]_q C_N}{r^2} 
     (dx_i x_j - {\hat{R}}^{k l}_{ij} x_k dx_l)
 = (\one - (\hat{R}^2)^{k l}_{ij}) \xi_k x_l
\label{reality_id}
\ee
which can be verified with some effort, see Section \ref{subsec:proofs-I}.
In particular, this shows that if one imposed
$x_i \xi_j = (\hat{R}^{-1})^{k l}_{ij} \xi_k x_l$ instead of 
(\ref{xxi_braid}), one would obtain an equivalent calculus.
This is unlike in the flat case, where one has two inequivalent
calculi \cite{Fiore:1993mb,Ogievetsky:1992qp}.
Moreover, one can show that this real form is consistent
with (\ref{xixi}).

\paragraph{\bf 2) $|q| =1$.}
In view of (\ref{x_star_phase}), it is easy to see that  
the star structure in this case is
\be
(\xi_i)^{\ast} =  q^{-4} \tw \xi_i \tw^{-1}, 
        %=  - \xi_j {L^-}^{j}_{k} q^{-6} g^{k i}, 
               \quad x_i^{\ast} = - \tw x_i \tw^{-1}.
\label{x_calc_phase}
\ee
Recall that $\tw$ is a particular unitary element of $\S_{q,N}$ introduced in 
Section \ref{subsec:real_form}. 

It is obvious using 
$({\hat R}^{k l}_{ij})^* = (\hat{R}^{-1})^{lk}_{ji}$
that this is an involution which is consistent with (\ref{xxi_braid}),
and one can verify that 
\be
\Theta^{\ast} = - \Theta.
\ee
This also implies 
$$
[\tw, \Theta] =0,
$$
hence 
\be
(d x_i)^{\ast} = - \tw d x_i \tw^{-1}.
\ee
Finally, $\ast_H$ is also compatible with the star structure:
\be
(\ast_H(\a))^{\ast} = \ast_H(\a^{\ast})
\ee
where $\a \in \Omega^1_{q,N}$,
for both $q \in \R$ and $|q| =1$. This is easy to see
for $\a = \xi_i$ in the latter
case, and for $\a = dx_i$ in the case $q \in \R$.
This implies that indeed
$(d\a^{(n)})^{\ast} = d(\a^{(n)})^{\ast}$
for all $n$.

We summarize the above results:
\begin{theorem}
The definitions (\ref{d_1}), (\ref{d_2}) define a 
covariant differential calculus 
on $\Omega^{*}_{q,N} = \oplus_{n=0}^3 \; \Omega^n_{q,N}$
over $\S_{q,N}$ with $dim(\Omega^n_{q,N}) = (1,3,3,1)$ for $n=(0,1,2,3)$. 
Moreover, this is a $*$--calculus with the 
star structures (\ref{x_calc_real}) 
and (\ref{x_calc_phase}) for $q \in \R$ and $|q|=1$, respectively.
\end{theorem}

\subsection{Frame formalism}
\label{subsec:frames}

On many noncommutative spaces \cite{Cerchiai:2000qu,Mad99c}, it is possible 
to find a particularly 
convenient set of one--forms (a ``frame'') $\theta_a \in \Omega^1$,
which commute with all elements in the function space $\Omega^0$.
Such a frame exists here as well, and in terms of the
$\xi_i$ variables, it takes a similar form to that of \cite{Cerchiai:2000qu}. 
Consider the elements
\berr
\theta^a &=&  \La_N \; S({L^+}^a_j)\; g^{jk} \xi_k \;\; \in \Omega^1_{q,N}, \\
\la_a     &=& \frac 1{\La_N}\; 
             x_i \; {L^+}^i_a \qquad \;\; \in \S_{q,N}.
\label{vielbein}
\err
where as usual
\berr
{L^+}^i_j &=& \RR_1 \pi^i_j(\RR_2),    \label{L_plus} \\
S({L^+}^i_j) &=& \RR_1^{-1} \pi^i_j(\RR^{-1}_2)      
\err
are elements of $U_q(su(2))$, which we
consider here as elements in $\S_{q,N}$ via (\ref{j_map}).
Then the following holds:
\begin{lemma}
\berr
[\theta^a, f] &=& 0,      \label{frame}  \\
d f              &=& [\la_a, f] \theta^a,   \label{d-frame}   \\
\Theta = x_i \xi_j g^{ij} &=& \la_a \theta^a.   \label{Theta_decomp}
\err
for any $f \in \S_{q,N}$. In this sense, 
the $\la_a$ are dual to the frame $\theta^b$. They satisfy the relations
\berr
\la_a \la_b g^{ba} &=& \frac 1{q^{4} \La_N^{2}} \; r^2, \nn\\
\la_a \la_b \eps_c^{ba} &=& -\frac 1{q^{2}} \; \la_c, \nn\\
\theta^a \theta^b  &=& -q^2 \;\hat R^{ba}_{cd}\theta^d \theta^c  \label{theta_AS}\\
d \theta^a &=& \la_b [\theta^a,\theta^b]_+ 
        + \frac 1{q^{2} [2]_{q^2}} \eps^a_{bc} \theta^c \theta^b \nn\\
\ast_H \theta^a    &=& - \frac 1{q^{2}[2]_{q^2}} 
                       \eps^a_{bc} \theta^c \theta^b \nn\\
\theta^a \theta^b \theta^c &=& -\La_N^2 \frac{q^6}{r^2} \;\eps^{cba} \Theta^3
\label{l_relations}
\err
\end{lemma}
In particular in the limit $q =1$, this becomes  
$\la_a = \frac 1{\La_N}\; x_a$,
and $dx_a = - \eps_{ab}^c x_c \theta^b$, using (\ref{dx_i}).

\begin{proof}
Using 
$$
S({L^+}^i_j) x_k = x_l ({\hat R}^{-1})^{ln}_{jk} S({L^+}^i_n)
$$
(which follows from (\ref{smash})) and $\Delta(S({L^+}^i_j)) = 
S({L^+}^n_j) \tens S({L^+}^i_n)$, 
it is easy to check that 
$[\theta^a, x_i] =0$ for all $i,a$, and (\ref{frame}) follows.
(\ref{Theta_decomp}) follows immediately from 
${L^+}^i_a S({L^+}^a_j) = \d^i_j$, and 
To see (\ref{l_relations}), one needs 
the well--known relation ${L^+}^{l}_r {L^+}^k_s g^{sr} = g^{kl}$, as
well as ${L^+}^{l}_r {L^+}^k_s  \eps_n^{sr} = \eps^{kl}_m {L^+}^m_n$;
the latter follows from the quasitriangularity of $U_q(su(2))$. 
The commutation relations among the $\theta$ are obtained 
as in \cite{Cerchiai:2000qu} by observing
\berr
\theta^a \theta^b  &=& 
         \La_N \theta^a  S({L^+}^b_n) g^{nl} \xi_l \nn\\
 &=&     \La_N S({L^+}^b_n) \theta^a g^{nl} \xi_l    \nn\\
 &=&     \La_N^2 S({L^+}^b_n) S({L^+}^a_j) g^{jk}  g^{nl}  \xi_k \xi_l,
\err
using the commutation relations
$\hat R^{kl}_{ij} S{L^+}^i_n  S{L^+}^j_m = 
S{L^+}^k_i  S{L^+}^l_j \hat R^{ij}_{nm}$, as well as (\ref{RRg_braid}).
The remaining relations can be checked similarly.
\end{proof}

\subsection{Integration of forms}
\label{subsec:integral_forms}

As classically, it is natural to define the integral over the forms of
the highest degree, which is $3$ here. Since any $\a^{(3)} \in \Omega^3_{q,N}$
can be written in the from $\a^{(3)} = f \Theta^3$, we define
\be
\int \a^{(3)} = 
\int f \Theta^3 := \int\limits_{\S_{q,N}} f
\ee
by (\ref{integral}), so that $\Theta^3$ is the volume form.
This definition is natural, since $[\Theta^3, f] = 0$.
Integrals of forms with degree $\neq 3$ will be set to zero. 

This integral satisfies an important cyclic property, as did
the quantum trace (\ref{cyclic_integral}).
To formulate it, we extend the map $S^2$ from $\S_{q,N}$ to $\Omega^*_{q,N}$
by 
$$
S^2(\xi_i) = q^{H} \trr \xi_i,
$$ 
extended as an algebra map. 
Then the following holds (see Section \ref{subsec:proofs-I}): 
\be
\int \a \; \b = \int S^{-2}(\b)\; \a
\label{cyclic_forms}
\ee
for any $\a, \b \in \Omega^*_{q,N}$ with $deg(\a) + \deg(\b) = 3$.
Now Stokes theorem follows immediately:
\be
\int d\a^{(2)} =  \int [\Theta,\a^{(2)}] = 0
\ee
for any $\a^{(2)} \in  \Omega^2_{q,N}$, because $S^2 \Theta = \Theta$.
This purely algebraic derivation is also valid on some other spaces 
\cite{Steinacker:1995jh}.

Finally we establish the compatibility of the integral with the star structure.
From $\Theta^{\ast} = -\Theta$ and (\ref{I_star_real}), 
we obtain 
\be
(\int \a^{(3)})^* = -\int (\a^{(3)})^{\ast} \qquad \mbox{for} \;\; q\in \R.
\ee 
For $|q|=1$, we have to extend the algebra map $\obar{f}$ (\ref{obar}) to 
$\Omega^*_{q,N}$. It turns out that the correct definition is
\be
\obar{\xi_i} = - q^{-4}\;g^{ij}\xi_j +  
            q^{-2} (q-q^{-1})\frac{[2]_q C_N}{r^2} g^{ij} d x_j,
\label{obar_xi}
\ee
extended as an antilinear algebra map; 
compare (\ref{xi_real}) for $q\in\R$. To verify that 
this is compatible with (\ref{xxi_braid}) and (\ref{xixi})
requires the same calculations as to verify the star structure 
(\ref{xi_real}) for $q \in \R$. Moreover one can check using 
(\ref{sxdx_id}) that 
\be
\obar{dx_i} = -g^{ij} dx_j,
\ee
which implies that $\obar{\Theta} = \Theta$, and
\berr
\obar{\ast_H(\a)} &=& \ast_H(\obar{\a}), \nn\\
\obar{d\a} &=& d\obar{\a},  \nn\\
\obar{\obar{\a}} &=& S^{-2} \a
\err
for any $\a \in \Omega^*_{q,N}$. Hence we have
\be
(\int \a^{(3)})^* = \int \obar{\a^{(3)}} \qquad \mbox{for} \;\; |q|=1.
\ee

\section{Actions and fields}

\subsection{Scalar fields}
\label{subsec:scalars}

With the tools provided in the previous sections, it is possible to 
construct actions for 2--dimensional euclidean field theories 
on the $q$--deformed fuzzy sphere. 

We start with scalar fields, which are simply elements 
$\psi \in \S_{q,N}$. 
The obvious choice for the kinetic term is 
\berr
S_{kin}[\psi] &=& i \frac {r^2}{\La_N^2}\int (d\psi)^{\ast} \ast_H d \psi 
                             \qquad \mbox{for} \;\; q\in \R, \nn\\
S_{kin}[\psi] &=&  \frac {r^2}{\La_N^2}\int \obar{d\psi} \ast_H d \psi 
                             \qquad \mbox{for} \;\; |q|=1,
\err
which, using Stokes theorem, can equivalently be written in the form
\berr
S_{kin}[\psi] &=& - i \frac {r^2}{\La_N^2}\int \psi^{\ast} (d \ast_H d) \psi 
  = - \frac {r^2}{\La_N^2} i\int\limits_{\S_{q,N}} \psi^{\ast} (\ast_H d \ast_H d) \psi 
                            \qquad \mbox{for} \;\; q\in \R, \nn\\
S_{kin}[\psi] &=& -  \frac {r^2}{\La_N^2}\int \obar{\psi} (d \ast_H d) \psi 
   =- \frac {r^2}{\La_N^2}\int\limits_{\S_{q,N}} \obar{\psi} (\ast_H d \ast_H d) \psi 
                            \qquad \mbox{for} \;\;  |q|=1. \nn\\
\err
They are real
\be
S_{kin}[\psi]^* = S_{kin}[\psi] 
\ee
for both $q \in \R$ and $|q| =1$, using the reality properties
established in the previous sections. 

The fields can be expanded in terms of the irreducible \reps
\be
\psi(x) = \sum_{K,n} a^{K,n} \; \psi_{K,n}(x) 
\label{psi_field-0}
\ee
according to (\ref{A_decomp}), with coefficients $a^{K,n} \in \C$;
this corresponds to the first--quantized case. 
However, in order to ensure invariance of the actions
under $U_q(su(2))$ (or a suitable subset thereof),
we must assume that $U_q(su(2))$ acts on products of fields
via the $q$--deformed coproduct. This can be implemented 
consistently only after a ``second quantization'', such that the coefficients
in (\ref{psi_field-0}) generate a $U_q(su(2))$--module algebra. 
This will be presented in Chapter \ref{chapter:qFSII}.

One can also consider
real fields, which have the form 
\berr
\psi(x)^* &=& \psi(x)  \qquad \mbox{for} \;\; q\in \R, \nn\\
\obar{\psi(x)} &=& \psi(x)   \qquad \mbox{for} \;\; |q|=1.
\err
This is preserved under the action of a certain real sector 
$\cG \subset U_q(su(2))$ (\ref{gauge_group}); 
the discussion is completely parallel to the one 
below (\ref{B_gauge}) in the next section, hence we will not give it here.
%This amounts to $a^{K,n} g^{nm}= a^{K,m}$, or something like that.

Clearly $\ast_H d \ast_H d$ is the analog of the Laplace operator for 
functions, which can also be written in the usual form
$d \d + \d d$, with $\d = \ast_H  d \ast_H$.
It is hermitian by construction. 
We wish to evaluate it on the irreducible \reps 
$\psi_K \in (2K+1)$, that is, on spin--$K$ representations. The result is the 
following:
\begin{lemma}
If $\psi_K \in \S_{q,N}$ is a spin $K$ representation, then 
\be
\ast_H d\ast_H d \psi_{K} = 
   \frac{2}{[2]_q C_N} [K]_q [K+1]_q \; \psi_{K}.
\label{scalar_laplace}
\ee
\label{laplace_lemma}
\end{lemma}
The proof is in Section \ref{subsec:proofs-I}.

It is useful to write down explicitly the hermitian forms associated to
the above kinetic action. Consider
\berr
S_{kin}[\psi,\psi'] &=& i \frac {r^2}{\La_N^2}\int (d\psi)^{\ast} 
                              \ast_H d \psi'
                             \qquad \mbox{for} \;\; q\in \R, \nn\\
S_{kin}[\psi,\psi'] &=&  \frac {r^2}{\La_N^2}\int \obar{d\psi} \ast_H d \psi'
                             \qquad \mbox{for} \;\; |q|=1.
\err
Using Lemma \ref{inner_product_lemma}, it follows immediately 
that they satisfy 
\berr
S_{kin}[\psi,\psi']^* &=& S_{kin}[\psi',\psi],  \nn\\
S_{kin}[\psi, u\trr \psi']  &=& S_{kin}[u^{\ast} \trr \psi, \psi']
\err
for both $q\in\R$ and $|q|=1$. 
To be explicit, let $\psi_{K,n}$ be an orthonormal basis of (2K+1). We
can be assume that it is a weight basis, so that $n$ labels the weights
from $-K$ to $K$. Then it follows that
\be
S_{kin}[\psi_{K,n},\psi_{K',m}] = c_{K}\; \d_{K,K'}  \;\d_{n,m}
\ee
for some $c_K \in \R$.
Clearly one can also consider interaction terms, which could be of the form
\be 
S_{int}[\psi] = \int\limits_{\S_{q,N}} \psi \psi \psi,
\ee
or similarly with higher degree.

\subsection{Gauge fields}
\label{subsec:gauge}

Gauge theories arise in a very natural way on $\S_{q,N}$. For simplicity,
we consider only the analog of the abelian gauge fields here. 
They are simply one--forms
\be
B = \sum B_a \theta^a r  \quad \in \Omega^1_{q,N},
\label{B_expand}
\ee 
which we expand in terms of the frames $\theta^a$ introduced in Section
\ref{subsec:frames}. Notice that they have 3 independent components, which 
reflects the fact that calculus is 3--dimensional. Loosely speaking, the 
fuzzy sphere does see a shadow of the 3--dimensional embedding space. 
One of the components is essentially radial and should be considered
as a scalar field, however it is naturally tied up with the 
other 2 components of $B$. 
We will impose the reality condition
\berr
B^* &=& B  \qquad \mbox{for} \;\; q\in \R, \nn\\
\obar{B} &=& q^{-\frac H2} \trr B  \qquad \mbox{for} \;\; |q|=1.
\label{B_real}
\err
Since only 3--forms can be integrated, the most simple candidates for 
Langrangians that can be written down have the form
\be
S_3 = \frac {1}{r^2\La_N^2} \int B^3, \quad 
S_2 = \frac {1}{r^2\La_N^2} \int B \ast_H B, 
    \quad S_4 = \frac {1}{r^2\La_N^2} \int B^2 \ast_H B^2.
\label{B_actions}
\ee
They are clearly real, with the reality condition (\ref{B_real});
the factor $i$ for real $q$ is omitted here. We also define 
\be
F := B^2 - \ast_H B,
\label{curvature}
\ee
for reasons which will become clear below.
The meaning of the field $B$ becomes obvious if one writes it in the form 
\be
B = \Theta + A, \qquad B_a = \frac 1{r}\la_a + A_a
\label{B_A_split}
\ee
While $B$ and $\Theta$ become singular in the limit $N \rightarrow \infty$, 
$A$ remains well--defined. Using 
\berr
F &=& dA + A^2, \nn\\
\int A \Theta^2 &=& \int d A \Theta = \int  \ast_H A \; \Theta,  \nn\\
\int A^2 \Theta &=& \frac 12 \int(A d A + A  \ast_H A)
\err
which follow from (\ref{d_1}), one finds
\berr
S_2 &=& \frac {1}{r^2\La_N^2} \int  A  \ast_H A + 2 A \Theta^2  \nn\\
S_3 &=& \frac {1}{r^2\La_N^2} \int  A^3 + \frac 32(A dA +  A  \ast_H A)  
        + 3 A \Theta^2 +  \Theta^3 
\err
and {\large
\be \fbox{$
S_{YM} := \frac {1}{r^2\La_N^2} \int F \ast_H F = 
        \frac {1}{r^2\La_N^2} \int (dA + A^2) \ast_H (dA + A^2). $}
\label{YM}
\ee}
The latter action (which is a linear combination of $S_2, S_3$, and $S_4$) 
is clearly the analog of the Yang--Mills action, 
which in the classical limit contains a gauge field an a scalar,
as we will see below.
In the limit $q \rightarrow 1$, it reduces to the action considered in 
\cite{Grosse:1992bm}.  

The actions $S_3$ and $S_2$ alone contain terms which are linear in $A$, 
which would indicate
that the definition of $A$ (\ref{B_A_split}) is not appropriate. 
However, the linear terms cancel in the following linear combination
{\large
\be\fbox{$
S_{CS} := \frac 13 S_3 - \frac 12 S_2 = - \frac{2\pi}{3\La_N^2}
    + \frac 12 \frac {1}{r^2\La_N^2} \int A dA + \frac 23 A^3. $}
\label{S_CS}
\ee}
Notice that the ``mass term'' $A \ast_H A$ has also disappeared. 
This form is clearly the 
analog of the Chern--Simons action. It is very remarkable that it
exists on $\S_{q,N}$, which is related to the fact that the calculus 
is 3--dimensional. In the case $q=1$, 
this is precisely what has been found recently in the context of
2--branes on the $SU(2)$ WZW model \cite{Alekseev:2000fd}.

In terms of the components (\ref{B_expand}), 
$B^2 = B_a B_b \theta^a \theta^b r^2$, and
$\ast_H B = -\frac {r}{q^2 [2]_{q^2}} B_a \eps^a_{bc}\theta^c\theta^b$. 
Moreover, it is easy to check that 
\berr
\ast_H(\theta^b \theta^c) &=& -q^2 \; \eps^{cb}_a \theta^a,\nn\\
\theta^a \ast_H \theta^b &=& \La_N^2 \frac{q^4}{r^2} g^{ba} \;\Theta^3, \nn\\
\theta^a \theta^b \ast_H \theta^c\theta^d &=& 
      [2]_{q^2}\La_N^2 \frac {q^8}{r^2} 
       (P^-)^{dc}_{a' b'}\; g^{b' b} g^{a' a}\; \Theta^3
   =  \La_N^2 \frac {q^8}{r^2} \eps^{dc}_n \eps^{ba}_m g^{nm} \Theta^3.
\err
Hence 
\berr
F &=&(B_a B_b + \frac 1{q^2 r [2]_{q^2}} B_c\; \eps^c_{ba}) 
          \theta^a \theta^b r^2=
     (\frac{\la_a}{r} A_b + A_a \frac{\la_b}{r} + A_a A_b 
           + \frac 1{q^2 r [2]_{q^2}} A_c \;\eps^c_{ba}) 
           \theta^a \theta^b r^2\nn\\
 &=& F_{ab} \; \theta^a \theta^b r^2,
\label{F_BB}
\err
where we define  $F_{ab}$ to be totally antisymmetric, i.e. 
$F_{ab} = (P^-)^{b' a'}_{ba} F_{a' b'}$ using (\ref{theta_AS}). 
This yields
\beq
S_{YM} = q^8 [2]_{q^2}\int\limits_{\S_{q,N}} F_{ab} F_{cd}\; 
                                (P^-)^{dc}_{a' b'}\; g^{b' b} g^{a' a}, 
\label{gauge_actions_components}
\eeq
To understand these actions better, we write the gauge fields in terms 
of ``radial'' and ``tangential'' components,
\be
A_a = \frac {x_a}{r}\phi + A^t_a 
\ee
where $\phi$ is defined such that 
\be
x_a A^t_b \; g^{ab} = 0; 
\label{A^t}
\ee
this is always possible. However to get a better insight,
we consider the case $q=1$, and take
the classical limit $N \rightarrow \infty$ in the 
following sense: for a given (smooth) field configuration in $\S_N$, we 
use the sequence of embeddings of $\S_{q,N}$ 
to approximate it for $N \rightarrow \infty$. 
Then terms of the form $[A^t_a, A^t_b]$
vanish in the limit (since the fields are smooth in the limit).
The curvature then splits into a tangential and radial part, 
$F_{ab} = F^t_{ab} + F^{\phi}_{ab}$, where\footnote{the pull--back of
$F$ to the 2--sphere in the classical case is unaffected by this split} 
\berr
F^t_{ab} &=& \frac 1{2r} \left([\la_a, A^t_b] - [\la_b, A^t_a]
             + A^t_c \;\eps^c_{ba}\right) , \nn\\
F^{\phi}_{ab} 
      &=& \frac 1{2r^2} \left(\eps_{ab}^c x_c \phi + [\la_a,\phi] x_b
          - [\la_b, \phi] x_a \right).
\err
Moreover,  
\berr
x^a F^t_{ab} &\rightarrow&  \frac 1{4r} [x^a \la_a, A^t_b] 
          - \frac 1{2r} [\la_b, x^a A^t_a] = 0 ,\nn\\
x^a [\la_a,\phi] &\rightarrow& \frac 12[x^a \la_a,\phi] = 0
\label{approx}
\err
in the classical limit, which implies that 
\berr
\int\limits_{S^2} F^t_{ab} {F^{\phi}}^{ab} &=& \int\limits_{S^2} 
          \frac1{2r^2} \eps_{ab}^n x_n \phi {F^t}^{ab}, \nn\\
\int\limits_{S^2} F^{\phi}_{ab} {F^{\phi}}^{ab}
       &=& \int\limits_{S^2} \frac 1{2r^2} 
         \left( \phi^2 + [\la_a, \phi] [\la^a, \phi] \right) \nn
\err
in the limit. Therefore we find
{\large
\be\fbox{$
S_{YM} =  -\int\limits_{S^2} \left( 2F^t_{ab} {F^t}_{ab}
         +  \frac2{r^2} \eps_{ab}^n x_n \phi {F^t}^{ab}
         + \frac 1{r^2} (\phi^2 + [\la_a, \phi] [\la^a, \phi] ) \right)$}
\ee}
in the limit, as in \cite{Grosse:1992bm}.
Notice that in the flat limit $r \rightarrow \infty$, the  
$F-\phi$ coupling term vanishes. 
Similarly, the Chern--Simons action (\ref{S_CS} ) becomes
\berr
S_{CS}  &\rightarrow& - \frac{2\pi}{3\La_N^2} + 
         \frac {1}{2r^2\La_N^2} \int dA^t (A^t + 2\La_N \Theta\phi)
                 - \La_N^2 \phi^2 \Theta^3   \nn\\
  &=& - \frac{2\pi}{3\La_N^2} + 
      \frac {1}{2r} \int\limits_{S^2} F^t_{ab} (A_c^t + 2\frac{x_c}r \phi)
      \eps^{abc} - \frac {1}{2r^2} \int\limits_{S^2} \phi^2 
\err
for $N \rightarrow \infty$. In the flat limit $r \rightarrow \infty$, the  
term $ F^t_{ab} A_c^t\eps^{abc}$ vanishes because of (\ref{approx}), 
leaving the $F-\phi$ coupling term (after multiplying with $r$).

Back to finite $N$ and $q\neq 1$.
To further justify the above definition of curvature (\ref{curvature}),
we consider the zero curvature condition, $F=0$. 
In terms of the $B$ fields, this is 
equivalent to 
\be
\eps^{ba}_c B_a B_b + \frac 1{q^2 r} B_c = 0
\ee
which is (up to rescaling) the same as equation (\ref{xx_sigma}) with 
{\em opposite} multiplication\footnote{this can be implemented e.g. using the 
antipode of $U_q(su(2))$}; in particular, the solutions 
$B_a \in \S_{q,N}$ are precisely all possible representations of 
$U_q^{op}(su(2))$
in the space of $N+1$--dimensional matrices. They are of course
classified by the number of partitions of $Mat(N+1)$  into blocks with
sizes $n_1, ..., n_k$ such that $\sum n_i = N+1$, as in the case $q=1$.

These solutions can be interpreted as fuzzy spheres of various 
sizes. Quite remarkably, this shows that gauge theory on the fuzzy sphere
($q$-deformed or not) is more that we anticipated: it describes
in fact a gauge theory on various (superpositions of) 
fuzzy spheres simultaneously, and one can expect that even transitions 
between different spheres can occur. This is indeed the case, and
will be described in Chapter \ref{chapter:fuzzyinst}
in the undeformed case. This is very much in accord with the 
picture of $D$-branes on group manifolds as described in 
\cite{Alekseev:2000fd}.

\paragraph{Gauge invariance.}

We have seen that actions which  describe gauge theories in the limit $q=1$
arise very naturally on $\S_{q,N}$ (as on certain other 
higher--dimensional $q$--deformed spaces \cite{Steinacker:1997za}).
However, it is less obvious in which sense they are actually 
gauge--invariant for $q \neq 1$. 
For $q=1$, the appropriate gauge transformation is
$B \rightarrow U B U^{-1}$, for a unitary element $U \in \S_{N}$. 
This transformation does not work for $q \neq 1$, because of 
(\ref{cyclic_forms}). Instead, we propose the following: let
\berr
\cH &=& \{\g \in U_q(su(2)): \;\; \eps(\g) = 0, \; \g^* = S\g \}, \nn\\
\cG &=& \{\g \in U_q(su(2)): \;\; \eps(\g) = 1, \; \g^* = S\g \} = e^{\cH},
\label{gauge_group}
\err
for $q \in \R$;
for $|q|=1$, the $S\g$ on the rhs should be replaced by 
$S_0(\g) = q^{-\frac H2} S(\g) q^{\frac H2}$.
Clearly $\cH$ is a subalgebra (without 1) of $U_q(su(2))$, and $\cG$
is closed under multiplication. Using the algebra map $j$ 
(\ref{j_map}), $\cH$ can be mapped to some real sector of the space of
functions on the fuzzy sphere.

Now consider the following ``gauge'' transformations:
\be
B \rightarrow j(\g_{(1)}) B j(S \g_{(2)}) \qquad \mbox{for} \;\; \g \in \cG.
\label{B_gauge}
\ee
It can be checked easily that these transformations preserve the reality
conditions (\ref{B_real}) for both real $q$ and $|q|=1$. 
In terms of components $B = B_a \theta^a r$, this transformation is 
simply (suppressing $j$)
\be
B_a \rightarrow \g_{(1)} B_a S \g_{(2)} = \g \trr B_a,
\ee
which is the rotation of the fields $B_a \in \S_{q,N}$ 
considered as scalar fields\footnote{notice that this is {\em not}
the rotation of the one--form $B$, because 
$\g_{(1)} \xi_i S \g_{(2)} \neq \g\trr \xi_i$}, 
i.e. the rotation $\g \in U_q(su(2))$
does not affect the index $a$ because of (\ref{frame}). In terms of 
the $A_a$ variables, this becomes
\be
A_a \theta^a \rightarrow \g_{(1)} A_a S \g_{(2)}\theta^a + 
\g_{(1)} d( S \g_{(2)})
 = (\g \trr A_a) \theta^a + \g_{(1)} d( S \g_{(2)}),
\label{A_gauge}
\ee
using (\ref{d_0}) and (\ref{frame}).
Hence these transformations are a mixture
of rotations of the components (first term) and 
``pure gauge transformations'' (second term).  Moreover,
the radial and tangential components get mixed. 

To understand these transformations better, consider 
$q=1$. Then we have two transformations of a given gauge field $B_a$, 
the first by conjugation with an unitary element $U \in \S_{q,N}$, and 
the second by (\ref{B_gauge}). We claim that the respective spaces of
inequivalent gauge fields are in fact equivalent. 
Indeed, choose e.g. $a=1$; then there exists a unitary $U \in \S_{q,N}$ 
such that $U^{-1} B_a U$ is a diagonal matrix with real entries.
On the other hand, using a suitable $\g \in \G$ and recalling (\ref{A_decomp}),
one can transform $B_a$ into the form $B_a = \sum_i b_i (x_0)^i$
with real $b_i$, which is again represented
by a diagonal matrix in a suitable basis. Hence at least generically,
the spaces of inequivalent gauge fields are equivalent.

One can also see more intuitively that 
(\ref{A_gauge}) corresponds to an abelian gauge transformation
in the classical limit. 
Consider again $\g(x) = e^{i h(x)}$ with $h(x)^* = -Sh(x)$, 
approximating a smooth function in the limit $N \rightarrow \infty$.
Using properly rescaled variables $x_i$, 
one can see using (\ref{U_A_relation}) that if viewed as an element in 
$U(su(2))$, $\g$ approaches the identity,
that is $\g \trr A_a(x) \rightarrow A_a(x)$ in the classical limit. 
Now write the functions on $\S_{N}$  
in terms of the variables $x_1$ and $x_{-1}$, for example. 
Then (\ref{U_A_relation}) yields
\be
(\one\tens S)\Delta(x_i) = x_i \tens 1 - 1 \tens x_i,
\ee
for $i = \pm 1$, and one can see that  
\be
\g_{(1)} [\la_i, S \g_{(2)}] \approx \partial_i h(x_i)
\ee
in the (flat) classical  limit. Hence (\ref{A_gauge}) indeed 
becomes a gauge transformation in the classical limit. 

To summarize, we found that the set of gauge transformations 
in the noncommutative case is a (real sector of a) quotient of $U_q(su(2))$, 
and can be identified with the space of (real) functions
on $\S_{q,N}$ using the map $j$.
However, the transformation of products of fields is different from the 
classical case. Classically, the gauge group acts on products 
``componentwise'', which means that the coproduct is trivial.
Here, we must assume that $U_q(su(2))$ acts on products of fields
via the $q$--deformed coproduct, so that 
the above actions are invariant under gauge transformations,
by (\ref{invariance_int}). 
In particular, the ``gauge group'' has become a real sector of
a Hopf algebra. Of course, this can be properly implemented on the fields 
only after a ``second quantization'', as in the case of rotation invariance
(see Section \ref{subsec:scalars}). This will be presented in 
Chapter \ref{chapter:qFSII}. This picture is also quite consistent with 
observations of a BRST--like structure in $U_q(so(2,3))$ at roots of unity,
see \cite{Steinacker:1997za,Steinacker:1997qz}.

Finally, we point out that the above actions are invariant under a global 
$U_q(su(2))$ symmetry, by rotating the frame $\theta^a$.

\section{Discussion}

In this chapter, we studied the $q$--deformed fuzzy sphere $\S_{q,N}$,
which is an associative algebra which is covariant under $U_q(su(2))$, 
for real $q$ and $q$ a phase. In the first 
case, this is the same as the ``discrete series'' of Podle\'s quantum spheres.
We developed the formalism of differential forms and frames, as well as 
integration. We then briefly considered
scalar and gauge field theory on this space. 
It appears that $\S_{q,N}$ is a nice and perhaps the simplest example of
a quantum space which is covariant under a quantum group.
This makes it particularly well suited for studying field theory,
which has proved to be rather difficult on other $q$--deformed
spaces. We were able to write hermitian actions for scalar and gauge fields,
including analogs of Yang--Mills and Chern--Simons actions. 
In particular, the form of the actions for gauge theories suggests a new
type of gauge symmetry, where the role of the gauge group is played by 
$U_q(su(2))$, which can be mapped onto the space of functions on $\S_{q,N}$.
This suggests that formulating field theory on 
quantized spaces which are not based on a 
Moyal-Weyl star product is qualitatively different, 
and may lead to interesting new insights. 

To put all this into perspective, we recall that
the main motivation for considering $\S_{q,N}$ was
the discovery \cite{Alekseev:1999bs} that 
a quasi--associative twist of $\S_{q,N}$
arises on spherical $D$--branes in the $SU(2)_k$ WZW model, for $q$
a root of unity. In view of this,  
we hope that the present formalism may be useful 
to formulate a low--energy effective field theory 
induced by open strings ending on the $D$--branes.
This in turn suggests to consider also the second quantization of 
field theories on $\S_{q,N}$, corresponding to a loop expansion and 
many--particle states. It is quite interesting that even from a purely 
formal point of view, a second quantization seems necessary 
for a satisfactory implementation of the $q$-deformed 
symmetries in such a field theory. 
This is the subject of  Chapter \ref{chapter:qFSII}.
Finally, while the question of using either the quasi--associative algebra
(\ref{quasiass-ope}) or the associative $\S_{q,N}$ 
(\ref{ass-ope}) may 
ultimately be a matter of taste, the latter does suggest 
certain forms for Lagrangians, induced by the
differential calculus. It would be very interesting 
to compare this with a low--energy effective action 
induced from string theory beyond the leading term in a $1/k$ expansion.

\section{Technical complements to Chapter \ref{chapter:qFSI}}

\subsection{Invariant tensors for $U_q(su(2))$}
\label{subsec:inv-su2}

The general properties of invariant tensors were explained in 
Section \ref{subsec:reps}. 
The $q$--deformed epsilon--symbol (``spinor metric'') for spin 1/2 \reps  
is given by
\be
\eps^{+ -} = q^{\haf}, \;\; \eps^{- +} = - q^{-\haf},
\ee
all other components are zero. The corresponding tensor with lowered indices
is $\eps_{\a \b} = -\eps^{\a \b}$ and satisfies 
$\eps^{\a \b} \eps_{\b \g} = \d^{\a}_{\g}$. In particular, 
$\eps^{\a \b} \eps_{\a \b} = -(q + q^{-1}) = -[2]_q$.

The  $q$--deformed sigma--matrices, i.e. the Clebsch--Gordon coefficients
for $(3) \subset (2) \tens (2)$, are given by 
\berr
\sigma^{+ +}_{1} &=& 1 = \;\sigma^{- -}_{-1}, \nn\\
\sigma^{+ -}_{0} &=& \frac{q^{-\haf}}{\sqrt{[2]_q}}, \;\;
\sigma^{- +}_{0} = \frac{q^{\haf}}{\sqrt{[2]_q}}
\err
in an orthonormal basis, and 
$\sigma^{\a \b}_{i} = \sigma_{\a \b}^{i}$.
They are normalized such that 
$\sum_{\a \b} \sigma_{\a \b}^{i} \sigma^{\a \b}_{j} = \d^i_j$.
That is, they define a unitary map (at least for $q \in \R$).

The $q$--deformed invariant tensor for spin 1 \reps  
is given by
\be
g^{1 -1} = -q, \;\; g^{0 0} = 1, \;\; g^{-1 1} = -q^{-1},
\ee
all other components are zero.
Then $g_{\a \b} = g^{\a \b}$ satisfies  
$g^{\a \b} g_{\b \g} = \d^{\a}_{\g}$, and 
$g^{\a \b} g_{\a \b} = q^2 +1 + q^{-2} = [3]_q$.

The  Clebsch--Gordon coefficients
for $(3) \subset (3) \tens (3)$, i.e. the $q$--deformed structure 
constants, are given by 
\be
\begin{array}{ll} 
\eps^{1 0}_{1} =  q, & \eps^{0 1}_{1} = -q^{-1},  \\
\eps^{0 0}_{0} = q-q^{-1}, & \eps^{1 -1}_{0} = 1 = -\eps^{-1 1}_{0}, \\
\eps^{0 -1}_{-1} = q, & \eps^{-1 0}_{-1} = -q^{-1}
\end{array}
\label{C_ijk}
\ee
in an orthonormal basis, and $\eps_{i j}^{k} = \eps^{i j}_{k}$.
They are normalized such that 
$\sum_{ij} \eps_{ij}^{n} \eps^{ij}_{m} = [2]_{q^2} \d^n_m$.
Moreover, the following identities hold:
\berr
\eps_{ij}^n g^{jk} &=& \eps^{nk}_i   \label{C_g}  \\
g_{ij} \eps^j_{kl} &=& \eps^j_{ik} g_{jl}  \label{Cg_id} \\
\eps_i^{nk} \eps_k^{lm} - \eps_i^{km} \eps_k^{nl} 
     &=& g^{nl} \d_i^m - \d^n_i g^{lm}   \label{CC_id}   
\err
which can be checked explicitly.
In view of (\ref{Cg_id}), the $q$--deformed totally ($q$--)antisymmetric 
tensor is defined as follws: 
\be
\eps^{ijk} = g^{in} \eps_n^{jk} = \eps^{ij}_n g^{nk}.
\label{q_epsilon}
\ee
It is invariant under the action of $U_q(su(2))$.

\subsection{Some proofs}
\label{subsec:proofs-I}

\paragraph{\em Proof of (\ref{dx_i}):}

Using the identity
\be
\one = q^{-2} \hat R + (1+q^{-4})P^- + (1-q^{-6}) P^0,
\label{one_id}
\ee
(\ref{C_g}), (\ref{RRg_braid}),
and the braiding relation (\ref{xxi_braid})
we can calculate the commutation relation of $\Theta$ 
with the generators $x_i$:
\berr
x_i \Theta &=& x_i (x_j \xi_t g^{jt}) \nn\\
  &=& q^{-2} {\hat R}^{kl}_{ij} x_k x_l \xi_t g^{jt} + 
      q^{-2} \La_N \eps^n_{ij} x_n \xi_t g^{jt}
      + \frac{r^2}{[3]} (1-q^{-6}) g_{ij} \xi_t g^{jt} \nn\\
  &=& q^{-2} \Theta x_i + q^{-2} \La_N \eps_{ij}^n x_n \xi_t g^{jt}
              + {r^2}\frac{(1-q^{-6})}{[3]_q} \xi_i \nn\\
  &=&  q^{-2} \Theta x_i + q^{-2} \La_N \eps^{nk}_i x_n \xi_k 
               + {r^2} q^{-3} (q-q^{-1}) \xi_i,  \nn
\err
which yields (\ref{dx_i}).

\paragraph{Proof of (\ref{theta_2}) and (\ref{dtheta}):}

Using (\ref{xixi}), one has
\be
\Theta \xi_i = -q^2 \xi_i \Theta,
\label{theta_xi}
\ee
which implies
$\Theta \Theta = \Theta (x \cdot \xi) = dx \cdot \xi - q^2 \Theta \Theta$, 
hence
$$
(1+q^2) \Theta^2 = dx \cdot \xi.   
$$
On the other hand, (\ref{dx_i}) yields
$$
dx \cdot \xi = - \La_N x_i \eps_j^{kl} \xi_k \xi_l g^{ij}
               - q^3 (q-q^{-1}) \Theta^2,
$$
and combining this it follows that
$$
\Theta^2 = - \frac{q^2 \La_N}{[2]_{q^2}} x_i \xi_k \xi_l \eps^{ikl}.
$$
We wish to relate this to $dx_i dx_j g^{ij}$, which is proportional
to $d \Theta$.
Using  the relations $\eps^{nk}_i x_n \xi_k = -q^{-2} \eps^{nk}_i \xi_n x_k$,
$\Theta = q^{-4} \xi\cdot x$,
(\ref{Cg_id}) and (\ref{CC_id}), one can show that
$$
\eps^{nk}_i x_n \xi_k \eps^{ml}_j x_m \xi_l g^{ij} 
  = \La_N  x_i  \xi_k \xi_l \eps^{ikl} + q^2 \Theta^2
$$
which using (\ref{dx_i}) implies
$$
dx_i dx_j g^{ij} = -  \frac 1{C_N}\;r^2\; \Theta^2.
$$

\paragraph{Proof of $\; d\circ d = 0$ on $\Omega^1_{q,N}$.}

First, we calculate
\berr
[\Theta, d\xi_i] &=& (q^2-1)(q^2+1)\left(\xi_i \Theta^2 + 
     \frac{q^{-2} \La_N}{[2]_{q^2}} \eps_i^{kl} \xi_k \xi_l \Theta\right) \nn\\
  &=& (q^2-1)(q^{2}+1)  
       \left(\xi_i (\ast_H \Theta) - (\ast_H \xi_i) \Theta\right) =0  \nn
\err
using (\ref{theta_xi}), (\ref{ast_theta}), and (\ref{star_adj}).
This implies that 
\berr
d(d(f \xi_i)) &=& [\Theta, df \xi_i + f d\xi_i] \nn\\
   &=& [\Theta,  df]_+ \; \xi_i - df [\Theta, \xi_i]_+
       + df d\xi_i + f [\Theta, d\xi_i] \nn\\
  &=& d df + \ast_H( df) \xi_i - 
                   df(d\xi_i + \ast_H(\xi_i)) + df d\xi_i  \nn\\
  &=&  - df \ast_H \xi_i + (\ast_H df) \xi_i = 0 \nn
\err
by (\ref{star_adj}) for any $f \in \S_{q,N}$. This proves 
$d\circ d = 0$ on $\Omega^1_{q,N}$.

\paragraph{Proof of (\ref{star_adj}).}

First, we show that 
\be
(\ast_H \xi_i) \xi_j = \xi_i (\ast_H \xi_j),
\ee
which is equivalent to
$$
\eps^{nk}_i \xi_n \xi_k \xi_j = \xi_i \eps^{nk}_j \xi_n \xi_k.
$$
Now $\Omega^3_{q,N}$ is one--dimensional as module over $\S_{q,N}$,
generated by $\Theta^3$ (\ref{om_3}), which in particular is a singlet
under $U_q(su(2))$. This implies that 
\berr
\eps^{nk}_i \xi_n \xi_k \xi_j &=& 
             (P^0)_{ij}^{rs} \eps^{nk}_r \xi_n \xi_k \xi_s\nn\\
  &=& -\frac{q^6 [2]_{q^2}}{\La_N r^2} \; g_{ij}\;  \Theta^3  \nn\\
  &=& (P^0)_{ij}^{rs} \xi_r \eps^{nk}_s \xi_n \xi_k =\xi_i\eps^{nk}_j \xi_n \xi_k,
\label{cyc_calc}
\err
as claimed. Now (\ref{star_adj}) follows immediately using the
fact that $\ast_H$ is a left--and right $\S_{q,N}$--module map.

\paragraph{Reality structure for $q \in \R$:}
These are the most difficult calculations, and they are needed 
to verify (\ref{obar_xi}) as well. First, we have to show that 
(\ref{xxi_braid}) is compatible with the star structure (\ref{xi_real}).
By a straightforward calculation, one can reduce the problem to
proving (\ref{reality_id}). We verify this by projecting this 
quadratic equation to its spin 0, spin 1, and spin 2 part. 
The first two are easy to check, using (\ref{sxdx_id}) in the spin 1 case. 
To show the spin 2 sector, it is enough to consider (\ref{reality_id})
for $i=j=1$, by covariance. This can be seen e.g. using
$[x_1, \eps_1^{ij} x_i \xi_j] = -q^{-2} \La_N x_1 \xi_1$, which 
in turn can be checked 
using (\ref{one_id}), (\ref{RRC_braid}) and (\ref{C_ijk}). 

Next, we show that (\ref{xixi}) is compatible with the star structure 
(\ref{xi_real}). This can be reduced to 
$$
(q^2 \hat{R} - q^{-2}\hat{R}^{-1})^{k l}_{ij}\; dx_k \; \xi_l
 = q^2(q -q^{-1})\frac{[2]_q C_N}{r^2} 
     (\one + q^2 \hat{R})^{k l}_{ij}) \; dx_k\; dx_l
$$
The spin 0 part is again easy to verify, and the spin 1 part 
vanishes identically (since then $\hat R$ has eigenvalue $-q^{-2}$). 
For the spin 2 part, one can again choose $i=j=1$, and verify it e.g. by 
comparing with the differential of equation (\ref{reality_id}).

\paragraph{Proof of (\ref{cyclic_forms}):}
Since $\Omega^*_{q,N}$ is finitely generated and because of 
(\ref{cyclic_integral}) and $[\Theta^3, f] = 0$, it is enough to consider
$\b = \xi_k$. In this case, the claim reduces to
$$
\xi_i \xi_j \xi_k = S^{-2}(\xi_k) \xi_i \xi_j.
$$
Now $ S^{-2}(\xi_k) = D_k^l \xi_l$, where $D_k^l = \d_k^l q^{2r_l}$
with $r_l = (-2,0,2)$ for $l= (1,0,-1)$, respectively. Since 
$\xi_i \xi_j = \frac 1{[2]_{q^2}}\; \eps_{ij}^n (\eps_{n}^{rs} \xi_r \xi_s)$, 
there remains to show that 
$(\eps_{n}^{rs} \xi_r \xi_s) \xi_k = S^{-2}(\xi_k) (\eps_{n}^{rs} \xi_r \xi_s)$.
By (\ref{cyc_calc}), this is equivalent to
$$
g_{nk} \Theta^3 = D_k^l g_{ln} \Theta^3,
$$
which follows from the definition of $ D_k^l$.

\paragraph{Proof of Lemma \ref{laplace_lemma}:}

Using (\ref{comm_theta2}), (\ref{theta3_f}) and 
$d \Theta^2 = 0$, we have
\berr
d\ast_H d \psi &=& d( \psi \Theta^2 - \Theta^2 \psi)
  = (d\psi)  \Theta^2 - \Theta^2 d\psi \nn\\
  &=& (d\psi)  \Theta^2 + [\Theta^2,\psi] \Theta \nn\\
  &=& (d\psi)  \Theta^2 + (\ast_H d\psi) \Theta.   \label{lapl_calc}
\err
To proceed, we need to evaluate $d\psi_{K}$. 
Because it is an irreducible representation, 
it is enough to consider $\psi_{K} = (x_1)^K$. From 
(\ref{reality_id}) and using 
$\xi_1 x_1 = q^{-2} x_1 \xi_1$, it follows that 
$$
dx_1 x_1 = q^2 x_1 dx_1 - \frac{q^{-2}}{C_N} r^2 x_1 \xi_1,
$$
since $\hat R$ can be replaced by $q^2$ here. By induction, one finds
\be
dx_1 x_1^k = x_1^k \left( q^{2k}  dx_1 - 
       [k]_{q^2} \frac{q^{-2}}{C_N}\;r^2 \xi_1 \right),
\ee
and by an elementary calculation it follows that
\be
d(x_1^{k+1}) = 
  [k+1]_q x_1^k 
  \left(q^k dx_1 - \frac{q^{-2}}{[2]_q C_N} [k]_q r^2 \xi_1 \right).
\ee
Moreover, we note that using (\ref{star_adj}) 
\be
(\xi_i \Theta + \ast_H \xi_i) \Theta =
  \xi_i (\ast_H\Theta) + (\ast_H\xi_i) \Theta = 2 (\ast_H\xi_i) \Theta 
  = -\frac 2{r^2} x_i \Theta^3.
\ee
The last equality follows easily from (\ref{cyc_calc}) and (\ref{theta_xi}).
Similarly
\be
(dx_i \Theta + \ast_H dx_i) \Theta = 2 \ast_H d x_i \Theta = 2 dx_i \Theta^2.
\ee
Now we can continue (\ref{lapl_calc}) as  
\berr
d\ast_H d x_1^{K} &=& (dx_1^{K-1} \;\Theta + \ast_H dx_1^{K-1}) \Theta \nn\\
  &=& [K]_q  x_1^{K-1} \left(2 q^{K-1} dx_1 \Theta^2 -    
     2 \frac{q^{-2}}{[2]_q C_N} [K-1]_q x_1 \Theta^3\right).
\err
Finally it is easy to check that
\be
dx_i \Theta^2 = -\frac 1{C_N} x_i \Theta^3,
\ee
and after a short calculation one finds (\ref{scalar_laplace}).

%\end{document}

%% fuzzy Instantons on  S^2_{N}

\chapter{Fuzzy instantons}
\label{chapter:fuzzyinst}

To some extent, 
the ideas behind the construction of instantons and solitons can be applied
to the noncommutative geometries under consideration here.
For our purpose, an instanton is 
a finite--action solution of Yang--Mills type equations of motion
with euclidean signature, such that the field 
equations reduce to self--duality conditions. The word soliton 
refers to a stable finite-energy solution.
In ordinary geometry they 
are both stable because of a topological obstruction to their decay.
The notion of topology now becomes somewhat ``fuzzy'';
nevertheless such solutions exist in our context, and it is 
interesting to study their structure on a fuzzy space.
This chapter is based on a joint work \cite{fuzzyinst}
with Harald Grosse, Marco Maceda and John Madore,
however with some changes especially in the scaling
behaviour of the sphere, adapting it to the $D$-brane scenario 
of the preceding sections.

We add in this chapter a commuting (euclidean) time coordinate
to the undeformed fuzzy sphere $S^2_N$, and consider 
a Maxwell--like $U(1)$ gauge theory ~\cite{Madore:1992bw,Grosse:1992bm}.
The spacetime is hence 3-dimensional, and 
it may seem impossible to have an instanton, 
which is a self-dual solution of some equation of motion.
Nevertheless, this is possible here. 
The reason is that the fuzzy sphere $S^2_N$ sees some trace of the 
3-dimensional embedding space. We already noticed this in Chapter
\ref{chapter:qFSI}, where the calculus on the fuzzy sphere 
(whether or not it is $q$-deformed) turned out to be 3-dimensional.
Hence with an extra time it is 4-dimensional, and admits a 
notion of self-duality. 

From a physical point of view, the most important use of instantons is 
to describe tunneling between different topological sectors 
of a gauge theory. In our situation, they turn out to play a similar role:
they interpolate between different geometrical vacua of the theory.
We already encountered these vacua in Chapter \ref{chapter:qFSI}, as
block-type solutions of the zero curvature condition 
in Section \ref{subsec:gauge}. They describe in the classical limit 
spheres of different sizes, or superpositions thereof. 
They can be viewed as `fuzzy' spherical $D2$-branes, and were
already found in ~\cite{Dubois-Violette:1989ps}. The instantons now
tunnel between two such $D2$-brane configurations, for example
one in which the branes coincide and the other in which they are
completely separated.
Clearly, these instantons disappear in the classical limit; 
however the quantum nature of fuzzy geometry admits such 
tunneling between different classical geometries.

We stress here the role of instantons as mediators between different
stable vacuum sectors of a matrix geometry.  Several other aspects of
instantons have also been carried over ~\cite{Baez:1998he,
Gopakumar:2000zd, Gross:2000ss, Harvey:2000jt, Connes:2000tj} into various
noncommutative geometries including the fuzzy
sphere~\cite{Iso:2001mg, Hashimoto:2001xy, Hikida:2000cp}. Similar
calculations have been carried out ~\cite{Krajewski:2001ch} on the torus.

\section{Abelian gauge theory on $S^2_N \times \R$}

\subsection{The noncommutative geometry.}

We recall very briefly some formulae from 
Section \ref{sec:diff-calc} %, \ref{subsec:frames} 
which will be useful  below.
The algebra $\cA$ of functions on $S^2_N \times \R$ is generated by 
generators $\lambda_a$ which satisfy the commutation relations
\beq
[\lambda_a, \lambda_b] = \eps^c{}_{ab} \lambda_c, 
\quad \lambda_a\lambda^a = \frac{N(N+2)}{4},
\label{lala-1}
\eeq
together with a commutative time coordinate $t$. 
Let $\theta^a$ be a real frame
~\cite{Dubois-Violette:1989at,madoreDim,FioMad98a}
for the differential calculus over the fuzzy sphere, and let 
$\theta^0 = dt$ be the standard de~Rham differential along the real line.
The differential $df$ of $f \in \cA$ can then be written (by definition) as 
$$
df = [\la_a, f] \theta^a + \dot f \theta^0,  \qquad \dot f = \partial_t f,
$$
The $\la_a$ are considered dimensionless here.
We use the basis of anti-commuting one-forms 
$\theta^\alpha = (\theta^0, \theta^a)$ for $a=1,2,3$.
The form $\theta = \la_a \theta^a$ generates the 
``spatial part'' of the exterior derivative,
and can be considered as a flat connection \refeq{dtheta}.
%$$
%d\theta - \theta^2 = 0.
%$$
%The fuzzy 2-solitons are in 1-1 correspondence with the projectors
%and their number is given by the (Hardy-Ramanujan) partition function $p(n)$.
The exterior derivative of a 2-form is given by the obvious adaptation of 
\refeq{d_1},
\berr
d: \; \Omega^1_{N} &\rightarrow& \Omega^2_{N}, \nn\\
                   \a &\mapsto& [\Theta,\a]_+ - \ast_H(\a).
\err
where $\ast_H(\theta^a) = \frac 1{2} \eps^a_{bc} \theta^b \theta^c$.
Finally, we can define a modified Hodge-star operator 
\beqa
\ast (\theta^a \theta^0) &=& \frac 1{2} \eps^a_{bc} \theta^b \theta^c \nn\\
\ast (\theta^a \theta^b) &=& \; \eps^{ab}_c \theta^c \theta^0,\nn
\eeqa
will play an essential role in the sequel. It is clearly a generalization 
of the definition in Section \ref{sec:diff-calc}, but now 
maps indeed 2-forms into 2-forms. 
In particular, 
\beq
\ast(\theta \theta^0) = \theta\theta.
\eeq
No length scale has been introduced so far,
only an integer $N$.

There are now different possibilities to introduce a scale
into the theory, leading to different classical limits. 
The first is to 
consider the fuzzy sphere to have a fixed radius as in \refeq{xx_square},
\berr
\eps_k^{i j}  x_i  x_j &=& 
            \La_N \; x_k, \label{xx_sigma-1}\\ 
 x\cdot x &:=& g^{i j} x_i x_j =  r^2
\label{xx_square-1}
\err
where
\beq
\La_N = r \; \frac{2}{\sqrt{N(N+2)}}.
\eeq
Another possibility is to fix 
\berr
\eps_k^{i j}  y_i  y_j &=& 
            \sqrt{\kbar} \; y_k, \label{yy_sigma}\\ 
 y\cdot y &=&  \frac{N(N+2)}{4} \kbar.
\label{yy_square}
\err
introducing a length scale $\sqrt{\kbar}$ \cite{fuzzyinst}.
Then the radius is basically $r = N/2\; \sqrt{\kbar}$ for large N. 
However, we will develop the formalism of gauge theories without
introducing any scale here, and use $\la_a$ which satisfy \refeq{lala-1}.
Space then acquires an onion-like structure with an infinite sequence of
concentric fuzzy spheres at the radii given by the above relation
(i.e. $\kbar = 1$).
This is also the scenario which is appropriate to study $D$-branes 
on $SU(2)$, where transitions between various such spheres
are indeed expected. 
There are further possibilities, which were discussed in some detail in 
\cite{fuzzyinst}.

We define the integral over functions as
$$
\int_{S^2_N} f =  4\pi \; \tr (f).
$$
Similarly, we define the integral over a 3-form $\a = f dV$ where 
$dV = \theta\theta\theta$
by
\beq
\int_{S^2_N} \a = \int_{S^2_N} f.
\eeq
This leads to a quantization condition on the area:
$$
\mbox{Area}[S_N^2] = \int_{S^2_N} dV \quad \approx\; 4\pi\; N
$$
Stokes theorem is easily seen to work for the 
present situation, noting that the sphere has no boundary. Therefore
$$
\int_{S^2_N} d\a^{(2)} = 0
$$
for any 2-form $\a^{(2)}$ on the fuzzy sphere, and
$$
\int_{S^2_N \times \R} d\a^{(3)}
= \int_{S^2_N} \a^{(3)}(t=\infty) - \int_{S^2_N} \a^{(3)}(t=-\infty) 
$$
for any 3-form $\a^{(3)}$ on $S^2_N \times \R$.

\subsection{Gauge theory.}

The definition of a gauge theory on the 
fuzzy sphere can be taken over from Section \ref{subsec:gauge},
setting $q=1$ and adding the time variable. 
The degrees of freedom of a $U(1)$ gauge theory are encoded in a 
one-form $B = \sum B_\a \theta^\a$. 
It is naturally written as $B = \theta + A$ defining the
``connection'' 1-form $A = A_\alpha \theta^\alpha$. Hence
$B_a = \la_a + A_a$ and $B_0 =  A_0$.
Notice that $A_\a$ is dimensionless here. 
The curvature is then defined formally as
$$
F = (B+ d_t) (B + d_t) - \ast_H(B) = 
  (dA + A^2) = \frac 12 F_{\alpha\beta}\theta^\alpha \theta^\beta,
$$
cp. \refeq{curvature}. 
Its components are simply
$$
F_{0a} = \dot B_a + [\la_a, B_0] + [B_0, B_a], \qquad
F_{ab} = [B_a, B_b] - \eps^c{}_{ab} B_c.
$$
The dynamics of the model is defined by the 
Yang-Mills action
\be \fbox{$
S_{YM} =  \int_{S_N^2 \times \R} F \ast F $}
\label{YM-1}
\ee
which is invariant under the gauge transformations
\beq
B \to U^{-1} B U, \quad A \to U^{-1} A U + U^{-1} d U
\eeq
One can now easily derive equations of motion \cite{fuzzyinst}.
Moreover, we can conclude immediately from the inequality
$$
\int_{S_N^2 \times \R} (F \pm \wm F) \wm (F \pm \wm F) \geq 0
$$
and
\beq
FF = dK = d(A dA + \frac 23 A^3)
\eeq
as well as 
$$
\int_{S^2_N \times \R} dK = \int_{S^2_N} K(t=\infty) - 
\int_{S^2_N} K(t=-\infty) 
$$
that the action $S[A]$ is bounded from below by 
$$
S_{YM} \geq |\int_{S^2_N \times \R} FF| 
= |\int_{S^2_N} K(t=\infty) - \int_{S^2_N} K(t=-\infty)| 
$$
for any configuration $A$ ``in the same sector''
determined by the 
boundary values of $\int_{S^2_N} K(t=\pm \infty)$.
This is exactly the same argument as in the undeformed case.

\subsection{Zero curvature solutions and multi-brane configurations.}

One particular solution of the zero-curvature condition $F=0$ 
is given by $F = \theta$. More generally, its time-independent
solutions have the form
\beq
B = \oplus_i \la_a^{(i)} \theta^a
\eeq
as in Section \ref{subsec:gauge},
where $\la_a^{(i)}$ are irreducible representations of the Lie algebra 
$su(2)$ acting on $n_i$-dimensional subspaces of $\C^n$ where $n=N+1$. 
The solutions to $F=0$  are hence determined by the partitions 
$[n] = [n_1, ..., n_k]$ of $n$ such that $\sum n_i = n$,
which can be interpreted as ``multi-brane'' configuration consisting of 
branes (fuzzy spheres) $S^2_{n_i-1}$. A better interpretation of our
gauge theory is therefore as describing $n$ ``$D0$-branes'', which can form
bound states as fuzzy spheres of various sizes. This fits nicely with the 
string-theoretical picture of $D0$-branes on group manifolds,
in the limit of infinite radius of the group.

\section{Fuzzy Instantons}

From the above considerations, it follows that solutions of the 
self-duality conditions 
\beq\fbox{$
F = \pm \wm F $}
\eeq
are certainly solutions of the equations of motions, and moreover they are
the solutions with the smallest action for any configuration $A$ with given
boundary values at $t = \pm \infty$. 
We shall now find these solutions, called instantons. 
First, it is useful to go to the Coulomb gauge $B_0 = 0$.
%this is always possible, by a suitable gauge transformation.
We then make the Ansatz $B_a = f(t) \lambda_a$, which
can be also written as
$$
B = f(t) \theta, \qquad A = (f(t)-1)\theta.
$$
The $f(t)$ is {\it a priori}
an arbitrary element of the algebra. 
The field strength is then
\begin{equation}
F = dA + A^2 = df \theta  + (f-1)\theta\theta + 
(f-1)\theta(f-1)\theta.    \label{curv}
\end{equation}
We shall compute only the simple case with $f(t)$
in the center. This is reasonable, 
because it describes the configurations which are invariant under $SU(2)$. 
In this case
$$
F = \dot f dt \theta + (f-1)f\theta\theta,
$$
and self-duality condition becomes
$$
\dot f dt \theta + (f-1)f\theta^2 = 
\pm(\dot f\; (-\theta^2) - (f-1)f dt \theta).
$$
Hence 
\beq
\dot f = \mp (f-1)f.
\eeq
The constant solutions
$f=0,1$ correspond to the two stable ground states of the
system, $B = 0$ and $B = \theta$.
The instanton solution 
interpolates between the first two, and is given by
\begin{equation}
f_+ = \frac 12 (1 + \tanh (\frac{t}{2} + b)),  
\qquad b \in \R                                          \label{t-sol}
\end{equation}
interpolates between $f(-\infty) = 0$ and $f(+\infty) = 1$. This has
the same form as the classical double-well instanton~\cite{Polyakov:1977fu}.
The ``anti-instanton'' is given by
\begin{equation}
f_- = \frac 12 (1 + \tanh (-\frac{t}{2} + b)),
\qquad b \in \R                                       
\end{equation}
connecting the vacua $f(-\infty) = 1$ and $f(+\infty) = 0$.

To calculate the value of the action
$S_{YM}$ for this instanton, it is again convenient to use the fact that
the 4-form $F^2$ is the exterior derivative of a 3-form,
$$
\int_{S_N^2 \times \R} F^2 = 
\int_{S_N^2 \times \R} dK, \qquad 
K = A dA + \frac 23 A^3.
$$
The action then becomes
$$
S =  \int_{S_N^2} K(+\infty) - \int_{S_N^2}K(-\infty).
$$
Our solution tunnels between $f=0$ at 
$t =-\infty$ to $f=1$ at $t=\infty$. That is $K(+\infty) = 0$ and
$$
K(-\infty) =  \frac 1{3} \theta^3 = \frac1{3}\; dV. 
$$
But
$$
\int_{S_N^2} \theta^3  = 4\pi (N+1)
$$
up to an arbitrary overall normalization constant. 
We find then that the action of the instanton solution~(\ref{t-sol}) is
$$
S = - \frac{1}{3}\int_{S_N^2} \theta^3 = 
\frac{4\pi}{3}\; (N+1).
$$ 
In particular, this is a lower bound for the action of any field
configuration. The precise
value is of course somewhat arbitrary, depending on the normalization of 
the integral. However, it acquires meaning in the case of multiple 
branes, which we will now discuss.

\subsection{Multi-brane instantons and Fock space}

We have found an instanton solution which tunnels between $f=0$ and
$f=1$. 
If we return to the definition of the curvature as a functional
of the fields $A$, we see that this corresponds to a transition from
the irreducible representation of dimension $n=N+1$ to the completely
reducible representation of the same dimension. The former corresponds
to the partition $[n]$ of $n$; the latter to the partition 
$[1, \dots 1]$.  
Now recall that the general solution of $F=0$ is given by multi-brane 
configurations labeled by partitions $[n_1, ..., n_k]$.  
By the same construction there is,
for each index $i$, an instanton which tunnels between $[n_i]$ and
$[1, \dots 1]$.  There are perhaps other instantons, those which
correspond to transitions between non-trivial projectors. 

Consider for example an instanton $T^n_l$ which tunnels as
$$
[l, \underset{n-l}{\underbrace{1, \cdots 1}}] 
\buildrel T^n_l \over \longrightarrow [\underset{n}{\underbrace{1, \cdots 1}}].
$$
The corresponding gauge field is $B_a = f(t) e_l\; \lambda_a$
where $e_l$ is the projector on $\C^l$. 
Using the same definitions of curvature and integral as before, 
the only thing that changes in the above calculation\footnote{note in 
particular that the frame $\theta^a$ is independent of $l$} is that 
now
$$
\int_{S_N^2} e_l dV  = 4\pi l
$$
and the action is given by
\beq\fbox{$
S[T^n_l] = - \frac{\pi}{3}\int_{S_N^2}  e_l \theta^3 = 
\frac{4\pi}{3}\; l.  $} \label{Bohr}
\eeq
This is the basic relation which we find. The action 
of such instantons is basically an integer $l$, or the sum 
$\sum l_i$ for multi-instantons of sizes $l_i$ interpolating between
the corresponding branes. This is quite remarkable: of course classically, 
the action of instantons is always an integer, given by the Chern number; 
the reason is that $F$ is a connection on a nontrivial bundle.
The physical and geometric meaning of the present situation 
is very different and entirely non-classical. Nevertheless, 
the integral over the corresponding action is still basically 
integer-valued, and measures certain ``topological'' classes.

\paragraph{Fock space.}

Consider a general partition $[n_1, \cdots, n_k]$.  Let $e_i$
be the projector onto the $i$th sector. An arbitrary projector $e$ can
be written in the form
$$
e = \sum_i \epsilon_i e_i, \qquad \epsilon_i = 0,1.
$$
The corresponding expression (\ref{curv}) for the field strength 
is given by
$$
F = \sum_i F_i
$$
with each $F_i$ of the form (\ref{curv}). Each corresponding
$f_i$ evolves independently according to its field equation. The
instantons tunnel therefore between different partitions.

If one quantizes the system one obtains a bosonic Fock space of
ordinary `vacuum modes'. Due to the gauge symmetry, different orderings
of the partition $[n_1, \cdots, n_k]$ are equivalent, and 
we can describe the vacuum states (branes) as a bosonic Fock space
labeled by the multiplicities of each brane,
$$
\ket{k_{1}, k_{2}, \cdots, k_n} = 
[\underset{k_1}{\underbrace{1, \cdots, 1}},
 \underset{k_2}{\underbrace{2, \cdots, 2}},...,
\underset{k_n}{\underbrace{n, \cdots, n}} ].
$$
The integer $k_{j}$ is the occupation number of the brane $[j]$,
which is a fuzzy sphere $S^2_{j-1}$. 
Besides these vacuum modes there is also a Fock space $\cF$ of `tunneling
modes', which transform a brane $[j]$ into $j$ branes $[1]$.
In Fock-space notation the basic transition is of the general form
$$
\ket{k_1,\cdots, k_i, \cdots, k_n} 
\buildrel T^n_{i} \over \longleftrightarrow 
\ket{k_1+i, \cdots, k_i - 1, \cdots, k_n}.
$$
One of the $k_i$ irreducible components of dimension $i$ in the given
representation `decays' into $i$ representations of dimension
one, or the inverse. The tunneling
modes interact with all the vacuum modes and change their energy
eigenvalues in a rather complicated way.  
Without the tunneling modes
the vacuum modes do not interact, so we consider the tunneling modes as
responsible for the dynamics. 
An instanton gas is an ensemble of tunneling modes,
considered as a Bose gas.

In physical terms, instantons describe quantum mechanical tunneling between
these brane configurations; see e.g. \cite{Coleman:1978ae} for a 
lucid discussion of the physical aspects of instantons.
One could therefore try to use them for calculating 
transition rates between such configurations. 
The probability of these transitions
is proportional to the barrier penetration rate~\cite{Landau:1987gn}
$$
p[T^n_{j}] = A[T^n_{j}] e^{-S[T^n_{j}]}.
$$  
The $A[T^n_{i}]$ is a $WKB$ amplitude difficult to
calculate in general. Furthermore, while 
all the different partitions are degenerate classically, 
the tunneling phenomena would lift this degeneracy and make some
partitions more favorable.  Consider the case $n=2$ with its two
partitions $[2]$ and $[1,1]$.  
The two degenerate levels split by an amount proportional
to the transition probability $p = A e^{-8\pi/3}$.
Such processes are indeed predicted by string theory. 

To discuss this situation in more detail,
one should consider an instanton gas consisting of many bounces, which
is to some extent discussed in \cite{fuzzyinst}. This is 
a complicated problem, and we will not pursue this here any further. 
One can see the complexity of the situation also in a different way,
noting that the degeneracy of the vacua is lifted by 
the zero-point fluctuations. For
example, the completely reducible configuration $[1,...,1]$
and the irreducible configuration $[n]$
will have vacuum energy given respectively by
$$
E_{[1,...,1]} \simeq \frac 12 \hbar \omega \cdot n^2,
\qquad E_{[n]} \simeq \frac 12 \hbar\omega \cdot \sum 2l(2l+1),
\qquad \omega \approx 1.
$$
In the first there are $n^2$ modes of equal frequency $\omega$,
because there is no kinetic term. In
the second, for each $l \lesssim n$ there are $2(2l+1)$ modes of
frequency $l\omega$, due to the kinetic energy of the gauge fields.
The remaining vacuum energies lie between these
two values.

%% scalar QFT on S^2_{N} at one loop

\chapter{One--loop effects on the fuzzy sphere}
\label{chapter:floops}

This chapter covers the first one-loop calculation of 
quantum effects on the fuzzy sphere, which was done in 
a joint work \cite{fuzzyloop}
with John Madore and Chong-Sun Chu. We consider scalar $\Phi^4$ theory,
and  calculate the two point function at one loop.
The fuzzy sphere $S^2_N$ is 
characterized by its radius $R$ and a ``noncommutativity'' parameter
$N$ which is an integer. It  approaches 
the classical sphere in the limit $N \rightarrow \infty$ for fixed $R$,
and can be thought of as consisting of $N$ ``quantum cells''.
The algebra of functions on $S^2_N$ is finite, with maximal angular 
momentum $N$. Nevertheless, it admits the full symmetry group $SO(3)$ 
of motions. The fuzzy sphere is closely related to several other
noncommutative spaces \cite{qFSI,qFSII}. In particular, it can be 
used as an approximation to the quantum plane $\R^2_\th$, by
``blowing up'' for example the neighborhood of the south pole. 
This is the quantum spaces with the basic commutation relations
$$
[x_i,x_j] = i \theta_{ij}
$$
for a constant antisymmetric tensor $\theta_{ij}$
which has been studied extensively in recent years.
Many of the problems that can arise in QFT on noncommutative spaces are 
illustrated in this much--studied example of the noncommutative plane
$\R^n_\th$; see \cite{Douglas:2001ba} for a recent
review. One of the most intriguing phenomena on that space
is the existence of an ultraviolet/infrared 
(UV/IR) mixing \cite{Minwalla:1999px} in the quantum effective 
action. Due to this
mixing, an IR singularity arises from integrating out the UV degrees of
freedom. This threatens the renormalizability and even the
existence of a QFT. Hence a better
understanding (beyond the technical level) of the mechanism of UV/IR mixing
and possible ways to resolve it
are certainly highly desirable. 
One possible approach is to approximate
$\R^n_\th$ in terms of a different noncommutative space.
This idea is realized here, 
approximating $\R^2_\th$ by a fuzzy sphere.
This will allow to understand the UV/IR mixing as an infinite variant of a
``noncommutative anomaly'' on the fuzzy sphere,
which is a closely related but different phenomenon
discussed below. This is one of our main results. 
A related, but less geometric approach was considered in \cite{Kinar:2001yk}.

The fuzzy sphere has the great merit that it is very clear how to 
quantize field theory on it, using a finite analog of the path
integral \cite{Grosse:1996ar}. Therefore QFT on this space is a priori 
completely well--defined, on a mathematical level. 
Nevertheless, it is not clear
at all whether such a theory makes sense from a
physical point of view, i.e. whether there exists a limiting theory 
for large $N$, which could be interpreted as a 
QFT on the classical sphere. There might be a similar UV/IR 
problem as on the quantum plane $\R^2_\th$, as was claimed
in a recent paper \cite{Vaidya:2001bt}. In other words, it is not clear if 
and in what sense such a QFT is renormalizable. 
As a first step, we calculate 
the two point function at one loop and find that it is well
defined and regular, without UV/IR mixing. 
Moreover, we find a closed formula for 
the two point function in the commutative limit, i.e. we calculate the
leading term in a $1/N$ expansion. 

It turns out that the 1--loop effective action on $S^2_N$
in the commutative limit differs 
from the 1--loop effective action on the commutative sphere $S^2$
by a finite term, which we call  ``noncommutative anomaly'' 
(NCA). It is a mildly nonlocal, ``small'' correction to the kinetic energy 
on $S^2$, and changes the dispersion relation. 
It arises from the nonplanar loop integration.
Finally, we consider the planar limit of the fuzzy sphere. 
We find that a IR singularity is
developed in the nonplanar two point function, and hence the UV/IR mixing
emerges in this limit. 
This provides an understanding of the UV/IR mixing for QFT on $\R^2_\th$
as a ``noncommutative anomaly'' which 
becomes singular in the planar limit of the fuzzy sphere dynamics.

This chapter is organized as follows.
In Section \ref{sec:fuzzylimits}, we consider different geometrical limits of
the classical (ie. $\hbar =0$) fuzzy sphere. In particular, we show how the
commutative sphere and the noncommutative plane $\R^2_\th$
can be obtained  in different corners of the moduli space of the fuzzy sphere.
In Section \ref{sec:one-loop}, we study the quantum 
effects of scalar $\Phi^4$ field theory on the fuzzy sphere at 1-loop. 
We show that the planar and nonplanar 2-point function are both regular in the
external angular momentum and no IR singularity is developed.
This means that there is no UV/IR mixing phenomenon on the fuzzy sphere.
We also find that the planar and nonplanar two point functions differ 
by a finite amount which is smooth in the external angular momentum,
and survives in the commutative limit. Therefore the commutative
limit of the $\Phi^4$ theory at one loop differs from the
corresponding one loop quantum theory on the commutative sphere 
by a finite term \refeq{NCA}.
In section 4, we consider the planar limit of this QFT, 
and recover the UV/IR mixing.

\section{More on the Fuzzy Sphere and its Limits}
\label{sec:fuzzylimits}

\subsection{The multiplication table for $S^2_N$}

%%%%%%%%%%%%%%%%%%%%%%%
%% maybe: change R -> r, point out different integral here

We need some more explicit formulas for the multiplication
on $S^2_N$. Recall from  Section \ref{sec:fuzzysphere} that
the algebra of functions on the fuzzy sphere is 
generated by Hermitian operators 
$x= (x_1, x_2, x_3)$ satisfying the defining relations
\beq
[x_i, x_j] = i \la_N \e_{ijk} x_k, \quad
x_1^2 + x_2^2 +x_3^2  = R^2. \label{defl2}
\eeq
The radius $R$ is quantized in units of $\la_N$ by \refeq{defl3-0}
\be\label{defl3}
\frac {R}{\la_N} = \sqrt{\frac{N}{2} \left( \frac{N}{2} +1\right)
} \; , 
\quad \mbox{$N = 1,2,\cdots$ } 
\ee
The algebra of ``functions'' $S_N^2$ is simply the
algebra $Mat(N+1)$ of $(N+1) \times (N+1)$ matrices. 
It is covariant under the adjoint action of $SU(2)$, under which it 
decomposes into the irreducible \reps with dimensions
$(1) \oplus (3) \oplus (5) \oplus ... \oplus (2N+1)$.
The integral of a function $F \in S_N^2$ over the fuzzy sphere is given by
\be
R^2 \int F  = \frac{4 \pi R^2}{N+1} \tr[ F(x)],
\ee
where we have introduced $\int$, the integral over the fuzzy sphere with 
unit radius.
It agrees with the integral $\int d \Omega$  on $S^2$ in the large $N$ limit. 
One can also introduce the inner product
\be
(F_1,F_2) = \int F_1\dag F_2.
\ee

A complete basis of functions on $S_N^2$ is given by
the $(N+1)^2$ spherical harmonics, 
$Y^J_j, (J = 0, 1, ..., N;  -J \leq j \leq J)$ \footnote{We will
use capital and small letter (e.g. $(J,j)$) to refer to the 
eigenvalue of the angular momentum operator $\bf{J^2}$ and $J_z$
respectively.}, which are the weight basis of the spin $J$ component
of $S_N^2$ explained above.
They correspond to the usual spherical harmonics, however 
the angular momentum has an upper bound $N$ here. This is a
characteristic feature of fuzzy sphere. 
The normalization and reality for these matrices can be taken to be
\be
(Y^J_j, Y^{J'}_{j'}) = \delta_{JJ'} \delta_{jj'}, \qquad
(Y^J_j)^\dagger = (-1)^J Y^J_{-j}.
\ee
They obey the ``fusion'' algebra
\beqa \label{YY}
Y^I_i Y^J_j &=& 
\sqrt{\frac{N+1}{4\pi}}
\sum_{K,k} (-1)^{2\a+I+J+K+k} \sqrt{(2I+1)(2J+1)(2K+1)} \cdot   \nn\\ 
&&\cdot \left( \begin{array}{ccc} I&J&K\\ i&j&-k \end{array} \right)      
\left\{\begin{array}{ccc} I&J&K\\ \a&\a&\a \end{array} \right\} \; Y^K_k ,
\eeqa
where the sum is over $0 \leq K \leq N, -K \leq k \leq K$, and 
\be
\a = N/2. 
\ee
Here the first bracket is the  Wigner $3j$-symbol
and the curly bracket is the $6j$--symbol of
$su(2)$, in the standard mathematical normalization
\cite{Varshalovich:1988ye}. Using the Biedenharn--Elliott identity \refeq{BE}, 
it is easy to show that \refeq{YY} is associative.
In particular, 
$Y^0_0 = \frac{1}{\sqrt{4\pi}}\; \bf{1}$.
The relation \refeq{YY} is independent of the radius $R$, but depends on
the deformation parameter $N$. It is  
a deformation of the algebra of product of the spherical harmonics on the
usual sphere.  We will need \refeq{YY} to derive the form of
the propagator and vertices in the angular momentum basis. 

Now we turn to various limits of the fuzzy sphere. By tuning the
parameters $R$ and $N$, one can obtain different limiting algebras of
functions. In particular, we consider the commutative sphere $S^2$ and 
the noncommutative plane $\R^2_\th$.

\subsection{The limit of the commutative sphere $S^2$}

The commutative limit is defined by 
\be \label{limit-s}
N \rightarrow \infty ,\quad \mbox{keeping $R$ fixed}.
\ee
In this limit, \refeq{defl1} reduces to $[x_i,x_j]
=0$ and
we obtain the commutative algebra of functions on the usual sphere $S^2$. 
Note that \refeq{YY} reduces to the standard product of spherical
harmonics, due to the asymptotic relation between the $6j$--symbol and 
the Wigner $3j$-symbol \cite{Varshalovich:1988ye}, 
\be
\lim_{\a\rightarrow \infty} (-1)^{2\a} \sqrt{2 \a}
\left\{\begin{array}{ccc} I&J&K\\ \a&\a&\a \end{array} \right\} 
= (-1)^{I+J+K}
\left( \begin{array}{ccc} I&J&K\\ 0&0&0 \end{array} \right) .
\ee
 
\subsection{The limit of the quantum plane $\bf{R}^2_\th$}

If the fuzzy sphere is blown up around a given point, it 
becomes an approximation of the quantum plane \cite{Madore:1992bw}. 
To obtain this planar limit, it is convenient to
first introduce an alternative representation of the fuzzy sphere
in terms of stereographic projection. Consider the generators
\be
y_+ = 2 R x_+ (R-x_3)^{-1}, \quad y_- = 2 R (R-x_3)^{-1} x_-,
\ee
where $x_\pm = x_1 \pm i x_2$. The generators 
$y_\pm$  are the coordinates of the stereographic projection 
from the north pole. $y=0$ corresponds to the south pole. Now we take
the large $N$ and large $R$ limit, such that
\be \label{limit-p}
N \to \infty, \quad R^2 = N \th /2 \to \infty, 
\quad \mbox{keeping $\th$ fixed}. 
\ee
In this limit, 
\be
\frac{\la_N}{\sqrt{\th}}\; \sim \;  \frac 1{\sqrt{N}}
\ee
and $[y_+,y_-] = -4 R^2 \la_N (R-x_3)^{-1} + o(\la_N^2)$.
Since $y_+ y_- = 4R^2 (R+x_3) (R-x_3)^{-1} + o(N^{-1/2})$, 
we can cover the whole $y$-plane with
$x_3 = -R + \b/R$ with finite but arbitrary $\b$.
The commutation relation of the $y$ generators takes the form
\be
[y_+,y_-] = -2 \th
\ee
up to corrections of order $\la_N^2$, or
\be
[y_1, y_2] = -i \th
\ee
with $y_\pm = y_1 \pm i y_2$. 

\section{One Loop Dynamics of $\Phi^4$ on the Fuzzy Sphere}
\label{sec:one-loop}

Consider a scalar $\Phi^4$ theory on the fuzzy sphere, with action
\be
S_0 = \int \frac 12\; \Phi (\Delta + \mu^2) \Phi + 
    \frac{g}{4 !} \Phi^4.
\ee
Here $\Phi$ is Hermitian, $\mu^2$ is the dimensionless  
mass square, $g$ is a dimensionless coupling
and $\Delta = \sum J_i^2$ is the Laplace operator.
The differential operator $J_i$ acts on function $F \in S_N^2$
as
\be
J_i F = \frac{1}{\la_N}[x_i, F].
\ee
This action is valid for any radius $R$, since $\mu$ and
$g$ are dimensionless. To quantize the theory, 
we will follow the path integral quantization procedure as explained in
\cite{Grosse:1996ar}. We expand $\Phi$ in terms of the modes, 
\be
\Phi = \sum_{L,l} a^L_{l} Y^L_l, \quad 
a^{L}_l{}\dag = (-1)^l a^L_{-l}.
\ee
The Fourier
coefficient $a^L_l$ are then treated as the dynamical variables, and 
the path integral
quantization is defined by integrating over all possible configuration of
$a^L_l$. Correlation functions are computed using \cite{Grosse:1996ar}
\be
\langle a^{L_1}_{l_1}   \cdots  a^{L_k}_{l_k} \rangle = 
\frac{\int [\cD \Phi] e^{-S_0}\; a^{L_1}_{l_1}   \cdots  a^{L_k}_{l_k}  }
{\int [\cD \Phi] e^{-S_0}}.
\ee
For example, the propagator is
\be \label{prop}
\langle a^{L}_{l}   a^{L'}_{l'}{}\dag \rangle =
(-1)^l\langle a^{L}_{l}   a^{L'}_{-l'} \rangle =
\d_{L L'} \d_{l l'} \frac{1}{L(L+1) + \mu^2 },
\ee
and the vertices for the $\Phi^4$ theory are given by
\be
a^{L_1}_{l_1} \cdots  a^{L_4}_{l_4}\;  V(L_1,l_1; \cdots; L_4,l_4)
\ee
where
\beqa 
 V(L_1,l_1; \cdots; L_4,l_4)&& = \frac{g}{4!} \frac{N+1}{4 \pi} 
 (-1)^{L_1+L_2+L_3+L_4} \prod_{i=1}^4 (2L_i+1)^{1/2}  \sum_{L,l}
 (-1)^{l}(2L+1) \cdot  \nn\\ 
&& \cdot\left( \begin{array}{ccc} L_1&L_2&L\\ l_1&l_2&l \end{array} \right)
\left( \begin{array}{ccc} L_3&L_4&L\\ l_3&l_4&-l \end{array} \right) 
\left\{\begin{array}{ccc} L_1&L_2&L\\ \a&\a&\a \end{array} \right\} 
\left\{\begin{array}{ccc} L_3&L_4&L\\ \a&\a&\a \end{array} \right\}. \nn\\
\label{V}
\eeqa
One can show that $V$ is symmetric with respect to cyclic
permutation of its arguments $(L_i,l_i)$.

The  $1PI$ two point function at one loop is obtained by
contracting 2 legs in \refeq{V} using the propagator \refeq{prop}.
The planar contribution is defined by contracting neighboring legs:
\be \label{aap}
(\Gamma^{(2)}_{planar})^{L L'}_{l l'}
= \frac{g}{4\pi} \frac 13\; \d_{L L'} \d_{l, -l'}(-1)^{l} 
\cdot I^P, 
\quad
I^P:= \sum_{J=0}^N \frac{2J+1}{J(J+1) +\mu^2} .
\ee
All 8 contributions are identical.
Similarly by contracting non--neighboring legs, we find 
the non--planar contribution
\beqa \label{aanp}
(\Gamma^{(2)}_{nonplanar})^{L L'}_{l l'}
&=&  \frac{g}{4 \pi} \frac 16\; \d_{L L'} \d_{l, -l'} (-1)^{l} \cdot I^{NP},
\nn\\
I^{NP}&:=& \sum_{J=0}^N (-1)^{L+J+2\a} \;\frac{(2J+1)(2\a+1)}{J(J+1) +\mu^2 } 
\left\{\begin{array}{ccc} \a&\a&L\\ \a&\a&J \end{array} \right\}.
\eeqa
Again the 4 possible contractions agree.
These results can be found using standard identities for the 
$3j$ and $6j$ symbols, see e.g. \cite{Varshalovich:1988ye} and Section \ref{sec:supp-loop}.

It is instructive to note that $I^{NP}$ 
can be written in the form  
\be
I^{NP}= \sum_{J=0}^N \frac{2J+1}{J(J+1) +\mu^2} \; f_J ,
\ee
where $f_J$ is obtained from the generating function
\be
f(x) = \sum_{J=0}^\infty f_J x^J
=\frac{1}{1-x} \;
{}_2F_1 (-L,L+1,2 \a+2, \frac{x}{x-1}) 
{}_2F_1 (-L,L+1, -2 \a, \frac{x}{x-1}).
\ee
Here the hypergeometric function ${}_2F_1(-L,L+1; c;z)$ is a
polynomial of degree $L$ for any $c$. 
For example, for $L=0$, one obtains 
\be
f_J =1, \quad \mbox{$0 \leq J \leq N$},
\ee 
and hence the planar and nonplanar two point functions coincides.
For $L=1$, we have
\be
f_J = 1-\frac{J(J+1)}{2 \a (\a+1)}, \quad \mbox{$0 \leq J \leq N$ },
\ee
and hence 
\be \label{L1}
I^{NP} = I^{P} - \frac{1}{2 \a(\a+1)}
\sum_{J=0}^{2\a} 
\frac{J (J+1) (2J+1)}{J(J+1) +\mu^2} .
\ee
Note that 
the difference between the planar and nonplanar two point functions is 
finite. It is easy to convince oneself that for any finite external
angular momentum $L$, the difference between the planar and nonplanar
two point function is finite and analytic in $1/ \a$. 
This fact is important as it implies that, unlike in the $\R^n_\th$ case, 
there is no infrared singularity developed in the nonplanar 
amplitude\footnote{It was argued in \cite{Vaidya:2001bt} that the nonplanar
two-point function has a different sign for even and odd 
external angular momentum $L$. This is misleading, however, because only 
$J=2\a$ was considered there. This is not significant 
in the loop integral which is a sum over all J.}.
We will have more to say about this later.  

\subsection{On UV/IR mixing and the commutative limit}

Let us recall that in the case of noncommutative space $\R^{n}_\th$,
the one--loop contribution to the effective action often
develops a singularity at $\th p =0$ 
\cite{Minwalla:1999px,Matusis:2000jf,Ruiz:2000hu,Landsteiner:2001ky}.
This infrared singularity is generated by
integrating out the infinite number of 
degree of freedom in the nonplanar loop. 
This phenomenon is referred to as ``UV/IR mixing'', and it implies 
in particular that  
(1)  the nonplanar amplitude is singular when the external 
momentum is zero in the noncommutative directions;
and 
(2) the quantum effective action in the commutative limit is different
from the quantum effective action of the commutative limit \cite{Chu:2000bz}.

\begin{itemize}
\item
\underline{Effective action on the fuzzy sphere.}
\end{itemize}

We want to understand the behavior of the corresponding planar and 
nonplanar two point functions on the fuzzy sphere, to see if there is 
a similar UV/IR phenomenon. We emphasize that this is not obvious a priori 
even though quantum field theory on the fuzzy sphere is always finite.
The question is whether the 2--point function is smooth at 
small values of $L$, or rapidly oscillating as was indeed claimed in 
a recent paper \cite{Vaidya:2001bt}. 
Integrating out all the degrees of freedom in the
loop could in principle generate a IR singularity, for large $N$.

However, this is not the case.  We found above  that the planar
and nonplanar two point function agree precisely with each 
other when the external angular momentum $L=0$. 
For general $L$, a closed expression for $g_J$ 
for general $L$ is difficult to obtain. We will derive below an approximate 
formula for the difference $I^{NP} - I^{P}$, which is found to 
be an excellent approximation for large $N$ by numerical tests, and 
becomes exact in the  commutative limit $N \rightarrow \infty$.

First, the planar contribution to the two point function 
\be
I^P= \sum_{J=0}^N \frac{2J+1}{J(J+1) +\mu^2} 
\ee
agrees precisely with the corresponding terms on the classical
sphere as $N \rightarrow \infty$, and it diverges
logarithmically
\be
I^P  \sim \log \a + o(1).
\ee
To understand the nonplanar contribution, 
we start with the following approximation formula \cite{Varshalovich:1988ye} 
for the $6j$ symbols due to Racah,
\be
\left\{\begin{array}{ccc} \a&\a&L\\ \a&\a&J \end{array} \right\} \approx
\frac{(-1)^{L+ 2 \a +J}}{2\a} P_L (1-\frac{J^2}{2 \a^2}),
\label{racah_app}
\ee
where $P_L$ are the Legendre Polynomials.
This turns out to be an excellent approximation for 
all $0 \leq J \leq 2\a$, provided
$\a$ is large and $L \ll \a$. 
Since this range of validity of this 
approximation formula is crucial for us, we shall derive it in 
Section \ref{sec:supp-loop}.
This allows then to rewrite the sum in \refeq{aanp} to a very good
approximation as
\be
I^{NP} - I^{P} = \sum\limits_{J = 0}^{2\a} \frac{2J+1}{J(J+1) + \mu^2}
\left(P_L(1-\frac{J^2}{2 \a^2}) -1 \right)
\ee
for large $\a$. Since $P_L(1) = 1$ for all $L$, only $J \gg 1$
contributes, and one can approximate the sum by the integral
\beqa
I^{NP} - I^{P} &&\approx \int\limits_0^{2} du \;
          \frac{2u + \frac{1}{\a}}{u^2 + \frac{u}{\a}+ \frac{\mu^2}{\a^2}}
  \left(P_L(1-\frac{u^2}{2}) -1 \right) \nn\\
 &&= \int\limits_{-1}^{1}dt \; \frac{1}{1-t} (P_L(t) - 1) + o (1/\a),
\eeqa
assuming $\mu \ll \a$.
This integral is finite for all $L$. 
Indeed using generating functions techniques,
it is easy to show that
\be
\int\limits_{-1}^{1}dt  \frac{1}{1-t} (P_L(t) - 1) = 
  -2\; (\sum_{k=1}^L \frac 1k) = -2 h(L),
\label{Legendre-int}
\ee 
where $h(L) = \sum_{k=1}^L \frac 1k$ is the harmonic number and $h(0) =0$. 
While $h(L) \approx log L$ for large $L$, it is finite and 
well--behaved for small $L$. Therefore
we obtain the effective action 
{\large
\be\fbox{$
 S_{one-loop} = S_0 + 
           \int \frac 12\;  \Phi (\delta \mu^2 - 
          \frac{g}{12\pi} h(\widetilde\Delta)) \Phi + o(1/\a)  $}
\ee}
to the first order in the coupling where
\be 
\delta \mu^2 = \frac{g}{8\pi} \sum_{J=0}^N \frac{2J+1}{J(J+1) +\mu^2}
\ee 
is the mass square renormalization, and $\widetilde\Delta$ 
is the function of the Laplacian which has eigenvalues
$L$ on $Y^L_l$. Thus we find that the effects due
to noncommutativity are analytic in the noncommutative parameter
$1/\a$.
This is a finite quantum effect with nontrivial, but mild $L$ dependence.
Therefore no IR singularity is
developed, and we conclude that there is no UV/IR problem on the
fuzzy sphere\footnote{
The author of \cite{Vaidya:2001bt} argued that the effective
action is not a smooth function of the external momentum and suggested
this to be a signature of UV/IR mixing. We disagree
with his result. }.

\begin{itemize}
\item
\underline{The commutative limit}
\end{itemize}

The commutative limit of the QFT is defined by the limit 
\be
\a \to \infty, \quad\mbox{keeping $R$, $g$, $\mu$ fixed}.
\ee
In this limit, the resulting
one-loop effective action differs from the effective action obtained
by quantization on the commutative sphere by an amount 
\be  \label{NCA}\fbox{$
\Gamma^{(2)}_{NCA} =  - \frac{g}{24\pi} \int \Phi h(\widetilde\Delta) \Phi.$}
\ee
We refer to this as a ``NonCommutative Anomaly'', 
since it is the piece of the quantum effective action which 
is slightly nonlocal and therefore not present in the classical action. 
``Noncommutative'' also
refers to fact that the quantum effective action depends on 
whether we quantize first or take the commutative limit first.

The new term $\Gamma^{(2)}_{NCA}$ modifies the dispersion relation on the 
fuzzy sphere.
It is very remarkable that such a ``signature'' of 
an underlying noncommutative space exists, even as the noncommutativity
on the geometrical level is sent to zero. A similar phenomena is the
induced Chern-Simon term in 3-dimensional gauge theory on $\R^3_\th$
\cite{Chu:2000bz}. 
This has important 
implications on the detectability of an underlying noncommutative structure.
The reason is that the vacuum fluctuations ``probe'' the structure
of the space even in the UV, 
and depend nontrivially on the external momentum  
in the nonplanar diagrams. 
Higher--order corrections may modify the result.
However since the theory is completely
well--defined for finite $N$, the above result  
\refeq{NCA} is meaningful for small
coupling $g$.

Summarizing, we find  
that quantization and taking the commutative limit does not commute 
on the fuzzy sphere, a fact which we refer to as 
``noncommutative anomaly''. A similar phenomenon also occurs on the
noncommutative quantum plane $\R^n_\th$. 
However, in contrast to the case of the quantum plane, the
``noncommutative anomaly'' here
is not due to UV/IR mixing since there is
no UV/IR mixing on the fuzzy sphere.
We therefore suggest that the existence of a ``noncommutative anomaly'' 
is a generic phenomenon and is independent of UV/IR 
mixing\footnote{However, 
as we will see, they are closely related.}. 
One can expect that the ``noncommutative anomaly'' does not occur for
supersymmetric theories on the 2--sphere. 

\section{Planar Limit of Quantum $\Phi^4$}

In this section, we consider the planar limit 
of the $\Phi^4$ theory on the fuzzy sphere at one loop. 
Since we
have shown that there is no UV/IR mixing on the fuzzy sphere, one may
wonder whether \refeq{sphere-reg} could provide a regularization for the
nonplanar two point function  \refeq{planeI2} on $\R^2_\th$ which does
not display an infrared singularity. This would be very nice,
as this would mean that UV/IR can be understood as an artifact that arises
out of a bad choice of variables. However, this is not the case.

To take the planar limit, we need in addition
to  \refeq{limit-p}, also 
\be
\mu^2 = m^2 R^2 \sim \a \to \infty, \quad 
     \mbox{keeping  $m$ fixed},
\ee
so that a massive scalar theory is obtained. 
We wish to identify in the limit of large $R$ the 
modes on the sphere with angular
momentum $L$ with modes on the plane with linear momentum $p$. This can
be achieved by matching the Laplacian on the plane 
with that on the sphere in the large radius limit, ie.
\be \label{Lp}
L (L+1)/R^2 = p^2.
\ee
It follows that
\be \label{pR}
p = \frac{L}{R}. 
\ee
Note that by (\ref{limit-p}), 
a mode with a fixed nonzero $p$ corresponds to a mode on the
sphere with large $L$:
\be \label{Lfinitep}
L \sim R \sim \sqrt{\a}. 
\ee
Since $L$ is bounded by $\a$, there is a UV cutoff $\La$ on the plane at
\be
\La = \frac{2\a}{R}.
\ee
Denote the external momentum of the two point function by $p$. 
It then follows that $\a \gg L \gg 1$ as long as $p \neq 0$. 

It is easy to see that the planar amplitude \refeq{aap} becomes 
\be
I^P = 2 \int_0^\La dk \frac{k}{k^2+m^2} 
\ee
in the quantum plane limit, with $k = J/R$. This is
precisely the planar contribution to the two point function on $\R^2_\th$. 

For the nonplanar two point function \refeq{aanp}, 
we can again use the formula (\ref{racah_app}) which is valid for all
$J$ and large $\a$, since the condition $\a \gg L$ is guaranteed by 
(\ref{Lfinitep}). 
Therefore 
\be \label{I-int-racah}
I^{NP}(p) = 2 \int_{0}^\La dk\; \frac{k}{k^2+m^2} 
       P_{p R}(1-2\frac{k^2}{\La^2})
\ee
For large $L = p R$, we can use the approximation formula
\be
P_L(\cos \phi) = \sqrt{\frac{\phi}{\sin \phi}}J_0((L+1/2)\phi) \quad
  +\; O(L^{-3/2}),
\label{PLJ-approx}
\ee
which is uniformly convergent \cite{Magnusbook} 
as $L \rightarrow \infty$ in the interval 
$0 \leq \phi \leq \pi - \epsilon$ for any small, but finite $\epsilon>0$.
Then one obtains
\beqa 
I^{NP}(p) &\approx&  2\int_{0}^\La dk \;\frac{k}{k^2+m^2} 
                   \sqrt{\frac{\phi_k}{\sin \phi_k}}\; J_0(p R \phi_k) \nn\\
          &\approx&  2\int_{0}^\La dk \;\frac{k}{k^2+m^2}\; J_0(\th p k), 
\label{sphere-reg}
\eeqa
where $\phi_k = 2 \arcsin(k/\La)$. The singularity at $\phi = \pi$ on the rhs 
of \refeq{PLJ-approx} (which is an artefact of the approximation and not 
present in the lhs) is integrable and does not contribute to \refeq{sphere-reg}
for large $p\La\th$. 
The integrals in \refeq{sphere-reg} are (conditionally) 
convergent for $p\neq 0$, and the approximations become exact
for $p\La\th \rightarrow \infty$. Therefore we recover
precisely the same form as the one loop nonplanar two point function on
$\R^2_\th$,
\be \label{planeI2}
\frac{1}{2 \pi} \int_{0}^\La d^2 k \; 
      \frac{1}{k^2+m^2} e^{i \th p \times k}.
\ee
For small $p\La\th$, i.e. in the vicinity of the induced infrared divergence
on $\R^2_\th$,
these approximations are less reliable. We can obtain
the exact form of the infrared divergence from \refeq{Legendre-int},
\be
I^{NP} = - 2 \log(p\sqrt{\th}) + (I^P - \log{\a}).
\ee
Hence we find the same logarithmic singularity in the infrared
as on $\R^2_\th$ \cite{Minwalla:1999px}. In
other words, we find that the UV/IR mixing phenomenon
which occurs in QFT on $\R^2_\th$ can be understood as the infinite 
limit of the noncomutative anomaly \refeq{NCA} on the fuzzy sphere.
Hence one could use the fuzzy sphere as a regularization
of $\R^2_\th$, where the logarithmic singularity 
$\log(p\sqrt{\th})$ gets ``regularized'' by \refeq{Legendre-int}.

\section{Discussion}

We have done a careful analysis of the one--loop dynamics 
of scalar $\Phi^4$ theory on the fuzzy sphere $S^2_N$. 
It turns out that the 
two point function is completely regular, without any UV/IR
mixing problem. We also give a closed expression for 
the two point function in the commutative limit, i.e. we find an exact form
for the leading term in a $1/N$ expansion.
Using this we discover a ``noncommutative anomaly'' 
(NCA), which characterizes the difference between  
the quantum effective action on the commutative sphere $S^2$ and
the commutative limit $N \rightarrow \infty$
of the quantum effective action on the fuzzy sphere. 
This anomaly is finite but mildly nonlocal on $S^2$, and changes the 
dispersion relation. 
It arises from the nonplanar loop integration. 

It is certainly intriguing and perhaps disturbing
that even an ``infinitesimal'' quantum structure of
(space)time has a finite, nonvanishing 
effect on the quantum theory. Of course this
was already found in the UV/IR phenomenon on $\R^n_\th$, 
however in that case
one might question whether the quantization procedure based on deformation
quantization is appropriate. On the fuzzy sphere, the
result is completely well-defined and unambiguous.
One might argue that a ``reasonable'' QFT should be free of such a
NCA, so that  the effective, macroscopic theory is insensitive to
small variations of the structure of spacetime at short distances. 
On the other hand, it is conceivable that 
our world is actually noncommutative, and the noncommutative dynamics 
should be taken seriously. Then  there is no reason to
exclude theories with NCA. In particular, one would like to 
understand better how sensitive these ``noncommutative anomalies''
are to the detailed quantum structure of spacetime.

By approximating the QFT on $\R^2_\th$ with the QFT on the fuzzy
sphere, we can explain the UV/IR mixing from the point of view of the
fuzzy sphere as a infinite variant of the NCA. 
In some sense, we have regularized $\R^2_\th$.
It would be interesting
to provide an explanation of the UV/IR mixing also for the higher
dimensional case $\R^4_\th$. To do this, the first step is to realize
$\R^4_\th$ as a limit of a ``nicer'' noncommutative
manifold. A first candidate is the product of two fuzzy spheres. Much
work remains to be done to clarify this situation. 

It would also be very desirable to include fermions and gauge fields in 
these considerations. In particular it will be interesting to determine 
the dispersion relation for ``photons'', depending on the 
``fuzzyness'' of the underlying geometry.
In the case of noncommutative QED on $\R^4_\th$, this question was studied 
in \cite{Matusis:2000jf,Ruiz:2000hu,Landsteiner:2001ky} 
where a nontrivial modification to the dispersion relation of the ``photon''
was found which makes the theory ill--defined.
In view of our results, 
one may hope that these modifications are milder on the fuzzy sphere 
and remain physically sensible.

\section{Technical supplements} 
\label{sec:supp-loop}

We derive the approximation formula (\ref{racah_app}) for large $\a$
and $0 \leq J \leq 2\a$, assuming $L \ll \a$. 

There is an exact formula for the $6j$ coefficients due to Racah
(see e.g. \cite{Varshalovich:1988ye}), which can be written in the form
\beqa
\left\{\begin{array}{ccc} \a&\a&L\\ \a&\a&J \end{array} \right\} &=&
(-1)^{2\a+J}\sum\limits_{n} (-1)^n \cdot  \nn\\
&&    \cdot    \left(\begin{array}{c}L \\ n\end{array}\right)^2
   \frac{(2\a-L)!(2\a+J+n+1)!(2\a-J)!(J!)^2}
   {(2\a+L+1)!(2\a+J+1)!(2\a-J-n)!((J-L+n)!)^2}.  \nn
\eeqa
The sum is from $n = \max\{0,L-J\}$ to $\min\{L,2\a-J\}$,
so that all factorials are non-negative.
Assume first that $L \leq J \leq 2\a-L$, so that the 
sum is from $0$ to $L$. 
Since $\a \gg L$, this becomes
\be
\left\{\begin{array}{ccc} \a&\a&L\\ \a&\a&J \end{array} \right\} \approx
(-1)^{2\a+J} \frac 1{(2\a)^{2L+1}} \sum\limits_{n=0}^L (-1)^n 
  \left(\begin{array}{c}L \\ n\end{array}\right)^2
  (4\a^2 - J^2)^n \left(\frac{J!}{(J-(L-n))!}\right)^2,
\label{6j-deriv}
\ee
dropping corrections of order $o(\frac L{\a})$.
Now there are 2 cases: either $J \gg L$, or otherwise $J \ll \a$
since $\a \gg L$. Consider first
\begin{enumerate}
\item  $J \gg L$:

Then $\frac{J!}{(J-(L-n))!}$ can be replaced by $J^{L-n}$, up to 
corrections of order $o(\frac LJ)$. Therefore 
\beqa
\left\{\begin{array}{ccc} \a&\a&L\\ \a&\a&J \end{array} \right\} &\approx&
  \frac{(-1)^{2\a+J}}{2\a} \left(\frac J{2\a}\right)^{2L}
  \sum\limits_{n=0}^L (-1)^n  \left(\begin{array}{c}L \\ n\end{array}\right)^2
  \left( (\frac{2\a}J)^2 - 1 \right)^n \nn\\
  &=& \frac{(-1)^{2\a+J}}{2\a} P_L(1- \frac{J^2}{2\a^2}),
\eeqa
as claimed.
\item $J \ll \a$: 

Then in the sum (\ref{6j-deriv}), 
the dominant term is $n = L$, because 
$\frac{J!}{(J-(L-n))!} \leq J^{L-n}$. Therefore one can safely
replace the term $\frac{J!}{(J-(L-n))!}$ in this sum by its value at $n=L$,
namely $J^{L-n}$. The remaining terms are smaller by a 
factor of $(\frac J{\a})^2$. Hence we can continue as in case 1.

\end{enumerate}

If  $J \leq L$, one 
can either use the same argument as in the 2nd case since the
term $n = L$ is dominant, or use the symmetry of the 
$6j$ symbols in $L,J$ together with 
$P_L(1- \frac{J^2}{2\a^2}) \approx  P_J(1- \frac{L^2}{2\a^2})$
for $J,L \ll \a$.
Finally if $J+L \geq 2\a$, then the term $n=0$ dominates, and 
one can proceed as in case 1. Therefore
(\ref{racah_app}) is valid for all $0 \leq J \leq 2\a$.

One can illustrate
the excellent approximation for the $6j$ symbols provided 
by (\ref{racah_app}) for all $0 \leq J \leq 2\a$ using numerical calculations.

\paragraph{Identities for $3j$ and $6j$ symbols.}

We quote here some identities of the $3j$ and $6j$ symbols which are 
used to derive the expressions (\ref{aap}) and (\ref{aanp})
for the one--loop corrections.
The $3j$ symbols satisfy the orthogonality relation
\be
\sum_{j,l}
\left( \begin{array}{ccc} J&L&K\\ j&l&k \end{array} \right)
\left( \begin{array}{ccc} J&L&K'\\ -j&-l&-k' \end{array} \right)
= \frac{(-1)^{K-L-J}}{2K+1} \d_{K, K'} \d_{k, k'},
\ee
assuming that $(J,L,K)$ form a triangle. 

The $6j$ symbols satisfy standard symmetry properties, and
the orthogonality relation
\be
\sum_{N} (2N+1)
    \left\{ \begin{array}{ccc} A&B&N\\ C&D&P \end{array} \right\}
    \left\{ \begin{array}{ccc} A&B&N\\ C&D&Q  \end{array} \right\}
  = \frac 1{2P+1}\; \d_{P, Q} ,
\ee
assuming that $(A,D,P)$ and $(B,C,P)$ form a triangle.
Furthermore, the following sum rule is used in \refeq{aanp}
\be
\sum_{N} (-1)^{N+P+Q} (2N+1) 
  \left\{ \begin{array}{ccc} A&B&N\\ C&D&P  \end{array} \right\}
  \left\{ \begin{array}{ccc} A&B&N\\ D&C&Q  \end{array} \right\}
 = \left\{ \begin{array}{ccc} A&C&Q\\ B&D&P  \end{array} \right\}.
\ee
The Biedenharn--Elliott relations are needed to verify associativity of
(\ref{YY}): 
\beqa \label{BE}
&& \sum_{N} (-1)^{N+S} (2N+1) 
 \left\{ \begin{array}{ccc} A&B&N\\ C&D&P  \end{array} \right\}
 \left\{ \begin{array}{ccc} C&D&N\\ E&F&Q  \end{array} \right\}
 \left\{ \begin{array}{ccc} E&F&N\\ B&A&R  \end{array} \right\} = \nn\\
&&\qquad \left\{ \begin{array}{ccc} P&Q&R\\ E&A&D  \end{array} \right\}
 \left\{ \begin{array}{ccc} P&Q&R\\ F&B&C  \end{array} \right\},
\eeqa
where $S = A+B+C+D+E+F+P+Q+R$.
All these can be found e.g. in \cite{Varshalovich:1988ye}.

%% QFT on S^2_{q,N}

\chapter{Second quantization on the $q$-deformed fuzzy sphere}
\label{chapter:qFSII}

We have seen in the example of the fuzzy sphere that field theory can be 
$q$-deformed, in a more or less straightforward manner. 
Other $q$--deformed field theories have been considered
before, see for example 
\cite{Fiore:1993mb,Cerchiai:1998ee,Hebecker:1992gc,Isaev:1993mt,Majid:1994cm,Meyer:1995wi,Podles:1996qy} 
and references therein, mainly for scalar fields though.
However, they 
should be considered as ``classical'' field theory on
some kind of nonlocal space, rather than a  
quantized field theory in the physical sense, where each mode of a field 
should be an operator on a Hilbert space or equivalently all
possible configurations should be integrated over via a Feynman path
integral. 
In this final chapter, we proceed to (euclidean) {\em quantum} 
field theory on the $q$--deformed fuzzy sphere $\S_{q,N}$. 
It is based on a paper \cite{qFSII} written in collabration with
Harald Grosse and John Madore.

The second quantization of $q$-deformed field theories
has proved to be difficult. The main 
problem is probably the apparent incompatibility between the 
symmetrization postulate of quantum field theory (QFT)
which involves the permutation group, 
and the fact that quantum groups are naturally related to the
braid group rather than the permutation group. One could of course 
consider theories with generalized statistics; however if 
$q$--deformation is considered as a ``deformation'' of ordinary 
geometry,
then it should be possible to define models with a smooth limit 
$q \rightarrow 1$. In particular, the degrees of freedom 
should be independent of $q$. This is the guiding principle 
of the present approach, together with covariance under the quantum group
of motions $U_q(su(2))$: our goal is to define 
a $q$--deformed (euclidean) 
quantum field theory which is essentially bosonic, and has 
a smooth limit $q \rightarrow 1$ as an ordinary quantum field theory.
Moreover, we would like to have a map from the $q$-deformed
quantum field theory to some undeformed, but nonlocal theory, i.e.
some kind of Seiberg-Witten map. This is also expected from the 
point of view of string theory \cite{Seiberg:1999vs}.
While some proposals have been given in the literature
\cite{Oeckl:1999zu,Chaichian:1999wy} how to define 
quantum field theories 
on spaces with quantum group symmetry, none of them seems to
satisfies these requirements.

We will show how to accomplish this goal using a path integral approach, 
integrating over all modes or harmonics of the field. To this end,
it turns out to be useful to define a 
quasiassociative star--product of the modes, based on the Drinfeld twist. 
Quasiassociativity appears only in intermediate steps of the mathematical
formalism, and is not at all in contradiction with the axioms of 
quantum mechanics. Indeed we provide also an equivalent
formulation which is entirely within the framework of associative algebras,
and in particular we give in Section \ref{subsec:2plus1} an operator 
formulation of scalar field theory in $2_q + 1$ dimensions. 
The latter is very instructive to understand how e.g. 
bosons and quantum groups
can coexist.

The models we find have a manifest $U_q(su(2))$ symmetry with 
a smooth limit $q \rightarrow 1$, and satisfy positivity and twisted bosonic 
symmetry properties. We also develop some of 
the standard tools of quantum field theory, in particular we
give a systematic way to calculate $n$--point correlators 
in perturbation theory. As applications of the formalism, the
4--point correlator of a free scalar field theory is calculated, 
as well as the planar contribution to the tadpole diagram in a $\phi^4$ theory.
Gauge fields are discussed in Section \ref{subsec:gaugefields-2}. 
Here the correct quantization is less clear at present,
and we only suggest 2 possible variants of a path integral quantization. 

We should point out that while only the 
$q$-deformed fuzzy sphere $\S_{q,N}$ is considered here,  
the proposed quantization prodecure 
is not restricted to this case, and not even to 2 dimensions.
This chapter is the result of a long period of trial-and-error
trying to $q$-deform
quantum field theory with the above requirements. 
$\S_{q,N}$ is particularly well suited to attack the problem of
second quantization, because there is only a finite number of modes.
This means that all considerations can be done on a purely algebraic 
level, and are essentially rigorous.
However, our constructions 
can be applied in principle to any other $q$-deformed space,
provided the decomposition of the fields in terms 
of irreducible representations of the underlying quantum group is 
known. Furthermore, it requires the knowledge of some rather involved
group-theoretical objects (such as coassociators) 
build from Drinfeld twists. While the latter
are not needed explicitly, much work is still needed before
for example a full loop calculation becomes possible.

\section{Why QFT on $q$-deformed spaces is difficult}

To understand the problem, consider scalar fields,
which are elements $\psi \in \S_{q,N}$. 
A reasonable action could have the form
\be
S[\psi] = - \int\limits_{\S_{q,N}}\hf\;\psi \Delta \psi + \lambda\psi^4,
\ee
where $\Delta$ is the Laplacian \cite{qFSI}. Such actions are
invariant under the quantum group $U_q(su(2))$ of rotations, and they are
real, $S[\psi]^* = S[\psi]$. 
They define a first--quantized euclidean scalar
field theory on the $q$--deformed fuzzy sphere. 

We want to study the second quantization of these models. 
On the undeformed fuzzy sphere, this is fairly straightforward 
\cite{Grosse:1996ar,Madore:1992bw}:
The fields can be expanded in terms of irreducible \reps of $SO(3)$,
\be
\psi(x) = \sum_{K,n} \psi_{K,n}(x) \;  a^{K,n} 
\label{psi_field_intro}
\ee
with coefficients $a^{K,n} \in \C$. The above actions then become 
polynomials in the variables $a^{K,n}$
which are invariant under $SO(3)$, 
and the ``path integral'' is naturally defined as 
the product of the ordinary integrals over the coefficients $a^{K,n}$.
This defines a quantum field theory which has a $SO(3)$ rotation symmetry, 
because the path integral is invariant. 

In the $q$--deformed case, this is not so easy.
The reason is that the coefficients $a^{K,n}$ in (\ref{psi_field_intro}) 
must be considered as \reps of $U_q(su(2))$ in order 
to have such a symmetry at the quantum level. This implies that
they cannot be ordinary complex numbers, because
a commutative algebra is not consistent with the action of $U_q(su(2))$,
whose coproduct is not cocommutative. Therefore an ordinary
integral over commutative modes $a^{K,n}$ would violate $U_q(su(2))$ 
invariance at the quantum level. On the other hand, no associative 
algebra with generators $a^{K,n}$ is known (except for some simple 
representations) which is both covariant under 
$U_q(su(2))$ and has the same Poincar\'e series as classically, 
i.e. the dimension of the space of polynomials at a given degree is 
the same as in the undeformed case. The 
latter is an essential physical requirement at least for 
low energies, in order to have the correct number of degrees of freedom,
and is usually encoded in a symmetrization postulate. It means that 
the ``amount of information'' contained in the $n$--point functions 
should be the same as for $q=1$. These issues will 
be discussed on a more formal level in Section \ref{sec:QFT}.
While some proposals have been given in the literature
\cite{Oeckl:1999zu,Chaichian:1999wy} how to define 
QFT on spaces with quantum group symmetry, none of them seems to
satisfies all these requirements.

One possible way out was suggested in \cite{Fiore:1996bu}, where it
was pointed out that a symmetrization can be achieved using a  
Drinfeld twist, at least in any given $n$--particle sector.
Roughly speaking, the Drinfeld twist relates the tensor product of \reps
of quantum groups to the tensor product of undeformed ones, 
and hence essentially allows to use the usual completely
symmetric Hilbert space. The problem remained, however, how to 
treat sectors with different particle number simultaneously,
which is essential for a QFT, and how to handle
the Drinfeld twists which are very difficult to calculate.

We present here a formalism which solves these problems, by
defining a star product of the modes $a^{K,n}$ which is 
covariant under the quantum group, and in the limit
$q \rightarrow 1$ reduces to the commutative algebra of functions in the
$a^{K,n}$. This algebra is quasiassociative, but satisfies all
the requirements discussed above. In particular, the number of 
independent polynomials in the $a^{K,n}$ is the same as usual.
One can then define an invariant path integral, which yields a 
consistent and physically reasonable definition of a second--quantized 
field theory with a quantum group symmetry. In particular, the 
``correlation functions'' will satisfy invariance, hermiticity, positivity
and symmetry properties. An essentially equivalent formulation in terms 
of a slightly extended associative algebra will be presented
as well, based on constructions by Fiore \cite{Fiore:1997fb}. It turns out 
to be related to the general considerations in \cite{Mack:1992tg}.
The appearance of quasiassociative algebras is also reminiscent of 
results in the context of $D$--branes on WZW models 
\cite{Alekseev:1999bs,Cornalba:2001sm}.

Our considerations are not restricted to 2 dimensions, 
and should be applicable to other spaces with quantum group symmetry as well.
The necessary mathematical tools will be developed in 
Sections \ref{sec:drinfeld_twist}, \ref{sec:twisting}, and 
\ref{sec:operator-form}. After discussing the definition and basic
properties of QFT on $\S_{q,N}$ in Section \ref{sec:QFT}, we
derive formulas to calculate
$n$--point functions in perturbation theory, and find
an analog of Wick's theorem. All diagrams on $\S_{q,N}$ are of course finite,
and vacuum diagrams turn out to cancel a usual.
The resulting models can also be interpreted as field theories
on the undeformed fuzzy sphere, with slightly ``nonlocal'' interactions.

As applications of the general method, we consider first the
case of a free scalar field theory, and calculate the 4--point functions.
The tadpole diagram for a $\phi^4$ model is studied as well, and
turns out to be linearly divergent as $N \rightarrow \infty$. 
We then discuss two possible quantizations of gauge models,
and finally consider scalar field theory on $\S_{q,N}$ with an extra 
time.

We should stress that our approach is 
quite conservative, as it aims to find a ``deformation'' of standard 
quantum field theory in a rather strict sense, with ordinary statistics.
Of course on can imagine other, less conventional approaches,
such as the one in \cite{Oeckl:1999zu}.
Moreover, we only consider the case $q \in \R$ in this paper. It should be 
possible to modify our methods so that the 
case of $q$ being a root of unity can also be covered. 
Then QFT on more realistic spaces such as
4--dimensional quantum Anti--de Sitter space \cite{Steinacker:1999xu} could 
be considered as well. There, the number of modes as well as the
dimensions of the relevant \reps are finite at roots of unity, 
as in the present paper.

\section{More on Drinfeld twists}
\label{sec:drinfeld_twist}

We first have to extend our knowledge of Drinfeld twists,
which were briefly introduced in Chapter \ref{chapter:qDbranes}.
In order to avoid confusions, the language
will be quite formal initially. To a given a finite--dimensional simple Lie 
algebra $g$ (for our purpose just $su(2)$), one can associate 
2 Hopf algebras \cite{Faddeev:1990ih,Drinfeld:1988in,Jimbo:1985zk}:
the usual $(U(g)[[h]],m,\eps,\Delta,S)$,
and the $q$--deformed $(U_q(g),m_q,\eps_q,\Delta_q,S_q)$. Here
$U(g)$ is the universal enveloping algebra, 
$U_q(g)$ is the $q$--deformed universal enveloping algebra, and
$U(g)[[h]]$ are the formal power series in $h$ with 
coefficients in $U(g)$. The symbol
$$
q=e^h
$$
is considered formal for now.
As already discussed in Section \ref{subsec:drinfeld_twist}, 
a theorem by Drinfeld states that there exists an algebra isomorphism
\be
\varphi: U_q(g) \rightarrow U(g)[[h]]
\label{phi}
\ee
and a `twist', i.e. an element 
$$
\F = \F_1\tens \F_2 \; \in \; U(g)[[h]]\tens U(g)[[h]]
$$
(in a Sweedler notation, where a sum is implicitly understood) 
satisfying
\berr
&&(\varepsilon\tens \id)\F=\one=(\id\tens \varepsilon)\F,
\label{cond2}\\
&&\F = \one\tens\one +o(h),
\label{cond2bis}
\err
which relates these two Hopf algebra $U_q(g)$ and $U(g)[[h]]$ 
as follows: if  $\F^{-1}=\F^{-1}_1\tens \F^{-1}_2$ is the 
inverse\footnote{it exists as a formal power series because of 
(\ref{cond2bis})} of $\F$, then  
\berr
\p (m_q)    &=& m \circ(\p\tens\p),                      \label{defm}\\
\eps_q      &=& \eps\circ \p,                             \label{defc}\\
\p(S_q(u))  &=& \g^{-1} S(\p(u))\g,                      \label{def1}\\
\p(S^{-1}_q(u))  &=&  \g' S(\p(u))\g'^{-1},              \label{def1'}\\
(\p\tens\p) \Delta_q(u)&=& \F\Delta(\p(u))\F^{-1},        \label{defd-1}\\
(\p\tens\p)\RR &=& \F_{21} q^{ \frac t2}\F^{-1}.        \label{defR}
\err
for any $u \in U_q(g)$.
Here $t:=\Delta(C)-\one\tens C-C\tens \one$ is 
the canonical invariant element 
in $U(g)\tens U(g)$, $C$ is the quadratic Casimir, and
\berr
\g  &=& S(\F_1^{-1}) \F_2^{-1},\qquad\qquad \g' =\F_2 S \F_1  %\nn\\
%\g^{-1} &=& \F_1  S \F_2 = S \g', \qquad 
%        \g'^{-1} = S(\F^{-1}_2) \F^{-1}_1 = S \g.
\label{gammas}
\err
%(from $S \phi_1 \phi_2 \tens \eps(\phi_3)$ ...).
Moreover, $\g^{-1} \g'$ is central in $U(g)[[h]]$.
The undeformed maps\footnote{we will suppress the multiplication maps 
from now on} $m,\eps,\Delta,S$ have
been linearly extended from $U(g)$ to $U(g)[[h]]$; notice that $S^2 = 1$.
$\F_{21}$ is obtained from $\F$ by flipping the tensor product. 
This kind of notation will be used throughout from now on.
Coassociativity of $\Delta_q$ follows from the fact that the  
(nontrivial) coassociator
\be
\phi:=[(\Delta\tens \id)\F^{-1}](\F^{-1}\tens \one)(\one \tens \F)
[(\id \tens \Delta)\F]
\label{defphi}
\ee
is $U(g)$--invariant, i.e. 
$$
[\phi,\Delta^{(2)}(u)]=0
$$
for $u \in U(g)$. Here $\Delta^{(2)}$ denotes the usual 2--fold coproduct.

In the present paper, we only consider finite--dimensional
representations, i.e. operator algebras rather than the abstract ones.
Then the formal parameter $q=e^h$ can be replaced by a real
number close to 1, and all statements in this section still hold 
since the power series will converge. One could then identify the algebras
$U(su(2))$ with $U_q(su(2))$ (but not as coalgebras!) via the 
the isomorphism $\varphi$. We will usually keep $\varphi$ explicit, however, 
in order to avoid confusions.

It turns out that the twist $\F$ is not determined uniquely, but there
is some residual ``gauge freedom'' \cite{Drinfeld:quasiGal,Drinfeld:quasi},
\be
\F \rightarrow \F T
\label{F_gauge}
\ee
with an arbitrary symmetric $T \in U(g)[[h]]^{\tens 2}$ which commutes with 
$\Delta(U(g))$ and satisfies 
(\ref{cond2}), (\ref{cond2bis}). The symmetry of $T$ guarantees that $\RR$ 
is unchanged, so that $\F$ remains a twist from $(U(g)[[h]],m,\eps,\Delta,S)$
to $(U_q(q),m_q,\eps_q,\Delta_q,S_q)$.
We will take advantage of this below.

While for the twist $\F$, little is known apart apart from 
its existence, one can show \cite{Fiore:1997fb} using results of 
Kohno \cite{kohno1,kohno2} and Drinfeld 
\cite{Drinfeld:quasiGal,Drinfeld:quasi} 
that the twists can be chosen 
such that the following formula holds:
\be
\phi=\lim_{x_0,y_0\rightarrow 0^+}\left\{x_0^{-\frac h{2\pi i} t_{12}}
\vec{P}\exp\left[-\frac h{2\pi i}\int\limits^{1-y_0}_{x_0}dx
  \left(\frac{t_{12}}{x}+\frac{t_{23}}{x-1}\right)\right] 
   y_0^{\frac h{2\pi i} t_{23}}\right\} 
=\one + o(h^2).
\label{phi-integral}
\ee
Here $\vec P$ denotes the path--ordered exponential. 
Such twists were called ``minimal'' by Fiore \cite{Fiore:1997fb}, who showed 
that they satisfy the following remarkable relations:
\berr
\one &=& \F \Delta \F_{1}(1 \tens (S \F_2)\gamma) \label{F_id_1},\\
 &=& (1\tens S\F_2 {\g'}^{-1})\F (\Delta \F_1)   \label{F_id_2} \\
 &=& \F \Delta \F_{2}((S \F_1)\gamma'^{-1}\tens 1) \label{F_id_3}\\
 &=&(\gamma^{-1}S\F_1^{-1}\tens 1)\Delta\F_{2}^{-1}\F^{-1} \label{F_id_4}\\
 &=&  \Delta \F^{-1}_2 \F^{-1}(\g' S\F^{-1}_1\tens 1)\label{F_id_5}\\
 &=&  (1\tens \gamma' S\F^{-1}_2)\Delta \F^{-1}_{1}\F^{-1} \label{F_id_6}
\err 
All coproducts here are undeformed. 
For such twists, one can write down inverses of the elements $\g, \g'$:
\be
\g^{-1} = \F_1  S \F_2 = S \g', \qquad 
        \g'^{-1} = S(\F^{-1}_2) \F^{-1}_1 = S \g.
\label{gammas'}
\ee
Furthermore, we add the following observation: let $(V_i,\trr)$ be \reps of 
$U(g)$ and $I^{(3)}\in V_1\tens V_2\tens V_3$ be an invariant tensor, so that 
$u \trr I^{(3)} \equiv \Delta^{(2)}(u) \trr I^{(3)} = \eps(u) I^{(3)}$
for $u \in U(g)$. 
Then the (component--wise) action of $\phi$ on $I^{(3)}$ is trivial:
\be
\phi \trr I^{(3)} = I^{(3)}.
\label{phi_trivial}
\ee 
This follows from (\ref{phi-integral}): observe that 
$t_{12}$ commutes with $t_{23}$ in the exponent,
because e.g. $(\Delta(C)\tens\one)$ can be replaced by $\one\tens\one\tens C$ 
if acting on invariant tensors. Therefore the 
path--ordering becomes trivial, and (\ref{phi_trivial}) follows.

\paragraph{Star structure.}

Consider on $U(su(2))[[h]]$ the (antilinear) star structure
\be
H^* = H ,\quad {X^{\pm}}^* = X^{\mp},
\label{U_star_cpct_1}
\ee
with $h^* = h$, since $q$ is real.
It follows e.g. from its explicit form 
\cite{Curtright:1991ri,Curtright:1990sw} that the algebra
map $\p$ is compatible with this star, 
$$
\p(u)^* = \p(u^*).
$$
It was shown in \cite{Jurco:1994te} that using a suitable gauge transformation
(\ref{F_gauge}), it is possible to choose $\F$
such that it is unitary,
\be
(\ast \tens \ast) \F = \F^{-1}.
\label{F_real}
\ee
Moreover, it was stated in  \cite{Fiore:1997fb} without proof
that the following stronger statement holds:

\begin{prop}
Using a suitable gauge transformation (\ref{F_gauge}), 
it is possible to choose a twist $\F$ which for $q \in \R$
is both unitary and minimal, so that (\ref{F_real}) and 
(\ref{F_id_1}) to (\ref{F_id_6}) hold. 
\label{F_minimal}
\end{prop} 
Since this is essential for us, we provide a proof in 
Section \ref{chapter:supp-II}.

\section{Twisted $U_q(g)$--covariant $\star$ --product algebras}
\label{sec:twisting}

Let $(\A,\cdot,\trr)$ be an associative $U(g)$--module algebra,
which means that there exists an action 
\berr
U(g) \times \A &\rightarrow& \A, \nn\\
  (u,a) &\mapsto& u \trr a   \nn
\err
which satisfies $u \trr (ab) = (u_{(1)} \trr a) (u_{(2)} \trr b)$
for $a, b \in \A$. Here $\Delta(u) = u_{(1)} \tens u_{(2)}$ denotes the 
undeformed coproduct. Using the map $\p$ (\ref{defphi}), 
we can then define an action of $U_q(g)$ on $\A$ by
\be
u \trr_q a:= \p(u) \trr a,
\ee
or $u \trr_q a_i = a_j \pi^j_i(\p(u))$ in matrix notation.
This does {\em not} define a $U_q(g)$--module algebra,
because the multiplication is not compatible with the coproduct of $U_q(g)$.
However, one can define a new multiplication on $\A$ as follows 
\refeq{star-0}: 
\be\fbox{$
a \star b := (\F^{-1}_1\trr a) \cdot(\F^{-1}_2\trr b) 
            = \cdot(\F^{-1} \trr(a\tens b)) $}
\label{star}
\ee 
for any $a,b \in \A$. 
It is well--known \cite{Majid:1996kd} that  $(\A, \star, \trr_q)$ 
is now a $U_q(g)$--module algebra: 
\berr
u \trr_q (a \star b) &=& 
      \varphi(u)\trr \lb(\F^{-1}_1\trr a)\cdot(\F^{-1}_2\trr b)\rb \nn\\ 
 &=& \cdot \lb(\Delta(\varphi(u)) \F^{-1}) \trr a \tens b\rb \nn\\
 &=& \cdot \lb (\F^{-1}(\varphi\tens\varphi)\Delta_q(u)) \trr a \tens b\rb \nn\\
 &=& \star \lb \Delta_q(u) \trr_q a \tens b\rb \nn
\err
for $u\in U_q(g)$. In general, this product
$\star$ is not associative, but it is {\em quasiassociative},
which means that
\be
(a\star b)\star c = (\tilde\phi_1\trr a)\star\lb(\tilde\phi_2\trr b)
         \star(\tilde\phi_3\trr c)\rb.
\label{left_assoc}
\ee
where 
\be
\tilde\phi := (\one \tens \F)
   [(\id \tens \Delta)\F ][(\Delta\tens \id)\F^{-1}](\F^{-1}\tens \one) = 
    U_{\F} \; \phi \; U_{\F}^{-1}
\label{defphitilde}
\ee
with
$$
U_{\F} = (\one \tens \F) [(\id \tens \Delta)\F]\;\; \in U(g)^{\tens 3},
$$
which satisfies
$$
[\tilde\phi, \Delta^{(2)}_q(u)] = 0
$$
for $u \in U_q(g)$.
All this follows immediately from the definitions.
Moreover, the following simple observation will be very useful:

\begin{lemma}
In the above situation, 
\be
(a\star b)\star c =  a\star (b\star c) 
\label{triv_assoc}
\ee
if one of the factors $a,b,c \in \A$ is invariant under $U(g)$.
If $(\A, \cdot)$ is commutative, then any element $S \in \A$  which is 
invariant under the action of $U(g)$, 
$u \trr S = \eps(u)\; S$, is central in  $(\A, \star, \trr_q)$
\label{inv_central}
\end{lemma}
Note that invariance of an element $a \in \A$ under 
$U(g)$ is the same as invariance under $U_q(g)$.
\begin{proof}
This follows immediately from (\ref{cond2}) together with the definition of 
$\tilde \phi$. To see the last statement,
assume that $S$ is invariant. Then 
\berr
S \star a &=&  (\F_1^{-1} \trr S)\cdot(\F_2^{-1} \trr a)\nn\\
           &=& \cdot\lb((\eps\tens \one)\F^{-1}) \trr(S \tens a)\rb  \nn\\
           &=& S \cdot a = a \cdot S \nn\\
           &=&  a \star S
\err
for any $a\in \A$.
\end{proof}

For actual computations, it is convenient to use a tensor notation
as follows: assume that the elements
$\{a_i\}$ of $\A$ form a \rep of $U(g)$.
Denoting $\tilde\phi^{rst}_{ijk} = \pi^r_i(\tilde\phi_1)$
$\pi^s_j(\tilde\phi_2)\pi^t_k(\tilde\phi_3)$, 
equation (\ref{left_assoc}) can be written as 
\berr
(a_i\star a_j)\star a_k &=& a_r \star(a_s\star a_t) \tilde\phi^{rst}_{ijk},
       \;\; \mbox{or} \nn\\ 
(a_1\star a_2)\star a_3 &=&  a_1 \star(a_2\star a_3) \; \tilde\phi_{123}.
\err
The last notation will always imply a matrix multiplication as above.

Conversely, given a $U_q(g)$--module algebra $(\A, \star, \trr_q)$,
one can twist it into a $U(g)$--module algebra $(\A,\cdot,\trr)$ 
by 
$$
a \cdot b := (\p^{-1}(\F^{(1)})\trr_q a) \star(\p^{-1}(\F^{(2)})\trr_q b) 
$$ 
where of course $u \trr a = \p^{-1}(u) \trr_q a$.
Now if $(\A, \star,\trr_q)$ was associative, then 
$(\A, \cdot,\trr)$ is quasiassociative,
$$
a\cdot (b\cdot c) = \phi \trr_q^{(3)} \lb(a\cdot b) \cdot c\rb 
 :=\lb(\phi_1\trr_q a)\cdot(\phi_2\trr_q b)\rb
   \cdot(\phi_3\trr_q c).
$$
Such a twist was used in \cite{qFSI} to obtain the associative
algebra of functions on the $q$--deformed fuzzy sphere from the 
quasi--associative algebra of functions on $D2$--branes in the $SU(2)$
WZW model found in \cite{Alekseev:1999bs}.

\paragraph{Commutation relations and $\RR$--matrices.}

These twisted algebras have a more intrinsic
characterization, which is much more practical. Consider 
a commutative $U(g)$--module algebra $(\A, \cdot, \trr)$, and the associated
twisted $U_q(g)$--module algebra $(\A, \star, \trr_q)$ as defined above.
Observe that the definition (\ref{star}) is equivalent to 
\berr
a \star b &=& (\F_1^{-1} \trr a) \cdot (\F_2^{-1} \trr b) = 
            (\F_2^{-1} \trr b)\cdot (\F_1^{-1} \trr a) \nn\\
    &=& \cdot\lb (\F^{-1} \F \F^{-1}_{21}) \trr (b\tens a) \rb \nn\\
    &=& (\TR_2 \trr_q b) \star (\TR_1\trr_q a)
\label{star_R_calc}
\err
where we define
\be
\TR := (\p^{-1}\tens\p^{-1}) \F_{21}\F^{-1} = \TR^{-1}_{21}.
\ee 
In a given representation, this can be written as
\be 
a_i \star a_j =  a_k \star a_l \; \TR_{ij}^{lk}, 
     \;\; \mbox{or}\;\;\;  a_1 \star a_2 = a_2 \star a_1 \; \TR_{12}
\label{aa_CR}
\ee
where
\be
\TR^{ji}_{kl} = (\pi_k^j \tens  \pi_l^i) (\TR).
\label{TR-matrix}
\ee
Now there is no more reference to the ``original'' $U(g)$--covariant 
algebra structure. $\TR$ does not satisfy the quantum Yang--Baxter
equation in general, which reflects the non--associativity of the 
$\star$ product. However it does satisfy
\berr
\TR \TR_{21} &=& \one,  \label{triang}\\
\TR_{(12),3} := (\Delta_q \tens 1) \TR  &=& 
                  \tp_{312} \TR_{13} \tp^{-1}_{132} \TR_{23}\tp_{123} \\
\TR_{1,(23)} := (1 \tens \Delta_q) \TR  &=&
              \tp^{-1}_{231} \TR_{13} \tp_{213} \TR_{12} \tp^{-1}_{123},
\label{quasitr}
\err
as can be verified easily. This means that we are working with
the quasitriangular quasi--Hopf algebra 
\cite{Drinfeld:quasiGal,Drinfeld:quasi} 
$(U_q(g),\Delta_q,\tp,\TR)$, which is obtained from the ordinary Hopf algebra 
$(U(g),\Delta,\one,\one)$ by the Drinfeld twist $\F$. 
In practice, it is much easier to work with $\TR$ than with $\F$. 
For $q \in \R$, one can in fact write 
\be
\TR = \RR \; \sqrt{\RR_{21} \RR_{12}}^{-1},
\ee
where $\RR$ is the usual universal $R$ --matrix (\ref{defR}) 
of $U_q(g)$, which does satisfy the quantum Yang--Baxter equation.
The product $(\RR_{21} \RR_{12})$ could moreover be expressed in terms 
of the  Drinfeld--Casimir 
\be
v = (S_q \RR_2) \RR_1 q^{-H},
\label{v}
\ee
which is central
in $U_q(g)$ and satisfies $\Del(v) = (\RR_{21} \RR_{12})^{-1} v\tens v$.
The square root is well--defined on all the \reps 
which we consider, since $q$ is real.

\paragraph{Twisted Heisenberg algebras.}

Consider the $U(g)$--module algebra 
$(\A_H, \cdot, \trr)$ with generators $a_i$ and $a^{\dagger}_j$ in
some given irreducible representation and commutation relations 
\berr
[a^{\dagger}_i, a^{\dagger}_j] &=& 0 = [a_i,a_j], \nn\\
\[a^{\dagger}_i, a_j\] &=& (g_c)_{ij} 
\err
where $(g_c)_{ij}$ is the (unique) invariant tensor
in the given representation of $U(g)$. 
We can twist $(\A_H, \cdot, \trr)$ as above, and
obtain the $U_q(g)$--module algebra $(\A_H, \star, \trr_q)$. 
The new commutation relations among the generators can be 
evaluated easily:
\berr
a_1 \star a_2 &=&  a_2 \star a_1 \;  \TR_{12} , \nn\\
a^{\dagger}_1 \star a^{\dagger}_2 &=& 
           a^{\dagger}_2 \star a^{\dagger}_1 \; \TR_{12}, \nn\\
a^{\dagger}_1 \star a_2 &=& g_{12} + \; a_2\star a^{\dagger}_1\;\TR_{12}.
\label{a_del_CR}
\err
Here 
\be
 {g}_{n m} = (g_c)_{r s} \; \pi^{r}_{n}(\F_1^{-1}) \pi^{s}_{m}(\F_2^{-1})
\label{g_twist}
\ee  
is the unique rank 2 tensor which is invariant under the action 
$\trr_q$ of $U_q(g)$. A similar relation holds for the invariant tensor 
with upper indices:
\be
{g}^{n m} = \pi_{r}^{n}(\F_1) \pi_{s}^{m}(\F_2)\; g_c^{r s},
\label{g_twist_up}
\ee  
which satisfies ${g}^{n m} {g}_{m l} = \d^n_l$. 
In particular, it follows that
\be
a\star a:=  a_1\star a_2 \; g^{12} =  a_1 \cdot a_2\; (g_c)^{12},
\label{twisted_square}
\ee
therefore the invariant bilinears remain undeformed. This
is independent of the algebra of the generators $a_i$.

It is sometimes convenient to use the $q$--deformed antisymmetrizer
\cite{Fiore:1996bu}
$$
P^-_{12} = \F_{12} (1-\d^{21}_{12}) \F^{-1}_{12} 
    = (\one - P \tilde \RR)_{12}
$$
acting on the tensor product of 2 identical representations, where 
$P$ is the flip operator. Then the commutation relations
(\ref{aa_CR}) can be written as
\be
a_1\star a_2\; P^-_{12} = 0.
\ee
For products of 3 generators, the following relations hold: 
\begin{lemma}
\berr
a_{1} \star (a_2\star a_3) &=&  (a_2 \star a_3) \star a_{ 1}\;\TR_{1,(23)},
                             \label{aaa_braid}\\
a^{\dagger}_1 \star (a_2\star a_3\; g^{23}) &=&
   2 a_1 + (a_2\star a_3\; g^{23})\star a^{\dagger}_1, \label{del_asquare}\\
a^{\dagger}_1 \star (a_2\star a_3\; P^-_{23}) &=&
   (a_2\star a_3\; P^-_{23}) \star a^{\dagger}_1\; \TR_{1,(23)}.
\label{aP-a_braid}
\err

\label{twist_qa_lemma}
\end{lemma}
The proof is in Section \ref{chapter:supp-II}.

\subsection{Integration}
\label{subsec:integral}

From now on we specialize to $g = su(2)$, even though much of the following
holds more generally.
Let  $\{a_i\}$ be a basis of the spin $K$ \rep of $U(su(2))$ with integer $K$,
and consider the (free) commutative algebra $\A$ generated by these variables.
Let $g_c^{ij}$ be the (real, symmetric)
invariant tensor, so that $a\cdot a :=a_i a_j g_c^{ij}$
is invariant under $U(su(2))$, and $g_c^{ij} g_c^{jk} = \d^{ik}$. 
We now impose on $\A$ the star structure
\be
a_i^* = g_c^{ij} a_j,
\label{a_class_star}
\ee
so that $\A$ can be interpreted as the algebra of  
complex--valued functions on $\R^{2K+1}$; in particular, $a \cdot a$ is 
real. Then the usual integral on $\R^{2K+1}$ defines a 
functional on (the subset of integrable 
functions in a suitable completion of) $\A$, which satisfies
\berr
\int d^{2K+1}\!a\; u \trr f &=& \eps(u) \int d^{2K+1}\!a\; f, \nn\\
\lb\int d^{2K+1}\!a\; f\rb^* &=& \int d^{2K+1}\!a\; f^*
\err
for $u \in U(su(2))$ and integrable $f \in \A$. 
More general invariant functionals on $\A$
can be defined as  
\be
\langle f \rangle := \int d^{2K+1}\!a\; \rho(a\cdot a)\; f 
\label{inv_functional}
\ee
for $f \in \A$, where $\rho$ is a suitable real weight function. 
They are invariant, real and positive:
\berr
\langle u \trr f\rangle &=& \eps(u) \langle f\rangle, \nn\\
\langle f\rangle ^* &=& \langle f^* \rangle \nn\\
\langle f^* f\rangle  &\geq& 0
\label{functional_props}
\err
for any $u \in U(su(2))$ and $f \in \A$. 
As usual, one can then define a Hilbert space of 
square--(weight--) integrable functions by
\be
\langle f, g \rangle := \langle f^* g\rangle = 
   \int d^{2K+1}\!a\; \rho(a\cdot a)\; f^* g.
\label{GNS}
\ee
Now consider the twisted $U_q(su(2))$--module algebra $(\A, \star, \trr_q)$
defined in the previous section. We want to find an integral on $\A$
which is invariant under the action $\trr_q$ of $U_q(su(2))$. Formally, this is 
very easy: since the {\em space} $\A$ is unchanged by the twisting, 
we can simply use the classical integral again, and verify invariance
$$
\int d^{2K+1}\!a\; u \trr_q f = \int d^{2K+1}\!a\; \p(u) \trr f
 = \eps(\varphi(u)) \int d^{2K+1}\!a\; f 
 = \eps_q(u) \int d^{2K+1}\!a\; f.
$$
Notice that the algebra structure of $\A$ does not enter here at all.
The compatibility 
with the reality structure will be discussed in the next section. 

Of course we have to restrict to certain classes of integrable functions.
However, this is not too hard in the cases of interest. 
Consider for example
the space of Gaussian functions, i.e. functions of the form 
$P(a_i) e^{-c (a\cdot a)}$ with suitable (polynomial, say) $P(a_i)$.
Using (\ref{twisted_square}), this is the same as 
the space of Gaussian functions in the sense of the star product,
$P_{\star}(a_i) e^{-c (a\star a)}$. This will
imply that all integrals occuring in perturbation theory are well--defined. 
Furthermore, one can obtain a twisted sphere by imposing 
the relation $a \star a = a\cdot a = R^2$. On this sphere, 
the integral is well--defined for any polynomial functions. 
The integral over the twisted $\R^{2K+1}$ 
can hence be calculated by first integrating over the sphere and then 
over the radius.
Finally, we point out the following obvious fact:
\be
\langle P(a) \rangle  = \langle P_0(a) \rangle 
\ee
where 
$P_0(a) \in \A$ is the singlet part of the decomposition 
of the polynomial $P(a)$ under the action $\trr_q$ of $U_q(su(2))$,
or equivalently under the action $\trr$ of $U(su(2))$.

\section{An $U_q(su(2))$--covariant operator formalism}
\label{sec:operator-form}

In the previous section, we defined quasi--associative algebras
of functions on arbitrary \rep spaces of $U_q(su(2))$.
We will apply this to the coefficients of the fields on $\S_{q,N}$ later.
However, there is an alternative approach 
within the framework of ordinary operators and representations,
which is essentially equivalent for our purpose. We shall follow here 
closely the constructions in \cite{Fiore:1997fb}.
It seems that both approaches have their own advantages, 
therefore we want to discuss them both.

We first recall the notion of the semidirect product (cross--product) 
algebra, which is useful here.
Let $(\A,\cdot,\trr)$ be an associative $U(su(2))$--module algebra.
Then $U(su(2)) \smash \A$ is the vector space $\A \tens U(su(2))$,
equipped with the structure of an associative algebra defined by
$ua = (u_{(1)}\trr a) u_{(2)}$.
Here $u_{(1)} \tens u_{(2)}$ is the undeformed coproduct of $U(su(2))$.

In the following, we shall be interested in \reps of $\A$ which 
have a ``vacuum'' vector $\rangle$ such that all elements can be 
written in the form $\A \rangle$, i.e. 
by acting with $\A$ on the vacuum vector. In particular, 
we will denote with $V_{\A}$ the free left $\A$--module $\A \rangle$  
which as a vector space is equal to $\A$. This will be called the 
the ``left vacuum representation'' (or left regular representation).
Now any\footnote{provided the kernel of the \rep
is invariant under $U(su(2))$, which we shall assume.}  
such \rep of $\A$ can naturally be viewed as a 
\rep of $U(su(2)) \smash \A$, if
one declares the vacuum vector $\rangle$ to be a singlet under $U(su(2))$,
$$
u \rangle = \eps(u) \rangle,
$$
and $u \trr (a\rangle) = (u \trr a) \rangle$. One can then verify the relations
of $U(su(2)) \smash \A$.

Inspired by \cite{Fiore:1997fb}, we define
for any $a \in \A$ the element
\be
\hat a:= (\F^{-1}_1 \trr a) \F^{-1}_2 \; \in U(su(2)) \smash \A.
\label{a_hat}
\ee
Using the definition of the Drinfeld twist, it is immediate 
to verify the following properties:
\berr
\hat a \rangle &=& a \rangle \nn\\
\hat a \hat b \rangle &=& (a \star b) \rangle
\err
where $(a\star b)$ is the twisted multiplication on $\A$ defined in 
(\ref{star}). More generally, 
\be \fbox{$
\hat a_1 \hat a_2 .... \hat a_k \rangle = 
(a_1 \star(a_2 \star (.... a_{k-1} \star a_k) ...)) \rangle $}
\label{hata_astar}
\ee
for any $a_i \in \A$. Hence the elements $\hat a$ realize the twisted
product (\ref{star}) on $\A$, with this particular bracketing. 
If $c \in \A$ is a singlet, or equivalently $[c, U(su(2))] = 0$
in $U(su(2)) \smash \A$, then
\be
\hat c = c.
\label{c_equation}
\ee
If in addition the algebra $\A$ is commutative, then 
$\hat c$ is central in  $U(su(2)) \smash \A$. 
Moreover, the new variables $\hat a_i$ are 
automatically covariant under the quantum group $U_q(su(2))$,
with the $q$--deformed coproduct: denoting 
$$
\hat u:= \p(u) \in U(su(2))
$$
for $u \in U_q(su(2))$, one easily verifies
\be
\hat u \hat a = \widehat{u_1 \trr_q a}\; \widehat{u_2},
\label{uhat_ahat}
\ee
where $u_1 \tens u_2$ denotes the $q$--deformed coproduct. 
In particular,
\berr
\hat u \hat a \rangle    &=& u \trr_q a \rangle  
  = \widehat{u \trr_q a}\rangle \nn\\
\hat u \hat a \hat b\rangle &=& 
    (\widehat{u_1 \trr_q a}) (\widehat{u_2 \trr_q b})\rangle. \nn
\err 
Therefore $U_q(su(2))$ acts correctly on the $\hat a$--variables in the
left vacuum representation. 
More explicitly, assume that $\A$ is generated (as an algebra) 
by generators $a_i$ transforming in the spin $K$ \rep $\pi$ of $U(su(2))$, 
so that $u a_i = a_j \pi^j_i(u_{(1)}) u_{(2)}$. Then (\ref{uhat_ahat}) 
becomes
\be
\hat u \hat a_i = \hat a_j \pi^j_i(\widehat u_{1}) \widehat u_{2}. 
\ee
In general, the generators $\hat a_i$ will not satisfy 
closed commutation relations, even if the $a_i$ do.
However if $[a_i, a_j] = 0$, then one can verify that 
(cp. \cite{Fiore:1997fb})
\be
\hat a_i \hat a_j = \hat a_k \hat a_l \; \mathfrak{R}^{lk}_{ij}
\label{aa_hat_CR}
\ee
where 
\be
\mathfrak{R}^{ij}_{kl} = (\pi^i_k \tens \pi^j_l\tens id) 
      (\tilde \phi_{213}\; \tilde R_{12}\; \tilde \phi_{123}^{-1})
     \quad  \in U(su(2)).
\label{R_phi}
\ee
Again, this involves only the coassociator and the universal 
$\RR$--matrix. Such relations for field operators 
were already proposed in \cite{Mack:1992tg} on general grounds;
here, they follow from the definition (\ref{a_hat}).
In the case of several variables, one finds 
\be
\hat a_i \hat b_j = \hat b_k \hat a_l \; \mathfrak{R}^{lk}_{ij}.
\ee
Indeed, no closed quadratic commutation relations for 
deformed spaces of function with generators $a_i$ in arbitrary \reps of 
$U_q(su(2))$ are known, which has has been a major obstacle for defining
QFT's on $q$--deformed spaces. 
In the present approach, the generators $\hat a_i$ satisfy quadratic 
commutation relations which close only in the bigger algebra
$U(su(2)) \smash \A$. In general, they are not easy to work with.
However some simplifications occur if we use minimal  
twists $\F$ as defined in Section \ref{sec:drinfeld_twist}, 
as was observed by Fiore \cite{Fiore:1997fb}:

\begin{prop}
For minimal twists $\F$ as in (\ref{phi-integral}), 
the following relation holds:
\be
g^{ij}\hat a_i \hat a_j = g_c^{ij} a_i  a_j.
\label{quadratic_hat}
\ee
Here
\be
g^{ij} = \pi_{r}^{i}(\F_1) \pi_{s}^{j}(\F_2)\; g_c^{rs} =
         g_c^{il} \pi^j_l(\g'),
\label{g_twist_3}
\ee
where $\g'$ is defined in (\ref{gammas}).
In particular if $\A$ is abelian, this implies
that $g^{jk}\hat a_j \hat a_k$ is central in $U(su(2)) \smash \A$.
\label{a2_central_prop}
\end{prop}
We include a short proof in Section \ref{chapter:supp-II} for convenience.
This will be very useful to define a quantized field theory.
From now on, we will always assume that the twists are minimal.

\paragraph{Derivatives.}

Let $\A$ be again the free commutative algebra with generators $a_i$
in the spin $K$ \rep of $U(su(2))$, and
consider the left vacuum \rep $V_{\A} = \A \rangle$ of $U(su(2)) \smash \A$.
Let $\del_i$ be the (classical) derivatives,
which act as usual on the functions in $\A$. 
They can be considered as operators acting on $V_{\A}$,
and as such they satisfy the relations of the classical Heisenberg algebra, 
$\del_i a_j = g^c_{ij} + a_j \del_i$. 
Now define
\be
\hat \del_i :=  (\F^{-1}_1 \trr \del_i) \F^{-1}_2,
\label{del_hat}
\ee
which is an operator acting on $V_{\A} = \A \rangle$; in particular, it
satisfies $\hat \del_i \rangle = 0$.
Then the following relations hold:

\begin{prop}
For minimal $\F$ as in (\ref{phi-integral}), the operators 
$\hat a_i$,  $\hat \del_j$ acting on the left vacuum representation satisfy
\berr
\hat \del_i (g^{jk}\hat a_j \hat a_k) &=&
       2 \hat a_i + (g^{jk}\hat a_j \hat a_k) \hat \del_i, \label{del_a2}\\
\hat \del_i \hat a_j &=& g_{ij} + \hat a_k \hat \del_l \; 
                \mathfrak{R}^{lk}_{ij}.                    \label{del_a}
\err
\label{derivatives-prop}
\end{prop}
The proof is given in Section \ref{chapter:supp-II}; 
the second relation (\ref{del_a})
is again very close to a result (Proposition 6) in \cite{Fiore:1997fb},
and it holds in fact in $U(su(2)) \smash \A$.
Of course, the brackets in (\ref{del_a2}) were
just inserted for better readability, 
unlike in Lemma \ref{twist_qa_lemma} where they were essential.
If we have algebras with several variables in the 
same representation, then for example
\be
\hat \del_{a_i} (g^{jk}\hat b_j \hat a_k) =
       \hat b_i + (g^{jk}\hat b_j \hat a_k) \hat \del_{a_i}
\label{del_mixed}
\ee
holds, in self--explanatory notation.

One advantage of this approach compared to the quasi--associative
formalism in the previous section
is that the concept of a star is clear,
induced from Hilbert space theory. This will be explained next.

\subsection{Reality structure.}
\label{subsec:star}

Even though the results of this section are more general, 
we assume for simplicity that 
$\A$ is the free commutative algebra generated by the elements
$\{a_i\}$ which transform in the spin $K$ \rep  of $U(su(2))$ with integer $K$,
i.e. the algebra of complex--valued functions on $\R^{2K+1}$
(or products thereof). Then the classical integral defines an
invariant positive functional on $\A$ which satisfies 
(\ref{functional_props}), and
$V_{\A}$ becomes a Hilbert space (\ref{GNS}) 
(after factoring out a null space if necessary). 
Hence we can calculate the operator adjoint of the generators
of this algebra. By construction, 
$$
a_i^* = g_c^{ij} a_j
$$
where $g_c^{ij}$ is the invariant tensor, normalized such that  
$g_c^{ij} g_c^{jk} = \d^{ik}$.
As discussed above, $V_{\A}$ is a \rep of the semidirect product 
$U(su(2)) \smash \A$, in particular it is a
unitary \rep of $U(su(2))$. Hence the star on the generators of $U(su(2))$ is
$$
H^* = H ,\quad {X^{\pm}}^* = X^{\mp}.
%\label{U_star_cpct}
$$
Now one can simply calculate the star of the twisted variables 
$\hat a_i \in U(su(2)) \smash \A$. The result is as expected: 

\begin{prop}
If $\F$ is a minimal unitary twist
as in Proposition \ref{F_minimal}, then
the adjoint of the operator $\hat a_i$ acting on
the left vacuum representation $\A\rangle$ is 
\be
\hat a_i^* = g^{ij} \hat a_j.
\ee
\label{a_star_real_prop}
\end{prop}
This is proved in Section \ref{chapter:supp-II}, 
and it was already found in \cite{Fiore:1997fb}.
It is straightforward to extend these results to the case of 
several variables $a^{(K)}_i, b^{(L)}_j, ...$ in different representations,
using a common vacuum $\rangle$. The star structure is always of the form
(\ref{a_star_real_prop}). 

If $\A$ is the algebra of functions on $\R^{2K+1}$, we have seen above
that the left vacuum \rep
$\A\rangle$ is also a \rep of the Heisenberg algebra
$\A_H$ with generators  $a_i, \del_j$.
Again we can calculate the operator adjoints, and the result is
\berr
\del_i^* &=& - g_c^{ij} \del_j, \nn\\
\hat \del_i^* &=& - g^{ij} \hat\del_j. \nn
\err
Of course, all these statements are on a formal level,
ignoring operator--technical subtleties.

\subsection{Relation with the quasiassociative $\star$ --product}
\label{subsec:QO_relation}

Finally, we make a simple but useful observation, which provides
the connection of the operator approach in this section with the 
quasiassociative approach of Section \ref{sec:twisting}.
Observe first that an invariant (real, positive (\ref{functional_props}))
functional $\langle \; \rangle$ on $\A$ extends trivially as a
(real, positive)
functional on $U(su(2)) \smash \A$, by evaluating the generators of
$U(su(2))$ on the left (or right) of $\A$ with the counit. Now
for any tensor $I^{i_1 ... i_k}$ of $U_q(su(2))$, denote
$$
I(\hat a) := I^{i_1 ... i_k} \hat a_{i_1} ... \hat a_{i_k}  \qquad
\in U(su(2)) \smash \A,
$$
and
\be
I_{\star}(a) := I^{i_1 ... i_k} a_{i_1}\star 
( ...  \star (a_{i_{k-1}} \star a_{i_k}) ... ) \qquad
\in \A.
\label{I_star_tensor}
\ee
Then the following holds:

\begin{lemma}
\begin{itemize}
\item[1)] If $I = I^{i_1 ... i_k}$ is an invariant tensor of $U_q(su(2))$, 
then $I(\hat a)$ as defined above commutes with $u \in U_q(su(2))$,
\be
[u, I(\hat a)] = 0 \qquad \mbox{in}\;\;U(su(2)) \smash \A.
\ee
\item[2)]
Let $s\rangle \in \A\rangle$ be invariant, i.e.
$u\cdot s\rangle  = \eps_q(u) s\rangle$, and 
$I$, ..., $J$ be invariant tensors of $U_q(su(2))$. Then
$$
I(\hat a) ... J(\hat a)\; s \rangle 
  = I_{\star}(a) \star .... \star J_{\star}(a)\; s \rangle.
$$
\item[3)]
Let $I,J$ be invariant, and $P = P^{i_1 ... i_k}$ be an arbitrary tensor 
of $U_q(su(2))$. Denote with 
$P_0$ the trivial component of $P$ under the action of $U_q(su(2))$. 
Then for any invariant functional $\langle \; \rangle$ on $\A$,
\berr
\langle I(\hat a) ... J(\hat a)\; P(\hat a)\rangle
 &=& \langle I(\hat a) ... J(\hat a)\; P_0(\hat a)\rangle \nn\\
  &=& \langle I_{\star}(a) \star .... \star J_{\star}(a)
           (P_0)_{\star}(a)\rangle  \nn \\
  &=& \langle I_{\star}(a) \star .... \star J_{\star}(a) \star
           P_{\star}(a)\rangle
\label{IJ_stuff}
\err
Moreover if $\A$ is abelian, 
then the $I(\hat a)$, $J(\hat a)$ etc. can be considered as central
in an expression of this form, for example
$$
\langle I(\hat a) ... J(\hat a)\; P(\hat a_i)\rangle
  = \langle P(\hat a_i)\; I(\hat a) ... J(\hat a) \rangle
 = \langle P(\hat a_i)\; J(\hat a) ... I(\hat a) \rangle
$$
and so on.
\end{itemize}
\label{star_hat_lemma}
\end{lemma}
The proof follows easily from (\ref{uhat_ahat}), (\ref{hata_astar}) and 
Lemma \ref{inv_central}. The stars between the invariant 
polynomials $I_{\star}(a), ... , J_{\star}(a)$ 
are of course trivial, and no brackets are needed.

\section{Twisted Euclidean QFT}
\label{sec:QFT}

These tools can now be applied to our problem of quantizing
fields on the $q$--deformed fuzzy sphere $\S_{q,N}$.  
Most of the discussion is not restricted to
this space, but it is on a much more rigorous level there
because the number of modes is finite.
We will present 2 approaches, the first based on twisted $\star$
--products as defined in Section \ref{sec:twisting}, 
and the second using an operator formalism as in Section 
\ref{sec:operator-form}. Both have their own merits which seem 
to justify presenting them both. Their equivalence will follow from
Lemma \ref{star_hat_lemma}.

First, we discuss some basic requirements for a quantum field theory 
on spaces with quantum group symmetry. 
Consider a scalar field, and expand it in its modes as
\be
\Psi(x) = \sum_{K,n} \psi_{K,n}(x)\;  a^{K,n}.
\label{psi_field}
\ee
Here the $\psi_{K,n}(x) \in \S_{q,N}$ are a basis of 
the spin $K$ \rep of $U_q(su(2))$,
\be
u \trr_q \psi_{K,n}(x) = \psi_{K,m}(x) \pi^{m}_{n}(u),
\label{psi_covar}
\ee
and the coefficients $a^{K,n}$ transform
in the dual (contragredient) representation of $\tilde U_q(su(2))$,
\be 
u \; \ttr_q a^{K,n} = \pi^{n}_{m}(\tilde S u) \; a^{K,m}.
\label{a_covar}
\ee  
It is important to distinguish the Hopf algebras
which act on the coefficients $a^{K,n}$ and on the functions $\psi_{K,n}(x)$,
respectively.
The Hopf algebra $\tilde U_q(su(2))$ is obtained from $U_q(su(2))$
by flipping the coproduct and using the opposite antipode $\tilde S = S^{-1}$.
In particular, the $\RR$--matrix and the invariant tensors are
also flipped: 
\be
\tilde g^K_{nm} = g^K_{mn}
\label{g_tilde}
\ee
where $\tilde g^K_{nm}$ is the invariant tensor of $\tilde U_q(su(2))$.
The reason for this will become clear soon.
Moreover, it is sometimes convenient to express the
contragredient generators in terms of ``ordinary'' ones,
\be 
a_{K,n} = \tilde g^K_{nm}  a^{K,m}.
\label{lower_index}
\ee 
Then $\tilde U_q(su(2))$ acts as 
$$
u\; \ttr_q a_{K,n} = a_{K,m} \;  \pi^{m}_{n}(u).
$$  
We assume that the coefficients $a^{K,n}$ generate 
some algebra $\A$. This is not necessarily the algebra of field operators, 
which in fact would not be appropriate in the Euclidean case even for $q=1$. 
Rather, $\A$ could be the algebra 
of coordinate functions on configuration space (space of modes)
for $q=1$, and an analog thereof for $q \neq 1$.
The fields $\Psi(x)$ can then be viewed as ``algebra--valued distributions''
in analogy to usual field theory, by defining
$$
\Psi[f]:= \int\limits_{\S_{q,N}} \Psi(x) f(x) \quad \in \A
$$
for $f(x) \in \S_{q,N}$. Then the covariance properties (\ref{psi_covar}) and 
(\ref{a_covar}) could be stated as 
\be
u\; \ttr_q \Psi[f] =  \Psi[u \trr_q f],
\label{Psi_covariance}
\ee
using the fact that $\int (u\trr_q f) g = \int f (S(u) \trr g)$.

Our goal is to define some kind of correlation functions
of the form 
\be\fbox{$
\langle \Psi[f_1] \Psi[f_2] ... \Psi[f_k] \rangle \quad \in \C $}
\label{correlation_function}
\ee
for any $f_1, ..., f_k \in \S_{q,N}$, in analogy to the undeformed case.
After ``Fourier transformation'' (\ref{psi_field}), this amounts to 
defining objects
\be
G^{K_1,n_1; K_2,n_2; ...;K_k, n_k} := 
   \langle a^{K_1,n_1} a^{K_2,n_2} ... a^{K_k,n_k}\rangle =: 
   \langle P(a) \rangle
\label{correlation_function_a}
\ee 
where $P(a)$ will denote some polynomial in the $a^{K,n}$ from now on,
perhaps by some kind of a ``path integral''
$ \langle P(a) \rangle = \frac 1{\cN} \int \D a\; e^{-S[\Psi]}  P(a)$.
We require that they should satisfy at least the following properties, 
to be made more precise later:
\begin{itemize}

\item[(1)] {\em Covariance:}
\be
\langle u\; \ttr_q P(a) \rangle = \eps_q(u) \; \langle P(a) \rangle,
\ee
which means that the $G^{K_1,n_1; K_2,n_2; ...;K_k, n_k}$ are invariant 
tensors of $\tilde U_q(su(2))$,

\item[(2)] {\em Hermiticity:} 
\be
\langle P(a) \rangle^* = \langle P^*(a) \rangle
\ee
for a suitable involution $*$ on $\A$,

\item[(3)] {\em Positivity:}
\be
\langle P(a)^* P(a)\rangle \geq 0,
\ee

\item[(4)] {\em Symmetry} 

under permutations of the fields, in a suitable sense discussed below.
\end{itemize}
This will be our heuristic ``working definition'' of a 
quantum group covariant Euclidean QFT.

In particular, the word ``symmetry'' in (4) needs some explanation.
The main purpose of a symmetrization axiom is that it
puts a restriction on the number of degrees 
of freedom in the model, which in the limit $q \rightarrow 1$
should agree with the undeformed case. More precisely, the amount of 
information contained in the correlation functions 
(\ref{correlation_function_a}) should be the same as for $q=1$, i.e. 
the Poincare series of $\A$ should be the same.
This means that the polynomials in the $a^{K,n}$
can be ordered as usual, i.e. they
satisfy some kind of ``Poincare--Birkhoff--Witt'' property.
This is what we mean with ``symmetry'' in (4). In more physical terms, it
implies the statistical properties of bosons\footnote{we do not 
consider fermions here.}.

However, it is far from trivial how to impose such a ``symmetry'' on tensors 
which are invariant under a quantum group.
Ordinary symmetry is certainly not consistent with 
covariance under a quantum group.
One might be tempted to replace ``symmetry'' by some kind of invariance 
under the braid group which is 
naturally associated to any quantum group. This group is generally
much bigger than the group of permutations, however, 
and such a requirement is qualitatively different and
leaves fewer degrees of freedom. The
properties (3) and (4) are indeed very nontrivial requirements
for a QFT with a quantum group spacetime symmetry,
and they are not satisfied in the proposals that have been given up to 
now, to the knowledge of the authors.

Covariance (1) suggests that the algebra $\A$ generated by the 
$a^{K,n}$ is a $U_q(su(2))$--module algebra.
This implies immediately that $\A$ cannot be commutative, 
because the coproduct of $U_q(su(2))$ is not cocommutative. 
The same conclusion can be reached by contemplating the meaning 
of invariance of an action $S[\Psi]$, which will be clarified below.
One could even say that a second quantization
is required by consistency. As a further guiding line, 
the above axioms (1) -- (4) should be verified easily
in a ``free'' field theory.
 
In general, there is no obvious candidate for an associative algebra
$\A$ satisfying all these requirements. 
We will construct a suitable quasiassociative algebra $\A$ 
as a star--deformation of the algebra of functions on configuration space
along the lines of Section \ref{sec:twisting}, 
which satisfies these requirements.
Our approach is rather general and should be applicable 
in a more general context, such as for
higher--dimensional theories. Quasiassociativity implies that the 
correlation functions (\ref{correlation_function}) 
make sense only after specifying 
the order in which the fields should be multiplied (by explicitly
putting brackets), however different ways of bracketing are always
related by a unitary transformation. Moreover, 
the correct number of degrees of freedom is guaranteed by construction. 
We will then define QFT's which satisfy the above
requirements using a path integral over the fields $\Psi(x)$, 
i.e. over the modes $a^{K,n}$. An associative approach will also be presented
in Section \ref{subsec:associative_quant}, 
which is essentially equivalent.

\subsection{Star product approach}
\label{subsec:quant_1}

The essential step is as follows.
Using the map $\p$ (\ref{phi}), the coefficients 
$a^{K,n}$ transform also under the spin $K$ representation
of $U(su(2))$, via $u\; \ttr a^{K,n} = \p^{-1}(u)\;\ttr_q  a^{K,n}$.
Hence we can consider the usual commutative 
algebra $\A^K$ of functions on $\R^{2K+1}$ generated
by the $a^{K,n}$, and view it as a left $U(su(2))$--module
algebra $(\A^K, \cdot, \ttr)$. As explained in the Section \ref{sec:twisting}, 
we can then obtain from it 
the left $\tilde U_q(su(2))$--module algebra $(\A^K, \star,\ttr_q)$, 
with multiplication $\star$ as defined in (\ref{star}). 
More generally, we consider the 
left $\tilde U_q(su(2))$--module algebra $(\A, \star,\ttr_q)$
where $\A = \otimes_{K=0}^N \A^K$.
Notice that
the twist $\tilde \F$ corresponding to the reversed coproduct must be 
used here, which is simply $\tilde \F_{12} = \F_{21}$.
The reality issues will be discussed in Section \ref{subsec:associative_quant}.

\paragraph{Invariant actions.}

Consider the following candidate for an invariant action,
\be
S_{int}[\Psi] = \int\limits_{\S_{q,N}} \Psi(x) \star (\Psi(x) \star \Psi(x)).
\label{S_int_3}
\ee
Assuming that the functions on 
$\S_{q,N}$ commute with the coefficients, $[x_i, a^{K,n}] = 0$, 
this can be written as
\berr
S_{int}[\Psi] &=& \int\limits_{\S_{q,N}} \psi_{K,n}(x)\; \psi_{K',m}(x)\; 
                \psi_{K'',l} (x) \; a^{K,n}\star (a^{K',m}\star a^{K'',l})\nn\\
   &=& I^{(3)}_{K,K',K'';\; n,m,l} \; a^{K,n}\star (a^{K',m}\star a^{K'',l})
       \quad \in \A.
\label{S_example}
\err
Here\footnote{note that the brackets are actually not necessary here because 
of (\ref{phi_trivial}). For higher--order terms they are essential, however.} 
$I^{(3)}_{K,K',K'';n,m,l} =$ 
$\int\limits_{\S_{q,N}} \psi_{K,n}\; \psi_{K',m}\; \psi_{K'',l}\;$
is by construction an invariant tensor of $U_q(su(2))$, 
\be
I^{(3)}_{K,K',K'';n,m,l} \; 
                 \pi^n_r(u_{1}) \pi^m_s(u_{2}) \pi^l_t(u_{3})
 = \eps_q(u) \; I^{(3)}_{K,K',K'';r,s,t}.
\label{invar_I}
\ee
We have omitted the labels on the various representations.
Hence $S_{int}[\Psi]$ is indeed an invariant element of $\A$:
\berr
u\; \ttr_q S_{int}[\Psi] &=&  I^{(3)}_{K,K',K'';\; n,m,l} \;
    \pi^{n}_r (\tilde S u_{\wtilde{1}}) 
    \pi^{m}_{s}(\tilde Su_{\wtilde{2}})
    \pi^{l}_t (\tilde Su_{\wtilde{3}})\; 
      a^{K,r}\star (a^{K',s}\star a^{K'',t}) \nn \\
 &=& \eps_q(u) \; S_{int}[\Psi] \nn
\err
using (\ref{invar_I}), 
where $u_{\wtilde{1}} \tens u_{\wtilde{2}}\tens u_{\wtilde{3}}$
is the 2--fold coproduct of $u\in \tilde U_q(su(2))$;
notice that the antipode reverses the coproduct. 
This is the reason for using $\tilde U_q(su(2))$. 

In general, our actions $S[\Psi]$ will be polynomials
in $\A$, and we shall only consider invariant actions, 
\be
u \;\ttr_q S[\Psi] = \eps_q(u)\;  S[\Psi] \qquad \in \A,
\label{invar_actions}
\ee
for $u \in \tilde U_q(su(2))$.
It is important to note that by Lemma \ref{inv_central}, 
the star product of any such invariant actions 
is commutative and associative, even though the full algebra of
the coefficients $(\A, \star)$ is not. 
Moreover we only consider actions which are obtained 
using an integral over $\S_{q,N}$ as in (\ref{S_int_3}), which 
we shall refer to as ``local''.

In particular, consider the quadratic action 
$$
S_{2}[\Psi] = \int\limits_{\S_{q,N}} \Psi(x) \star \Psi(x),
$$
which can be rewritten as
\berr
S_{2}[\Psi]  &=& \int\limits_{\S_{q,N}} \psi_{K,n}(x)\; \psi_{K,m}(x)\; 
               a^{K,n}\star a^{K,m}
   = \sum_{K=0}^N \;  g^K_{nm} \; a^{K,n}\star a^{K,m} \nn\\
   &=& \sum_{K=0}^N \;  \tilde g^K_{mn} \; a^{K,n}\star a^{K,m}.
\label{S_mass}
\err
Here we assumed that the
basis $\psi_{K,n}(x)$ is normalized such that
\be
\int\limits_{\S_{q,N}} \psi_{K,n}(x) \psi_{K',m}(x)
= \d_{K,K'} \; g^K_{n,m}.
\label{psi_normaliz}
\ee 
This action is of course invariant,
$
u \;\ttr_q S_{2}[\Psi] = \eps_q(u) \; S_{2}[\Psi].
$
Moreover, the invariant quadratic actions agree
precisely with the classical ones. Indeed, the most general invariant
quadratic action has the form
\berr
S_{free}[\Psi]  &=& \hf \sum_{K=0}^N D_K g^K_{nm} \; a^{K,n}\star a^{K,m}\nn\\
     &=& \hf \sum_{K=0}^N \; D_K (g_c^K)_{nm}\; a^{K,n}\cdot a^{K,m} 
\label{quadratic_action}
\err
using (\ref{twisted_square}), for some $D_K \in \C$. 
This will allow to derive Feynman rules from Gaussian integrals as usual.

\paragraph{Quantization: path integral.}

We will define the quantization by 
a (configuration space) path integral, i.e. some kind of integration
over the possible values of the coefficients $a^{K,m}$.
This integral should be invariant under $\tilde U_q(su(2))$.
Following Section \ref{subsec:integral}, we 
consider $\A^K$ as the vector space of complex--valued functions on 
$\R^{2K+1}$, and use the usual classical integral over $\R^{2K+1}$.
Recall that the algebra structure of $\A^K$ does not enter here at all.
The same approach was used in 
\cite{Grosse:1996ar} to define the quantization of the 
undeformed fuzzy sphere, and an analogous approach is usually taken on 
spaces with a star product \cite{Seiberg:1999vs}.
Notice that $K$ is an integer, since we do not consider fermionic fields here.
Explicitly, let $\int d^{2K+1}\!a^{K}\; f$ be the integral 
of an element $f \in \A^K$ over $\R^{2K+1}$. It is 
invariant under the action of $\tilde U_q(su(2))$ 
(or equivalently under $U(su(2))$) 
as discussed in Section \ref{subsec:integral}:
$$
\int d^{2K+1}\!a^K\; u \;\ttr_q f = \eps_q(u)\; \int d^{2K+1}\!a^K\; f.
$$
Now we define
$$
\int \cD \Psi \; f[\Psi] := \int \prod_K d^{2K+1}\!a^{K}\; f[\Psi],
$$
where $f[\Psi] \in \A$ is any integrable function (in the usual sense)
of the variables $a^{K,m}$.
This will be our path integral, which is by construction 
invariant under the action $\ttr_q$ of $\tilde U_q(su(2))$.

Correlation functions can now be defined as 
functionals of ``bracketed polynomials'' 
$P_{\star}(a) = a^{K_1,n_1} \star (a^{K_2,n_2} \star (...\star a^{K_l,n_l}))$
in the  field coefficients by
{\large
\be\fbox{$
\langle P_{\star}(a)\rangle := 
           \frac{\int  \cD \Psi\; e^{-S[\Psi]} P_{\star}(a)}
                                    {\int  \cD \Psi\; e^{-S[\Psi]}}.$}
\label{correlation}
\ee}
This is natural, because all invariant actions $S[\Psi]$
commute with the generators $a^{K,n}$. 
Strictly speaking there should be a factor $\frac 1{\hbar}$ in front
of the action, which we shall omit. In fact there are now 3 different
``quantization'' parameters: $\hbar$ has the usual meaning, while 
$N$ and  $q-q^{-1}$ determines a quantization or deformation of space. 

Invariance of the action $S[\Psi] \in \A$ implies
that 
\be
\langle u\; \ttr_q P_{\star}(a) \rangle  = 
             \eps_q(u)\; \langle P_{\star}(a) \rangle,
\label{P_invariance}
\ee
and therefore
\be
\langle P_{\star}(a) \rangle  = \langle (P_0)_{\star}(a) \rangle
\label{P_P_0}
\ee
where $P_0$ is the singlet part of the polynomial $P$, as in
Lemma \ref{star_hat_lemma}.
These are the desired invariance properties, and they would not hold
if the $a^{K,n}$ were commuting variables.
By construction, the number of
independent modes of a polynomial $P_{\star}(a)$ with given degree
is the same as for $q=1$. One can in fact order them, using
quasiassociativity together with the commutation relations (\ref{aa_CR}) 
which of course also hold under the integral:
\be
\langle P_{\star}(a) \star((a_i \star a_j - a_k \star a_l \; \TR_{ij}^{lk}) 
\star Q_{\star}(a))\rangle  = 0,
\label{aa_CR_corr}
\ee
for any polynomials $P_{\star}(a),  Q_{\star}(a) \in \A$. This can also be 
verified using the perturbative formula (\ref{correlator_Z}) below.
Therefore the symmetry requirement (4) of Section \ref{sec:QFT} is satisfied.
Moreover, the following cyclic property holds:
\be
\langle a_i \star P_{\star}(a) \rangle = 
\langle  P_{\star}(a) \star a_k \rangle\;  \tilde D^k_i, 
  \qquad \tilde D^k_i = \tilde g^{kn} \tilde g_{in}
\ee
for any $P_{\star}(a)$. This follows using (\ref{P_P_0}) and
the well--known cyclic property of the $q$--deformed invariant tensor
$\tilde g_{ij}$. 

In general, the use of quasiassociative algebras for QFT
is less radical than one might think, and it is consistent with results of
\cite{Alekseev:1999bs} on boundary correlation functions in BCFT.
Before addressing the issue of reality, we develop some tools
to actually calculate such correlation functions in perturbation theory.

\paragraph{Currents and generating functionals.}

One can now introduce the usual tools of quantum field theory. 
We introduce (external) currents $J(x)$ by
\be
J(x) = \sum_{K,n}  \psi_{K,n}(x) \; j^{K,n},
\label{currents}
\ee
where the new generators $j^{K,n}$ are included into the 
$\tilde U_q(su(2))$--module algebra $\A$, again 
by the twisted product (\ref{star}).
We can then define a generating functional 
\be\fbox{$
Z[J] = \frac 1{\cN} \int  \cD \Psi \; e^{-S[\Psi] + \int \Psi(x)\star J(x)},$}
\label{generating_Z}
\ee
which is an element of $\A$ but depends only on the current variables.
Here $\cN = \int  \cD \Psi \; e^{-S[\Psi]}$. Note that
$$
\int \Psi(x) \star J(x) =  \int J(x) \star \Psi(x),
$$
which follows e.g. from (\ref{quadratic_action}).
Invariance of the functional integral implies that 
\be
u\; \ttr_q Z[J] =  \eps_q(u)\; Z[J]
\label{Z_invar}
\ee
for any $u \in \tilde U_q(su(2))$, 
provided the actions $S[\Psi]$ are invariant.

It is now useful to introduce derivatives $\dl_{(j)}^{K,n}$ 
similar to (\ref{a_del_CR}), which together with the currents form a 
twisted (quasiassociative)
Heisenberg algebra as explained in the previous section:
$$
\dl_{(j) n}^K \star j^{K'}_m =  \d_{K,K'}\;  \tilde g^K_{nm} +  
         \; j^{K'}_r \star \dl_{(j) s}^{K} \;\TR_{nm}^{sr}
$$
By a calculation analogous to (\ref{del_asquare}), it follows that
$$
\dl_{(j)}^{K,n} \Big(\int \Psi(x)\star J(x)\Big) = a^{K,n} + 
  \Big( \int \Psi(x)\star J(x) \Big)  \dl_{(j)}^{K,n}.
$$ 
Recall that it is not necessary to put a star if one of the 
factors is a singlet.

This is exactly what we need. We conclude immediately that 
$[\dl_{(j)}^{K,n}, \exp(\int \Psi\star J)] =$ 
$a^{K,n} \exp(\int \Psi\star J)$, and 
by an inductive argument it follows that
the correlation functions (\ref{correlation}) can be written as 
\be 
\langle P_{\star}(a)\rangle  = 
      {{}_{{}_{J=0}}}\langle  P_{\star}(\dl_{(j)})\; Z[J] \rangle_{\dl=0}.
%_{|_{J=\dl=0}}.
\label{correlator_Z}
\ee
Here %$ |_{J=\dl=0}$ 
$ {{}_{{}_{J=0}}}\langle ...  \rangle_{\dl=0}$ 
means ordering the derivatives to the right of the currents
and {\em then} setting $J$ and $\dl_{(j)}$ to zero.
The substitution of derivatives into the bracketed polynomial $P_{\star}$ is 
well--defined, because the algebra of the generators $a$ is the same 
as the algebra of the derivatives $\dl_{(j)}$. 

The usual perturbative expansion can now be obtained easily. 
Consider a quadratic action of the form
$$
S_{free}[\Psi] = \int\limits_{\S_{q,N}} \hf \Psi(x) \star D \Psi(x),
$$
where $D$ is an invariant (e.g. differential) operator on $\S_{q,N}$, so that 
$ D \Psi(x) = \sum\psi_{K,n}(x)\; D_K a^{K,n}$
with $D_K \in \C$. It then follows as usual that 
\be
Z_{free}[J] := \frac 1{\cN_{free}}\; \int  \cD \Psi \; e^{- S_{free}[\Psi] +
             \int \Psi(x)\star J(x)} = 
            e^{ \frac 12  \int J(x)\star D^{-1} J(x)}.
\label{Z_free}
\ee
This implies that after writing the full action in the form 
$S[\Psi] =  S_{free}[\Psi] +  S_{int}[\Psi]$, one has
\berr
Z[J] &=& \frac 1{\cN}\;
             \int  \cD \Psi \; e^{- S_{int}[\Psi]} e^{- S_{free}[\Psi]
                   + \int \Psi(x)\star J(x)}  \nn\\
   &=& \frac 1{\cN'}\; e^{- S_{int}[\dl_{(j)}]}\;  Z_{free}[J] \rangle_{\dl=0}.
\label{Z_perturb}
\err
This is the starting point for a perturbative evaluation.
In the next section, we shall cast this into a form which is even more 
useful, and show that the ``vacuum diagrams''
cancel as usual.

\paragraph{Relation with the undeformed case.}

There is a conceptually simple relation of all the above models which are
invariant under $\tilde U_q(su(2))$ with models on the 
undeformed fuzzy sphere which are invariant under $U(su(2))$, at the 
expense of ``locality''.
First, note that the space of invariant actions (\ref{invar_actions})
is independent of $q$. More explicitly,
consider an interaction term of the form (\ref{S_example}).
If we write down explicitly the definition of the $\star$ product
of the $a^{K,n}$ variables, then it can be viewed as an interaction
term of $a^{K,n}$ variables with a tensor which is invariant under 
the {\em undeformed} $U(su(2))$, obtained from $I^{(3)}_{K,K',K'';\; n,m,l}$
by multiplication with representations of $\F$.
In the limit  $q = 1$, this $\F$ becomes trivial. 
In other words, the above actions can also be viewed
as actions on undeformed fuzzy sphere $S^2_{q=1,N}$, with interactions
which are ``nonlocal'' in the sense of  $S^2_{q=1,N}$, i.e.
they are given by traces of products of matrices only to the lowest 
order in $(q-1)$. Upon spelling out the $\star$ product in the 
correlation functions (\ref{correlation}) as well, they can be considered as
ordinary correlation functions of a slightly nonlocal field theory
on $S^2_{q=1,N}$, disguised by the transformation $\F$.

In this sense, $q$--deformation simply amounts to some kind of 
nonlocality of the interactions. 
A similar interpretation is well--known in the context of
field theories on spaces with a Moyal product \cite{Seiberg:1999vs}.
The important point is, however, that one can 
calculate the correlation functions for $q \neq 1$ {\em without}
using the twist $\F$ explicitly, using only $\hat R$ --matrices and the 
coassociators $\tilde \phi$, which are much easier to work with. This 
should make the $q$--deformed point of view useful.
It is also possible to generalize these results to other $q$--deformed spaces.

\subsection{Associative approach}
\label{subsec:associative_quant}

In order to establish the reality properties of the field theories introduced
above, it is easier to use an alternative formulation, using the results
of Section \ref{sec:operator-form}. The equivalence of the two 
formulations will follow from Section \ref{subsec:QO_relation}.
This will also allow to define field operators
for second--quantized models in 2+1 dimensions in Section \ref{subsec:2plus1}.

Consider the left vacuum \reps $V_{\A} = \A\rangle$ of $\A = \tens_K \A^K$
introduced in Section \ref{sec:operator-form}, and define the 
operators\footnote{they should not be considered as field operators.}
\be
\hat a^{K,n} = (\tilde \F^{-1}_1 \trr a^{K,n}) \tilde \F^{-1}_2 \; \;
          \in \; U(su(2)) \smash \A
\ee
acting on $\A\rangle$. We can then more or less repeat all the constructions
of the previous section with $a^{k,n}$ replaced by $\hat a^{K,n}$,
omitting the $\star$ product. The covariance property (\ref{Psi_covariance}) 
of the field
\be
\hat \Psi(x) = \sum_{K,n} \psi_{K,n}(x)\;  \hat a^{K,n}
\label{psi_field-op}
\ee
can now be written in the form
$$
 \Psi[u \trr_q f] = u_{\tilde 1} \Psi[f] \tilde Su_{\tilde 2}.
$$
Invariant actions can be obtained by contracting the $\hat a^{K,n}$ 
with invariant tensors of $\tilde U_q(su(2))$, and satisfy
$$
[u, S[\hat\Psi]] = 0
$$
for\footnote{recall that as algebra, there is no difference between
$U(su(2))$ and $\tilde U_q(su(2))$.}
$u \in \tilde U_q(su(2))$. For example, any actions of the form 
$$
S[\hat \Psi] = 
\int\limits_{\S_{q,N}} \hf \hat \Psi(x) D \hat \Psi(x) + 
    \lambda \hat \Psi(x)\hat \Psi(x) \hat \Psi(x)\; = 
  S_{free}[\hat \Psi] + S_{int}[\hat \Psi]\;  \quad 
          \in \;U(su(2)) \smash \A
$$
are invariant, where $D$ is defined as before. 
Using Proposition \ref{a2_central_prop}, the quadratic invariant actions 
again coincide with the undeformed ones. In general, higher--order 
actions are elements of $U(su(2)) \smash \A$ but not of $\A$. Nevertheless
as explained in Section \ref{subsec:QO_relation},
all such invariant actions $S[\hat \Psi]$ are in one--to--one correspondence
with invariant actions in the $\star$--product approach, 
with brackets as in (\ref{I_star_tensor}). This will be understood from now on.

Consider again the obvious (classical) functional 
$\int \prod_K d^{2K+1}\!a^K\;$ on $\A$ (or $V_{\A}$) as in the previous 
section,
and recall from Section \ref{subsec:QO_relation} that it extends trivially
to a functional on $U(su(2)) \smash \A$, by evaluating $U(su(2))$ with $\eps$.
We will denote this functional by 
$\int \cD \hat \Psi$. Define correlation functions of polynomials 
in the $\hat a^{K,n}$ variables as 
\be
\langle P(\hat a)\rangle := 
   \frac{\int \cD \hat \Psi e^{-S[\hat \Psi]} P(\hat a)}
        {\int \cD \hat \Psi e^{-S[\hat \Psi]}}
       = \langle P_0(\hat a)\rangle.
\ee
Here $P_0$ is again the singlet part of the polynomial $P$. 
Then  Lemma \ref{star_hat_lemma} implies
\be
\langle P(\hat a)\rangle = \langle P_{\star}(a)\rangle,
\ee
always assuming that the actions $S[\hat \Psi]$ are invariant
under $U_q(su(2))$. This shows the equivalence with the approach of 
the previous section. Moreover, 
\be
\langle P(\hat a)\;\hat a_i \hat a_j \;Q(\hat a)\rangle 
    = \langle P(\hat a) \;\hat a_k \hat a_l \; 
       \mathfrak{R}^{lk}_{ij} \; Q(\hat a) \rangle,
\label{aa_hat_CR_corr}
\ee
follows from (\ref{aa_hat_CR}), or from (\ref{correlator_Z_op}) below
on the perturbative level.

\paragraph{Currents and generating functionals.}
We can again extend $\A$ by other variables such as currents
\be
\hat J(x) = \sum_{K,n}  \psi_{K,n}(x) \; \hat j^{K,n} 
            \qquad \in U(su(2)) \smash \A,    \label{currents_op}
\ee
and consider the generating functional
\be
Z[\hat J] = \frac 1{\cN}\; \int  \cD \hat \Psi \; e^{-S[\hat\Psi] + 
             \int \hat \Psi(x)\hat J(x)} \rangle
\label{generating_Z_op}
\ee
with $Z[0] = 1$.
This is defined as the element of $\A \rangle \cong \A$ obtained
after integrating over the $a^K$--variables; 
the result depends on the currents only. 
The brace $\rangle$ indicates that the explicit $U(su(2))$
factors in $U(su(2)) \smash \A$ are evaluated by $\eps$.
Again, Lemma \ref{star_hat_lemma}
implies that $Z[\hat J]$ agrees precisely with 
the previous definition (\ref{generating_Z}).

As explained in Section \ref{sec:operator-form}, one can consider also the
twisted derivative operators $\hat \dl_{(j)}^{K,n}$, 
which act on $\A\rangle$. Using Proposition \ref{derivatives-prop},
we can derive essentially the same formulas as in the previous section,
omitting the star product. In particular, (\ref{del_mixed}) implies that
$$
\hat \dl_{(j)}^{K,n} \Big(\int \hat \Psi(x) \hat J(x)\Big) = \hat a^{K,n} + 
  \Big(\int \hat \Psi(x) \hat J(x) \Big)  \hat \dl_{(j)}^{K,n}. 
$$  
Since invariant elements of $\A$ are central as was pointed out 
below (\ref{c_equation}), we obtain as usual 
\berr
\langle P(\hat a)\rangle  &=& 
     {{}_{{}_{J=0}}}\langle P(\hat \dl_{(j)})\; Z[\hat J] \rangle_{\dl=0}  
                                              \label{correlator_Z_op} \\
Z[\hat J] &=& \frac 1{\cN}\;\int  \cD \hat \Psi \; e^{-( S_{free}[\hat\Psi]
         + S_{int}[\hat\Psi]) + \int \hat \Psi(x)\hat J(x)}\rangle \; 
   = \frac 1{\cN'}\; e^{- S_{int}[\hat\dl_{(j)}]}\;  
                            Z_{free}[\hat J] \rangle_{\dl=0}
                                             \nn \\
Z_{free}[\hat J] &=& \frac 1{\cN_{free}}\; 
                 \int  \cD \hat \Psi \; e^{- S_{free}[\hat\Psi] +
             \int \hat \Psi(x) \hat J(x)}\rangle \; = 
         e^{ \frac 12  \int \hat J(x) D^{-1} \hat J(x)}\rangle.
                                             \label{Z_free_op}
\err
Even though these formulas can be used to calculate correlators 
perturbatively, there is a form which is more convenient 
for such calculations. To derive it, observe that
(\ref{del_a2}) implies
\be
\hat \dl_{(j)}^{K,n}  \; e^{ \frac 12  \int \hat J(x) D^{-1} \hat J(x)}
   = e^{ \frac 12  \int \hat J(x) D^{-1} \hat J(x)} \;
     (D_K^{-1}\;  \hat j^{K,n} + \hat \dl_{(j)}^{K,n});
\ee
one can indeed verify that the algebra of 
\be
\hat b^{K,n} =  D_K^{-1}\; \hat j^{K,n} + \hat \dl_{(j)}^{K,n}
\label{b_def}
\ee
is the same as the algebra of $\hat a^{K,n}$. 
Therefore (\ref{correlator_Z_op}) can be rewritten as 
\berr
\langle P(\hat a)\rangle &=& \frac 1{\cN'}\;
 {{}_{{}_{J=0}}}\big\langle P(\hat \dl_{(j)})\; e^{- S_{int}[\hat\dl_{(j)}]}\; 
       e^{ \frac 12  \int \hat J(x) D^{-1} \hat J(x)}\big\rangle_{\dl=0} \nn\\
 &=& \frac 1{\cN'}\; 
    {{}_{{}_{J=0}}}\langle e^{ \frac 12  \int \hat J D^{-1} \hat J}
         P(\hat b)\; e^{- S_{int}[\hat b]}\rangle_{\dl=0}\nn\\
&=& \frac{{{}_{{}_{J=0}}}\langle P(\hat b)\; 
                e^{- S_{int}[\hat b]}\rangle_{\dl=0}}
       {{{}_{{}_{J=0}}}\langle e^{- S_{int}[\hat b]}\rangle_{\dl=0}}.
\label{J_del_Wick}
\err 
To evaluate this, one reinserts the definition (\ref{b_def}) 
of $\hat b$ as a sum of derivative operators $\hat \dl$ and current generators 
$\hat j$. Each $\hat \dl$ must be ``contracted'' with a $\hat j$ 
to the right of it using the commutation relations (\ref{del_a}),
which gives the inverse propagator $D_K^{-1}$,
and the result is the sum of all possible complete contractions. 
This is the analog of Wick's theorem. The contractions can be indicated 
as usual by pairing up the $\hat b$ variables with a line, before 
actually reordering them. Then each contribution can be 
reconstructed uniquely from a given complete contraction; this
could be stated in terms of Feynman rules. 

One can also show that the denominator exactly cancels the 
``vacuum bubbles'' in the numerator, as usual. 
Indeed, consider any given complete contraction of a term
$$
\hat b ... \hat b\; \frac 1{n!} (S_{int}[\hat b])^n.
$$
Mark the set of vertices which are connected (via a series of contractions)
to some of the explicit $\hat b$ generators on the left with blue, 
and the others with red. Then 2 neighboring red and blue vertices
can be interchanged keeping the given contractions, 
without changing the result. This is because only the homogeneous 
part of the commutation relations\footnote{associativity helps here.}
(\ref{del_a}) applies, and all
vertices are singlets (cp. Lemma \ref{star_hat_lemma}). 
Therefore the red vertices
can be moved to the right of the blue ones, and their contractions are
completely disentangled. Then the usual combinatorics yields
\be
\langle P(\hat a)\rangle = 
    {{}_{{}_{J=0}}}\langle P(\hat b)\; e^{- S_{int}[\hat b]}
     \rangle_{\dl=0,\;\mbox{\scriptsize no vac}}
\label{Wick_novac}
\ee
in self-explanatory notation. Of course this also holds in the 
quasiassociative version, but the derivation is perhaps less transparent.

In general, it is not easy to evaluate these expressions explicitly, 
because of the coassociators. However the lowest--order 
corrections $o(h)$ where $q=e^{h}$ are easy to obtain, using the fact 
that $\tilde \phi = \one + o(h^2)$ for minimal twists (\ref{phi-integral}).
If we write
$$
R_{12} = \one + h r_{12}\; + o(h^2),
$$
then 
$$
\tilde R_{12} = R_{12} \sqrt{R_{21} R_{12}}^{-1} 
              = \one + \frac h2 (r_{12} - r_{21}) \; + o(h^2),
$$
which allows to find the leading $o(h)$ corrections to the undeformed 
correlation functions explicitly.

\paragraph{Reality structure.}
One advantage of this formalism is that 
the reality structure is naturally induced from the Hilbert
space $V_{\A}$, as explained in Section \ref{subsec:star}.
Using  Proposition \ref{a_star_real_prop} and noting that the $a^{K,m}$ are in 
the contragredient \rep of $U(su(2))$, it follows that
\be
(\hat a^{K,n})^* = \tilde g^K_{nm} \hat a^{K,m}.
\label{a_reality}
\ee
We shall assume that all the actions are real, 
$$
S[\hat\Psi]^*  = S[\hat\Psi];
$$
this will be verified in the examples below.
Moreover, the classical integral defines a real functional 
on $U(su(2)) \smash \A$. Hence we conclude that
the correlation functions satisfy 
\be
\langle P(\hat a)\rangle^* = \langle P(\hat a)^*\rangle.
\ee
One can also show that
\be
\psi_{I,i}(x)^* = g^I_{ij} \psi_{I,j}(x),
\label{psi_reality}
\ee
where $g^I_{ij}$ is normalized such that $g^I_{ij} = (g^I)^{ij}$.
Therefore
\be
\hat \Psi(x)^* %= g^I_{ij} Y^I_j \tilde g^K_{im} \hat a^{K,m}
 = \hat \Psi(x),
\label{psi_hat_reality}
\ee
using (\ref{g_tilde}). This is useful to establish the reality of actions.
Of course, one could also consider complex scalar fields.
Finally, the correlation functions satisfy the positivity property
\be
\langle P(\hat a)^* P(\hat a)\rangle \geq 0,
\ee
provided the actions are real.
This is a simple consequence of the fact that $ P(\hat a)^* P(\hat a)$
is a positive operator acting on the left vacuum representation, together 
with the positivity of the functional integral.
It is one of the main merits of the present approach.

\section{Examples}

\subsection{The free scalar field}

Consider the action
\be
S_{free}[\Psi] = -\int\limits_{\S_{q,N}} 
    \hf \hat \Psi(x) \star \Delta \hat \Psi(x). %+ m\Psi(x)^{\ast}\star\Psi(x) 
\ee
Here the Laplacian was defined in \cite{qFSI} using a differential 
calculus as $\Delta = \ast_H d \ast_H d$, and 
satisfies\footnote{it is rescaled from the one in \cite{qFSI} 
so that its eigenvalues are independent of $N$.} 
$$
\Delta \psi_{K,n}(x)\; = %\frac{2[2]_q}{[N]_q [N+2]_q} 
     \frac 1{R^2}\; [K]_q [K+1]_q \;\psi_{K,n}(x) \equiv D_K \;\psi_{K,n}(x),
$$
where $[K]_q = \frac{q^K - q^{-K}}{q-q^{-1}}$.
The basis $\psi_{K,n}(x)$ is normalized as in (\ref{psi_normaliz}).
The action is real by (\ref{psi_hat_reality}), and can be rewritten as 
$$
S_{free}[\hat\Psi] = - \sum_{K,n} \hf D_K \;\tilde g^K_{nm}
               \hat a^{K,m} \hat a^{K,n}  
  = - \sum_{K,n} \hf D_K \; (\tilde g^K)^{mn} \hat a_{K,m} \hat a_{K,n}, 
$$
using (\ref{lower_index}).
As a first exercise, we calculate the 2--point functions. 
From (\ref{correlator_Z_op}) and (\ref{Z_free_op}), one finds
\berr
\langle \hat a^K_n \hat a^{K'}_{n'} \rangle &=& 
 {{}_{{}_{J=0}}}\langle \hat \dl_n^K \hat \dl_{n'}^{K'} 
           Z_{free}[\hat J]\rangle_{\dl=0} \nn\\
  &=&  {{}_{{}_{J=0}}}\langle\frac 12\;\hat \dl_{n}^K \hat \dl_{n'}^{K'} 
  \;(\sum (\tilde g^K)^{rs}\;\hat j^K_r D_K^{-1}\hat j^K_s)\rangle_{\dl=0}\nn\\
  &=& D_K^{-1}\;{}_{J=0}\langle\hat\dl_{n}^K  
        \hat j^{K'}_{n'} \rangle_{\dl=0}  
  \; = D_K^{-1}\; \d^{K K'} \; \tilde g^K_{n n'},  \nn
\err
where (\ref{del_a2}) was used in the last line.
This result is as expected, and it could also be obtained by using 
explicitly the definition of the twisted operators $\hat a^K$.

The calculation of the 4--point functions is more complicated, 
since it involves the coassociator. To simplify the notation, 
we consider the (most complicated) case where
all generators $a^K$ have the same spin $K$, which will be suppressed.
The result for the other cases can then be deduced easily. 
We also omit the prescriptions $(\dl=0)$ etc. 
Using first the associative formalism, (\ref{J_del_Wick}) yields
\berr
\langle \hat a_{n} \hat a_{m}\hat a_{k} \hat a_{l}  \rangle  
 &=& \langle(D^{-1} \hat j_n + \hat\dl_n)
     (D^{-1} \hat j_m + \hat\dl_m) 
    (D^{-1} \hat j_k + \hat \dl_k)
    (D^{-1} \hat j_l + \hat \dl_l)   \rangle \nn\\
 &=& \langle\hat\dl_n (D^{-1} \hat j_m + \hat\dl_m) 
    (D^{-1} \hat j_k + \hat\dl_k) D^{-1} \hat j_l  \rangle \nn\\
 &=& D^{-2}\;\langle\hat\dl_n \hat j_m  
     \tilde g_{kl} \; + \; \hat\dl_n \hat\dl_m
     \hat j_k \hat j_l  \rangle  \nn\\
 &=& D^{-2}\;\langle\tilde g_{nm} \tilde g_{kl} 
     + \hat\dl_n \hat\dl_m \hat j_k \hat j_l\rangle  \nn
\err
To evaluate this, consider
\berr
\langle\hat\dl_n \hat\dl_m \hat j_k \hat j_l \rangle 
&=& \langle\hat\dl_n (\tilde g_{mk} + \hat j_a \hat\dl_b\;
      \mathfrak{R}^{ba}_{mk}) \hat j_l \rangle \nn\\
&=& \tilde g_{mk} \tilde g_{nl}  + 
     \langle \hat\dl_n \hat j_a \hat\dl_b\;\hat j_s\pi^s_l
     (\mathfrak{R}^{ba}_{mk}) \rangle \nn\\
&=& \tilde g_{mk} \tilde g_{nl} 
    + \tilde g_{na} \tilde g_{bs}
    (\tilde\phi_{213} \tilde R_{12} \tilde\phi^{-1})_{mkl}^{bas} \nn
\err
Collecting the result, we recognize 
the structure of Wick contractions which are given by the invariant tensor
for neighboring indices, but involve the $\tilde R$--matrix and the 
coassociator $\tilde \phi$ for ``non--planar'' diagrams.

To illustrate the quasiassociative approach, we calculate the same 
4--point function using the $\star$ product. Then
\berr
\langle a_{n} \star (a_{m}\star(a_{k}\star a_{l})) \rangle 
  &=&  \langle \dl_n \star \big((D^{-1} j_m + \dl_m)\star 
    ((D^{-1} j_k + \dl_k)\star  D^{-1}  j_l)\big)\rangle \nn\\
  &=& D^{-2}\; \tilde g_{nm} \tilde g_{kl} 
    + D^{-2}\; \langle \dl_n \star (\dl_m \star
     (j_k\star  j_l)) \rangle  \nn
\err
using an obvious analog of (\ref{J_del_Wick}). Now
\berr
 \!\! \langle\dl_n \star (\dl_m \star (j_k\star j_l))\rangle \!\!\!
 &=& \langle\dl_n \star ((\dl_{m'} \star j_{k'})\star j_{l'})\rangle
     \;(\tilde\phi^{-1})^{m'k'l'}_{mkl} \nn\\
 &=& \tilde g_{nl'}\tilde g_{m'k'}  (\tilde\phi^{-1})^{m'k'l'}_{mkl} + 
    \langle\dl_n \star ((j_{m''} \star \dl_{k''} \tilde R^{k''m''}_{m'k'})
    \star j_{l'})\rangle\;(\tilde\phi^{-1})^{m'k'l'}_{mkl} \nn\\
 &=& \tilde g_{nl}\tilde g_{mk} +
     \langle\dl_n \star (j_{m'} \star (\dl_{k'} \star j_{l'}))\rangle\;
    (\tilde\phi_{213}\tilde R_{12}\tilde\phi^{-1})^{k'm'l'}_{mkl} \nn\\
 &=& \tilde g_{nl}\tilde g_{mk} +
     \tilde g_{n m'} \tilde g_{k'l'}\;
    (\tilde\phi_{213}\tilde R_{12}\tilde\phi^{-1})^{k'm'l'}_{mkl}, \nn
\err
in agreement with our previous calculation; here
the identity (\ref{gphi_id}) was used.
As pointed out before, the corrections to order $o(h)$ can now be obtained
easily.

\subsection{Remarks on $N \rightarrow \infty$ and $\phi^4$ theory.}

The above correlators for the free theory
are independent of $N$, as long as 
the spin of the modes is smaller than $N$. Therefore one can define 
the limit $N \rightarrow \infty$ in a straightforward way,
keeping $R$ constant. In this limit, the algebra of functions on
the $q$--deformed fuzzy sphere becomes
\be
\eps^{ij}_k x_i x_j = R\;(q-q^{-1})\; x_k, \qquad g^{ij} x_i x_j = R^2,
\ee
which defines $S^2_{q,N=\infty}$. It 
has a unique faithful (infinite--dimensional) 
Hilbert space representation \cite{Podles:1987wd}.

In an interacting theory, the existence of the limit $N \rightarrow \infty$
is of course a highly nontrivial question.
Consider for example the $\phi^4$ model, with action
$$
S[\Psi] = \int\limits_{\S_{q,N}} 
    \hf \hat \Psi(x) \Delta \hat \Psi(x) + \hf m^2 \hat\Psi(x)^2
   + \lambda \hat \Psi(x)^4 \; = S_{free} + S_{int}
$$
which is real, using (\ref{psi_hat_reality}).
We want to study the first--order corrections in $\lambda$ 
to the 2--point function
$\langle \hat a^K_i \hat a^{K}_{j} \rangle$ using 
(\ref{Wick_novac}):
$$
\langle \hat a^K_i \hat a^{K}_{j} \rangle = {{}_{{}_{J=0}}}\langle 
   \hat b^K_i\; \hat b^K_j\; \Big(1 - \lambda \int\limits_{\S_{q,N}} 
      \psi^{I,k}(x)\psi^{J,l}(x)\psi^{L,m}(x)\psi^{M,n}(x)\;
     \hat b^I_k \hat b^J_l \hat b^L_m \hat b^M_n\Big)
      \rangle_{\dl=0,\;\mbox{\scriptsize no vac}} .
$$
We only consider the ``leading'' planar tadpole diagram.
It is given by any contraction of the $\hat b^K_i$ and $\hat b^K_j$ with 
$\hat b$'s in the interaction term, which does not involve ``crossings''. 
All of these contributions are the same, hence we assume that 
$j$ is contracted with $k$ and $i$ with $l$. Then $\hat b^L_m $
is contracted with $\hat b^M_n$, which gives 
$D_L^{-1} \tilde g^L_{mn}\; \d^{LM}$. Now 
$\psi^{L,m}(x)\psi^{L,n}(x)\; \tilde g^L_{mn}\; \in \S_{q,N}$ 
is invariant under $U_q(su(2))$ and therefore proportional to the 
constant function. The numerical factor can be 
obtained from (\ref{psi_normaliz}):
$$
\int \psi^{L,m}(x)\psi^{L,n}(x)\; \tilde g^L_{mn} = 
\tilde g_L^{mn}\tilde g^L_{mn} = [2L+1]_q = {}_q \dim(V^L).
$$
Here $V^L$ denotes the spin $L$ \rep of $U_q(su(2))$. 
Using $\int 1 = 4\pi R^2$,
the contribution to $\langle \hat a^K_i \hat a^{K}_{j} \rangle$ is 
$$
\tilde g^K_{il} \tilde g^K_{jk}\; \lambda\int\psi^{K,k}(x)\psi^{K,l}(x)
       \sum_{L = 0}^N\;  D_L^{-1}\; \frac 1{4\pi R^2} [2L+1]_q = 
\tilde g^K_{ij}\; \frac{\lambda}{4\pi}
     \sum_{L=0}^N\; \frac{[2L+1]_q}{[L]_q [L+1]_q + m^2 R^2},
$$
up to combinatorial factors of order 1.
Unfortunately this diverges linearly in $N$ for $N \rightarrow \infty$,
whenever $q \neq 1$. This is worse that for $q=1$, where the divergence is
only logarithmic. This is in contrast to a result of \cite{Oeckl:1999zu}, 
which is however in the context of a different concept of (braided) 
quantum field theory which does not satisfy our requirements in 
Section \ref{sec:QFT}, and hence is not a ``smooth deformation'' of
ordinary QFT. 
The contributions from the ``non--planar'' tadpole diagrams 
are expected to be smaller, because the
coassociator $\tilde \phi$ as well as $\tilde R$ are unitary.
At least for scalar field theories, this behavior could be 
improved by choosing another Laplacian such as $\frac{v-v^{-1}}{q-q^{-1}}$
which has eigenvalues $[2L(L+1)]_q$, where $v$ is the Drinfeld Casimir 
(\ref{v}). Then all diagrams are convergent 
as $N \rightarrow \infty$. 
Finally, the case $q$ being a root of unity is more subtle, 
and we postpone it for future work.

\subsection{Gauge fields}
\label{subsec:gaugefields-2}

The quantization of gauge fields $\S_{q,N}$ is less clear at present,
and we will briefly indicate 2 possibilities.
Gauge fields were introduced in Section \ref{subsec:gauge} as
one--forms $B \in \Omega^1_{q,N}$.
%Here $\Omega^1_{q,N}$ is the subspace of one--forms in the 
%$U_q(su(2))$--module algebra %$\Omega^*_{q,N}$ 
%of differential forms on 
%$\S_{q,N}$. It turns out that there is a basis of 3 independent
%one--forms $\theta^a$ which commute with all functions on $\S_{q,N}$. 
It is natural to expand the gauge fields in terms of the 
frame $\theta^a$,
\be
B = \sum B_a \theta^a.
\label{B_expand-1}
\ee 
The fact that there are 3 independent one--forms $\theta^a$  means that 
one component is essentially radial and should be considered
as a scalar field on the sphere; however, it is impossible to find
a (covariant) calculus with ``tangential'' forms only. 
Therefore gauge theory on $\S_{q,N}$ as presented here 
is somewhat different from the conventional picture, but
may nevertheless be very interesting physically \cite{Alekseev:2000fd}.

Actions for gauge theories are expressions in $B$ which involve 
{\em no} explicit derivative terms.
We recall the simplest examples from Section \ref{subsec:gauge},
\be
S_3 = \int B^3, \quad 
S_2 = \int B \ast_H B,  \quad S_4 = \int B^2 \ast_H B^2,
\label{B_actions-1}
\ee
where $\ast_H$ is the Hodge star operator. 
The curvature was defined as $F = B^2 - \ast_H B$.
The meaning of the field $B$ becomes more obvious if it is written  
in the form 
\be
B = \Theta + A%, \qquad B_a = \l_a + A_a
%\label{B_A_split}
\ee
where $\Theta \in \Omega^1_{q,N}$ is the 
generator of exterior derivatives.
While $B$ and $\Theta$ become singular in the limit $N \rightarrow \infty$, 
$A$ remains well--defined. In these variables, a more standard form 
of the actions is recovered, including Yang--Mills 
\be 
S_{YM} :=  \int F \ast_H F =  \int (dA + A^2) \ast_H (dA + A^2)
%\label{YM}
\ee
and Chern--Simons 
\be
S_{CS} := \frac 13 \int B^3 - \frac 12 \int B \ast_H B = - \mbox{const}
    + \frac 12  \int A dA + \frac 23 A^3
%\label{S_CS}
\ee
terms.

Even though these actions (in particular the prescription 
``no explicit derivatives'') are very convincing and have the correct
limit at $q=1$, the precise meaning of gauge invariance is not clear. 
In the case $q=1$, gauge transformations have the form
$B_a \rightarrow U^{-1} B_a U$ for any unitary
matrix $U$, and actions of the above type are invariant. 
For $q \neq 1$, the integral is a quantum trace which contains an explicit 
``weight factor''$q^{-H}$, breaking this symmetry. 
There is however another symmetry of the above actions where  
$U_q(su(2))$ acts on the gauge fields $B_a$ as \cite{qFSI}
\be
B_a \rightarrow u_1 B_a S u_2
\label{gaugetransform}
\ee
or equivalently $B \rightarrow u_1 B S u_2$. 
This can be interpreted as a gauge transformation, 
leaving the actions
invariant for any $u \in U_q(su(2))$ with $\eps_q(u) = 1$, and it is
distinct from the rotations of $B$. 
There is no obvious extension to a deformed $U(su(N))$ invariance, however.
There is yet another $\tilde U_q(su(2))$
symmetry, rotating the frames $\theta^a$ only, i.e. mixing the 
components $B_a$. The rotation of the field $B$ is rather complicated 
if expressed in terms of the $B_a$, however.

The significance of all these different symmetries is not clear,
and we are not able to preserve them simultaneously at the quantum level.
We will therefore indicate 
two possible quantization schemes, leaving different symmetries manifest.

\paragraph{1) Quantization respecting rotation--invariance.}

First, we want to preserve the $U_q(su(2))$ symmetry corresponding
to rotations of the one--forms $\Omega^1_{q,N}$, which is 
underlies their algebraic properties \cite{qFSI}.
We shall moreover impose the constraint 
$$
d \ast_H B = 0,
$$
which can be interpreted as gauge fixing. It is invariant under rotations,
and removes precisely the null--modes in the Yang--Mills and Chern--Simons 
terms. We expand the field $B$ into irreducible \reps under this
action of $U_q(su(2))$:
\be
B = \sum_{K,n;\a} \;  \Xi^{\a}_{K,n}(x)\; b_{\a}^{K,n}.
\label{B_irreps}
\ee
Here $\Xi^{\a}_{K,n}(x) \in \Omega^1_{q,N}$ are one--forms 
which are spin $K$ \rep of $U_q(su(2))$ (``vector spherical harmonics''). 
The multiplicity is now generically $2$
because of the constraint, labeled by $\a$. 
%More precisely, 
%$mult(K) = \{1,3,3,...,3,2,1\}$ for $K=\{0,1,2,...,N-1, N, N+1\}$,
%since the maximal spin in $\S_{q,N}$ is $N$.
%Recall that $B$ contains a scalar field component 
%$\varphi(x) \Theta$, where $\Theta$ is the Dirac ``operator''
%which is the unique invariant one--form. 

To quantize this, we can use the same methods as in Section \ref{sec:QFT}.
One can either define a $\star$ product of the coefficients
$b_{\a}^{K,n}$ as discussed there, or introduce the operators
$\hat b_{\a}^{K,n}$ acting on a left vacuum representation. 
Choosing the star product approach to be specific, one
can then define correlation functions as 
\be
\langle P_{\star}(b)\rangle = \frac 1{\cN}\int  \D B \; 
                      e^{-S[B]} P_{\star}(b)
\label{correlation_B}
\ee
where $\D B$ is the integral over all $b_{\a}^{K,n}$, 
write down generating functions etc.
This approach has the merit that the remarkable solution $B = \Theta$
of the equation $F = 0$ in \cite{qFSI} 
survives the quantization, because the corresponding mode 
is a singlet (so that $\hat b_{\a}^{0,0} = b_{\a}^{0,0}$ is undeformed). 
Incidentally, observe that the bracketings 
$\int (B B)\ast_H (B B)$ and $\int B (B\ast_H(BB))$ in the star--product 
approach are equivalent, because of (\ref{phi-integral}).

\paragraph{2) Quantization respecting ``gauge invariance''.}

First, notice that there is no need for gauge fixing before quantization
even for $q=1$, since the group of gauge transformations is compact.
To preserve the symmetry (\ref{gaugetransform}) as well as the rotation
of the $\theta^a$, we expand $B$ into irreducible \reps under these
2 symmetries $U_q(su(2))$ and $\tilde U_q(su(2))$:
\be
B = \sum_{K,n;a} \;\psi_{K,n}(x)\theta^a\; \b_{a}^{K,n}.
\label{B_irreps_2}
\ee
Now $\b_{a}^{K,n}$ is a spin $K$ \rep of $\tilde U_q(su(2))$ and a spin 1
\rep of $U_q(su(2))$. These are independent and commuting
symmetries, hence
the quantization will involve their respective Drinfeld twists
$\tilde\F$ and $\F$. In the associative approach of Section \ref{sec:QFT}
we would then introduce
$$
\hat \b_{a}^{K,n} = \b_{a'}^{K,n'} \pi^{a'}_a(\F^{-1}_1) 
\pi^n_{n'}(\tilde S \tilde\F^{-1}_1)  \F^{-1}_2 \tilde\F^{-1}_2 \;  
            \in \; (\tilde U(su(2)) \tens U(su(2))) \smash \A.
$$
To avoid confusion, we have used an explicit matrix notation here.
The rest is formally as before, and will be omitted.
One drawback of this approach is that the above--mentioned
solution $B = \Theta$ is somewhat obscured now: the corresponding
mode is part of $\b_{a}^{1,n}$, but not easily identified. 
Moreover, ``overall'' rotation invariance is not manifest in this 
quantization.

%these modes will have a nontrivial star product, therefore 
%$B = \Theta$ may not solve $F=0$ any more. In fact, the definition of $F$ 
%becomes ``quantized'' as well, perhaps $F_{\star} = B\star B - \ast_H B$.

%(however: the $\tilde U_q(su(2))$ is quite simple and should be expressible
%as nice quantum algebra, by Fiore. Even the $\hat \b_{a'}^{1,n'}$ should
%be....in fact, should be just associative $Fun_q(SU(2))$, by Taktadjan???!!!!)

%Note: if use $\hat B$, then there is a very nice reality structure:
%\be
%\hat B^* = \hat B 
%\ee
%preserved under $U_q(su(2))$, I think. Real gauge fields.

\subsection{QFT in $2_q +1$ dimensions, Fock space}
\label{subsec:2plus1}

So far, we considered 2--dimensional $q$--deformed
Euclidean field theory. In this section,
we will add an extra (commutative) time and define a 2+1--dimensional 
scalar quantum field theory on $\S_{q,N}$ with manifest 
$\tilde U_q(su(2)) \times \R$ symmetry, 
where $\R$ corresponds to time 
translations. This will be done using an operator approach,
with $q$--deformed creation and anihilation operators acting on a 
Fock space. The purpose is mainly to elucidate the meaning of the 
Drinfeld twists as ``dressing transformations''.

We consider real scalar field operators of the form
\be
\hat \Psi(x,t) = \sum_{K,n} \;  \psi^{K,n}(x)\;\hat a_{K,n}(t)
  \; + \; \psi^{K,n}(x)^* \; \hat a^+_{K,n}(t)
\label{psi_field_t-op}
\ee
where 
\be
a^{(+)}_{K,n}(t)= U^{-1}(t)\; a^{(+)}_{K,n}(0)\; U(t)
\ee
for some unitary time--evolution operator 
$U(t) = e^{-iH t/\hbar}$; we will again put $\hbar = 1$.
The Hamilton--operator $H$ acts on some Hilbert space $\cH$.
We will assume that $H$ is invariant under rotations, 
$$
[H,u] = 0
$$
where $u \in U(su(2))$ is an operator acting on $\cH$;
recall that as (operator) algebra, $\tilde U_q(su(2))$ 
is the same as $U(su(2))$.
Rather than attempting some kind of quantization 
procedure, we shall assume that 
\be
\hat a^{(+)}_{K,n}(t) = \tilde \F^{-1}_1 \trr a^{(+)}_{K,n}(t)\; 
   \tilde \F^{-1}_2 \;  =  \; a^{(+)}_{K,m}(t)\; \pi^m_n (\tilde \F^{-1}_1)\;
        \tilde \F^{-1}_2  
\label{a_t_hat}
\ee
as in (\ref{a_hat}), where $a^{(+)}_{K,n}= a^{(+)}_{K,n}(0)$ 
are ordinary creation--and anihilation 
operators generating a oscillator algebra $\A$, 
\berr
[a_{K,n}, a^+_{K',n'}] &=& \d_{K K'}\; (g_c)_{n n'}, \nn\\
\;[a_{K,n}, a_{K',n'}] &=& [a^+_{K,n}, a^+_{K',n'}] = 0 \nn
\err
and act on the usual Fock space\footnote{note that
this is the same as the ``left vacuum \rep'' of the subalgebra
generated by the $a^+_{K,n}$, in the notation of Section 
\ref{sec:operator-form}.}
\be
\cH = \oplus\; (a^+_{K,n} ...\; a^+_{K',n'} |0\rangle).
\label{Fock_a}
\ee
$\cH$ is in fact a \rep of $U(su(2)) \smash \A$, and 
the explicit $U(su(2))$--terms in (\ref{a_t_hat}) are now understood 
as operators acting on $\cH$. 
Hence the $\hat a^{(+)}_{K,n}(t)$ are some kind of dressed 
creation--and anihilation operators, whose equal--time commutation 
relations follow from (\ref{aa_hat_CR}), (\ref{del_a}):
\berr
\hat a_{K,n}\; \hat a^+_{K',n'} &=& \d_{K,K'} \; g_{n n'} + 
     \hat a^+_{K',l'}\; \hat a_{K,l} \; \mathfrak{R}^{l l'}_{n n'}, \nn\\
\hat a^+_{K,n}\; \hat a^+_{K',n'} &=&
     \hat a^+_{K',l'}\; \hat a^+_{K,l} \; \mathfrak{R}^{l l'}_{n n'}, \nn\\
\hat a_{K,n}\; \hat a_{K',n'} &=&
     \hat a_{K',l'}\; \hat a_{K,l} \; \mathfrak{R}^{l l'}_{n n'}  \nn
\err
where $\hat a^{(+)}_{K,n} = \hat a^{(+)}_{K,n}(0)$. The Fock space 
(\ref{Fock_a}) can equivalently be written as
\be
\cH = \oplus \; \hat a^{+K,n} ... \;\hat a^{+K',n'} |0\rangle.
\label{Fock_ahat}
\ee
Here the main point of our construction of a quantum group covariant 
field theory is most obvious, namely that a symmetrization postulate 
has been implemented which restricts the number of states in the Hilbert 
space as in the undeformed case. This is the meaning of the postulate (4) in 
the introductory discussion of Section \ref{sec:QFT}.
One could even exhibit a (trivial) action of the symmetric group $S_n$
on the $n$--particle space, using the unitary transformation induced
by the Drinfeld twist $\F$, as in \cite{Fiore:1996bu}. Moreover,
using (\ref{psi_reality}) and an analog of (\ref{a_reality})
it follows that
$$
\hat \Psi(x,t) ^* = \hat \Psi(x,t).
$$
One can also derive the usual formulas for time--dependent 
perturbation theory, if we assume that the Hamilton operator has the form
$$
H = H_{free} + V
$$
where
\be
H_{free} = \sum_{K=0}^N \;  D_K\; (\tilde g^K)^{nm} \;
        \hat a^+_{K,n} \hat a_{K,m}
  = \sum_{K=0}^N \;  D_K\; (g_c^K)^{nm} \; a^+_{K,n} a_{K,m},
\ee
and $V$ may have the form 
$$
V = \int\limits_{\S_{q,N}} 
    \hat \Psi(x) \hat \Psi(x) ... \hat \Psi(x).
$$
Using (\ref{del_mixed}), one can see that 
$$
[H_{free}, \hat a^+_{K,l}] = D_K\;  \hat a^+_{K,l}
$$
and similarly for $\hat a_{K,l}$. 
Therefore the eigenvectors of $H_{free}$  have the form 
$\hat a^{+K,n} ... \;\hat a^{+K',n'} |0\rangle$ with 
eigenvalues $(D_K + ... + D_{K'}) \in \R$, and if $V=0$, then
the time evolution is given as usual by
$$
\hat a^+_{K,n}(t)= e^{-i D_K t/\hbar}\; \hat a^+_{K,n}, \qquad
\hat a_{K,n}(t)= e^{i D_K t/\hbar}\; \hat a_{K,n}.
$$
One can then go to the interaction picture if $V \neq 0$ and
derive the usual formula involving time--ordered products.
However one must now keep the time--ordering explicit, and 
there seems to be no nice formula for contractions of time--ordered
products. We shall not pursue this any further here.

The main point here is that the above
definitions are entirely within the framework of ordinary
quantum mechanics, with a smooth limit $q \rightarrow 1$ where
the standard quantum field theory on the 
fuzzy sphere is recovered. Again, one could also consider the limit 
$N \rightarrow \infty$ while keeping $q$ constant. The existence 
of this limit is far from trivial.
Moreover there is nothing special about 
the space $\S_{q,N}$ as opposed to other, perhaps higher--dimensional
$q$--deformed spaces, except the technical simplifications because
of the finite number of modes. 
This shows that there is no obstacle in principle 
for studying deformations of quantum field theory on such spaces.

\section{Technical complements to Chapter \ref{chapter:qFSII}}
\label{chapter:supp-II}

\paragraph{\em Proof of Proposition \ref{F_minimal}:}

Assume that $\F$ is minimal, so that  (\ref{phi-integral}) holds.
We must show that it can be chosen such that $\F$ is unitary as well.
Define
$$
A:=\F_{23}(1\tens \Delta)\F, \qquad B:= \F_{12} (\Delta\tens 1) \F,
$$
so that $\phi = B^{-1} A$. From (\ref{phi-integral}) it follows that
$(*\tens *\tens *)\phi = \phi^{-1}$, 
hence $A A^* = B B^*$, and more generally
$$
f(A A^*) = f(B B^*)
$$
for functions $f$ which are defined by a power series. This also implies that
$$
A f(A^* A) A^* = B f(B^* B) B^*
$$
for any such $f$, hence
$$
\phi f(A^* A) = f(B^* B) \phi^{*-1} = f(B^* B) \phi.
$$
In particular we can choose $f(x) = \sqrt{x}$ which makes sense
because of (\ref{cond2bis}), and obtain
\be
\sqrt{B^* B}^{-1} \phi \sqrt{A^*A} = \phi.
\label{BA_id}
\ee
On the other hand, the element $T:= ((* \tens *)\F) \; \F$ commutes with 
$\Delta(u)$ because $(* \tens *) \Delta_q(u) = \Delta_q(u^*)$,
and so does $\sqrt{T}$, which is well--defined in 
$U(su(2))[[h]]$ since $\F = 1 + o(h)$. Moreover, $T$ is symmetric,
noting that
\be
(* \tens *) (\F_{21} \F^{-1}) = \F \F_{21}^{-1}
\ee
which follows from the well-known relation 
$(* \tens *) \RR = \RR_{21}$ for $q \in \R$.
Therefore $T$ is an admissible gauge transformation, and 
$\F':= \F \sqrt{T}^{-1}$ is easily seen to be
unitary (this argument is due to \cite{Jurco:1994te}).
In particular, since $\F^* \F$ commutes with $\Delta(u)$, it follows that
$A^* A = (\F_{23}^* \F_{23})(1\tens \Delta)(\F^* \F)$
and  $B^* B = (\F_{12}^* \F_{12})(\Delta\tens 1)(\F^* \F)$. 
Looking at the definition (\ref{defphi}), this means that
the left--hand side of (\ref{BA_id}) is the gauge transformation of $\phi$ 
under a gauge transformation
$\F \rightarrow \F':= \F \sqrt{T}^{-1}$, which makes $\F$ unitary. 
Therefore the coassociator is unchanged under this gauge transformation, 
hence it remains minimal.

\paragraph{\em Proof of Lemma \ref{twist_qa_lemma}:}

We simply calculate
\berr
a^{\dagger}_1 \star (a_2\star a_3)  &=&  
        (a^{\dagger}_1 \star a_2) \star a_3 \; \tilde\phi_{123}^{-1}\nn\\
  &=& (g_{12} + a_2 \star a^{\dagger}_1 \; \TR_{12}) 
              \star a_3\; \tilde\phi_{123}^{-1} \nn\\
  &=& g_{12}\; a_3 \;\tilde\phi_{123}^{-1} + 
        a_2 \star (a^{\dagger}_1 \star a_3)\;\tilde\phi_{213}\; 
        \TR_{12} \tilde\phi_{123}^{-1} \nn\\
 &=& g_{12}\; a_3 \;\tilde\phi_{123}^{-1} + 
        a_2 \star (g_{13} + \; a_3\star a^{\dagger}_1\;\TR_{13})\;
       \tilde\phi_{213}\; \TR_{12} \tilde\phi_{123}^{-1} \nn\\
&=& g_{12}\; a_3 \;\tilde\phi_{123}^{-1} +  
    a_2 g_{13} \tilde\phi_{213}\; \TR_{12} \tilde\phi_{123}^{-1} 
   +\;( a_2 \star a_3) \star a^{\dagger}_1\;\tilde\phi_{231}^{-1}\;
     \TR_{13}\; \tilde\phi_{213}\;  
     \TR_{12} \tilde\phi_{123}^{-1} \nn\\
 &=&  g_{12}\; a_3 \;\tilde\phi_{123}^{-1} 
      + a_2 \; g_{31} \tilde\phi_{231} \TR_{1,(23)} 
    + (a_2 \star a_3) \star a^{\dagger}_1\; \TR_{1,(23)}, \nn
\err
where (\ref{quasitr}) and $g_{31} \TR_{13} = g_{13}$ 
was used in the last step. Now
the first identity (\ref{aaa_braid}) follows immediately along these lines, 
omitting the inhomogeneous terms. 
To see the last one (\ref{aP-a_braid}), observe that
$$
g_{12} \tilde\phi_{123}^{-1} = (g_c)_{12} \F_{1,(23)}^{-1} \F_{23}^{-1} 
$$
because $g_{12} \F_{(12),3} = g_{12}$, and similarly
$$
g_{31} \tilde\phi_{231} \TR_{1,(23)} 
  =  (g_c)_{13}  \F_{1,(23)}^{-1} \F_{23}^{-1} .
$$
This implies that 
$$ 
\Big(g_{12}\; a_3 \;\tilde\phi_{123}^{-1} 
   + a_2\; g_{31} \tilde\phi_{231} \TR_{1,(23)}\Big) P^-_{23}
 = ((g_c)_{12} \; a_3+ (g_c)_{13}\; a_2) (1-\d^{23}_{32})
    \F_{1,(23)}^{-1}\F_{23}^{-1}  = 0
$$
where we used the fact that the undeformed coproduct is symmetric.
The second (\ref{del_asquare}) follows as above using  
$$
g_{31} \tilde\phi_{231} \TR_{1,(23)} g^{23} = \d_1^2,
$$
or simply from (\ref{twisted_square}).

\paragraph{\em Proof of Proposition \ref{a2_central_prop}:}

Relation (\ref{g_twist_3}) follows easily from 
\be
\pi^j_s(u) (g_c)^{rs} = \pi^r_l(Su) (g_c)^{lj}
%\pi^i_j(Sx) = \pi^m_l(x) (g_c)^{li} (g_c)_{mj}.
\label{pi_S}
\ee
To prove (\ref{quadratic_hat}), consider
\berr
g^{ij}\hat a_i \hat a_j &=& g^{ij} \F_1^{-1} \trr a_i (\F_{2,1}^{-1}\F_a^{-1}) 
        \trr a_j \F_{2,2}^{-1} \F_b^{-1} \nn\\
  &=& a_k a_l g^{ij} \pi^k_i(\F^{-1}_1) \pi^l_j(\F_{2,1}^{-1}\F_a^{-1}) 
           \F_{2,2}^{-1} \F_b^{-1} \nn\\
  &=& a_k  a_l \pi^j_n(\g') (g_c)^{in} \pi^k_i(\F^{-1}_1)
         \pi^l_j(\F_{2,1}^{-1}\F_a^{-1}) \F_{2,2}^{-1} \F_b^{-1}. \nn
\err
Now we use
$\pi^k_i(\F^{-1}_1) (g_c)^{in} = (g_c)^{kr} \pi^n_r(S \F^{-1}_1)$, 
therefore
\berr
g^{ij}\hat a_i \hat a_j &=& a_k a_l (g_c)^{kr} 
 \pi^l_r(\F_{2,1}^{-1}\F_a^{-1}\g' S \F^{-1}_1)\F_{2,2}^{-1} \F_b^{-1} \nn\\
 &=& a_k a_l (g_c)^{kl}  \nn
\err
because of (\ref{F_id_5}).

\paragraph{\em Proof of Proposition \ref{derivatives-prop}:}

(\ref{del_a2}) follows easily from (\ref{quadratic_hat}):
\berr
\hat \del_i (g^{jk}\hat a_j \hat a_k)  &=&
       \hat\del_i ((g_c)^{jk} a_j a_k)  \nn\\
  &=& \del_n \pi^n_i(\F^{-1}_1) \F^{-1}_2  ((g_c)^{jk} a_j a_k)  \nn\\
  &=& \del_n \pi^n_i(\F^{-1}_1) ((g_c)^{jk} a_j a_k) \F^{-1}_2 \nn\\
  &=& 2 a_n \pi^n_i(\F^{-1}_1) \F^{-1}_2 
      +((g_c)^{jk} a_j a_k) \del_n \pi^n_i(\F^{-1}_1)\F^{-1}_2  \nn\\
  &=& 2 \hat a_i  
      +(g^{jk}\hat a_j \hat a_k) \hat \del_i, \nn
\err
as claimed. Next, consider
\berr
\hat \del_i \hat a_j &=& \del_n \pi^n_i(\F^{-1}_1) a_l 
    \pi^l_j(\F^{-1}_{2,1}\F^{-1}_a) \F^{-1}_{2,2} \F^{-1}_b \nn\\
    &=& (g_c)_{nl} \;\pi^n_i(\F^{-1}_1) 
         \pi^l_j(\F^{-1}_{2,1}\F^{-1}_a) \F^{-1}_{2,2} \F^{-1}_b 
    + a_l  \pi^l_j(\F^{-1}_{2,1}\F^{-1}_a)\del_n \pi^n_i(\F^{-1}_1)
             \F^{-1}_{2,2} \F^{-1}_b.   \nn
\err
The second term becomes 
$\hat a_k \hat \del_l \; \mathfrak{R}^{lk}_{ij}$ as in 
(\ref{aa_hat_CR}), and the first is
\berr
 (g_c)_{nl} \;\pi^n_i(\F^{-1}_1)
         \pi^l_j(\F^{-1}_{2,1}\F^{-1}_a) \F^{-1}_{2,2} \F^{-1}_b % \nn\\
  &=& \pi^t_l(S\F^{-1}_1) (g_c)_{ti}\pi^l_j(\F^{-1}_{2,1}\F^{-1}_a) 
      \F^{-1}_{2,2} \F^{-1}_b  \nn\\
  &=&  (g_c)_{ti}\; \pi^t_j(S\F^{-1}_1 \F^{-1}_{2,1}\F^{-1}_a) 
      \F^{-1}_{2,2} \F^{-1}_b \nn\\
  &=&  (g_c)_{ti}\; \pi^t_j(\g) = 
      (g_c)_{ti}\; \pi^t_l(S \F_1^{-1}) \pi^l_j(\F_2^{-1}) \nn\\
  &=& (g_c)_{tl}\; \pi^t_i(\F_1^{-1})  \pi^l_j(\F_2^{-1}) = g_{ij} 
\label{gphi_id}
\err
using (\ref{F_id_4}).

\paragraph{\em Proof of Proposition \ref{a_star_real_prop}:}
Since $\pi$ is a unitary representation, we have
\berr
\hat a_i^* &=& \F_2 a_j^* \pi^i_j(\F_1) = 
             \F_2 a_k (g_c)^{kj}\pi^i_j(\F_1) \nn\\
  &=& \F_2 a_k (g_c)^{ni}\pi^k_n(S\F_1) \nn\\ 
  &=& a_l \pi^l_k(\F_{2,1}) (g_c)^{in} \pi^k_n(S\F_1)\F_{2,2} \nn\\
  &=& a_l \pi^l_k(\F_{2,1}S\F_1)\F_{2,2} g^{it} \pi^k_t(\g'^{-1})\nn\\
  &=& a_l \pi^l_t(\F_{2,1}S\F_1 \g'^{-1})\F_{2,2} g^{it} \nn\\
  &=& a_l \pi^l_t(\F^{-1}_1)\F^{-1}_2 g^{it} \nn\\
  &=& \hat a_t g^{it} \nn
\err
where (\ref{F_id_3}) was essential.

%\begin{appendix}
%% rep-theory etc.
%\input{math.tex}
%\end{appendix}
\newpage
\addcontentsline{toc}{chapter}{Bibliography}

%\nocite{*}                   %this uses *everything* in the .bib file
  
\bibliographystyle{utphys} 
\bibliography{Main}        %or whatever your .bib file is

\providecommand{\href}[2]{#2}\begingroup\raggedright\begin{thebibliography}{10%
0}

\bibitem{Heisenberg:1996bv}
W.~Heisenberg, ``The universal length appearing in the theory of elementary
  particles,''. Ann. Phys. 32 (1938) 20-33. In Miller, A.I.: Early quantum
  electrodynamics 244-253.

\bibitem{Connes:book}
A.~Connes, ``Noncommutative geometry,''. Academic Press, 1994.

\bibitem{Douglas:1998fm}
M.~R. Douglas and C.~M. Hull, ``{D-}branes and the noncommutative torus,'' {\em
  JHEP} {\bf 02} (1998) 008,
\href{http://arXiv.org/abs/hep-th/9711165}{{\tt hep-th/9711165}}.
%%CITATION = HEP-TH 9711165;%%.

\bibitem{Chu:1998qz}
C.-S. Chu and P.-M. Ho, ``Noncommutative open string and {D-}brane,'' {\em
  Nucl. Phys.} {\bf B550} (1999) 151--168,
\href{http://arXiv.org/abs/hep-th/9812219}{{\tt hep-th/9812219}}.
%%CITATION = HEP-TH 9812219;%%.

\bibitem{Seiberg:1999vs}
N.~Seiberg and E.~Witten, ``String theory and noncommutative geometry,'' {\em
  JHEP} {\bf 09} (1999) 032,
\href{http://arXiv.org/abs/hep-th/9908142}{{\tt hep-th/9908142}}.
%%CITATION = HEP-TH 9908142;%%.

\bibitem{Schomerus:1999ug}
V.~Schomerus, ``{D-}branes and deformation quantization,'' {\em JHEP} {\bf 06}
  (1999) 030,
\href{http://arXiv.org/abs/hep-th/9903205}{{\tt hep-th/9903205}}.
%%CITATION = HEP-TH 9903205;%%.

\bibitem{Connes:1998cr}
A.~Connes, M.~R. Douglas, and A.~Schwarz, ``Noncommutative geometry and matrix
  theory: Compactification on tori,'' {\em JHEP} {\bf 02} (1998) 003,
\href{http://arXiv.org/abs/hep-th/9711162}{{\tt hep-th/9711162}}.
%%CITATION = HEP-TH 9711162;%%.

\bibitem{Alekseev:1999bs}
A.~Y. Alekseev, A.~Recknagel, and V.~Schomerus, ``Non-commutative world-volume
  geometries: Branes on {SU(2)} and fuzzy spheres,'' {\em JHEP} {\bf 09} (1999)
  023,
\href{http://arXiv.org/abs/hep-th/9908040}{{\tt hep-th/9908040}}.
%%CITATION = HEP-TH 9908040;%%.

\bibitem{Myers:1999ps}
R.~C. Myers, ``Dielectric branes,'' {\em JHEP} {\bf 12} (1999) 022,
\href{http://arXiv.org/abs/hep-th/9910053}{{\tt hep-th/9910053}}.
%%CITATION = HEP-TH 9910053;%%.

\bibitem{Jurco:2001rq}
B.~Jurco, L.~Moller, S.~Schraml, P.~Schupp, and J.~Wess, ``Construction of
  non-abelian gauge theories on noncommutative spaces,'' {\em Eur. Phys. J.}
  {\bf C21} (2001) 383--388,
\href{http://arXiv.org/abs/hep-th/0104153}{{\tt hep-th/0104153}}.
%%CITATION = HEP-TH 0104153;%%.

\bibitem{Jurco:2000dx}
B.~Jurco, P.~Schupp, and J.~Wess, ``Nonabelian noncommutative gauge fields and
  {Seiberg}-{Witten} map,'' {\em Mod. Phys. Lett.} {\bf A16} (2001) 343--348,
\href{http://arXiv.org/abs/hep-th/0012225}{{\tt hep-th/0012225}}.
%%CITATION = HEP-TH 0012225;%%.

\bibitem{Jurco:2001my}
B.~Jurco, P.~Schupp, and J.~Wess, ``Nonabelian noncommutative gauge theory via
  noncommutative extra dimensions,'' {\em Nucl. Phys.} {\bf B604} (2001)
  148--180,
\href{http://arXiv.org/abs/hep-th/0102129}{{\tt hep-th/0102129}}.
%%CITATION = HEP-TH 0102129;%%.

\bibitem{Alekseev:1998mc}
A.~Y. Alekseev and V.~Schomerus, ``{D-}branes in the {WZW} model,'' {\em Phys.
  Rev.} {\bf D60} (1999) 061901,
\href{http://arXiv.org/abs/hep-th/9812193}{{\tt hep-th/9812193}}.
%%CITATION = HEP-TH 9812193;%%.

\bibitem{Alekseev:2002rj}
A.~Y. Alekseev, S.~Fredenhagen, T.~Quella, and V.~Schomerus, ``Non-commutative
  gauge theory of twisted {D-}branes,''
\href{http://arXiv.org/abs/hep-th/0205123}{{\tt hep-th/0205123}}.
%%CITATION = HEP-TH 0205123;%%.

\bibitem{Pawelczyk:2000ah}
J.~Pawelczyk, ``{SU(2)} {WZW} {D-}branes and their noncommutative geometry from
  {DBI} action,'' {\em JHEP} {\bf 08} (2000) 006,
\href{http://arXiv.org/abs/hep-th/0003057}{{\tt hep-th/0003057}}.
%%CITATION = HEP-TH 0003057;%%.

\bibitem{Bordalo:2001ec}
P.~Bordalo, S.~Ribault, and C.~Schweigert, ``Flux stabilization in compact
  groups,'' {\em JHEP} {\bf 10} (2001) 036,
\href{http://arXiv.org/abs/hep-th/0108201}{{\tt hep-th/0108201}}.
%%CITATION = HEP-TH 0108201;%%.

\bibitem{Schweigert:2000iv}
C.~Schweigert, ``{D-}branes in group manifolds and flux stabilization,''
\href{http://arXiv.org/abs/hep-th/0012045}{{\tt hep-th/0012045}}.
%%CITATION = HEP-TH 0012045;%%.

\bibitem{Bachas:2000ik}
C.~Bachas, M.~R. Douglas, and C.~Schweigert, ``Flux stabilization of
  {D-}branes,'' {\em JHEP} {\bf 05} (2000) 048,
\href{http://arXiv.org/abs/hep-th/0003037}{{\tt hep-th/0003037}}.
%%CITATION = HEP-TH 0003037;%%.

\bibitem{Felder:1999ka}
G.~Felder, J.~Frohlich, J.~Fuchs, and C.~Schweigert, ``The geometry of {WZW}
  branes,'' {\em J. Geom. Phys.} {\bf 34} (2000) 162--190,
\href{http://arXiv.org/abs/hep-th/9909030}{{\tt hep-th/9909030}}.
%%CITATION = HEP-TH 9909030;%%.

\bibitem{Maldacena:2001xj}
J.~M. Maldacena, G.~W. Moore, and N.~Seiberg, ``{D-}brane instantons and
  {K-}theory charges,'' {\em JHEP} {\bf 11} (2001) 062,
\href{http://arXiv.org/abs/hep-th/0108100}{{\tt hep-th/0108100}}.
%%CITATION = HEP-TH 0108100;%%.

\bibitem{Fredenhagen:2000ei}
S.~Fredenhagen and V.~Schomerus, ``Branes on group manifolds, gluon
  condensates, and twisted {K-}theory,'' {\em JHEP} {\bf 04} (2001) 007,
\href{http://arXiv.org/abs/hep-th/0012164}{{\tt hep-th/0012164}}.
%%CITATION = HEP-TH 0012164;%%.

\bibitem{Figueroa-O'Farrill:2000kz}
J.~M. Figueroa-O'Farrill and S.~Stanciu, ``{D-}brane charge, flux quantization
  and relative (co)homology,'' {\em JHEP} {\bf 01} (2001) 006,
\href{http://arXiv.org/abs/hep-th/0008038}{{\tt hep-th/0008038}}.
%%CITATION = HEP-TH 0008038;%%.

\bibitem{Stanciu:2000fz}
S.~Stanciu, ``A note on {D-}branes in group manifolds: Flux quantization and
  {$D0$}-charge,'' {\em JHEP} {\bf 10} (2000) 015,
\href{http://arXiv.org/abs/hep-th/0006145}{{\tt hep-th/0006145}}.
%%CITATION = HEP-TH 0006145;%%.

\bibitem{Faddeev:1990ih}
L.~D. Faddeev, N.~Y. Reshetikhin, and L.~A. Takhtajan, ``Quantization of {Lie}
  groups and {Lie} algebras,'' {\em Leningrad Math. J.} {\bf 1} (1990)
193--225.
%%CITATION = 00064,1,193;%%.

\bibitem{Woronowicz:1987vs}
S.~L. Woronowicz, ``Compact matrix pseudogroups,'' {\em Commun. Math. Phys.}
  {\bf 111} (1987)
613--665.
%%CITATION = CMPHA,111,613;%%.

\bibitem{Mezincescu:1991ui}
L.~Mezincescu and R.~I. Nepomechie, ``Integrability of open spin chains with
  quantum algebra symmetry,'' {\em Int. J. Mod. Phys.} {\bf A6} (1991)
5231--5248.
%%CITATION = IMPAE,A6,5231;%%.

\bibitem{Kulish:1992qb}
P.~P. Kulish and E.~K. Sklyanin, ``Algebraic structures related to the
  reflection equations,'' {\em J. Phys.} {\bf A25} (1992) 5963--5976,
\href{http://arXiv.org/abs/hep-th/9209054}{{\tt hep-th/9209054}}.
%%CITATION = HEP-TH 9209054;%%.

\bibitem{PS:G}
J.~Pawelczyk and H.~Steinacker, ``A quantum algebraic description of {D-}branes
  on group manifolds,''
\href{http://arXiv.org/abs/hep-th/0203110}{{\tt hep-th/0203110}}.
%%CITATION = HEP-TH 0203110;%%.

\bibitem{Madore:1992bw}
J.~Madore, ``The fuzzy sphere,'' {\em Class. Quant. Grav.} {\bf 9} (1992)
69--88.
%%CITATION = CQGRD,9,69;%%.

\bibitem{Alekseev:2000fd}
A.~Y. Alekseev, A.~Recknagel, and V.~Schomerus, ``Brane dynamics in background
  fluxes and non-commutative geometry,'' {\em JHEP} {\bf 05} (2000) 010,
\href{http://arXiv.org/abs/hep-th/0003187}{{\tt hep-th/0003187}}.
%%CITATION = HEP-TH 0003187;%%.

\bibitem{Madore:2000en}
J.~Madore, S.~Schraml, P.~Schupp, and J.~Wess, ``Gauge theory on noncommutative
  spaces,'' {\em Eur. Phys. J.} {\bf C16} (2000) 161--167,
\href{http://arXiv.org/abs/hep-th/0001203}{{\tt hep-th/0001203}}.
%%CITATION = HEP-TH 0001203;%%.

\bibitem{fuzzyloop}
C.-S. Chu, J.~Madore, and H.~Steinacker, ``Scaling limits of the fuzzy sphere
  at one loop,'' {\em JHEP} {\bf 08} (2001) 038,
\href{http://arXiv.org/abs/hep-th/0106205}{{\tt hep-th/0106205}}.
%%CITATION = HEP-TH 0106205;%%.

\bibitem{Kirillov-book}
A.~Kirillov, ``Elements of the theory of representations,''. Die Grundlehren
  der mathematischen Wissenschaften 220, Springer 1976 ({\"Ubersetzung aus dem
  Russischen}).

\bibitem{Kostant-book}
B.~Kostant, ``Quantization and unitary representations,''. Lecture Notes in
  Math. 170, Springer, 1970.

\bibitem{Alekseev:2002km}
A.~Y. Alekseev, Y.~Kosmann-Schwarzbach, and E.~Meinrenken, ``Quasi-{Poisson}
  manifolds,'' {\em Canad. J. Math.} {\bf 54} (2002) 3--29,
\href{http://arXiv.org/abs/math.DG/0006168}{{\tt math.DG/0006168}}.
%%CITATION = MATH-DG 0006168;%%.

\bibitem{qFSI}
H.~Grosse, J.~Madore, and H.~Steinacker, ``Field theory on the {q-}deformed
  fuzzy sphere. {I},'' {\em J. Geom. Phys.} {\bf 38} (2001) 308--342,
\href{http://arXiv.org/abs/hep-th/0005273}{{\tt hep-th/0005273}}.
%%CITATION = HEP-TH 0005273;%%.

\bibitem{fuzzyinst}
H.~Grosse, M.~Maceda, J.~Madore, and H.~Steinacker, ``Fuzzy instantons,'' {\em
  Int. Jour. Mod. Phys. A} {\bf 17 No. 15} (2002) 2095,
\href{http://arXiv.org/abs/hep-th/0107068}{{\tt hep-th/0107068}}.
%%CITATION = HEP-TH 0107068;%%.

\bibitem{qFSII}
H.~Grosse, J.~Madore, and H.~Steinacker, ``Field theory on the {q-}deformed
  fuzzy sphere. {II}: Quantization,'' {\em J. Geom. Phys.} {\bf 43} (2002)
  205--240,
\href{http://arXiv.org/abs/hep-th/0103164}{{\tt hep-th/0103164}}.
%%CITATION = HEP-TH 0103164;%%.

\bibitem{Chari:1994pz}
V.~Chari and A.~Pressley, ``A guide to quantum groups,''. Cambridge, UK:
  University Press (1994).

\bibitem{Klimyk:1997eb}
A.~Klimyk and K.~Schmudgen, ``Quantum groups and their representations,''.
  Berlin, Germany: Springer (1997).

\bibitem{Majid:1996kd}
S.~Majid, ``Foundations of quantum group theory,''. Cambridge, UK: University
  Press (1995).

\bibitem{Jimbo:1985zk}
M.~Jimbo, ``A q-difference analog of {$U(g)$} and the {Yang-Baxter} equation,''
  {\em Lett. Math. Phys.} {\bf 10} (1985)
63--69.
%%CITATION = LMPHD,10,63;%%.

\bibitem{Drinfeld:1988in}
V.~G. Drinfeld, ``Quantum groups,''. Proceedings of the International Congress
  of Mathematicians, Berkeley, 1986 A.M. Gleason (ed.), p. 798, AMS,
  Providence, RI.

\bibitem{Douglas:2001ba}
M.~R. Douglas and N.~A. Nekrasov, ``Noncommutative field theory,'' {\em Rev.
  Mod. Phys.} {\bf 73} (2001) 977--1029,
\href{http://arXiv.org/abs/hep-th/0106048}{{\tt hep-th/0106048}}.
%%CITATION = HEP-TH 0106048;%%.

\bibitem{PS:su2}
J.~Pawelczyk and H.~Steinacker, ``Matrix description of {D-}branes on
  3-spheres,'' {\em JHEP} {\bf 12} (2001) 018,
\href{http://arXiv.org/abs/hep-th/0107265}{{\tt hep-th/0107265}}.
%%CITATION = HEP-TH 0107265;%%.

\bibitem{Alexanian:2001qj}
G.~Alexanian, A.~P. Balachandran, G.~Immirzi, and B.~Ydri, ``Fuzzy {$CP^2$},''
  {\em J. Geom. Phys.} {\bf 42} (2002) 28--53,
\href{http://arXiv.org/abs/hep-th/0103023}{{\tt hep-th/0103023}}.
%%CITATION = HEP-TH 0103023;%%.

\bibitem{Grosse:1999ci}
H.~Grosse and A.~Strohmaier, ``Noncommutative geometry and the regularization
  problem of {4D} quantum field theory,'' {\em Lett. Math. Phys.} {\bf 48}
  (1999) 163--179,
\href{http://arXiv.org/abs/hep-th/9902138}{{\tt hep-th/9902138}}.
%%CITATION = HEP-TH 9902138;%%.

\bibitem{Maldacena:2001ky}
J.~M. Maldacena, G.~W. Moore, and N.~Seiberg, ``Geometrical interpretation of
  {D-}branes in gauged {WZW} models,'' {\em JHEP} {\bf 07} (2001) 046,
\href{http://arXiv.org/abs/hep-th/0105038}{{\tt hep-th/0105038}}.
%%CITATION = HEP-TH 0105038;%%.

\bibitem{Fuchs:1992nq}
J.~Fuchs, ``Affine {Lie} algebras and quantum groups: An introduction, with
  applications in conformal field theory,''. Cambridge, UK: University Press
  (1992) (Cambridge monographs on mathematical physics).

\bibitem{DiFrancesco:1997nk}
P.~Di~Francesco, P.~Mathieu, and D.~Senechal, ``Conformal field theory,''. New
  York, USA: Springer (1997).

\bibitem{Alvarez-Gaume:1990aq}
L.~Alvarez-Gaume, C.~Gomez, and G.~Sierra, ``Duality and quantum groups,'' {\em
  Nucl. Phys.} {\bf B330} (1990)
347.
%%CITATION = NUPHA,B330,347;%%.

\bibitem{KL}
D.~Kazhdan and G.~Lusztig, ``Affine {Lie} algebras and quantum groups.,'' {\em
  Internat. Math. Res. Notices} {\bf no.2} (1991) 21--29.

\bibitem{Moore:1989ni}
G.~W. Moore and N.~Reshetikhin, ``A comment on quantum group symmetry in
  conformal field theory,'' {\em Nucl. Phys.} {\bf B328} (1989)
557.
%%CITATION = NUPHA,B328,557;%%.

\bibitem{UnitaryReps}
H.~Steinacker, ``Unitary representations of noncompact quantum groups at roots
  of unity,'' {\em Rev. Math. Phys.} {\bf 13} (2001) 1035,
\href{http://arXiv.org/abs/math.qa/9907021}{{\tt math.qa/9907021}}.
%%CITATION = MATH.QA 9907021;%%.

\bibitem{Drinfeld:quasi}
V.~G. Drinfeld, ``{Quasi-Hopf} algebras,'' {\em Leningrad Math. J.} {\bf 1,
  No.6} (1991) 1419.

\bibitem{Drinfeld:quasiGal}
V.~G. Drinfeld, ``On quasitriangular {quasi-Hopf} algebras and a group closely
  connected with {$Gal(\obar{Q}/Q)$},'' {\em Leningrad Math. J.} {\bf 2, No.4}
  (1991) 829.

\bibitem{Kulish:1993ep}
P.~P. Kulish, R.~Sasaki, and C.~Schwiebert, ``Constant solutions of reflection
  equations and quantum groups,'' {\em J. Math. Phys.} {\bf 34} (1993)
  286--304,
\href{http://arXiv.org/abs/hep-th/9205039}{{\tt hep-th/9205039}}.
%%CITATION = HEP-TH 9205039;%%.

\bibitem{Kulish:1993pb}
P.~P. Kulish and R.~Sasaki, ``Covariance properties of reflection equation
  algebras,'' {\em Prog. Theor. Phys.} {\bf 89} (1993) 741--762,
\href{http://arXiv.org/abs/hep-th/9212007}{{\tt hep-th/9212007}}.
%%CITATION = HEP-TH 9212007;%%.

\bibitem{Schupp:1993mt}
P.~Schupp, P.~Watts, and B.~Zumino, ``Bicovariant quantum algebras and quantum
  {Lie} algebras,'' {\em Commun. Math. Phys.} {\bf 157} (1993) 305--330,
\href{http://arXiv.org/abs/hep-th/9210150}{{\tt hep-th/9210150}}.
%%CITATION = HEP-TH 9210150;%%.

\bibitem{Kirillov:1991ec}
A.~N. Kirillov and N.~Y. Reshetikhin, ``Representations of the algebra
  {$U_q(sl(2)), q$}- orthogonal polynomials and invariants of links,''. In
  *Kohno, T. (ed.): New developments in the theory of knots* 202-256. (see Book
  Index).

\bibitem{Grosse:1996ar}
H.~Grosse, C.~Klimcik, and P.~Presnajder, ``Towards finite quantum field theory
  in noncommutative geometry,'' {\em Int. J. Theor. Phys.} {\bf 35} (1996)
  231--244,
\href{http://arXiv.org/abs/hep-th/9505175}{{\tt hep-th/9505175}}.
%%CITATION = HEP-TH 9505175;%%.

\bibitem{AspqFS}
H.~Steinacker, ``Aspects of the {q-}deformed fuzzy sphere,'' {\em Mod. Phys.
  Lett.} {\bf A16} (2001) 361--366,
\href{http://arXiv.org/abs/hep-th/0102074}{{\tt hep-th/0102074}}.
%%CITATION = HEP-TH 0102074;%%.

\bibitem{KhoroshkinTol}
S.~M. Khoroshkin and V.~N. Tolstoy, ``Universal {$R$}-matrix for quantized
  (super)algebras,'' {\em Comm. Math. Phys.} {\bf 141} (1991) 599--617.

\bibitem{Kirillov:1990vb}
A.~N. Kirillov and N.~Reshetikhin, ``q-{Weyl} group and a multiplicative
  formula for universal {R} matrices,'' {\em Commun. Math. Phys.} {\bf 134}
  (1990)
421--432.
%%CITATION = CMPHA,134,421;%%.

\bibitem{Faddeev:1996ie}
L.~D. Faddeev and P.~N. Pyatov, ``The differential calculus on quantum linear
  groups,'' {\em Am. Math. Soc. Transl.} {\bf 175} (1996) 35--47,
\href{http://arXiv.org/abs/hep-th/9402070}{{\tt hep-th/9402070}}.
%%CITATION = HEP-TH 9402070;%%.

\bibitem{Reshetikhin:1988ix}
N.~Y. Reshetikhin, ``Quantized universal enveloping algebras, the {Yang-Baxter}
  equation and invariants of links. {I},''. LOMI-E-4-87.

\bibitem{Podles:1987wd}
P.~Podles, ``Quantum spheres,'' {\em Lett. Math. Phys.} {\bf 14} (1987)
193--202.
%%CITATION = LMPHD,14,193;%%.

\bibitem{Dubois-Violette:1989at}
M.~Dubois-Violette, J.~Madore, and R.~Kerner, ``Gauge bosons in a
  noncommutative geometry,'' {\em Phys. Lett.} {\bf B217} (1989)
485--488.
%%CITATION = PHLTA,B217,485;%%.

\bibitem{madoreDim}
J.~Madore and A.~Dimakis, ``Differential calculi and linear connections,'' {\em
  J. Math. Phys.} {\bf 37, no. 9} (1996) 4647.

\bibitem{Mad99c}
J.~Madore, ``An introduction to noncommutative differential geometry and its
  physical applications,''. No. 257 in London Mathematical Society Lecture Note
  Series. Cambridge University Press, second edition, 1999.

\bibitem{Grosse:1997pr}
H.~Grosse, C.~Klimcik, and P.~Presnajder, ``Field theory on a supersymmetric
  lattice,'' {\em Commun. Math. Phys.} {\bf 185} (1997) 155--175,
\href{http://arXiv.org/abs/hep-th/9507074}{{\tt hep-th/9507074}}.
%%CITATION = HEP-TH 9507074;%%.

\bibitem{Wess:1991vh}
J.~Wess and B.~Zumino, ``Covariant differential calculus on the quantum
  hyperplane,'' {\em Nucl. Phys. Proc. Suppl.} {\bf 18B} (1991)
302--312.
%%CITATION = NUPHZ,18B,302;%%.

\bibitem{Carow-Watamura:1991zp}
U.~Carow-Watamura, M.~Schlieker, and S.~Watamura, ``{$SO_q(N)$} covariant
  differential calculus on quantum space and quantum deformation of
  {Schrodinger} equation,'' {\em Z. Phys.} {\bf C49} (1991)
439--446.
%%CITATION = ZEPYA,C49,439;%%.

\bibitem{Rosso:1988gg}
M.~Rosso, ``Finite dimensional representations of the quantum analog of the
  enveloping algebra of a complex simple {Lie} algebra,'' {\em Commun. Math.
  Phys.} {\bf 117} (1988)
581--593.
%%CITATION = CMPHA,117,581;%%.

\bibitem{Cerchiai:2000qu}
B.~L. Cerchiai, G.~Fiore, and J.~Madore, ``Geometrical tools for quantum
  euclidean spaces,'' {\em Commun. Math. Phys.} {\bf 217} (2001) 521--554,
\href{http://arXiv.org/abs/math.qa/0002007}{{\tt math.qa/0002007}}.
%%CITATION = MATH.QA 0002007;%%.

\bibitem{Cerchiai:2000tc}
B.~L. Cerchiai, J.~Madore, S.~Schraml, and J.~Wess, ``Structure of the
  three-dimensional quantum euclidean space,'' {\em Eur. Phys. J.} {\bf C16}
  (2000) 169--180,
\href{http://arXiv.org/abs/math.qa/0004011}{{\tt math.qa/0004011}}.
%%CITATION = MATH.QA 0004011;%%.

\bibitem{Keller:1990tg}
G.~Keller, ``Fusion rules of {$U_q(sl(2,C)), q^m = 1$},'' {\em Lett. Math.
  Phys.} {\bf 21} (1991) 273.

\bibitem{Steinacker:1999xu}
H.~Steinacker, ``Quantum anti-de {Sitter} space and sphere at roots of unity,''
  {\em Adv. Theor. Math. Phys.} {\bf 4} (2000) 155--208,
\href{http://arXiv.org/abs/hep-th/9910037}{{\tt hep-th/9910037}}.
%%CITATION = HEP-TH 9910037;%%.

\bibitem{lev-soib}
Y.~Soibelman and S.~Levendorskii, ``Some applications of the quantum {Weyl}
  group,'' {\em Journ. Geom. Phys.} {\bf 7 (2)} (1990) 241.

\bibitem{Mack:1992tg}
G.~Mack and V.~Schomerus, ``Quasihopf quantum symmetry in quantum theory,''
  {\em Nucl. Phys.} {\bf B370} (1992)
185--230.
%%CITATION = NUPHA,B370,185;%%.

\bibitem{Woronowicz:1989rt}
S.~L. Woronowicz, ``Differential calculus on compact matrix pseudogroups
  (quantum groups),'' {\em Commun. Math. Phys.} {\bf 122} (1989)
125--170.
%%CITATION = CMPHA,122,125;%%.

\bibitem{Podlescalc}
P.~Podles, ``Differential calculus on quantum spheres,'' {\em Lett. Math.
  Phys.} {\bf 18} (1989) 107.

\bibitem{Apel:1994du}
J.~Apel and K.~Schmuedgen, ``Classification of three dimensional covariant
  differential calculi on {Podles'} quantum spheres and related spaces,'' {\em
  Lett. Math. Phys.} {\bf 32} (1994)
25--36.
%%CITATION = LMPHD,32,25;%%.

\bibitem{Reshetikhin:1988iw}
N.~Y. Reshetikhin, ``Quantized universal enveloping algebras, the {Yang-Baxter}
  equation and invariants of links. {II},''. LOMI-E-17-87.

\bibitem{Fiore:1993mb}
G.~Fiore, ``The {$SO_q(N,R)$} symmetric harmonic oscillator on the quantum
  euclidean space {$R_q^N$} and its {Hilbert} space structure,'' {\em Int. J.
  Mod. Phys.} {\bf A8} (1993) 4679--4729,
\href{http://arXiv.org/abs/hep-th/9306030}{{\tt hep-th/9306030}}.
%%CITATION = HEP-TH 9306030;%%.

\bibitem{Ogievetsky:1992qp}
O.~Ogievetsky and B.~Zumino, ``Reality in the differential calculus on
  {q-}euclidean spaces,'' {\em Lett. Math. Phys.} {\bf 25} (1992) 121--130,
\href{http://arXiv.org/abs/hep-th/9205003}{{\tt hep-th/9205003}}.
%%CITATION = HEP-TH 9205003;%%.

\bibitem{Steinacker:1995jh}
H.~Steinacker, ``Integration on quantum euclidean space and sphere in {N}
  dimensions,'' {\em J. Math. Phys.} {\bf 37} (1996) 7438,
  \href{http://arXiv.org/abs/q-alg/9710016}{{\tt q-alg/9710016}}.

\bibitem{Grosse:1992bm}
H.~Grosse and J.~Madore, ``A noncommutative version of the {Schwinger} model,''
  {\em Phys. Lett.} {\bf B283} (1992)
218--222.
%%CITATION = PHLTA,B283,218;%%.

\bibitem{Steinacker:1997za}
H.~Steinacker, ``Quantum groups, roots of unity and particles on quantized
  anti-de {Sitter} space,'' \href{http://arXiv.org/abs/hep-th/9705211}{{\tt
  hep-th/9705211}}.
Ph.D. Thesis, Berkeley, May 1997.
%%CITATION = HEP-TH 9705211;%%.

\bibitem{Steinacker:1997qz}
H.~Steinacker, ``Unitary representations and {BRST} structure of the quantum
  anti-de {Sitter} group at roots of unity,''
  \href{http://arXiv.org/abs/q-alg/9710016}{{\tt q-alg/9710016}}. Proceedings
  to WigSym5, Vienna, Austria, 25-29 August, 1997.

\bibitem{Dubois-Violette:1989ps}
M.~Dubois-Violette, J.~Madore, and R.~Kerner, ``Classical bosons in a
  noncommutative geometry,'' {\em Class. Quant. Grav.} {\bf 6} (1989)
1709.
%%CITATION = CQGRD,6,1709;%%.

\bibitem{Baez:1998he}
S.~Baez, A.~P. Balachandran, B.~Idri, and S.~Vaidya, ``Monopoles and solitons
  in fuzzy physics,'' {\em Commun. Math. Phys.} {\bf 208} (2000) 787--798,
\href{http://arXiv.org/abs/hep-th/9811169}{{\tt hep-th/9811169}}.
%%CITATION = HEP-TH 9811169;%%.

\bibitem{Gopakumar:2000zd}
R.~Gopakumar, S.~Minwalla, and A.~Strominger, ``Noncommutative solitons,'' {\em
  JHEP} {\bf 05} (2000) 020,
\href{http://arXiv.org/abs/hep-th/0003160}{{\tt hep-th/0003160}}.
%%CITATION = HEP-TH 0003160;%%.

\bibitem{Gross:2000ss}
D.~J. Gross and N.~A. Nekrasov, ``Solitons in noncommutative gauge theory,''
  {\em JHEP} {\bf 03} (2001) 044,
\href{http://arXiv.org/abs/hep-th/0010090}{{\tt hep-th/0010090}}.
%%CITATION = HEP-TH 0010090;%%.

\bibitem{Harvey:2000jt}
J.~A. Harvey, P.~Kraus, F.~Larsen, and E.~J. Martinec, ``{D-}branes and strings
  as non-commutative solitons,'' {\em JHEP} {\bf 07} (2000) 042,
\href{http://arXiv.org/abs/hep-th/0005031}{{\tt hep-th/0005031}}.
%%CITATION = HEP-TH 0005031;%%.

\bibitem{Connes:2000tj}
A.~Connes and G.~Landi, ``Noncommutative manifolds: The instanton algebra and
  isospectral deformations,'' {\em Commun. Math. Phys.} {\bf 221} (2001)
  141--159,
\href{http://arXiv.org/abs/math.qa/0011194}{{\tt math.qa/0011194}}.
%%CITATION = MATH.QA 0011194;%%.

\bibitem{Iso:2001mg}
S.~Iso, Y.~Kimura, K.~Tanaka, and K.~Wakatsuki, ``Noncommutative gauge theory
  on fuzzy sphere from matrix model,'' {\em Nucl. Phys.} {\bf B604} (2001)
  121--147,
\href{http://arXiv.org/abs/hep-th/0101102}{{\tt hep-th/0101102}}.
%%CITATION = HEP-TH 0101102;%%.

\bibitem{Hashimoto:2001xy}
K.~Hashimoto and K.~Krasnov, ``{D-}brane solutions in non-commutative gauge
  theory on fuzzy sphere,'' {\em Phys. Rev.} {\bf D64} (2001) 046007,
\href{http://arXiv.org/abs/hep-th/0101145}{{\tt hep-th/0101145}}.
%%CITATION = HEP-TH 0101145;%%.

\bibitem{Hikida:2000cp}
Y.~Hikida, M.~Nozaki, and T.~Takayanagi, ``Tachyon condensation on fuzzy sphere
  and noncommutative solitons,'' {\em Nucl. Phys.} {\bf B595} (2001) 319--331,
\href{http://arXiv.org/abs/hep-th/0008023}{{\tt hep-th/0008023}}.
%%CITATION = HEP-TH 0008023;%%.

\bibitem{Krajewski:2001ch}
T.~Krajewski and M.~Schnabl, ``Exact solitons on noncommutative tori,'' {\em
  JHEP} {\bf 08} (2001) 002,
\href{http://arXiv.org/abs/hep-th/0104090}{{\tt hep-th/0104090}}.
%%CITATION = HEP-TH 0104090;%%.

\bibitem{FioMad98a}
G.~Fiore and J.~Madore, ``Leibniz rules and reality conditions,''.
math.QA/9806071.
%%CITATION = math-QA/9806071;%%.

\bibitem{Polyakov:1977fu}
A.~M. Polyakov, ``Quark confinement and topology of gauge groups,'' {\em Nucl.
  Phys.} {\bf B120} (1977)
429--458.
%%CITATION = NUPHA,B120,429;%%.

\bibitem{Coleman:1978ae}
S.~R. Coleman, ``The uses of instantons,''. Lecture delivered at 1977 Int.
  School of Subnuclear Physics, Erice, Italy, Jul 23-Aug 10, 1977.

\bibitem{Landau:1987gn}
L.~D. Landau and Lifshitz, ``Quantum mechanics,''. vol. 3 of Course of
  Theoretical Physics, Butterworth-Heinemann, third~ed., 1997.

\bibitem{Minwalla:1999px}
S.~Minwalla, M.~Van~Raamsdonk, and N.~Seiberg, ``Noncommutative perturbative
  dynamics,'' {\em JHEP} {\bf 02} (2000) 020,
\href{http://arXiv.org/abs/hep-th/9912072}{{\tt hep-th/9912072}}.
%%CITATION = HEP-TH 9912072;%%.

\bibitem{Kinar:2001yk}
Y.~Kinar, G.~Lifschytz, and J.~Sonnenschein, ``{UV/IR} connection: A matrix
  perspective,'' {\em JHEP} {\bf 08} (2001) 001,
\href{http://arXiv.org/abs/hep-th/0105089}{{\tt hep-th/0105089}}.
%%CITATION = HEP-TH 0105089;%%.

\bibitem{Vaidya:2001bt}
S.~Vaidya, ``Perturbative dynamics on fuzzy {$S^2$} and {$RP^2$},'' {\em Phys.
  Lett.} {\bf B512} (2001) 403--411,
\href{http://arXiv.org/abs/hep-th/0102212}{{\tt hep-th/0102212}}.
%%CITATION = HEP-TH 0102212;%%.

\bibitem{Varshalovich:1988ye}
D.~A. Varshalovich, A.~N. Moskalev, and V.~K. Khersonsky, ``Quantum theory of
  angular momentum: irreducible tensors, spherical harmonics, vector coupling
  coefficients, {3NJ} symbols,''. Singapore, Singapore: World Scientific (1988)
  514p.

\bibitem{Matusis:2000jf}
A.~Matusis, L.~Susskind, and N.~Toumbas, ``The {IR/UV} connection in the
  non-commutative gauge theories,'' {\em JHEP} {\bf 12} (2000) 002,
\href{http://arXiv.org/abs/hep-th/0002075}{{\tt hep-th/0002075}}.
%%CITATION = HEP-TH 0002075;%%.

\bibitem{Ruiz:2000hu}
F.~R. Ruiz, ``Gauge-fixing independence of {IR} divergences in non- commutative
  {U(1)}, perturbative tachyonic instabilities and supersymmetry,'' {\em Phys.
  Lett.} {\bf B502} (2001) 274--278,
\href{http://arXiv.org/abs/hep-th/0012171}{{\tt hep-th/0012171}}.
%%CITATION = HEP-TH 0012171;%%.

\bibitem{Landsteiner:2001ky}
K.~Landsteiner, E.~Lopez, and M.~H.~G. Tytgat, ``Instability of non-commutative
  {SYM} theories at finite temperature,'' {\em JHEP} {\bf 06} (2001) 055,
\href{http://arXiv.org/abs/hep-th/0104133}{{\tt hep-th/0104133}}.
%%CITATION = HEP-TH 0104133;%%.

\bibitem{Chu:2000bz}
C.-S. Chu, ``Induced {Chern-Simons} and {WZW} action in noncommutative
  spacetime,'' {\em Nucl. Phys.} {\bf B580} (2000) 352--362,
\href{http://arXiv.org/abs/hep-th/0003007}{{\tt hep-th/0003007}}.
%%CITATION = HEP-TH 0003007;%%.

\bibitem{Magnusbook}
W.~Magnus, F.~Oberhettinger, and R.~Soni, ``Formulas and theorems for the
  special functions of mathematical physics,''. 3rd Edition, Springer-Verlag,
  Berlin Heidelberg New York 1966.

\bibitem{Cerchiai:1998ee}
B.~L. Cerchiai and J.~Wess, ``{q-}deformed {Minkowski} space based on a
  {q-Lorentz} algebra,'' {\em Eur. Phys. J.} {\bf C5} (1998) 553--566,
\href{http://arXiv.org/abs/math.qa/9801104}{{\tt math.qa/9801104}}.
%%CITATION = MATH.QA 9801104;%%.

\bibitem{Hebecker:1992gc}
A.~Hebecker and W.~Weich, ``Free particle in q-deformed configuration space,''
  {\em Lett. Math. Phys.} {\bf 26} (1992)
245--258.
%%CITATION = LMPHD,26,245;%%.

\bibitem{Isaev:1993mt}
A.~P. Isaev and Z.~Popowicz, ``Quantum group gauge theories and covariant
  quantum algebras,'' {\em Phys. Lett.} {\bf B307} (1993) 353--361,
\href{http://arXiv.org/abs/hep-th/9302090}{{\tt hep-th/9302090}}.
%%CITATION = HEP-TH 9302090;%%.

\bibitem{Majid:1994cm}
S.~Majid, ``Introduction to braided geometry and {q-Minkowski} space,''
\href{http://arXiv.org/abs/hep-th/9410241}{{\tt hep-th/9410241}}.
%%CITATION = HEP-TH 9410241;%%.

\bibitem{Meyer:1995wi}
U.~Meyer, ``Wave equations on {q-Minkowski} space,'' {\em Commun. Math. Phys.}
  {\bf 174} (1995) 457--476,
\href{http://arXiv.org/abs/hep-th/9404054}{{\tt hep-th/9404054}}.
%%CITATION = HEP-TH 9404054;%%.

\bibitem{Podles:1996qy}
P.~Podles, ``Solutions of {Klein-Gordon} and {Dirac} equations on quantum
  {Minkowski} spaces,'' {\em Commun. Math. Phys.} {\bf 181} (1996) 569--586,
\href{http://arXiv.org/abs/q-alg/9510019}{{\tt q-alg/9510019}}.
%%CITATION = Q-ALG 9510019;%%.

\bibitem{Oeckl:1999zu}
R.~Oeckl, ``Braided quantum field theory,'' {\em Commun. Math. Phys.} {\bf 217}
  (2001) 451--473,
\href{http://arXiv.org/abs/hep-th/9906225}{{\tt hep-th/9906225}}.
%%CITATION = HEP-TH 9906225;%%.

\bibitem{Chaichian:1999wy}
M.~Chaichian, A.~Demichev, and P.~Presnajder, ``Quantum field theory on the
  noncommutative plane with {$E_q(2)$} symmetry,'' {\em J. Math. Phys.} {\bf
  41} (2000) 1647--1671,
\href{http://arXiv.org/abs/hep-th/9904132}{{\tt hep-th/9904132}}.
%%CITATION = HEP-TH 9904132;%%.

\bibitem{Fiore:1996bu}
G.~Fiore and P.~Schupp, ``Identical particles and quantum symmetries,'' {\em
  Nucl. Phys.} {\bf B470} (1996) 211--235,
\href{http://arXiv.org/abs/hep-th/9508047}{{\tt hep-th/9508047}}.
%%CITATION = HEP-TH 9508047;%%.

\bibitem{Fiore:1997fb}
G.~Fiore, ``Drinfeld twist and {q-}deforming maps for {Lie} group covariant
  {Heisenberg} algebrae,'' {\em Rev. Math. Phys.} {\bf 12} (2000) 327--359,
\href{http://arXiv.org/abs/q-alg/9708017}{{\tt q-alg/9708017}}.
%%CITATION = Q-ALG 9708017;%%.

\bibitem{Cornalba:2001sm}
L.~Cornalba and R.~Schiappa, ``Nonassociative star product deformations for
  {D-}brane worldvolumes in curved backgrounds,'' {\em Commun. Math. Phys.}
  {\bf 225} (2002) 33--66,
\href{http://arXiv.org/abs/hep-th/0101219}{{\tt hep-th/0101219}}.
%%CITATION = HEP-TH 0101219;%%.

\bibitem{kohno1}
T.~Kohno, ``Monodromy representations of braid groups and {Yang-Baxter}
  equations,'' {\em Ann. Inst. Fourier} {\bf 37} (1987) 139.

\bibitem{kohno2}
T.~Kohno, ``One-parameter family of linear representations of {Artin's} braid
  groups,'' {\em Adv. Stud. in Pure. Math.} {\bf 12} (1987) 189.

\bibitem{Curtright:1991ri}
T.~L. Curtright, G.~I. Ghandour, and C.~K. Zachos, ``Quantum algebra deforming
  maps, {Clebsch-Gordan} coefficients, coproducts, {U} and {R} matrices,'' {\em
  J. Math. Phys.} {\bf 32} (1991)
676--688.
%%CITATION = JMAPA,32,676;%%.

\bibitem{Curtright:1990sw}
T.~L. Curtright and C.~K. Zachos, ``Deforming maps for quantum algebras,'' {\em
  Phys. Lett.} {\bf B243} (1990)
237--244.
%%CITATION = PHLTA,B243,237;%%.

\bibitem{Jurco:1994te}
B.~Jurco, ``More on quantum groups from the quantization point of view,'' {\em
  Commun. Math. Phys.} {\bf 166} (1994) 63--78,
\href{http://arXiv.org/abs/hep-th/9301092}{{\tt hep-th/9301092}}.
%%CITATION = HEP-TH 9301092;%%.

\end{thebibliography}\endgroup

%\bibliographystyle{plain}
%\bibliography{Main}

%\newpage
%\addcontentsline{toc}{chapter}{Index}
%\printindex
\end{document}